\documentclass[11pt]{article}

\usepackage{epsfig}
\usepackage{psfig}
\def\laq{\raise 0.4ex\hbox{$<$}\kern -0.8em\lower 0.62
ex\hbox{$\sim$}}
\def\gaq{\raise 0.4ex\hbox{$>$}\kern -0.7em\lower 0.62
ex\hbox{$\sim$}}
\def\vk{\vec{k}}
\def\vp{\vec{p}}
\def\vx{\vec{x}}
\def\vy{\vec{y}}
\begin{document}

\begin{titlepage}
\begin{flushright}
CERN-PH-TH/2004-140
\end{flushright}
\vspace*{1cm}

\begin{center}
{\LARGE {\bf Theoretical tools for CMB physics}}
\vskip2.cm
\large{Massimo Giovannini \footnote{e-mail address: massimo.giovannini@cern.ch}}
\vskip 1.cm
{\it Department of Physics, Theory Division, CERN, 1211 Geneva 23, Switzerland}
\end{center}
\begin{abstract}
This review presents, in a self-consistent manner, those 
analytical tools that are relevant for the analysis of the physics 
of CMB anisotropies  generated in different theoretical models of the 
early Universe. After introducing the physical foundations 
of the Sachs-Wolfe effect,  the origin and evolution  
of the scalar, tensor and vector modes of the geometry is treated
in both  gauge-invariant and gauge-dependent descriptions.
Some of the recent progresses in the theory 
of cosmological perturbations are scrutinized with particular attention 
to their implications for the adiabatic and isocurvature paradigms, whose 
description is reviewed both within conventional fluid approaches and within 
the  the Einstein-Boltzmann treatment. Open problems and theoretical challenges 
for a unified theory of the early Universe are outlined in light of their 
implications for the generation of large-scale anisotropies in the CMB sky and in light 
of the generation of stochastic backgrounds of relic gravitons between few Hz and the GHz.
\end{abstract}
\vskip 2.cm

\centerline{\em This article is dedicated to Sergio Fubini}

\end{titlepage}

\pagenumbering{arabic}

\tableofcontents

\newpage

\renewcommand{\theequation}{1.\arabic{equation}}
\section{An inhomogeneous Universe}
\setcounter{equation}{0}
\subsection{Formulation of the problem}
 What is the origin of the inhomogeneities we observe  in the sky?
What is their typical wavelength? How come that we are able 
to observe these inhomogeneities?
The term inhomogeneity is rather generic and it indicates, for 
the purpose of the present introduction,  fluctuations both 
in the background geometry and in the energetic  content of the
 Universe.  For simplicity,  these inhomogeneities can be  
represented, for the purposes of the present introduction, by  plane waves characterized by a 
comoving wave-number $k$. Since the evolution of the 
Universe is characterized by a scale factor $a(t)$, a 
physical wave-number $\omega = k/a$ is also customarily defined. The comoving 
wave-number  does not feel the expansion  
while the physical momentum is different at different 
epochs in the life of the Universe. Conversely the value of a given 
physical frequency is fully specified only by stating the time 
at which the physical frequency is ``measured''. 

 If a given fluctuation has 
a momentum comparable with the present value of the Hubble parameter
 \footnote{The definition of the Planck mass 
adopted here is $ M_{\rm P} = ( 8\pi G)^{-1/2} = 2.4\times 10^{18} \,\,{\rm GeV}$ }, i.e. 
$\omega_0 \simeq H_{0} \simeq 5.6 \times 10^{-61} \,\, M_{\rm P}$, 
we will have that $\omega_{0} \simeq 2.3\times 10^{-18}$ Hz.
Fluctuations with momentum smaller that $H_{0}$ have a wave-length which 
is larger than the present value of the Hubble radius and, therefore, 
they seem to be impossible to detect directly since the distance 
between two maxima (or two minima) of the wave will be, in this case, 
larger than our observable Universe. 

At the decoupling epoch the Hubble rate is $H_{\rm dec} \simeq 6.7 \times 10^{-56} M_{\rm P}$.
The decoupling frequency red-shifted today will be 
$\omega_{\rm dec} \simeq 3\times10^{-16} {\rm Hz}$.   Fluctuations 
of this typical frequency can be ``detected" through the study of the 
of the temperature fluctuations of the Cosmic Microwave 
Background (CMB). The rationale for this statement is that 
the fluctuations in the background geometry and in the density 
contrasts of the various species composing the primeval plasma 
may induce tiny spatial variations in the CMB temperature.
The standard lore for the formation of the structures in the 
Universe, is that the fluctuations detected by means of CMB 
temperature inhomogeneities will eventually collapse by gravitational 
instability to form galaxies and clusters of galaxies.

At the  time of decoupling and, anyway prior to the recombination 
of free electrons with protons, the temperature 
of the Universe was of the order of a fraction of the ${\rm eV}$. 
The Universe, all along its history, reached much higher temperature 
(or, equivalently, much higher values of the expansion rate). 
At the epoch of the formation of light nuclei the temperature 
of the Universe was of the order of a fraction of the ${\rm MeV}$ and the corresponding 
value of the Hubble expansion rate  was of the order of 
$8\times10^{-44} \,\, M_{\rm P}$.
When (and if) the electroweak (EW) phase transition took place, the temperature of the Universe was 
of the order of $100\,\,{\rm GeV}$ leading, approximately, to $H_{\rm ew} \sim 10^{-32} M_{\rm P}$.
Comoving wave-numbers of the order of the Hubble rate at the EW or at the BBN epoch 
correspond, today, to physical frequencies of the order of
$\omega_{\rm ew} \sim 2 \times 10^{-5} \,\,\,{\rm Hz}$  
 and $\omega_{\rm N} \simeq 9 \times  10^{-11}\,\,\, {\rm Hz}$.
Finally, in the context of the conventional inflationary paradigm the value of the 
Hubble rate during inflation was $H_{\rm inf} \sim 10^{-6} M_{\rm P}$. Under the assumption the 
Universe was dominated by radiation right after inflation, the present value of the correspoonding 
physical frequency  was of the order of $10^{8}$ Hz.
 
To ``observe'' fluctuations corresponding to present 
frequencies $ 10^{-3} \,\,\, {\rm Hz} < \omega < {\rm MHz}$, CMB 
anisotropy experiments cannot be used. CMB anisotropy experiments 
are sensitive to present frequency scale only slightly higher 
than $10^{-16}$ Hz. For frequencies in the interval $ 10^{-3} \,\,\, 
{\rm Hz} < \omega < {\rm MHz}$, the only hope of direct observations are 
the experiments on direct detection of stochastic backgrounds of 
gravitational radiation. In some specific configurations these 
instruments could detect not only tensor fluctuations of the geometry 
but also scalar and vector modes. 

The cosmological inhomogeneities are usually characterized 
by their correlation functions. The simplest 
correlation function encoding informations on the nature 
of the fluctuations is the two-point function  whose 
Fourier transform is usually called power spectrum.  The two 
point function is computed by taking averages of the fluctuation amplitude at two 
spatially separated points but at the same time. From this 
quantity, with appropriate mathematical manipulations, it is also possible
to deduce other relevant physical quantities like, for instance, the 
energy density carried by these fluctuations. 

From the physical scales  
discussed above, one is led to conclude that the power 
spectrum of cosmological inhomogeneities 
is defined over a  huge interval of frequencies. For instance, the 
tensor fluctuations of the geometry which only couple to the curvature 
and not to the matter sources of the Universe have a spectrum 
ranging, in some specific models, from $10^{-18}$ Hz up to 
the GHz for, roughly 27 powers of ten.

From the figures obtained so far, legitimate questions may be raised:
\begin{itemize}
\item{}  given that the precise thermodynamical history of the Universe
above $10$ MeV is unknown, how it is possible to have a reliable framework 
describing the evolution of the cosmological fluctuations?
\item{}  does it make sense 
 to apply GR also up to curvature scales which are almost Planckian?
\item{} how is it possible 
to give sensible initial conditions to the various 
fluctuations of the geometry without having some handle 
on the theory of the initial singularity, i.e. the theory of the big-bang?
\end{itemize}

We do not have any (even indirect)  test of the thermodynamical state 
of the Universe for temperatures larger than the MeV. The four light isotopes 
${\rm D}$, $^{3}{\rm He}$, $^{4}{\rm He}$ and $^{7} {\rm Li}$ 
are mainly produced at the big-bang nucleosynthesis 
 below a typical temperature of $0.8$ 
MeV when neutrinos decouple from the plasma and the neutron abundance 
evolves via free neutron decay. The abundances calculated in the 
simplest (homogeneous and isotropic) 
 big-bang nucleosythesis model agree fairly well with the 
astronomical observations. This is the last indirect test of the 
thermodynamical state of the Universe. For temperature larger 
than $10$ MeV, collider physics and the success of the electroweak 
standard model offer a precious handle up to the epoch of the 
electroweak phase transition occurring for temperatures 
of the order of $100$ GeV. There is then the hope that our knowledge 
of particle physics could help us to fill the gap between the 
nucleosynthesis epoch and the physics beyond above the electroweak 
curvature scale. 

The evolution of inhomogeneities
over cosmological distances may well be  unaffected by our ignorance of the 
specific history of the expansion rate. The key concept is the one of conservation
 law. In the framework of Einsteinian theories of gravity (and 
in particular in GR) its is possible to show, on a rather general 
ground, that there exist conservation laws for the evolution of the 
fluctuations. These conservation laws are exact in the limit 
of vanishing comoving frequency, i.e. $ k\to 0$.
This conclusion holds under the assuming that general relativity 
is a valid description throughout the whole evolution of the Universe. Similar properties also 
hold, with some modifications, in scalar-tensor theories 
of gravity (of Brans-Dicke type) like the ones suggested 
by the low-energy limit of superstring theory. The conservation 
laws of cosmological perturbations apply 
strictly in the case of the tensor modes of the geometry. In
the case of the scalar fluctuations, the precise form of the 
conservation law depends upon the matter content of the early Universe since 
scalar fluctuations couple directly to the (scalar) matter sources. 
There is the hope that experiments aimed at the direct detection 
of stochastic backgrounds of relic gravitons will provide informations 
on the high-frequency behaviour of cosmological fluctuations. 
These informations could be, in principle, converted into 
valuable clues on the early evolution of the Hubble rate.

Finally, there is no way 
of addressing the singularity problem in general terms.  For instance, the 
conventional inflationary paradigm postulates that the evolution of 
the early Universe was driven by the potential energy of 
 a single (scalar) degree of freedom named 
the inflaton.  During inflation, the scale factor expands in 
an accelerated way 
(i.e. $\ddot{a} >0$ and $\dot{a}>0$). 
Inflationary cosmology does 
``predict'' a singularity in the far past of the Universe, however, the 
 effects of the singularity may well be invisible since the 
accelerated expansion of a quasi-de Sitter space-time is able to dilute 
all the spatial gradients which may eventually arise at the time 
of the cosmological singularity, prior to the beginning of inflation.
In conventional inflationary models the singularity is not addressed but it is 
just removed beyond our observational capabilities. 

As a general remark, one can say that, in modern cosmology, when the 
description of the early Universe is approached, many authors like to use 
a simplicity principle. This means, for instance, that instead of 
introducing many degrees of freedom in the description of the early Universe, 
we are often biased towards the model containing the fewest number 
of degrees of freedom (like in the case of single-field inflationary models). 
This is, however, not the typical situation arising in the context of the 
low-energy string effective action where various massless modes are simultaneously present.
It is here useful to recall the opinion 
of R. Feynman questioning the notion that simplicity should be a guiding 
principle in the search for truth about Nature \cite{feynman}:
``...the simplest solution,  by far, would be nothing, that 
there should be nothing at all in the Universe. Nature is more 
inventive than that, so I refuse to go along thinking that it always 
has to be simple.'' 

\subsection{A short introduction to a long history}

The history of the studies on the CMB 
anisotropies and on the cosmological 
fluctuations is closely linked to the the history of the standard 
cosmological model (see, for instance, the 
the textbooks by Weinberg \cite{weinberg} and 
by Kolb and Turner \cite{KT}). 
In this introduction only few interesting points 
will be swiftly mentioned since the main scope of this review is not historical.
Furthermore, the  second chapter of the excellent book \cite{patridge} of B. Patridge
describes in detail the excitement of experimental 
cosmology in the early sixties at the time when CMB was firstly 
discovered.  In the he issue number 81 of the ``Uspekhi Fizicheskikh
Nauk'' , on the occasion of the seventy-fifth 
anniversary of the birth of A. A. Friedmann, a number of rather 
interesting papers were published. Among them there is a 
review article of the development of Friedmannian 
cosmology by Ya. B. Zeldovich \cite{zel1} and  
the inspiring paper of Lifshitz and Khalatnikov \cite{LK} on the 
relativistic treatment of cosmological perturbations.

Reference \cite{zel1} describes mainly Friedmann's contributions \cite{friedmann}.
Due attention should also be paid 
to the work of G. Lema\^itre \cite{lemaitre1,lemaitre2} that 
was also partially motivated by the debate with A. Eddington
\cite{eddington}. According to the idea of Eddington 
the world evolved from an Einstein static Universe and so developed 
``infinitely slowly from a primitive uniform distribution in 
unstable equilibrium'' \cite{eddington}. The point of view of Lema\^itre 
was, in a sense, more radical since he suggested, in 1931, that the expansion 
really did start with the beginning of the entire Universe. Unlike the 
Universe of some modern big-bang cosmologies, the description of Lema\^itre 
did not evolve from a true singularity but from a material 
pre-Universe, what Lema\^itre liked to call ``primeval atom'' \cite{lemaitre2}.
The primeval atom was a unique atom whose atomic weight was the total 
mass of the Universe. This highly unstable atom would have 
experienced some type of fission and would have divided into smaller 
and smaller atoms by some kind of super-radioactive processes. 
The perspective  of Lema\^itre was that the early expansion of the Universe 
could be a well defined object of study for natural sciences even in the 
absence of a proper understanding of the initial singularity.

During the development of the standard cosmological model, 
a recurrent theme has been  the explanation of structures 
we observe in the Universe. This problem goes under the 
name of the structure formation problem. The  reviews 
of Kodama and Sasaki \cite{KS} and of Mukhanov, Feldman and 
Brandenberger \cite{MFB} contain some historical surveys that are a valuable 
introduction to this subject.

Among the various ideas put forward through the years we wish to mention 
two contributions that are suitable in order to introduce some of the 
problems treated in the present article: 
the classic paper of Sachs and Wolfe on the possible generation 
of anisotropies in the CMB \cite{sachs} and a paper of Sakharov \cite{sakharov} 
where the concept of Sakharov (i.e. acoustic) oscillations
has been introduced. 

In \cite{sachs} the basic general relativistic effects 
leading to the anisotropy of the CMB were estimated assuming 
that the CMB itself had a cosmological origin. This 
effect is now one of the cornerstones of the physics of CMB anisotropies 
and, in light of its importance, it will be one of the first 
derivations of the present review. In \cite{sachs} the authors 
argued that galaxy formation would imply $\Delta T /T \sim 10^{-2}$. However,
subsequent calculations \cite{silk1,PeeblesYu,doro,wilson} brought 
the estimate down to $10^{-4}$, closer to what experimentally observed.

The paper of Sakharov \cite{sakharov} is rather intriguing if one thinks that 
it was written in 1965, i.e. just at the time when the possible 
cosmological origin of CMB was recognized. Sakharov was speculating 
on a model proposed earlier by Zeldovich. The origin and details of the model
are, today, rather unimportant. Reference \cite{sakharov}
contained two ideas whose legacy, somehow unexpectedly, 
survived until the present day. The first 
idea is summarized in the abstract of Ref. \cite{sakharov}:
`` A hypothesis of the creation of astronomical bodies as a result 
of gravitational instability of the expanding Universe is 
investigated. It is assumed that the initial inhomogeneities arise 
as a result of {\em quantum fluctuations}...''. The model 
of Zeldovich indeed assumed that the initial state of the baryon-lepton 
fluid was cold. So the premises were wrong, but, still, it is 
amusing to notice that today quantum fluctuations are what it is assumed 
as initial conditions in a variety of  models 
present at a totally different curvature (or energy) scale as initial 
condition for the fluctuations of the geometry. 

In a theoretical perspective,  Refs. \cite{sachs} and \cite{sakharov}
underline three important problems:
\begin{itemize}
\item{} the problem of initial conditions of the fluctuations of the gravitational inhomogeneities (sections 4 and 6);
\item{} the problem of their evolution (sections 5 and 7);
\item{} the problem of their imprint on the CMB temperature fluctuations (sections 3, 8).
\end{itemize}
The sections quoted in connection with each of the previous items 
 indicate where each of the mentioned topic is treated in the present article.
 Due to the historical development of the subject some semantic ambiguities 
 may also arise. For instance, the  ``initial conditions of the CMB anisotropies"
 could be described either in terms of the amplitude of metric fluctuations around the time of
 matter-radiation equality (i.e. in the late Universe), or in terms of the normalization of the 
 cosmological perturbations   in the early Universe.  
 
\subsection{CMB experiments}

Even if the present review is theoretically oriented, it is appropriate to give a swift 
account of our recent experimental knowledge of CMB anisotropies.  A new twist 
in the study of CMB anisotropies came from the observations of the COBE satellite 
and, more precisely,  from the DMR (Differential Microwave Radiometer) instrument. 
The DMR was able to probe the angular power spectrum\footnote{While the precise definition of angular power 
spectrum will be given in section 8, here it suffices to recall that $\ell (\ell +1) C_{\ell}/(2\pi)$ measures the 
degree of inhomogeneity in the temperature distribution per logarithmic interval of $\ell$. Consequently,
a given multipole $\ell$ can be related to a given spatial structure in the microwave sky: small $\ell$ will correspond 
to low wavenumbers, high $\ell$ will correspond to larger wave-numbers 
but always around a present physical frequency of the order of $10^{-16} $ Hz.}
  $C_{\ell }$ spectrum (see section 8) up to $\ell \simeq 25$ 
(see, for instance, \cite{cobe,cob2} and references therein).  As the name says, DMR 
was a differential instrument measuring temperature differences in the microwave 
sky. The angular separation $\vartheta$ of the two horns of the antenna is related 
to the maximal multipole probed in the sky according to the approximate relation 
$\vartheta \simeq  \pi/\ell$.  This will give the angular separation in radians that can be easily 
related to the angular separation in degrees.

The angular separation explored by the COBE experiment were $\vartheta \gaq 7^{0}$.
After the COBE mission, various experiments attempted the exploration of smaller 
angular separation, i. e. larger multipoles.  
A definite convincing evidence of the existence and location of the first peak 
in the $C_{\ell}$ spectrum came from the Boomerang \cite{boom1,boom2} , Maxima \cite{maxima}
and Dasi \cite{dasi} experiments.  Both Boomerang and Maxima were balloon borne (bolometric) 
experiments. Dasi was a ground based interferometer. The data points of these last 
three experiments explored multipoles up to $1000$, determining the first acoustic oscillation 
(in the jargon the first Doppler peak) for $\ell \simeq 220$. 
Another important balloon borne experiments was Archeops \cite{archeops} providing 
interesting data for the region characterizing the first rise of the $C_{\ell}$ spectrum.
Some other useful references on earlier CMB experiments can be found in \cite{silk}.

\begin{figure}[t!]
\centering
\includegraphics[height=10cm,angle=90]{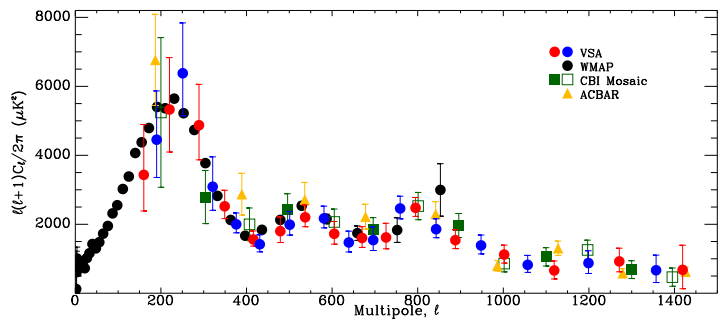}
\caption{Some of the most recent  CMB anisotropy data are reported (figure adapted from \cite{DICK}): 
 WMAP data (filled circles);
VSA data (shaded circles) \cite{DICK}; 
CBI data  (squares) \cite{mason,pearson};
ACBAR data (triangles) \cite{kuo}.}
\label{F13} 
\end{figure}

The $C_{\ell}$ spectrum, as measured by different recent experiments is reported in 
Fig. \ref{F13} (adapted from Ref. \cite{DICK}).
The WMAP (Wilkinson Microwave Anisotropy Probe)  data (filled circles in Fig. \ref{F13}) 
provided, among other important evidences: the precise determination of the position 
of the first peak (i.e. $\ell = 220.1 \pm 0.8$ \cite{WMAP02});  the clear evidence of the second
peak; a perspective (once all the four-years data will be collected) for good resolution
up to $\ell \sim 1000$, i.e. around the third peak. The WMAP experiment also 
measured temperature-polarization correlations (see section 8) providing a  distinctive 
signature (the so-called anticorrelation peak in the temperature-polarization 
power spectrum for $\ell \sim 150$) of primordial adiabatic fluctuations (see sections 4, 8).  
In Fig. \ref{F13} on top of the WMAP data we illustrated the results of  the 
Arcminute Cosmology Bolometer Array Receiver (ACBAR) as well as the data points of the 
Very Small Array (VSA) and of the  Cosmic Background Imager (CBI).

To have a more detailed picture of the evolution and relevance 
of CMB experiments we refer the reader to  Ref. \cite{white} (for review of the pre-1994 status of the art) 
 and Ref.  \cite{dodelson} for a review of the pre-2002 situation). The rather broad set of lectures by 
Bond \cite{bond} may also be usefully consulted.

In recent years, thanks to combined observations 
of Cosmic Microwave background (CMB) anisotropies \cite{WMAP01}, 
Large scale structure  \cite{LSS01,CL01}, supernovae of type Ia, big-bang nucleosyntheis \cite{BBN01},
some kind of paradigm for the evolution of the late time (or even present)
Universe emerged. It is  normally called by practitioners $\Lambda$CDM 
model or even, sometimes, ``concordance model". The terminology 
of $\Lambda$CDM refers to the fact that, in this model, the dominant 
(present) component of the energy density of the Universe 
is given by a cosmological constant $\Lambda$ and a fluid of cold dark matter 
particles interacting only gravitationally with the other (known) 
particle species such as baryons, leptons, photons.
According to this paradigm, our understanding of the Universe can be summarized 
in two sets of cosmological parameters: the first set of parameters refers to the homogeneous 
background, the second set of parameters to the inhomogeneities.
As far as the homogeneous background is concerned, on top of the indetermination 
on the (present) Hubble expansion rate\footnote{Recall that the present value of the 
Hubble expansion rate is customarily parametrized as $H_{0} = h \times 100 {\rm km}/[{\rm sec} \times {\rm Mpc}]$.
In the concordance model $h \simeq 0.72$.}, 
i.e. $h$, there are various other parameters such as:
the dark and CDM energy densities 
 in critical units, i.e. respectively $\Omega_{\rm cdm} \simeq 0.224$ and $ \Omega_{\Lambda} \simeq 0.73$;
 the radiation and baryon energy density in critical units, i.e. respectively $\Omega_{\rm r} \simeq 8\times 10^{-5}$ 
 and $\Omega_{\rm b} \simeq 0.046$; the number of neutrino species and their masses; the 
 possible contribution of the spatial curvature ($\Omega_{\kappa}=0$ in the 
 concordance model) the optical depth of the plasma at recombination (of the order of $0.16$ in the 
 concordance model)...
 
 The ellipses stand for various other parameters which are normally either not included in the analysis
 or just set to a fiducial value such as the equation of state for the dark energy component, the possible 
 effect of the dark energy sound speed; the possible variation (with the redshift) of the fine structure constant;
 some possible extra (warm) dark matter component and so on.
 The second set of parameters refers to the inhomogeneities. In the $\Lambda$CDM model 
 the initial conditions for the inhomogeneities belong to a specific class of scalar fluctuations of 
 the geometry that  are called adiabatic and are characterized by the Fourier transform of their 
 two-point function, i.e. the power spectrum which is usually parametrized in terms of an amplitude 
 and of a spectral slope. To these two parameters one usually adds the ratio between 
 the amplitudes of the scalar and tensor modes of the geometry. Depending on how we count 
 the minimal list includes from $10$ to $12$ parameters. Furthermore, few optional choices raise 
 easily the number of parameters to $16$ or even $18$.
 
 Having said this it is important to stress that the present article will not deal with the problem 
 of data analysis (or parameter extraction from the CMB data). The purpose of the 
 present review, as underlined before in this introduction, will be to describe the theoretical tools allowing 
 the calculation of large-scale temperature fluctuations in a broad range of cosmological 
 models whose predictions may not always fall in the concordance model. 
 
Finally a word of care concerning references. In reviewing this subject  attention has been given 
to ideas emerged in the last ten years. For this reasons, various important works will only be indirectly 
quoted and the reader may refer to earlier review articles on the subject that have been already quoted 
and will be quoted again when needed. 
Among the theoretical ideas reported here, for reasons of space 
two important topics have been omitted, i.e. the problem of cosmological inhomogeneities 
induced by cosmic defects and the problem of the (late time) fluctuations induced by 
inhomogeneous quintessence models. 

 \renewcommand{\theequation}{2.\arabic{equation}}
\section{FRW Universes and their inhomogeneities}
\setcounter{equation}{0}

\subsection{Friedmann-Robertson-Walker Universes}

A Friedmann-Robertson-Walker (FRW)  metric can be written in terms of the conformal time 
parametrization, leading to the line element 
\begin{equation}
\overline{g}_{\mu\nu} dx^{\mu} dx^{\nu} = 
a^2(\tau) \biggl\{ d\tau^2  - \biggl[\frac{d r^2}{1 - \kappa r^2} + r^2 ( d\theta^2 + \sin^2{\theta} d\varphi^2)\biggr]\biggr\}
\label{FRW}
\end{equation}
where $\kappa$ can take values $1$, $0$ and $-1$ corresponding, respectively, to  spherical, eucledian or hyperbolic 
spatial sections: $\tau$ is called conformal time coordinate.  Throughout 
this paper the signature of the metric will be, consistently,  mostly minus. Therefore, the flat Minkowski
metric will be  $\eta_{\mu\nu} = {\rm diag}(+,\,-,\,-,\,-)$.

In the standard cosmological model 
the evolution of the geometry is driven by the evolution of the matter sources according to Einstein equations
\begin{equation}
R_{\mu}^{\nu} - \frac{1}{2} \delta_{\mu}^{\nu} R = 8\pi G T_{\mu}^{\nu}, 
\label{ein1}
\end{equation}
where $R_{\mu\nu}$ and $R$ are the Ricci tensor and Ricci scalar; $T_{\mu\nu}$ is the energy-momentum tensor 
of the sources. The  form of the energy-momentum tensor 
may vary depending on the evolutionary stage of the Universe. 

After  electroweak 
interactions have fallen out of thermal equilibrium (i.e. temperatures 
$T \ll 1 $ MeV),  the energy-momentum tensor 
of the plasma contains photons, baryons, electrons, 
neutrinos, cold dark matter particles. If the temperature of the plasma is 
around the eV, then the electrons and baryons will be in thermal equlibrium at a common 
temperature and photons will be tightly coupled with the baryons because 
of Thompson scattering.  Neutrinos will be 
collisionless. 

To these species, a dark energy component is also usually added. 
The dark energy is customarily  
parametrized either in terms of a cosmological constant  $\Lambda$ \cite{carroll,pad} or in terms 
of some scalar degrees of freedom generically named quintessence (see for instance \cite{sahni} for 
a review).  As the CDM particles, the dark
energy fluid is supposed to have
negligible non-gravitational interactions with baryons, leptons and photons. 

For temperatures $ {\rm eV} < T < {\rm MeV}$ the 
Friedmann equations are
\begin{eqnarray}
&&{\cal H}^2 + \kappa = \frac{8\pi G}{3} a^2 \rho\sum_{\lambda} \Omega_{\lambda}  + \frac{\Lambda}{3} a^2,
\label{b1}\\
&& {\cal H}^2 - {\cal H}' + \kappa = 4 \pi G a^2 \rho\sum_{\lambda} \Omega_{\lambda} (1 + w_{\lambda}) ,
\label{b2}\\
&& \rho_{\lambda}' + 3 {\cal H}( 1 + w_{\lambda}) \rho_{\lambda} =0,
\label{b3}
\end{eqnarray}
where the summation index $\lambda$ refers to each component of the plasma and $w_{\lambda} = p_{\lambda}/\rho_{\lambda}$ is the 
corresponding barotropic index. Eqs. (\ref{b1}) and (\ref{b2}) depend on $\Omega_{\lambda}$, i.e. 
 the energy density, in critical units, for each component 
of the fluid.
In Eqs. (\ref{b1})--(\ref{b3}) ${\cal H} = (\ln{a})'$ and the prime denotes a derivation 
with respect to the conformal time coordinate $\tau$. The cosmic time parametrization $t$ is related to the
 conformal one by $dt = a(\tau) d \tau$. In the $t$-parametrization, 
$H = \dot{a}/a$  is the Hubble expansion rate; the dot denotes a derivation with respect to $t$. The relation 
between the Hubble factors in the two parametrizations is $ H a = {\cal H}$.
In the present article we will switch, when needed,  from one parametrization to the other.  Using the 
relations between the two parametrizations, Eqs. (\ref{b1})--(\ref{b3}) can be written 
in terms of  the cosmic time coordinate:
\begin{eqnarray}
&& H^2 + \frac{\kappa}{a^2} = \frac{8\pi G}{3} \rho\sum_{\lambda} \Omega_{\lambda} + \frac{\Lambda}{3},
\label{b1ct}\\
&& \dot{H} = - 4\pi G \rho\sum_{\lambda} \Omega_{\lambda} (1 + w_{\lambda})  + \frac{\kappa}{a^2},
\label{b2ct}\\
&&  \dot{\rho}_{\lambda} + 3 H( 1 + w_{\lambda}) \rho_{\lambda} =0,
\label{b3ct}
\end{eqnarray}
where, now, $\rho$, is the total energy density.
 Equations (\ref{b1})--(\ref{b3}) assume that,  for each 
species, the energy-momentum tensor can be written as 
\begin{equation}
T_{\mu\nu}^{(\lambda)} =( p_{\lambda} + \rho_{\lambda}) u^{(\lambda)}_{\mu} u^{(\lambda)}_{\nu} - p_{\lambda} g_{\mu\nu},
\label{formenmom}
\end{equation}
with the normalization condition $g^{\mu\nu} u^{(\lambda)}_{\mu} u^{(\lambda)}_{\nu} =1$ valid 
for each species $\lambda$.
 Since the thermodynamical history of the Universe 
is progressively less understood as we go back in time, also the 
energy-momentum tensor of the plasma may take different forms.  

\subsection{Fluctuations of the geometry in FRW Universes}
Given a conformally flat ($\kappa =0$) background metric of FRW type  
\begin{equation}
g_{\mu\nu}(\tau) = a^2(\tau) \eta_{\mu\nu}, 
\end{equation}
its first-order  fluctuations can be written as 
\begin{equation}
\delta g_{\mu\nu}(\tau,\vec{x}) = \delta_{\rm s} g_{\mu\nu}(\tau,\vec{x})
+  \delta_{\rm v} g_{\mu\nu}(\tau,\vec{x}) +
\delta_{\rm t} g_{\mu\nu}(\tau,\vec{x}),
\end{equation}
where the subscripts define, respectively, the scalar, vector and  tensor
perturbations classified according to rotations in the three-dimensional
Euclidean sub-manifold. Being a symmetric rank-two tensor in four-dimensions,  
the perturbed metric $\delta g_{\mu\nu}$  has, overall, 10 independent 
components whose explicit form will be parametrized as
\begin{eqnarray}
&& \delta g_{00} = 2 a^2 \phi,
\label{g001}\\
&& \delta g_{ij} = 2 a^2 ( \psi \delta_{ij} - \partial_{i} \partial_{j}E) 
- a^2 h_{ij} + a^2 ( \partial_{i} W_{j} + \partial_{j} W_{i}),
\label{gij1}\\
&& \delta g_{0i} = - a^2 \partial_{i} B - a^2 Q_{i},
\label{g0i1}
\end{eqnarray}
together with the conditions
\begin{equation}
\partial_{i} Q^{i} = \partial_{i} W^{i} =0 ,\,\,\,\,\,\,\,\,
h_{i}^{i} = \partial_{i}h^{i}_{j} =0.
\label{div1}
\end{equation}
The decomposition expressed by Eqs. (\ref{g001})--(\ref{g0i1}) and (\ref{div1}) 
 is the one normally employed in  the Bardeen 
formalism \cite{bardeen} (see also \cite{MFB}).

According to Eqs. (\ref{g001})--(\ref{g0i1}), the scalar fluctuations of 
the geometry are parametrized by the 4 scalar functions, i. e. 
$\phi$, $\psi$, $B$ and $E$. The vector fluctuations 
are described by the two (divergenceless) vectors in three (spatial) dimensions
 $W_{i}$ and $Q_{i}$, i.e. 
by $4$ independent degrees of freedom. Finally the tensor modes 
are described by $h_{ij}$, leading, overall, to 2 independent components 
because of the last two conditions of Eq. (\ref{div1}). 

Slightly different parametrizations 
exist in the literature. For instance, instead of denoting 
$\delta_{\rm s} g_{i j}(k,\tau) = 2 a^2(\tau) [\psi(k,\tau) \delta_{ij} + k_{i} k_{j} E(k,\tau)]$
the Fourier component of the spatial (scalar) fluctuation, the authors of
Ref. \cite{KS}  prefers to separate explicitely the longitudinal $H_{L}(k,\tau)$ 
and transverse, $H_{T}(k,\tau)$ perturbations: 
\begin{equation}
\delta_{\rm s} g_{ij}(k,\tau) = 2 a^2(\tau) \biggl[ H_{L}(k,\tau) \delta_{ij} + H_{T}(k,\tau) \biggl(
\hat{k}_{i} \hat{k}_{j} -\frac{1}{3}\delta_{ij}\biggl) \biggr],
\end{equation}
where $\hat{k}_{i} = k_{i}/k$.

In the applications 
related to the synchronous gauge description the fluctuations are 
customarily denoted, in Fourier space, by \cite{PV1,PV2,MB}
\begin{equation}
\delta_{\rm s} g_{i j}(k,\tau) = 
a^2(\tau)\biggl[ \hat{k}_{i} \hat{k}_{j} h(k,\tau) + 6 \xi(k,\tau)\biggl(  
\hat{k}_{i} \hat{k}_{j} - \frac{1}{3} \delta_{ij}\biggr)\biggr].
\label{synpar}
\end{equation}
Under infinitesimal coordinate transformations,
\begin{equation}
\tau \to \tilde{\tau} = \tau + \epsilon_{0},\,\,\,\,\,\,\,\,\,\,\, {x}^{i} \to \tilde{x}^{i} = x^{i} + \epsilon^{i}
\label{T1}
\end{equation}
the fluctuations of the geometry defined in Eqs. (\ref{g001})--(\ref{g0i1}) transform as 
\begin{equation}
\delta {g}_{\mu\nu} \to \delta\tilde{g}_{\mu\nu} = \delta g_{\mu\nu} - \nabla_{\mu} \epsilon_{\nu} - \nabla_{\nu} \epsilon_{\mu},
\label{T2}
\end{equation}
where $\nabla_{\mu}$ is the covariant derivative with respect to the background goemetry and 
$\epsilon_{\mu} = a^2(\tau)( \epsilon_{0}, - \epsilon_{i})$. The functions $\epsilon_{0}$ and 
$\epsilon_{i}$ are often called gauge parameters since the infinitesimal 
coordinate transformations of the type (\ref{T1}) form a group which is 
in fact the gauge group of gravitation. 
The gauge-fixing procedure, amounts, in four dimensions, to fix the four independent functions 
$\epsilon_{0}$ and $\epsilon_{i}$. The gauge parameters  $\epsilon_{i}$ can be separated 
into their divergenceless and divergencefull parts, i.e.
\begin{equation}
\epsilon_{i} = \partial_{i} \epsilon + \zeta_{i},
\end{equation}
where $ \partial_{i} \zeta^{i} =0$. The gauge transformations 
involving $\epsilon_{0}$ and $ \epsilon$ preserve the scalar nature of 
the fluctuations while the gauge transformations parametrized by $\zeta_{i}$ 
preserve the vector nature of the fluctuation.

Since the covariant derivatives appearing in Eq. (\ref{T2}) 
 (i.e. $\nabla_{\mu} \epsilon_{\nu} = \partial_{\mu} \epsilon_{\nu} - 
\overline{\Gamma}_{\mu\nu}^{\sigma} \epsilon_{\sigma}$) are computed in terms 
of the unperturbed connections (see Eqs. (\ref{Chback}) of the appendix),
from  Eqs. (\ref{g001})--(\ref{g0i1}),
the fluctuations in the tilded coordinate system, defined  by the 
transformation of  Eq. (\ref{T1}), can be written as 
\begin{eqnarray}
&& \phi \to \tilde{\phi} = \phi - {\cal H} \epsilon_0 - \epsilon_{0}' ,
\label{phi}\\
&& \psi \to \tilde{\psi} = \psi + {\cal H} \epsilon_{0},
\label{psi}\\
&& B \to \tilde{B} = B +\epsilon_{0} - \epsilon',
\label{B}\\
&& E \to \tilde{E} = E - \epsilon,
\label{E}
\end{eqnarray}
in the case of the scalar modes of the geometry. 
Under a coordinate transformation preserving the 
vector nature of the fluctuation, i.e. $x^{i} \to \tilde{x}^{i} = x^{i} 
+ \zeta^{i}$ (with  
$\partial_{i} \zeta^{i} =0$), the rotational  
modes of the geometry transform as 
\begin{eqnarray}
&& Q_{i} \to \tilde{Q}_{i} = Q_{i} - \zeta_{i}',
\label{Q}\\
&& W_{i} \to \tilde{W}_{i}= W_{i} + \zeta_{i}.
\label{W}
\end{eqnarray}
The tensor fluctuations, in the parametrization of Eq. (\ref{gij1})
are automatically invariant under infinitesimal diffeomorphisms.

In the appendix useful technical results concerning the fluctuations 
of the geometry and of the sources are reported: to derive the evolution equations
of the fluctuations in the different gauges employed in the present paper 
it is mandatory to have the perturbed form of the Christoffel connections and of the 
Ricci tensors to first order in the metric fluctuations of Eqs. (\ref{g001})--(\ref{g0i1}).
The fluctuations of the mentioned quantities have to be performed in general terms 
without a specific gauge choice.

The perturbed components of the energy-momentum tensor can be written, for a single species, as:
\begin{equation}
\delta T_{0}^{0} = \delta \rho_{\lambda}, \,\,\,\,\,  \delta T_{i}^{j} = - \delta p_{\lambda} \delta_{i}^{j},\,\,\,\,\,\,\,\,
\delta T_{0}^{i} =  ( p_{\lambda} + \rho_{\lambda})\partial^{i} v^{(\lambda)},
\label{enmomf}
\end{equation}
where we defined $\delta u_{i}^{(\lambda)} = \partial_{i} v^{(\lambda)}$.
Notice that the perturbed velocity field can be also written in a different way (see appendix). The 
convention adopted here differs from the one of \cite{MFB}. Under the infinitesimal 
coordinate transformations of Eq. (\ref{T1}) the fluctuations given in Eq. (\ref{enmomf}) 
 transform as 
\begin{eqnarray}
&& \delta \rho_{\lambda} \to \delta \tilde{\rho}_{\lambda} = \delta \rho_{\lambda} - \rho'_{\lambda} \epsilon_{0},
\label{drho}\\
&&  \delta p_{\lambda} \to \delta \tilde{p}_{\lambda} = \delta p_{\lambda} - w_{\lambda}\rho_{\lambda}' \epsilon_{0},
\label{dp}\\
&& v^{(\lambda)}\to \tilde{v}^{(\lambda)} = v^{(\lambda)} + \epsilon'.
\label{vL}
\end{eqnarray}
Using the covariant conservation equation (\ref{b3}) for the background fluid density, 
the gauge transformation for the density contrast, 
i.e. $\delta_{(\lambda)} = \delta \rho_{(\lambda)}/\rho_{(\lambda)}$, follows easily from Eq. (\ref{drho}): 
\begin{equation}
\tilde{\delta}_{(\lambda)} = \delta_{(\lambda)} - 
3 {\cal H} ( 1 + w_{(\lambda)}) \epsilon_{0}.
\label{denscontr}
\end{equation}
There are now, schematically,  three possible strategies
\begin{itemize}
\item{} a specific gauge can be selected  by fixing (completely or partially) 
the coordinate system;
\item{}  gauge-invariant fluctuations of 
the sources and of the geometry can be separately defined;
\item{} gauge-invariant fluctuations mixing the perturbations  
of the sources and of the geometry can be employed.
\end{itemize}

\subsection{Gauge choices and gauge-invariant variables}
The choice of the coordinate system and of the relevant 
gauge-invariant quantities  is determined by the peculiar 
features of the problem to be solved.   The tensor modes of the geometry, as 
introduced in the present section, 
are invariant under infinitesimal diffeomorphisms.
The vector and the scalar modes do change 
for infinitesimal coordinate transformations.  
The gauge fixing of the vector fluctuations is closely analog (but technically simpler) 
to the one of the scalar modes and that is why vectors will be discussed before scalars.
 
\subsubsection{Gauge fixing and gauge-invariant fluctuations for vector modes}

Two different gauge choices appear naturally from the analysis of the vector modes.  
The first choice could be to set 
$\tilde{Q}_{i} =0$. In this case, from Eq. (\ref{Q}),
 the gauge function $\zeta_{i}$ is determined to be 
\begin{equation}
\zeta_{i}(\tau,\vec{x})  = \int^{\tau} Q_{i}(\tau',\vec{x}) d \tau' + C_{i}(\vec{x}).
\end{equation}
Since, in this case, the gauge function is determined up to an arbitrary (space-dependent) 
constant, the coordinate system is not completely fixed. 
This occurrence is reminiscent of what happens in the synchronous 
coordinate system of scalar fluctuations. The gauge choice 
$\tilde{Q}_{i} =0$  has been used, for instance, by the authors of Refs. \cite{LK,GRV}.

Another equally useful choice is the one for which $ \tilde{W}_{i}=0$. 
Starting from an arbitrary coordinate system, according to Eq. (\ref{W}), 
the gauge function is determined as
 $\zeta_{i} = - W_{i}$ and the gauge freedom, in this case, is completely 
fixed.  The gauge $\tilde{W}_{i} =0$ has been exploited in \cite{br1,maxvec1,maxvec2}.

Instead of fixing a gauge, one may choose to work directly 
with gauge-invariant quantities.  By inspecting Eqs. (\ref{Q}) and (\ref{W}) 
it is easy to argue that  the quantity $W_{i}' + Q_{i}$   is invariant under 
infinitesimal coordinate transformations preserving the vector nature 
of the fluctuation. This is the first example of a gauge-invariant potential and it 
is the vector counterpart of the  Bardeen potentials
defined originally by Bardeen in \cite{bardeen} (see also \cite{bardeen2}).

\subsubsection{Gauge fixing and gauge-invariant fluctuations for scalar modes}

If the gauge freedom is used in order to eliminate all the off-diagonal 
entries of the perturbed metric (i.e. $E$ and $B$ according to the parametrization of Eqs. (\ref{g001})--
(\ref{g0i1})),  
we are led to the {\em conformally Newtonian} (or simply {\em longitudinal}) gauge (see \cite{MFB,harr1}).
In the longitudinal gauge $E_{\rm L} = B_{\rm L} =0$. 
To reach the conformally Newtonian gauge from an arbitrary coordinate
system characterized by arbitrary values of $E$ and $B$ 
one has to require, according to Eqs. (\ref{B}) and (\ref{E}) , that the gauge functions 
are determined as $\epsilon =  E$ and $ \epsilon_{0} = E' - B$. In this gauge the only non-vanishing 
entries of the perturbed metric are, according to the parametrization of Eqs. (\ref{g001})--(\ref{g0i1})
 $\phi$ and $\psi$.  

In a related perspective, a set of gauge-invariant generalization 
of the longitudinal degrees of freedom can be constructed. This set of gauge-invariant 
fluctuations can be written as  \cite{bardeen,MFB}
\begin{eqnarray}
&& \Phi = \phi + ( B - E') ' + {\cal H}( B- E'),
\label{phigi}\\
&& \Psi = \psi - {\cal H} ( B - E'),
\label{psigi}
\end{eqnarray}
for the metric perturbations and 
\begin{eqnarray} 
&& \delta \rho_{\rm g} = \delta \rho + \rho' (B - E'),
\label{drhogi}\\
&& \delta p_{\rm g} = \delta p + p' ( B - E'),
\label{dpgi}\\
&& V_{\rm g}^{i} = v^{i} + \partial^{i} E',
\label{vgi}
\end{eqnarray}
for the fluid inhomogeneities, written, for simplicity in the case of a single species.
Using the transformations derived in  
 Eqs. (\ref{phi})--(\ref{E}) and (\ref{drho})--(\ref{vL}),  it can be verified directly that 
 the quantities defined in
 Eqs. (\ref{phigi})--(\ref{vgi}) are gauge-invariant under infinitesimal 
 coordinate transformations preserving the scalar nature of the fluctuation.
Equation (\ref{vgi}) can also be written as $ V_{\rm g} = v + E'$ since, for 
scalar fluctuations, $V_{\rm g}^{i} = \partial^{i} V_{\rm g} $ and 
$ v^{i} = \partial^{i} v$. It 
is sometimes parctical to discuss not the peculiar velocity 
but its divergence, i.e. $ \theta = \partial_{i} v^{i}$ whose associates gauge-invariant 
quantity is $\Theta = \partial_{i} V_{\rm g}^{i} $.

The  {\em off-diagonal gauge } \cite{long2} (also correctly called
{\em uniform curvature gauge}  \cite{hwang1,hwang2,hwang3,hwang4})
 is characterized by the conditions $E_{\rm od} =0$  and $\psi_{\rm od} =0$. Starting 
from and arbitrary coordinate system with non vanishing 
$E$ and $\psi$,  the off-diagonal coordinate system is selected by fixing\footnote{To demonstrate this 
point, Eqs. (\ref{psi}) and (\ref{E}) have to be used in an arbitrary coordinate system.}
 $\epsilon_0 = - \psi /{\cal H}$ and $\epsilon= E$. The two relevant metric 
  fluctuations are then, in the off-diagonal gauge, $B$ and $\phi$.  Since $\psi_{\rm od} =0$, the 
  fluctuation of the spatial curvature is vanishing.

The {\em synchronous } coordinate system stipulates that $\phi_{\rm S} = B_{\rm S} =0$.
Unlike the longitudinal and the off-diagonal gauges, the 
synchronous description does not fix completely the coordinate system. This 
aspect can be appreciated from Eqs. (\ref{phi}) and (\ref{B}), implying that, in the 
synchronous description, the gauge functions are fixed by the following two differential relations 
\begin{equation}
(\epsilon_{0} a)' = a \phi,\,\,\,\,\,\,\,\,\,\,\,\,\, \epsilon' =  ( B + \epsilon_{0}),
\label{syncgaugecond}
\end{equation}
leaving, after integration over $\tau$, two undetermined space-dependent functions. 
This  remaining 
gauge freedom is closely linked with the presence, in the synchronous 
description, of unphysical gauge modes whose features have been, however, completely 
understood \cite{PV1,PV2}.  The parametrization 
usually employed in the literature is not the one based on $\psi$ and $E$ 
but rather the one already mentioned in Eq. (\ref{synpar}).

An important gauge choice is the so-called {\em comoving orthogonal gauge}.
In the comoving gauge the quantity $v_{\rm C} + B_{\rm C}$ is set to zero. In this 
gauge the expression of the curvature fluctuations coincides with a 
relevant gauge-invariant expression whose evolution obeys a conservation law which will be discussed 
extensively in section 5.

Finally, in the {\em uniform density gauge} the total matter density is unperturbed. The 
uniform density gauge is then defined from the relation $\delta \rho_{\rm D} =0$. The latter  implies 
 that, starting from an arbitrary coordinate system where the total 
density fluctuations are non vanishing, the gauge function has to be fixed as 
$\epsilon_0= (\delta \rho) /\rho'$ (see Eq. (\ref{drho})). The curvature fluctuations on constant density 
hypersurfaces  also obey a simple conservation law which will be discussed in section 5.
If the total energy-momentum tensor of the sources is represented by a single 
scalar field one can also define the {\em uniform field gauge}, i.e. the gauge 
in which the scalar field fluctuation vanishes (see section 5).

 \renewcommand{\theequation}{3.\arabic{equation}}
\section{Sachs-Wolfe effect}
\setcounter{equation}{0}

One of the first concepts relevant for the theory of the CMB anisotropies 
is the so-called Sachs-Wolfe (SW) effect \cite{sachs}\footnote{For a pedagogical derivation of the 
SW effect in the case of the standard adiabatic mode (see section 4), Ref. \cite{whitehu1} can be usefully consulted.
Here, however, the relativistic derivation will be presented.}. To introduce the SW effect, 
we recall that in the late-time Universe there are three rather close, but physically separate, 
time scales:  the time of matter-radiation equality (denoted by $\tau_{\rm eq}$);
 the time at which photons decouple from matter (denoted by  $\tau_{\rm dec}$);
the time at which electrons and protons recombine to form Hydrogen  (denoted by $ \tau_{\rm rec}$).
Recalling the definition of redshift, i.e. $z = a_{0}/a -1$, the redshift of equality is $z_{\rm eq} \sim 
2.4 \times 10^{4} \Omega_{\rm m} h^2$ while the redshift of decoupling is 
$z_{\rm dec} \sim 1100$. This occurrence implies that, at least, $\tau_{\rm dec} \sim 4.6 \tau_{\rm eq}$
so that the decoupling time takes place during the matter dominated epoch.

The geodesics of the photons that last-scattered 
at the decoupling time are  sensitive 
to the degree of homogeneity of the geometry after equality.
If, for sufficiently large length-scales after equality 
(but before decoupling) the geometry of the Universe was slightly
inhomogeneous the energy of a photon will experience a shift.

The shift in the 
energy of a photon as defined in the reference frame of a fluid 
moving at some velocity relative to the background space-time should then be computed:
this is the physical rationale for what is usually called the SW effect.  
 The tensor modes, being only coupled to the curvature, will 
behave differently from the scalar modes which are 
sensitive to the evolution of the (scalar) matter sources. These differences 
will be reflected in the SW effect that has to  be separately 
treated for scalar, vector and tensor fluctuations.

\subsection{Scalar Sachs-Wolfe effect}
It is practical  to recall that the geodesics of a photon
are invariant under conformal rescalings. This observation implies 
that by defining the appropriately rescaled affine parameter and   the appropriately 
rescaled metric tensor,  some parts of the derivation of the SW effect can be conducted 
as if the photons would propagate in a (slightly inhomogeneous)  flat space-time. 

Consider then the line element of a spatially flat FRW metric and notice 
that it can be written as 
\begin{equation}
g_{\alpha\beta} dx^{\alpha} dx^{\beta} = a^2(\tau) \tilde{g}_{\alpha\beta} d x^{\alpha} d x^{\beta},
\label{g1}
\end{equation}
where $\tau$ is the conformal time coordinate; $\tilde{g}_{\alpha\beta}$ is 
the flat space-time  (Minkowski) metric supplemented by the appropriate inhomogeneous part, i.e. 
in formulae,  $\tilde{g}_{\alpha\beta} = \eta_{\alpha\beta} + \delta \tilde{g}_{\alpha\beta}$.  

Introducing the affine parameter $s$, the geodesics of a photon, $x^{\mu}(s)$
in a spatially flat FRW metric ($\kappa =0$ in Eq. (\ref{FRW}))
are the same as those in the perturbed Minkowski metric $g_{\alpha\beta}$ with null geodesics 
$x^{\alpha}(\tau)$  and with affine parameters related by $ ds  = a^2(\tau) d\tau$.
The latter property can be checked from the action of the geodesic of a massless particle
\begin{equation}
S= \int d s  g_{\alpha\beta}[x(s)] \frac{dx^{\alpha}}{d s} \frac{dx^{\beta}}{d s}, 
\label{g2}
\end{equation}
whose specific form implies that under the rescalings
\begin{equation}
g_{\alpha\beta} \to a^2(\tau) \overline{g}_{\alpha\beta} ,\,\,\,\,\,\,\,\,\,\,
 d s \to  a^2(\tau)  d\tau,
 \label{resc}
\end{equation}
$S\to \tilde{S} = S$, i.e. the action is invariant.  As a simple application 
of this observation consider the geodesic equation in the inhomogeneous 
Minkowski metric. As discussed in section 2,  it is always possible 
to perform the derivation of a given physical quantity in a specific gauge.
Being the final results physical, they will also be gauge-invariant. 

The gauge we ought to select is the longitudinal  gauge
which stipulates that, from Eqs. (\ref{B}) and (\ref{E}) , $\tilde{E} =0$ and $\tilde{B} =0$, 
i.e. $\epsilon = E$ and $\epsilon_{0} = (E' -B)$.  Thus the perturbed Minkowski metric 
will be characterized by two independent scalar functions, i.e. 
\begin{equation}
 \delta_{\rm s} \tilde{g}_{00} = 2 \phi,\,\,\,\,\,\,\,\, \delta_{\rm s} \tilde{g}_{ij} = 2 \psi \delta_{ij}.
\label{CNG1}
\end{equation} 
The perturbation of the photon geodesic can be written as 
$x^{\mu}\to x^{\mu}(\tau) + \delta_{\rm s} x^{\mu}$, where 
$ x^{\mu} (\tau) = n^{\mu} \tau$ is the unperturbed photon geodesic; 
$n^{\mu}= ( 1, n^{i}) $ is the four-vector giving the direction of the 
photon in a coordinate system where the observer, located at 
the end of a photon world-line, is at $\vec{x} =0$.
The perturbed photon geodesic, obtained by perturbing 
to first order in the amplitude of the scalar fluctuations Eq. (\ref{g2}), can be 
written as 
\begin{equation}
\frac{d^{2}\delta  x^{\mu}}{d\tau^2 } + \delta_{\rm s} 
\tilde{\Gamma}^{\mu}_{\alpha\beta} \frac{d x^{\alpha}}{d\tau} \frac{d x^{\beta}}{d\tau} = 0,
\label{geod0}
\end{equation}
where $\delta_{\rm s} \tilde{\Gamma}_{\alpha\beta}^{\mu}$ are the perturbed connections of the inhomogeneous 
Minkowski metric $\tilde{g}_{\alpha\beta}$.  Since the connections vanish
when evaluated on the background geometry (Minkowski space), in the second term at the left hand side 
of Eq. (\ref{geod0}) possible terms linear in $\delta x^{\mu}$ are absent. The explicit
form of $  \delta_{\rm s} \tilde{\Gamma}_{\alpha\beta}^{\mu}$ can be directly inferred  from
Eqs. (\ref{SCHR}) of the appendix by setting $ {\cal H} =0$ (Minkowski space) and $ E =B=0$ 
(longitudinal gauge condition).  For the ends of the present derivation, the relevant 
component of the perturbed geodesic equation will then read 
\begin{equation}
\frac{d}{d\tau} \biggl[ \frac{d\delta x^{0}}{d\tau}\biggr] = - \delta_{\rm s} \tilde{\Gamma}_{00}^{0} n^{0} n^{0} - \delta_{\rm s} \tilde{\Gamma}_{ij}^{0} n^{i} n^{j}- 
2 \delta_{\rm s} \tilde{\Gamma}_{i0}^{0} n^{i} n^{0} \equiv \psi' - \phi' - 2 \partial_{i}\phi n^{i} n^{0},
\label{geod1}
\end{equation}
where the second equality follows from $\delta_{\rm s}\tilde{\Gamma}_{00}^{0} = \phi'$,
 $\delta_{\rm s} \tilde{\Gamma}_{i j}^{0} = - \psi' \delta_{i j}$,
$\delta_{\rm s} \Gamma_{0 i}^{0} = \partial_{i} \phi$. Notice, furthermore that the relation
 $x^{\mu} = n^{\mu} \tau$ has been used. Equation (\ref{geod1}) can be easily integrated once\footnote{Notice 
 that integration by parts is necessary in order to integrate the term $2 \partial_{i}\phi n^{i}$. Recall, in fact 
 that $d\phi / d\tau = \phi' + \partial_{i} \phi n^{i}$.}
with respect to $\tau$ between the time $\tau_{i}$ (coinciding with the decoupling time) and the 
time $\tau_{f}$ (coinciding with the present time) 
\begin{equation}
 \frac{d\delta x^{0}}{d\eta} = \int_{\tau_{i}}^{\tau_{f}} (\psi' + \phi') d\eta - 2 \phi.
\label{deltax0}
\end{equation}

The quantity to be computed, as previously anticipated, 
 is the photon energy as measured in the frame of reference 
of the fluid. Defining $u^{\mu}$ as the four-velocity 
of the fluid and $P^{\nu}$ as the photon four-momentum the photon energy to be computed is 
noting but  
\begin{equation}
{\cal E} = g_{\mu\nu} u^{\mu} P^{\nu}.
\label{calE}
\end{equation}
The four-momentum of a photon can then be written by generalizing the known special relativistic 
relation 
\begin{equation}
P^{\mu} =  P^{0} \frac{dx^{\mu}}{d\lambda} = \frac{P^0}{a^2} \frac{dx^{\mu}}{d\tau} = 
\frac{E}{a^2} \biggl[ n^{\mu} + \frac{d}{d\tau} \delta x^{\mu} \biggr]
\label{conj}
\end{equation}
where $E$ is a parameter (not to be confused with one of the off-diagonal 
entries of the perturbed metric) defining the red-shifting photon energy.  
The four-velocity of the fluid satisfies, within the conventions adopted in this 
review, the condition
\begin{equation}
g_{\mu\nu} u^{\mu} u^{\nu} =1.
\label{umu1}
\end{equation}
The logic will now be to determine $u^{\mu}$, $g_{\mu\nu}$ and $P^{\mu}$ to first 
order in the fluctuations of the geometry. This will allow to compute the 
right hand side of Eq. (\ref{calE})  as a function of the inhomogeneities of the metric.
The first-order variation of Eq. (\ref{umu1}) leads to
\begin{equation}
\delta_{\rm s}  g_{00} u^{0} = -2 \delta_{\rm s} u^{0} g_{00},
\label{umu2}
\end{equation}
so that in the longitudinal coordinate system 
Eq. (\ref{umu2}) gives, to first-order in the metric fluctuations, 
\begin{equation}
u^{0} = \frac{1}{a}( 1 - \phi).
\label{u0}
\end{equation}
Following the conventions of section 2 and of the appendix, the 
(divergencefull) peculiar  velocity field is given by
\begin{equation}
\delta_{\rm s} u^{i} = \frac{v^{i}}{a} \equiv \frac{1}{a} \partial^{i} v.
\label{ui}
\end{equation}
The relevant peculiar velocity field will be, in this derivation, the baryonic 
peculiar velocity since this is the component emitting and observing (i.e. absorbing)
the radiation. In the following this identification will be undesrtood and, hence,
$v^{i} = v^{i}_{\rm b}$. 

The energy of the photon in the frame of reference of the fluid becomes, then 
\begin{equation}
{\cal E} =  g_{\mu\nu} u^{\mu} P^{\nu} = g_{00} u^{0} P^{0} + g_{i j} u^{i} P^{j},
\label{photen}
\end{equation}
Inserting now Eqs. (\ref{conj}) and (\ref{u0}) into Eq. (\ref{photen}) and recalling 
the explicit forms of $g_{00}$ and $g_{ij}$ to first order in the 
metric fluctuations we have 
\begin{equation}
{\cal E} = \frac{E}{a}\biggl[ 1 + \phi - n_{i}v^{i}_{\rm b} + \frac{d \delta x^{0}}{d\tau} \biggr].
\label{SW1}
\end{equation}
Assuming, as previously stated,  that the observer, located at 
the end of a photon geodesic, is  at $\vec{x}=0$, Eq. (\ref{SW1}) can be expressed as 
\begin{equation}
{\cal E} = \frac{E}{a} \biggl\{ 1 - \phi - n_{i} v^{i}_{\rm b}+ \int_{\tau_{i}}^{\tau_{f}} (\psi' + \phi') d\tau\biggr\}
\label{SW2}
\end{equation}
The temperature fluctuation can be expressed by taking the difference between the final 
and initial energies, i.e. 
\begin{equation}
\frac{\delta T}{T} = \frac{ a_{f} {\cal E}(\eta_{f}) - a_{i} E_{i} }{a_{i} E_{i}}.
\label{SW3}
\end{equation}
In Eq. (\ref{SW3}) $E_{i}$ is the initial photon energy which can be expressed as 
\begin{equation}
E_{i} = E_{0} \biggl[ 1 + \biggl(\frac{\delta T}{T}\biggr)_{i}\biggr] \equiv  E_{0} \biggl[ 1 + \frac{\delta_{\rm r}(\eta_{i})}{4} \biggr]
\label{Ei}
\end{equation}
where $\delta_{\rm r} = \delta \rho_{\rm r}/\rho_{r}$ and $ \rho_{\rm r} \propto T^4$ is the radiation 
density contrast.
The final expression for the SW effect induced by scalar fluctuations can be written as 
\begin{equation}
\biggl(\frac{\Delta T}{T}\biggr)_{\rm s} = \frac{\delta_{\rm r}(\tau_{i})}{4} - [\phi]_{\tau_{i}}^{\tau_{f}} -
 [n_{i} v^{i}_{\rm b}]_{\tau_{i}}^{\tau_{f}} + 
\int_{\tau_{i}}^{\tau_{f}} ( \psi' + \phi') d\tau.
\label{SW0}
\end{equation}
Sometimes, for simplified esitmates, the temperature fluctuation can then be written, in explicit terms,  as
\begin{equation}
\biggl(\frac{\Delta T}{T}\biggr)_{\rm s} = \biggr[\frac{\delta_{\rm r}}{4} + \phi + n_{i} v^{i}_{\rm b}\biggr]_{\tau_{i}} + 
\int_{\tau_{i}}^{\tau_{f}}( \psi' +\phi') d\tau. 
\label{scalSW}
\end{equation}

Equation  (\ref{scalSW}) has three contribution
\begin{itemize}
\item{} the ordinary SW effect given by the first two terms at the righ hand side of Eq. (\ref{scalSW}) 
i.e. $\delta_{\rm r}/4$ and $\phi$;
\item{} the Doppler term (third term in Eq, (\ref{scalSW}));
\item{} the integrated SW effect (last term in Eq, (\ref{scalSW})).
\end{itemize}
The ordinary SW effect is due both to the intrinsic temperature  inhomogeneities on the last 
scattering surface and to the inhomogeneities of the metric.  On large angular scales 
the ordinary SW contribution dominates. The Doppler term arises thanks to the relative 
velocity of the emitter and of the receiver. At large angular scales its contribution is subleading 
but it becomes important at smaller scales, i.e. in multipole space, for $\ell \sim 200$
corresponding to the first peak in Fig. \ref{F13}.
The SW integral contributes to the temperature anisotropy if $\psi$ and $\phi$ depend on time.
It will be shown that this may be the case for some classes of initial conditions (see section 4) 
but it is not the case for the standard adiabatic mode. In quintessence models off dark energy 
the longitudinal fluctuations may acquire a time-dependence in spite of the adiabaticity 
inducing a late time integrated SW effect.  In section 8 Boltzmann equation will be used in order 
to derive, in a more refined way, the same formula given in Eq. (\ref{scalSW}).

\subsection{Vector Sachs-Wolfe effect}
In the case of the vector Sachs-Wolfe effect the shift 
in the photon geodesics is induced by the vector modes of the 
geometry defined in Eqs. (\ref{gij1}) and (\ref{g0i1}) and named 
$Q_{i}$ and $W_{i}$ (see also Eqs. (\ref{VF})--(\ref{VRICCI}) of the appendix).
Since $W_{i}$ and $Q_{i}$ are both divergenceless the photon energy 
in the frame of reference of the baryonic fluid will be, to first-order 
in the amplitude of the vector fluctuations:
\begin{equation}
{\cal E} = \frac{E}{a} \biggl[ - n_{i} {\cal V}^{i}_{\rm b} + \frac{ d \delta_{\rm v}
 x^{0}}{d\tau} \biggr]
\end{equation}
where ${\cal V}^{i}_{\rm b}$ is the rotational component
of the baryonic peculiar velocity and, as anticipated in section 2,
$\partial_{i} {\cal V}_{\rm b}^{i}=0$; the notation $\delta_{\rm v}$, as 
defined in the appendix, denotes the fluctuation induced by the vector modes 
of the geometry.  

From the geodesic equation we have 
\begin{equation}
 \frac{ d \delta_{\rm v} x^{0}}{d\tau} = \frac{1}{2} (\partial_{i} V_{j} + \partial_{j} V_{i}) n^{i} n^{j},
\end{equation}
where, as before, the perturbed connections are obtained by setting ${\cal H}=0$ 
in Eqs. (\ref{VCHR}); the quantity $V_{i}$ is defined as 
\begin{equation}
V_{i} = Q_{i} + W_{i}'.
\label{GIV}
\end{equation}
Recalling the transformation rules for $Q_{i}$ and $W_{i}$ as derived in Eqs. (\ref{Q}) and (\ref{W}),
it can be appreciated that $V_{i}$ is invariant under infinitesimal diffeomeorphisms.
Thus,  the induced temperature fluctuations induced by the vector 
modes of the geometry can be written as 
\begin{equation}
\biggl(\frac{\Delta T}{T}\biggr)_{\rm v} = [ -\vec{{\cal V}}\cdot \vec{n}]_{\tau _{i}}^{\tau_{f}} + \frac{1}{2} \int_{\tau_{i}}^{\tau_{f}}
(\partial_{i} V_{j} + \partial_{j} V_{i}) n^{i} n^{j} d\tau.
\end{equation} 
In the vector case there is no ordinary SW effect but the only terms are the ones 
connected with the Doppler effect and with the integrated SW effect.

\subsection{Tensor Sachs-Wolfe effect}

The tensor Sachs-Wolfe contribution, as anticipated,  corresponds only to the integrated Sachs-Wolfe effect.
In full analogy with the derivation of the scalar Sachs-Wolfe effect, the tensor contribution can be also 
obtained. Recalling that the tensor fluctuations of the metric are 
parametrized by a rank-two tensor in three-dimensions, $h_{ij}$  which is divergenceless 
and traceless, from Eqs. (\ref{gij1}) and (\ref{TCHR}) the perturbed metric and the perturbed 
connections can be obtained. As before, the perturbed Minkowski metric will be 
considered. The photon energy is then  
\begin{equation}
{\cal E} = \frac{E}{a}\biggl[ 1 + \frac{d \delta_{\rm t} x^0}{d\tau} \biggr].
\end{equation}
The geodesic equation to be integrated is then
\begin{equation}
\frac{d}{d\tau} \biggl[ \frac{d \delta_{\rm t} x^0}{d\tau}\biggr] = - \delta_{\rm t}
 \Gamma_{ij}^{0} n^{i} n^{j}.
\end{equation}
The explicit value of $\delta_{\rm t} \tilde{\Gamma}_{ij}^{0}$ can be obtained from 
Eqs. (\ref{TCHR}) by setting ${\cal H} =0$ and, consequently,
\begin{equation}
\biggl(\frac{\Delta T}{T}\biggr)_{\rm t} = - \frac{1}{2} \int_{\tau_i}^{\tau_{f}} h_{ij}' n^{i} n^{j} d\tau. 
\end{equation}
The only contribution to the tensor  Sachs-Wolfe effect is given by the Sachs-Wolfe integral.

\renewcommand{\theequation}{4.\arabic{equation}}
\section{Initial conditions for CMB anisotropies}
\setcounter{equation}{0}
In the derivation of the SW the possible existence of metric inhomogeneities after 
equality has been simply assumed. However, the metric inhomogeneities 
have to be solutions of the Einstein equations and this 
restricts their possible form.

The effect of metric fluctuations on the temperature 
anisotropies can be discussed within two complementary 
approaches. The first approach could be called, in broad 
terms, model independent. In this case,  the spatial and temporal 
variations of the metric inhomogeneities are determined 
as functions of a number of unknown initial conditions 
parametrized in terms of a suitable number of arbitrary 
constants. The second approach could be called 
 model dependent. In this case the initial conditions 
of the fluctuations are given at an early stage in the life of the 
Universe (for instance during conventional inflation or in the 
context of other cosmological models inspired by superstring theory). The fluctuations 
will then be evolved in time using the techniques to be
described in section 5 and the result of this procedure will 
then allow to compute the SW effect induced by  the different models.

The two approaches are complementary since the metric inhomogeneities 
 computed in a specific model of evolution will always fall in one of the 
 general solutions described in the model-independent approach. 
 In other words, once a specific model of early evolution 
 is adopted,  the arbitrary constants appearing in the model-independent approach 
 will be fixed to a specific value.  Sometimes, in the literature, 
 the model-independent study of the metric inhomogeneities relevant 
 for the SW effect, is swiftly phrased as the problem of initial 
 conditions. This terminology will be also adopted here with the caveat 
 that there could be ambiguities. In fact, there is another problem of 
 initial conditions which will be treated later on (in section 6): namely
 the problem of initial conditions of the fluctuations in a specific 
 model of the physics of the early Universe.

To set initial conditions for the CMB 
anisotropies means to find large-scale (i.e. $ k\tau < 1$) solutions 
for the evolution of the metric inhomogeneities and 
of the relevant plasma quantities.  When a given mode satisfies 
the relation $k \tau <1$, it is said to be outside the horizon.
If this is the case, the associated physical frequency is smaller than the 
Hubble rate $H$,  i.e. $ \omega = k/a(\tau) < H$. In the opposite case, i.e. 
$k \tau > 1$ the mode is said to be inside the horizon and the 
corresponding physical frequency will be larger than the Hubble rate. 
For $ k\tau \simeq 1$ the mode $k$ is said to be crossing the horizon. The 
initial conditions of CMB anisotropies can be set, for numerical purposes, 
deep within the radiation dominated epoch. However, if swift estimates of the SW 
effect have to be obtained, one must recall that, since decoupling 
occurs already in the matter-dominated phase, the solutions of the evolution 
equations have to be modified accordingly.

A relevant distinction playing a key r\^ole in the theory of the CMB 
anisotropies is the one between adiabatic and isocurvature \footnote{In order 
to avoid misunderstandings  it would be more appropriate to use the 
terminology non-adiabatic since the term isocurvature may be interpreted 
as denoting a fluctuation giving rise to a uniform curvature. In the 
following the common terminology will be however used.}
initial conditions. Consider, for simplicity, the idealized 
case of a plasma where the only relevant fluid variables are the ones 
associated with CDM particles and radiation. Incidentally, this 
is a rather good approximation for $\tau< \tau_{\rm eq}$. 
The entropy per dark matter particle will then be given by 
$ \varsigma = T^3/n_{\rm c}$ where $n_{\rm c}$ is the number density 
of CDM particles and $\rho_{\rm c} = m_{\rm c} n_{\rm c}$ is the associated
 energy density.
Recalling that $\delta_{\rm r} = \delta \rho_{\rm r}/\rho_{\rm r}$ and $\delta_{\rm c} = \delta \rho_{\rm c}/\rho_{\rm c}$ 
are, respectively the density contrast in radiation and in CDM, the fluctuations of the specific entropy will then be 
\begin{equation}
{\cal S} = \frac{\delta{\varsigma}}{{\varsigma}} = 3 \frac{\delta T}{T} - \delta_{\rm c}=  \frac{3}{4} \delta_{\rm r} 
- \delta_{\rm c},
\label{SE}
\end{equation}
where the second equality follows recalling that $\rho_{\rm r} \propto T^4$. 
If the fluctuations in the specific entropy vanish, at large-scales, then a 
chacteristic relation 
between the density contrasts of the various plasma quantities appears, i.e. 
for a baryon-photon-lepton fluid with CMD particles,
\begin{equation}
\delta_{\gamma }\simeq \delta_{\nu} \simeq \frac{4}{3} \delta_{\rm c} 
\simeq \frac{4}{3} \delta_{\rm b}.
\label{ad}
\end{equation}
Eq. (\ref{SE}) can be generalized to the case of a mixture of different fluids with
arbitrary equation of state as specifically discussed by Kodama and Sasaki \cite{KS}. For 
instance, in the case of two fluids ${\rm a}$ and ${\rm b}$ with arbitrary barotropic indices 
$w_{\rm a}$ and $w_{\rm b}$ the fluctuations in the specific entropy are
\begin{equation}
{\cal S}_{{\rm a\,b}} = \frac{\delta_{\rm a}}{1 + w_{{\rm a}}}  - \frac{\delta_{\rm b}}{1 +  w_{{\rm b}}},
\label{SEgen}
\end{equation}
where $\delta_{\rm a}$ and $\delta_{\rm b}$ are the density contrasts of the two species.
It is relevant to stress that, according to Eq. (\ref{denscontr}), giving the gauge variation 
of the density contrast of a given species,  ${\cal S}_{{\rm a\,b}}$ is gauge-invariant.

If the solution of the coupled system of the pertured  Einstein and fluid 
equations satisfies Eq. (\ref{ad}), then the initial conditions are said to be 
adiabatic. This 
possibility is the one favoured by conventional inflationary models. 
If the condition (\ref{ad}) is not satisfied by the initial conditions 
are said to be 
{\em isocurvature}. 
On top of the adiabatic mode, there are, in principle, four well 
characterized non-adiabatic modes: a baryon isocurvature mode, a  CDM isocurvature mode and
two neutrino isocurvature modes. While some of these modes are 
divergent in the longitudinal gauge description, they are all 
finite and well behaved in the synchronous gauge which, therefore, will be used to 
introduce them.  In the following section after discussing 
the problem of the initial conditions in the longitudinal gauge, we will 
move to the synchronous gauge and complete the analysis.

\subsection{Longitudinal gauge description} 
From the results derived in Eqs. (\ref{dg00})--(\ref{dg0i}), the perturbed components 
of the Einstein tensor in the longitudinal gauge are obtained by setting $ E =0$ and $ B=0$.   
The perturbed Einstein equations can then  be written as:
\begin{eqnarray}
&& \delta_{\rm s} {\cal G}_{0}^{0} = 8 \pi G \delta_{\rm s} T_{0}^{0},
\label{l00a}\\
&& \delta_{\rm s} {\cal G}_{i}^{j} = 8 \pi G \delta_{\rm s} T_{i}^{j},
\label{lija}\\
&& \delta_{\rm s} {\cal G}_{0}^{i} = 8\pi G \delta_{\rm s} T_{0}^{i}.
\label{l0ia}
\end{eqnarray}
The fluctuations of total the energy-momentum tensor are 
written as the sum of the fluctuations over the various 
species composing the plasma, i.e. according to Eq. (\ref{enmomf}),
\begin{eqnarray}
&& \delta_{\rm s} T_{0}^{0} = \delta \rho = \sum_{\lambda} \rho_{\lambda} \delta_{\lambda},
\label{dT00la}\\
&& \delta_{\rm s} T_{i}^{j} = -\delta_{i}^{j} \delta p + \Pi_{i}^{j}= 
- \delta_{i}^{j} \sum_{\lambda} w_{\lambda} \rho_{\lambda} \delta_{\lambda} + \Pi_{i}^{j},
\label{dTija}\\
&& \delta_{\rm s} T_{0}^{i} = (p + \rho) v^{i} = \sum_{\lambda} ( 1 + w_{\lambda}) \rho_{\lambda} v^{i}_{\lambda} .
\label{dToia}
\end{eqnarray}
In Eqs. (\ref{dT00la})--(\ref{dToia}) the sum over $\lambda$ run over the different 
species of the plasma.  We will 
be interested in the situation where the universe is deep within 
the radiation-dominated epoch for $T< 1 $ MeV when neutrinos 
have already decoupled and form a quasi-perfect (collisionless) fluid. 
In this case the species are baryons, photons, neutrinos and CDM particles. 
Notice that if the energy-momentum tensor may include a dark-energy component
parametrized, for instance, in terms of a cosmological term. In this case 
the fluctuations of the dark-energy are negligible. This situation may change, after 
decoupling, if the dark energy is parametrized in terms of scalar field(s).  In this case 
contributions to the integrated SW effect are naturally expected.

In Eq. (\ref{dTija}) an anisotropic stress $\Pi_{i}^{j}$ has been included 
to account for the possible presence of collisionless 
particles, like, in the present case, neutrinos. 
Equations (\ref{l00a})--(\ref{l0ia}) lead then to the following system 
of equations:
\begin{eqnarray}
&&\nabla^2 \psi- 3 {\cal H} ( {\cal H} \phi + \psi') = 4\pi G a^2 \delta \rho,
\label{p00l}\\
&& \nabla^2( {\cal H} \phi + \psi') =- 4 \pi G a^2 (p + \rho) \theta,
\label{p0il}\\
&& \biggl[ \psi'' + {\cal H} ( 2\psi' +\phi') + ( 2 {\cal H}' + {\cal H}^2) \phi + 
\frac{1}{2} \nabla^2(\phi - \psi) \biggr] \delta_{i}^{j}  
\nonumber\\
&& - \frac{1}{2} \partial_{i}\partial^{j} ( \phi - \psi) = 
 4\pi G a^2 (\delta p \delta_{i}^{j} - \Pi_{i}^{j}).
\label{pijl}
\end{eqnarray}
where the divergence of the total velocity field has been defined as:
\begin{equation}
( p + \rho) \theta = \sum_{\lambda} ( p_{\lambda} + \rho_{\lambda})\theta_{\lambda}
\label{thetadefinition}
\end{equation}
with $\theta = \partial_{i} v^{i}$ and 
$\theta_{\lambda}= \partial_{i} v^{i}_{\lambda}$. Equations  (\ref{p00l}) and 
 (\ref{p0il}) are, respectively, the Hamiltonian and the 
momentum constraint. The enforcement of these two constraints is
crucial for the regularity  of the initial conditions. 
Taking the trace of Eq. (\ref{pijl}) and recalling that 
the anisotropic stress is, by definition,  traceless (i.e. $\Pi_{i}^{i}=0$) 
it is simple to obtain 
\begin{equation}
\psi'' + 2 {\cal H} \psi' + {\cal H} \phi' + ( 2 {\cal H}' + {\cal H}^2)\phi + 
\frac{1}{3} \nabla^2 (\phi - \psi) = 4 \pi G a^2 \delta p.
\label{B1}
\end{equation}
The difference of Eqs. (\ref{pijl}) and (\ref{B1}) leads to 
\begin{equation}
\frac{1}{6} \nabla^2( \phi - \psi)\delta_{i}^{j}  - \frac{1}{2} \partial_{i} 
\partial^{j} ( \phi - \psi) = - 4 \pi a^2 G \Pi_{i}^{j}.
\label{B2}
\end{equation}
Going then to  Fourier space and defining (see appendix)
\begin{equation}
 \biggl( \frac{k_{j}k^{i}}{k^2} - \frac{1}{3} \delta_{j}^{i} \biggr)
 \Pi_{i}^{j} = (p + \rho)\sigma = \sum_{\lambda} ( p_{\lambda} + 
\rho_{\lambda}) \sigma_{\lambda},
\end{equation}
Then, Eq. (\ref{B2}) becomes : 
\begin{equation}
\nabla^2 ( \phi - \psi) = 12 \pi G a^2 ( \rho + p) \sigma =  12 \pi G a^2 \sum 
( \rho_{\lambda} + p_{\lambda}) \sigma_{\lambda}.
\label{B3}
\end{equation}
Since, as elaborated before, the only collisionless species for 
$T< 1 \,\,{\rm MeV}$ are neutrinos, 
the relevant contribution to Eq. (\ref{B3}) will mainly come from the 
neutrinos. 

Equations (\ref{p00l})--(\ref{pijl}) may be 
supplemented with the perturbation of the 
covariant conservation of the energy-momentum tensor, i.e. Eqs. (\ref{T0a}) 
and (\ref{T0b}); by selecting the longitudinal coordinate system (as 
explained in the appendix), Eqs. (\ref{T0a})--(\ref{T0b}) become, for a single species,
\begin{eqnarray}
&&\delta_{\lambda}' = ( 1 + w_{\lambda}) ( 3 \psi' - \theta_{\lambda}) + 
3 {\cal H} \biggl[ w_{\lambda} - 
\frac{\delta p_{\lambda}}{\delta \rho_{\lambda}}\biggr] \delta_{\lambda}, 
\label{delta}\\
&&\theta_{\lambda}' = ( 3 w_{\lambda} -1) {\cal H} \theta_{\lambda} - 
\frac{w_{\lambda}'}{w_{\lambda} + 1} 
\theta_{\lambda} - \frac{1}{w_{\lambda}+1} \frac{\delta 
p_{\lambda}}{\delta \rho_{\lambda}} \nabla^2 \delta_{\lambda} + \nabla^2 
\sigma_{\lambda} - \nabla^2 \phi.
\label{theta}
\end{eqnarray}
In the following, the evolution equations of the fluctuations in the 
longitudinal gauge will be 
exploited for the derivation of  the large-scale adiabatic  mode.
\subsubsection{The adiabatic mode}
Defining as $\delta_{\nu}$, $\delta_{\gamma}$, $\delta_{\rm b}$ and 
$\delta_{\rm c}$ the neutrino, photon, baryon and CDM 
density contrasts, the Hamiltonian constraint of Eq. (\ref{p00l}) 
can be written as 
\begin{equation}
- 3 {\cal H} ( {\cal H} \phi + \psi') - k^2 \psi = \frac{3}{2} {\cal H}^2 
[ ( R_{\nu} \delta_{\nu} + ( 1 - R_{\nu}) \delta_{\gamma})  + 
\Omega_{\rm b} \delta_{\rm b} + \Omega_{\rm c} \delta_{\rm c}],
\label{p00lex}
\end{equation}
where, for $N_{\nu}$ species of massless neutrinos,   
\begin{equation}
R= \frac{7}{8} N_{\nu} \biggl( \frac{4}{11}\biggr)^{4/3},\,\,\,\,\,\,\, 
R_{\nu}  = \frac{R}{1 + R},\,\,\,\,\,\,\,\, R_{\gamma} = 1 - R_{\nu},
\label{DEFRNU}
\end{equation}
so that $R_{\nu}$ and $R_{\gamma}$ represent the fractional 
contributions of photons and neutrinos to the total density at early times 
deep within the radiation-dominated epoch.
Eq. (\ref{b1}) has been used into Eq. (\ref{p00lex}) in order to 
eliminate the explicit dependence upon the total energy density 
of the background.  Notice that, in the following {\em the index $k$ appearing in the 
Fourier transformed quantities will be omitted since it could be confused 
with the other indices defining the various species present in the plasma}.

From the momentum constraint of Eq. (\ref{p0il}), 
and from Eq. (\ref{B1}) the following pair of equations can be derived:
\begin{eqnarray}
&& k^2 ( {\cal H} \phi + \psi') = \frac{3}{2} {\cal H}^2
 \biggl[ \frac{4}{3} (R_{\nu} \theta_{\nu} + R_{\gamma} 
\theta_{\gamma}) + 
\theta_{\rm b} \Omega_{\rm b} + \theta_{\rm c} \Omega_{\rm c} \biggr],
\label{p0ilex}\\
&& \psi'' + ( 2 \psi' + \phi') {\cal H} + (2 {\cal H}' + {\cal H}^2) \phi
 - \frac{k^2}{3} ( \phi - \psi) = \frac{{\cal H}^2}{2} 
( R_{\nu} \delta_{\nu} +
\delta_{\gamma}  R_{\gamma}),
\label{pijlex}
\end{eqnarray}
where following Eq. (\ref{thetadefinition})  the divergence of the (total) peculiar velocity field 
has been separated for the different species. Furthermore, 
in Eqs (\ref{p0ilex})--(\ref{pijlex}),  Eq. (\ref{b2}) has 
been used in order to eliminate the explicit dependence upon the 
(total) energy and pressure densities. 
Finally, according to Eq. (\ref{B3}) the neutrino anisotropic stress fixes 
the difference between the two longitudinal fluctuations of the 
geometry, i.e. 
\begin{equation}
k^2 (\phi - \psi) = -6 {\cal H}^2 \sigma_{\nu},
\label{pineqjex}
\end{equation}
completing the set of  perturbed Einstein equations written in explicit form.

Equations (\ref{delta}) and (\ref{theta}) should now be written in the case 
of the various species.
In the case of CDM particles (i.e.  $w_{\rm c}=0$), Eqs. (\ref{delta}) and (\ref{theta}) imply
\begin{eqnarray}
&&\theta_{\rm c}' + {\cal H} \theta_{c} = k^2 \phi,
\label{CDM1}\\
&& \delta_{\rm c}' = 3 \psi' - \theta_{\rm c}.
\label{CDM2} 
\end{eqnarray}
Photons an baryons are tightly coupled through Thompson 
scattering and, therefore,  are treated as collisional species. The fluid equations
for photons 
\begin{eqnarray}
&& \delta_{\gamma}' = - \frac{4}{3} \theta_{\gamma} + 4 \psi',
\label{phot1}\\
&& \theta_{\gamma}' = \frac{k^2}{4} \delta_{\gamma} + k^2 \phi  
+ a n_{\rm e} \sigma_{\rm T} ( \theta_{\rm b} - \theta_{\gamma} ).
\label{phot2}
\end{eqnarray}
and baryons 
\begin{eqnarray}
&& \delta_{\rm b}' = 3 \psi' - \theta_{\rm b},
\label{baryon1}\\
&& \theta_{\rm b}' = - {\cal H} \theta_{\rm b} + c_{{\rm s, b}}^2 k^2 
\delta_{\rm b} + k^2  \phi +  
\frac{4}{3} \frac{\Omega_{\gamma}}{\Omega_{\rm b}} a n_{\rm e} 
x_{\rm e} \sigma_{\rm T} ( \theta_{\gamma} - \theta_{\rm b}),
\label{baryon2} 
\end{eqnarray}
are again obtained from Eqs. (\ref{delta}) and (\ref{theta}) by recalling 
that 
\begin{equation}
w_{\gamma} = \frac{1}{3},\,\,\,\,\,\,\,\, w_{\rm b} =0,\,\,\,\,\,\,\,\,
c_{\rm s, b}^2 = \frac{\delta p_{\rm b}}{\delta\rho_{\rm b}}.
\end{equation}
In Eqs. (\ref{phot2}) and (\ref{baryon2}) 
\begin{equation}
\sigma_{\rm T} = \frac{8 \pi}{3} r_{0}^2 = 0.665246 \,\,{\rm barn}, \,\,\,\,\,\,\,\,\,\,\,\,\,\,r_{0} = 
\frac{e^2}{m_{\rm e}c^2}
\label{thompson}
\end{equation}
is the Thompson scattering cross section \cite{jackson}, $x_{\rm e}$ is the ionization 
fraction of the plasma and 
\begin{equation}
n_{\rm e} = 1.13 \times 10^{-5} \times (\Omega_{\rm b} h^2) \,\,(1 + z)^3 \,\,\,{\rm cm}^{-3}
\label{electrondensity}
\end{equation}
is the electron density.   From Eqs. (\ref{thompson}) and (\ref{electrondensity})  the 
mean free path of CMB photons can be computed prior to recombination as a function 
of the redshift $z$:
\begin{equation}
\ell_{\gamma} \sim \frac{1}{ x_{\rm e}\sigma_{\rm T} n_{\rm e}(z+ 1)^{-1}} \simeq 5 \times 10^{4}
(\Omega_{\rm b} h^2)^{-1} ( 1 + z )^{-2}\,\,\,{\rm Mpc}.
\label{MFPphot}
\end{equation}
For comoving scales shorter than the photon mean free path photons and baryons 
are  tightly coupled and  the characteristic time for the synchronization of the photon 
and baryon velocities is small with respect to the expansion time and to the oscillation period. 
 In the first approximation,  $\sigma_{\rm T} \to \infty$ and
$ \theta_{\gamma}\simeq \theta_{\rm b}$. This result holds to 
zeroth order in the tight-coupling expansion. 
Higher orders are also interesting in the context of the CMB polarization 
and will be treated after having introduced the full Boltzmann hierarchy 
for the brightness functions (see section 8). 

The last species to be introduced is given by neutrinos. 
Unlike baryons and photons 
neutrinos are not collisional and therefore the fluid approximation, for this 
species is not strictly justified. The usual approach is 
to characterize neutrinos with a set of fluid equations where, however, the 
anisotropic stress contributes directly to the evolution equation of 
the peculiar velocity.
Since, in this approach, the neutrino anisotropic stress is 
 dynamical, to be fully consistent the evolution of $\sigma_{\nu}$  has to be 
included. The full set of relevant equations will then be 
\begin{eqnarray}
&& \delta_{\nu}' = - \frac{4}{3} \theta_{\nu} + 4\psi',
\label{nu1}\\
&& \theta_{\nu}' = \frac{k^2}{4} \delta_{\nu} - k^2 \sigma_{\nu} + k^2 \phi,
\label{nu2}\\
&& \sigma_{\nu}' = \frac{4}{15} \theta_{\nu} - \frac{3}{10} k {\cal F}_{\nu 3}.
\label{nu3}
\end{eqnarray}
Equations (\ref{nu1}) and (\ref{nu2}) are directly obtained from Eqs. 
(\ref{delta}) and (\ref{theta}) 
in the case $ w_{\nu} = 1/3$ and $\sigma_{\nu} \neq 0$. Eq. (\ref{nu3}) 
is not obtainable in the fluid approximation and the full Boltzmann hierarchy has to be introduced.
The quantity  ${\cal F}_{\nu 3}$   introduced in 
 Eq. (\ref{nu3}),  
is the octupole term of the neutrino phase space distribution (see section 8). 

The system of the above equations will now be solved for the case 
of the adiabatic mode deep within the radiation dominated epoch. 
Inspection of the Hamiltonian  and momentum 
constraints (\ref{p00lex})--(\ref{p0ilex})  shows that if the 
adiabaticity relation of Eq. (\ref{ad}) is satisfied, then the fluctuations, in the 
conformally Newtonian gauge, should be constant in the limit $ k \tau <1$. 
Inspection of the momentum constraint also implies that the peculiar velocity 
fields should be of order $k^2\tau$ and that they should all be equal 
to lowest order. Thus, the solution of the system can be found as a 
perturbative expansion in $k\tau$ and the starting trial solution can be written as:
\begin{eqnarray}
&& \delta_{\nu}=  - 2 \phi_{0} + A k^2 \tau^2 = \delta_{\gamma} = 
\frac{4}{3} \delta_{\rm b} = \frac{4}{3} \delta_{\rm c}
\label{deltasol1}\\
&& \phi = \phi_0 + C_{\phi} k^2\tau^2,
\label{phisol1}\\
&& \psi = \psi_{0} + C_{\psi} k^2 \tau^2,
\label{psisol1}\\
&&\theta_{\gamma} = \frac{\phi_{0}}{2} k^2 \tau  + D_{\gamma} k^4 \tau^3,
\label{thgammasol1}\\
&& \theta_{\nu} =  \frac{\phi_{0}}{2} k^2 \tau  + D_{\nu} k^4 \tau^3,
\label{thnusol1}\\
&& \theta_{\rm b} =\frac{\phi_{0}}{2} k^2 \tau   + D_{\rm b} k^4 \tau^{3},
\label{thbsol1}\\
&& \theta_{\rm c} = \frac{\phi_{0}}{2} k^2 \tau  + D_{\rm c} k^4 \tau^{3},
\label{thcsol1}\\
&&\sigma_{\nu} = \sigma_{0} k^2 \tau^2 + \sigma_{1} k^4 \tau^4.
\label{sigsol1}
\end{eqnarray}
Notice that the equality between $\delta_{\gamma}$ and $\delta_{\nu}$ up to ${\cal O}(k^2 \tau^2)$
has not been postulated but it 
is a direct consequence of Eqs. (\ref{nu1}) and (\ref{phot1}).  Furthermore, 
 the leading term of the velocity fields has been determined by 
consistency of the trial solution with Eqs. (\ref{CDM2}), (\ref{phot2}, 
(\ref{baryon2})  and (\ref{nu2}).

Consider now  the evolution equations for the neutrinos as written in 
Eqs. (\ref{nu1})--(\ref{nu3}), and  assume, for the moment,  that the octupole 
moment of the neutrino phase-space density vanishes, i.e. ${\cal F}_{\nu 3} =0$ to 
the wanted order 
in $k\tau$. In this case, Eqs. (\ref{thnusol1}) and (\ref{sigsol1}) imply, to lowest order,
\begin{equation}
 \sigma_{0} = \frac{\phi_{0}}{15},
\label{cond1}
\end{equation}
Since, from Eq. (\ref{pineqjex}), the value of the anisotropic stress is fixed in 
terms of the difference between the two longitudinal fluctuations of the metric to 
\begin{equation}
\sigma_{0} = \frac{\psi_{0} - \phi_{0}}{6 R_{\nu}},
\label{sigma01}
\end{equation}
Eq. (\ref{cond1}) implies that, to lowest order, the relation between  $\psi_{0}$ and $\phi_0$ is:
\begin{equation}
\psi_{0} = \biggl(1 + \frac{2}{5} R_{\nu}\biggr) \phi_{0}.
\label{25R}
\end{equation}
The remaining  arbitrary constants  appearing in Eqs. (\ref{deltasol1})--(\ref{sigsol1})  
are fixed by requiring the compatibility with the Einstein-fluid system of equations
to the  relevant order  $k\tau$. From the Hamiltonian and 
momentum constraints we get, respectively,:
\begin{eqnarray}
&& C_{\phi} + 2 C_{\psi} = - \biggl( \frac{A}{2} + \frac{\psi_{0}}{3}\biggr),
\nonumber\\
&& C_{\phi} + 2 C_{\psi} = 2 R_{\nu} D_{\nu} + 2 ( 1 - R_{\nu}) D_{\gamma}.
\label{comp1}
\end{eqnarray}
Equation (\ref{pijlex}) implies, after the use of Eq. (\ref{25R}), 
\begin{equation}
6 C_{\psi} + C_{\phi} + \frac{2}{15} R_{\nu} \phi_0 - \frac{A}{2}=0.
\label{comp2}
\end{equation}
From Eqs. (\ref{nu2}) and (\ref{phot2}) the following algebraic relations can respectively be obtained:
\begin{eqnarray}
&& 3 D_{\nu} - \frac{A}{4} - C_{\phi}  + \sigma_0 =0,
\nonumber\\
&& 3 D_{\gamma} - \frac{A}{4} - C_{\phi}=0.
\label{comp3}
\end{eqnarray}
From Eqs. (\ref{nu1}) and (\ref{phot1})  the relation between the constants is 
\begin{equation}
2 A = 8 C_{\psi} - \frac{2}{3} \phi_0.
\label{comp4}
\end{equation}
Finally, from Eqs. (\ref{nu3}) and (\ref{pineqjex}), we do get
\begin{eqnarray}
&& C_{\psi} - C_{\phi} = \frac{2}{5} R_{\nu} D_{\nu},
\nonumber\\
&& \sigma_{1} = \frac{D_{\nu}}{15}.
\label{comp5}
\end{eqnarray}
The equations for the baryon and CDM velocity fields, i.e. 
Eqs. (\ref{baryon1}) and (\ref{CDM1}), simply imply that 
$4 D_{\rm b, c} = C_{\phi}$ both for baryons and CDM particles.
By solving the algebraic system of equations provided by all 
the conditions obtained so far in Eqs. (\ref{comp1})--(\ref{comp5})
the various constants are determined to be
\footnote{Notice that not all the conditions obtained from Eq. (\ref{comp1}) 
to Eq. (\ref{comp5}) 
are independent.}
\begin{eqnarray}
&& A = - \frac{\left( 525 + 188\,R_{\nu} + 16\,R_{\nu}^2 \right) }{45\,\left( 25 + 2\,R_{\nu} \right) }\phi_{0}
\nonumber\\
&& C_{\phi}   = 4 D_{\rm b,c} =-\frac{\left( 75\, + 14\,\,R_{\nu} - 8\,\,R_{\nu}^2 \right) }{90\,\left( 25 + 2\,R_{\nu} \right) }\phi_0,
\nonumber\\
&& C_{\psi}   = -\frac{\left( 75 + 79\,R_{\nu} + 8\,R_{\nu}^2 \right) }{90\,\left( 25 + 2\,R_{\nu} \right) }\phi_{0},
\nonumber\\
&& D_{\gamma} = -\frac{\left( 25 + 8\,R_{\nu} \right) }{20\,\left( 25 + 2\,R_{\nu} \right) }\phi_{0},
\nonumber\\
&& D_{\nu}    = 15 \sigma_{1} = - \frac{\left( 65 + 16\,R_{\nu} \right) }{36\,\left( 25 + 2\,R_{\nu} \right) }\phi_0,
\label{cond3}
\end{eqnarray}
which leads to the solution 
\begin{eqnarray}
&& \overline{\delta}_{\rm b} = 
\overline{\delta}_{\rm c} = - \frac{3}{2} \phi_{0} - \frac{\left( 525 + 188\,R_{\nu} + 16\,R_{\nu}^2 \right) }{60\,\left( 25 + 2\,R_{\nu} \right) }\phi_{0} k^2 \tau^2,
\label{ad1}\\
&& \overline{\delta}_{\gamma} = \overline{\delta}_{\nu} 
= - 2 \phi_{0} - \frac{\left( 525 + 188\,R_{\nu} + 16\,R_{\nu}^2 \right) }{45\,\left( 25 + 2\,R_{\nu} \right) }\phi_{0} k^2 \tau^2,
\label{solde1}\\
&& \overline{\phi} = \phi_0  -\frac{\left( 75\, + 14\,\,R_{\nu} - 8\,\,R_{\nu}^2 \right) }{90\,\left( 25 + 2\,R_{\nu} \right) }\phi_0 k^2 \tau^2,
\label{solphi1}\\
&& \overline{\psi} 
= \biggl( 1 + \frac{2}{5} R_{\nu} \biggr)\phi_{0}  -\frac{\left( 75 + 79\,R_{\nu} + 8\,R_{\nu}^2 \right) }{90\,\left( 25 + 2\,R_{\nu} \right) }\phi_{0} k^2 \tau^2,
\label{solpsi1}\\
&& \overline{\theta}_{\nu} = \frac{\phi_0}{2}k^2 \tau  - \frac{\left( 65 + 16\,R_{\nu} \right) }{36\,\left( 25 + 2\,R_{\nu} \right) }\phi_0 k^4 \tau^3,
\label{solthnu1}\\
&& \overline{\theta}_{\rm b} = \frac{\phi_{0}}{2} k^2\tau  -\frac{\left( 75\, + 14\,\,R_{\nu} - 8\,\,R_{\nu}^2 \right) }{360\,\left( 25 + 2\,R_{\nu} \right) }\phi_0 k^4 \tau^3,
\label{solthb1}\\
&&\overline{\theta}_{\rm c} = \frac{\phi_{0}}{2} k^2 \tau -\frac{\left( 75\, + 14\,\,R_{\nu} - 8\,\,R_{\nu}^2 \right) }{360\,\left( 25 + 2\,R_{\nu} \right) }\phi_0 k^4 \tau^3,
\label{solthc1}\\
&& \overline{\theta}_{\gamma} =  \frac{\phi_0}{2}k^2 \tau -\frac{\left( 25 + 8\,R_{\nu} \right) }{20\,\left( 25 + 2\,R_{\nu} \right) }\phi_{0} k^2 \tau^2,
\label{solthgam1}\\
&& \overline{\sigma}_{\nu} = \frac{\phi_{0}}{15} k^2 \tau^2  - \frac{\left( 65 + 16\,R_{\nu} \right) }{540\,\left( 25 + 2\,R_{\nu} \right) }\phi_0 k^4 \tau^4,
\label{solsig1}
\end{eqnarray}
This is the expression of the adiabatic mode. Clearly, the adiabatic mode is 
determined,  by fixing the numerical value and spectral dependence of $ \phi_{0}(k)$. Having fixed 
the value of $\delta g_{00}$, the value of $\psi$ follows from the value of the 
anisotropic stress which is proportional to the fractional contribution of the neutrinos, i.e. Eq. (\ref{25R}).
It should also be noticed that, to lowest order, all the velocity field are equal. However, as $ k\tau$ 
approaches $1$ the values of $\theta_{\gamma}$, $\theta_{\nu}$, $\theta_{\rm b}$ and $\theta_{\rm c}$ 
will start becoming different.

\subsection{Synchronous gauge description}
There are examples in the literature of isocurvature modes that are divergent, at early times, 
in the longitudinal gauge, but which are perfectly physical and regular in the 
synchronous coordinate system. Moreover, it is relevant to consider 
the formulation of the problem of initial conditions in the synchronous gauge, since 
various numerical codes solving the Boltzmann hierarchy use, indeed, the synchronous description.

The synchronous gauge condition stipulates that $ \phi =0$ and $B=0$.  The perturbed 
form of the FRW line element is then parametrized in terms of $\psi$ and $E$.  For the problem 
at hand, the parametrization given in Eq. (\ref{synpar}) is customarily  chosen. The relation between 
the two parametrizations is simply, in Fourier space \footnote{Notice that sometimes 
in the literature the variable called $\xi$ is sometimes called $\eta$. This notation 
will be avoided since $\eta$ will be reserved for one of the slow-roll parameters in the 
context of conventional inflationary modes.}, 
\begin{equation}
2 k^2 E = h + 6 \xi,\,\,\,\,\,\,\,\,\,\,\,\,\,\,\,\, \psi = - \xi.
\label{syncrel}
\end{equation}
Inserting  Eq. (\ref{syncrel}) into the perturbed Einstein tensors of Eqs. (\ref{dg00})--(\ref{dg0i}) and 
setting $\phi= B=0$ the fluctuation of the left hand side of Einstein equations can be 
obtained in the synchronous gauge parametrized as in Eq. (\ref{synpar}).

By performing an infinitesimal gauge transformation the 
$\delta g_{00}$ and $\delta g_{0i}$ parts of the metric (which vanish in the synchronous gauge) 
also transform. The gauge modes can be exactly identified by requiring 
that $ \delta g_{00} =0$ and $\delta g_{ij}=0$ in the transformed 
coordinate system, i.e.  for gauge transformations that preserve the synchronous nature 
of the coordinate system.  
The first gauge mode corresponds to a spatial reparametrization of the constant-time hypersurfaces. 
As a consequence, in this mode the metric perturbation is constant and the matter density unperturbed.
The second gauge mode corresponds to a spatially dependent shift in the time direction. 

With the notation of Eqs. (\ref{syncrel}), the perturbed Einstein equations become 
\begin{eqnarray}
&& k^2 \xi - \frac{{\cal H}}{2} h' = \frac{3}{2} {\cal H}^2[ R_{\nu} \delta_{\nu} + ( 1 - R_{\nu}) \delta_{\gamma}  
+ \Omega_{\rm b} \delta_{\rm b} + \Omega_{\rm c} \delta_{\rm c}],
\label{00s}\\
&& k^2 \xi' = - \frac{3}{2} {\cal H}^2 \biggl[  \frac{4}{3} ( R_{\nu} \theta_{\nu} + 
( 1 - R_{\nu}) \theta_{\gamma}) + \Omega_{\rm b} \theta_{\rm b} + \Omega_{\rm c} \theta_{\rm c}\biggr],
\label{0is}\\
&& h'' + 2 {\cal H} h' - 2 k^2 \xi = 3 {\cal H}^2 [ R_{\nu} \delta_{\nu} + ( 1 - R_{\nu}) \delta_{\gamma}],
\label{ijs}\\
&& h'' + 6 \xi'' + 2 {\cal H} h' + 12 {\cal H} \xi' - 2 k^2 \xi = 12 {\cal H}^2  R_{\nu} \sigma_{\nu} ,
\label{shs}
\end{eqnarray}
where the background equations (\ref{b1})--(\ref{b3}) have  already been used to eliminate the energy and pressure 
densities\footnote{In this section the density contrasts 
and the peculiar velocity field are named in the same way as in the longitudinal gauge. 
It is understood that the density 
contrasts and the peculiar velocity fields are {\em not} equal in the two gauges and 
are related by the transformations listed below (see Eqs. (\ref{Tr1}) and (\ref{Tr2})).}. 

By combining appropriately Eqs. (\ref{00s}) and (\ref{ijs}), it is possible to obtain a further useful equation
\begin{equation}
h'' + {\cal H} h' = 3 {\cal H}^2 [ 2 R_{\nu} \delta_{\nu} + 2 ( 1 - R_{\nu}) \delta_{\gamma} + 
\Omega_{\rm b} \delta_{\rm b} + \Omega_{\rm c} \delta_{\rm c}].
\end{equation}
The evolution equations of the peculiar velocities and density contrasts of the various species of the 
plasma can be obtained by perturbing the covariant conservation. The result of this calculation is reported 
in Eqs. (\ref{gencov1}) and (\ref{gencov2}) without any gauge fixing. Hence, setting $\phi = B=0$ in 
Eqs. (\ref{gencov1}) and (\ref{gencov2}) and using Eq. (\ref{syncrel}) the following set of equations 
can be readily obtained:
\begin{eqnarray}
&& \delta_{\nu}' = - \frac{4}{3} \theta_{\nu} + \frac{2}{3} h',
\,\,\,\,\,\,\,\,\,\,\,\,\,\,\,\,\,\,\,\delta_{\gamma}' = - \frac{4}{3} \theta_{\gamma} + \frac{2}{3} h', 
\label{dgs}\\
&& \delta_{\rm b}' = - \theta_{\rm b} + \frac{1}{2} h' ,\,\,\,\,\,\,\,\,\,\,\,\,\,\,\,\,\,\,\,\,\,\,\,\,
\delta_{\rm c}' = - \theta_{\rm c} + \frac{h'}{2},
\label{dcs}\\
&& \theta_{\nu}' = -k^2 \sigma_{\nu} + \frac{k^2}{4} \delta_{\nu},\,\,\,\,\,\,\,\,\,\,\,\,\,\,\,\,\,
 \theta_{\gamma}' = \frac{k^2}{4} \delta_{\gamma},
\label{thgs}\\
&& \theta_{\rm b}' = - {\cal H} \theta_{\rm b},\,\,\,\,\,\,\,\,\,\,\,\,\,\,\,\,\,\,\,\,\,\,\,\,\,\,\,\,\,\,\,\,\,\,\,
\theta_{\rm c}' = - {\cal H} \theta_{\rm c}
\label{thcs}\\
&& \sigma_{\nu}' = \frac{4}{15} \theta_{\nu} - \frac{3}{10} k {\cal F}_{\nu 3} + \frac{2}{15} h' + \frac{4}{5} \xi'.
\label{signus}
\end{eqnarray}
As anticipated the synchronous gauge modes can be made harmless by
 eliminating the constant solution 
for $h$ and by fixing, for instance, the CDM velocity field to  zero. These two requirements  specify 
the coordinate system completely.

The adiabatic solution can be also obtained in the synchronous gauge, and in praticular, 
\begin{eqnarray}
&& \xi = - 2 C + \biggl[ \frac{5 + 4 R_{\nu}}{6( 15 + 4 R_{\nu})} C \biggr]k^2 \tau^2,
\label{xisol1}\\
&& h = - C k^2 \tau^2.
\label{hsol1}
\end{eqnarray}
The constant 
introduced in Eqs. (\ref{xisol1}) and (\ref{hsol1}) has been defined in order 
to match the standard notation usually employed in the literature (see for instance \cite{MB}) 
to characterize the adiabatic (inflationary) mode.
The solution  for the density contrasts will be 
\begin{eqnarray}
&& \delta_{\gamma} = \delta_{\nu} =- \frac{2}{3} C  k^2 \tau^2,
\label{dnusols1}\\
&& \delta_{\rm b} =  \delta_{\rm c} - \frac{C}{2} k^2 \tau^2.
\label{dbsols1}
\end{eqnarray}
Setting, as explained, $\theta_{\rm c}=0$, the peculiar velocity satisfy instead:
\begin{eqnarray}
&& \theta_{\gamma} = - \frac{C}{18} ,\,\,\,\,\,\,\,\,\,\,\,\,\,\,\, \theta_{\nu} = - \frac{23 + 4 R_{\nu}}{15 + 4 R_{\nu}}.
\label{thnusols1}
\end{eqnarray}
Any solution obtained in the synchronous coordinate system can be studied
in the conformally Newtonian gauge by using the appropriate transformation, i. e.
\begin{eqnarray}
&&  \phi_{\rm L} = - \frac{1}{2 k^2} [ ( 6 \xi + h)'' + {\cal H} ( 6\xi + h)'],
\nonumber\\
&& \psi_{\rm L} = - \xi + \frac{{\cal H}}{2 k^2} ( 6 \xi' + h'),
\nonumber\\
&& \delta^{(\lambda)}_{\rm L} = \delta^{(\lambda)}_{\rm s} + 
3 ( w_{\lambda} + 1) \frac{{\cal H}}{2 k^2} ( h' + 6 \xi'), 
\nonumber\\
&& \theta^{(\lambda)}_{\rm L} = \theta^{(\lambda)}_{\rm s} - \frac{1}{2} ( h' + 6 \xi'), 
\label{Tr1}
\end{eqnarray}
where the subscripts refer to the quantities evaluated either in the longitudinal or in the synchronous gauge.
Using Eq. (\ref{Tr1}), from Eqs. (\ref{xisol1})--(\ref{hsol1}) and (\ref{dnusols1}) --(\ref{thnusols1}),
we recover the standard adiabatic mode  with the constant $C$ determined as  
\begin{equation}
C = \frac{15 + 4 R_{\nu}}{20} \phi_{0},
\end{equation}
where $\phi_{0}$ is the value of the adiabatic mode in the longitudinal gauge discussed previously.
Conversely, the solutions of the longitudinal gauge can also be directly 
transformed into the synchronous gauge 
using the appropriate transformation, which can be easily derived:
\begin{eqnarray}
&& \xi = - \psi_{L} - \frac{{\cal H}}{a} \int a \phi_{L} d\tau,
\nonumber\\
&& h = 6 \psi_{L} + 6 \frac{{\cal H}}{a} \int a \phi_{L} d\tau, 
- 2 k^2 \int \frac{d\tau''}{a(\tau'')} \int^{\tau''} a(\tau') \phi_{L}(\tau') d\tau', 
\nonumber\\
&& \delta^{(\lambda)}_{s} = \delta^{(\lambda)}_{L} + 3 {\cal H} ( w + 1)\frac{1}{a} \int \phi_{L} a d\tau,
\nonumber\\
&& \theta^{(\lambda)}_{s} = \theta^{(\lambda)}_{L} - \frac{k^2}{a} \int a \phi_{L} d\tau.
\label{Tr2}
\end{eqnarray}

\subsubsection{Isocurvature modes}
It is useful to recall that, in looking for perturbative solutions (both in the 
longitudinal and in the synchronous gauge),
there are various small parameters. One is certainly $k\tau$, which is small outside the horizon. However, 
also $\Omega_{\rm b}$ and $\Omega_{\rm c}$ are small parameters deep within the 
radiation-dominated epoch. Let us make this statement more precise by considering the 
scale factor 
\begin{equation}
a(\tau) = \biggl[ \biggl(\frac{\tau}{\tau_{ 1}}\biggl) + \biggl(\frac{\tau}{\tau_{1}}\biggl)^2 \biggr],
\label{interpolation}
\end{equation}
interpolating between the radiation-dominated phase for $ \tau 
\ll  \tau_{1}$ and the matter dominated  epoch for $ \tau  \gg \tau_{1}$.
In this case we can also write 
\begin{equation}
\Omega_{\rm b,c} = \overline{\Omega}_{\rm b,c} \frac{a(\tau) }{a(\tau) + 1},
\label{ombar1}
\end{equation}
where the subscripts in Eq. (\ref{ombar1}) refer either to baryons or to CDM and where we took, for simplicity, 
$\tau_{1} =1$.

Eqs. (\ref{interpolation}) and (\ref{ombar1})  can be inserted into the evolution equations 
for the perturbations in the synchronous gauge, regarding $\Omega_{\rm b, c}$ as small paramters
deep within the radiation-dominated epoch ($\tau \to 0$). The result of this procedure is summarized by the 
following solutions:
\begin{eqnarray}
h \simeq ( - 4 \overline{\Omega}_{\rm b} \tau + 6 \overline{\Omega}_{\rm b} \tau^2) ,\,\,\,\,\,\,\,\,\,\,\,\,\,\,\,\,\,\,\,\,\,\,
 \xi \simeq \frac{2}{3} \overline{\Omega}_{\rm b} \tau- \overline{\Omega}_{\rm b} \tau^2, 
\label{isoxih}
\end{eqnarray}
for the metric perturbations and 
\begin{eqnarray}
&& \delta_{\gamma} \simeq \biggl( - \frac{8}{3} \overline{\Omega}_{\rm b} \tau + 4 \overline{\Omega}_{\rm b} ^2 \biggr), \,\,\,\,\,\,\,\,\,\,\,\,\,\,\,\,\,\,
\delta_{\rm b} \simeq  ( 1 - 2 \overline{\Omega}_{\rm b} \tau+ 3 \overline{\Omega}_{\rm b} \tau^2),
\label{isodbgam}\\
&& \delta_{\nu} \simeq \biggl( - \frac{8}{3} \overline{\Omega}_{\rm b}  \tau+ 4 \overline{\Omega}_{\rm b} 
\tau^2 \biggr),\,\,\,\,\,\,\,\,\,\,\,\,\,
 \delta_{\rm c} \simeq  2 \overline{\Omega}_{\rm b} \tau+ 3 \overline{\Omega}_{\rm b} \tau^2,
\label{isodcnu}\\
&& \theta_{\gamma} \simeq -\frac{1}{3} \overline{\Omega}_{\rm b} k^2\tau^2 ,\,\,\,\,\,\,\,\,\,
\,\,\,\,\,\,\,\,\,\,\,\,\,\,\,\,\,\,\,\,\,\,\,\,\,\,
\theta_{\nu} \simeq -\frac{1}{3} \overline{\Omega}_{\rm b} k^2\tau^2 ,
\label{isonu}\\
&& \theta_{\rm c} =0,\,\,\,\,\,\,\,\,\,\,\,\,\,\,\,\,\,\,\,\,\,\,\,\,\,\,\,\,\,\,\,\,\,\,\,\,\,\,\,\,\,\,\,\,\,\,\,\,\,\,\,\,\,\,\,\,\,\,\,\,
 \sigma_{\nu} \simeq - \frac{2}{3} \frac{\overline{\Omega}_{\rm b}}{2 R_{\nu} + 15} k^2 \tau^3 .
\label{isosigma}
\end{eqnarray}
for the fluid quantities. Clearly, for this mode, the adiabaticity 
condition is not satisfied. Furthermore, by transforming 
the solution to the Newtonian gauge, it is easy to check that the longitudinal 
fluctuations of the metric vanish for $\tau \to 0$.  This is 
 the baryon isocurvature mode already discussed in \cite{turok1}. 
A similar solution can 
be obtained by changing 
$\overline{\Omega}_{\rm b} \to \overline{\Omega}_{\rm c}$, i.e. 
the CDM isocurvature mode. The baryon isocurvature mode has been 
studied in detail in Refs. \cite{Peebles1,Peebles2, Peebles3,Peebles4}.

On top of the baryon and CDM isocurvature modes, there are other two modes which 
are physical and which may parametrize, together with the modes 
discussed so far, the most general initial conditions 
for the CMB anisotropies. They are the neutrino isocurvature density mode and the 
neutrino isocurvature velocity mode. These modes were analyzed in \cite{turokiso} 
and partially introduced in \cite{challinoriso}. Following the same procedure described 
in the case of the baryon and CDM density modes it can be shown that, for the 
neutrino isocurvature density mode,  the metric inhomogeneities are given by 
\begin{equation}
h \simeq - \frac{\overline{\Omega}_{\rm b} }{10} \frac{R_{\nu}}{R_{\gamma} } k^2 \tau^3, 
\,\,\,\,\,\,\,\,\,\,\,\, \xi \simeq \frac{R_{\nu}}{6 ( 4 R_{\nu} + 15)} k^2 \tau^2,
\end{equation}
while, setting $\theta_{\rm c}=0$, the fluid quantities are determined to be 
\begin{eqnarray}
&& \delta_{\rm c} \simeq  - \frac{\overline{\Omega}_{\rm b}}{20} \frac{R_{\nu}}{R_{\gamma} } k^2 \tau^3,
\,\,\,\,\,\,\,\,\,\,\,\,\,\,\, \delta_{\rm b} \simeq \frac{1}{8} \frac{R_{\nu}}{R_{\gamma}} k^2 \tau^2, 
\\
&&  \delta_{\nu} \simeq 1 - \frac{k^2 \tau^2}{6}, \,\,\,\,\,\,\,\,\,\,\,\,\,\,\,\,\,\,\, \delta_{\gamma} \simeq - \frac{R_{\nu}}{R_{\gamma}} + \frac{1}{6} \frac{R_{\nu}}{R_{\gamma}} k^2 \tau^2,
\\
&& \theta_{\gamma}  \simeq \theta_{\rm b} \simeq - \frac{1}{4} \frac{R_{\nu}}{R_{\gamma}} k^2 \tau 
+ \frac{3}{4} \overline{\Omega}_{\rm b} \frac{R_{\nu}}{R_{\gamma}^2}  k^2 \tau^2,
\\
&& \sigma_{\nu} \simeq \frac{k^2 \tau^2}{2 ( 15  + 4 R_{\nu})}, \,\,\,\,\,\,\,\,\,\,\,\,\,\,\,  \theta_{\nu} \simeq 
\frac{k^2 \tau}{4}.
\label{nidm}
\end{eqnarray}
From the solution of Eq. (\ref{nidm})  we can see that $R_{\nu} \delta_{\nu} + R_{\gamma} \delta_{\gamma}\simeq 0$.
So, the initial conditions are such that the total energy density is essentially unperturbed. However, as the mode 
enters the horizon, the neutrinos free-stream  while photons are tightly coupled with baryons.

In the case of the neutrino isocurvature velocity mode, the solution for the metric inhomogeneities 
can be written as 
\begin{eqnarray}
&& h \simeq -\frac{3}{2} \overline{\Omega}_{\rm b} \frac{R_{\nu}}{R_{\gamma}} k \tau^2,
\\
&& \xi  \simeq \frac{ 4 R_{\nu}}{3 ( 5 + 4 R_{\nu})} k\tau  + \biggl( \overline{\Omega}_{\rm b} \frac{R_{\nu}}{R_{\gamma}} - \frac{ 20 R_{\nu}}{( 5 + 4 R_{\nu}) ( 15 + 4 R_{\nu})}\biggr) k\tau^2.
\label{metrivm}
\end{eqnarray}
The fluid variables will instead be given by:
\begin{eqnarray}
&& \delta_{\rm c} \simeq - \frac{3}{4} \overline{\Omega}_{\rm b} \frac{R_{\nu}}{R_{\gamma}} k\tau^2, 
\,\,\,\,\,\,\,\,\,\,\,\,\,\,\, \,\,\,\,\,\,\,\,\delta_{\rm b} \simeq \frac{R_{\nu}}{R_{\gamma}} k\tau - 3 \overline{\Omega}_{\rm b} 
\frac{R_{\nu} ( R_{\gamma} + 2 )}{4 R_{\gamma}^2} k\tau^2,
\\
&& \delta_{\nu} \simeq - \frac{4}{3} k\tau - \overline{\Omega}_{\rm b} \frac{R_{\nu}}{R_{\gamma}} k\tau^2 
 \,\,\,\,\,\,\,\,\,\,\,\,\,\,\, \delta_{\gamma} \simeq \frac{4}{3} \frac{R_{\nu}}{R_{\gamma}} k\tau - 
 \overline{\Omega}_{\rm b} \frac{ R_{\nu} (R_{\gamma} + 2)}{R_{\gamma}^2} k\tau^2,\,\,\,\,\,\,\,\,
 \\
&&  \theta_{\gamma} \simeq  \theta_{\rm b} \simeq - \frac{R_{\nu}}{R_{\gamma}}k 
 + 3 \overline{\Omega}_{\rm b} \frac{R_{\nu}}{R_{\gamma}^2}  k\tau +3 \frac{R_{\nu}}{R_{\gamma}^3} \overline{\Omega}_{\rm b}( 3 R_{\gamma} - 3 \overline{\Omega}_{\rm b} ) k\tau^2  + \frac{R_{\nu}}{6 R_{\gamma}} 
 k^3 \tau^2 ,
 \\
&& \theta_{\nu} \simeq k - \frac{ ( 9 + 4 R_{\nu})}{6( 5 + 4 R_{\nu})} k^3 \tau^2,
 \\
 &&\sigma_{\nu} \simeq \frac{ 4}{ 3 ( 4 R_{\nu} + 5)} k\tau 
 + \frac{16 R_{\nu}}{(5 + 4 R_{\nu}) (15 + 4 R_{\nu})} k\tau^2,
 \\
 &&{\cal F}_{\nu 3} \simeq \frac{4}{7 (5 + 4 R_{\nu})} k^2 \tau^2.
 \end{eqnarray}
 As in the previous examples of non-adiabatic modes, the CDM peculiar velocities are set to zero (i.e. 
 $\theta_{\rm c}=0$) to fix completely the coordinate system.
 
 In the longitudinal coordinate system the neutrino isocurvature velocity mode 
 leads to divergent perturbations in the limit $k \tau \to 0$. This point can be easily demonstrated 
 either by solving directly the equations in the longitudinal gauge or by transforming 
 the metric fluctuations of the neutrino isocurvature velocity mode 
 from the synchronous to the longitudinal gauge. Inserting then the two 
 expression of Eq. (\ref{metrivm}) at the right hand side of the first two 
 transformation rules given in Eq. (\ref{Tr1}) it can be easily obtained that 
 \begin{equation}
 \psi_{\rm L} = -\phi_{\rm L} = \frac{1}{|k \tau|} \frac{  4 R_{\nu}}{(  5 + 4 R_{\nu})},
 \label{divlong1}
 \end{equation}
 which diverge in the limit $ k \tau \to 0$.
 The fact that metric fluctuations diverge in a specific description should 
 not be regarded as a physical problem but rather as a mathematical 
 difficulty.  We shall get back to this point later on in  section 6.
 
 The possibility of having a mixture of modes (adiabatic and isocurvature)
 as initial condition for the CMB anisotropies led to a number of mode-independent 
 analysis of various sets of data. 
 In \cite{hannu2} it was argued, on the basis of Maxima data, that mixture 
 of isocurvature and adiabatic modes could not be excluded and in \cite{hannu1} 
 Enqvist and Kurki-Suonio pointed out that the Planck could detect 
 isocurvature modes thanks to its foreseen sensitivity to polarization. 
 CMB polarization is basically induced by scattering processes
 (either last Thompson scattering or late reionization scattering). This 
 occurrence allows to eliminate possible contamination with line-of-sight (integrated) effects and 
 makes polarization a valuable tool in order to constrain isocurvature modes.
 In \cite{hannu3}  CDM isocurvature modes in open and closed 
 FRW  backgrounds have also been considered.
 
 In \cite{turok1,turok2} the importance of polarization was also stressed in view 
 of the Planck planned sensitivities.   A general (model independent) 
 analysis of the recent WMAP data (combined with other sets of data including 2dF galaxy redhift
 survey) was recently performed in \cite{moodley1,moodley2}. 
 The results can be roughly summarized by saying that the isocurvature 
 fraction allowed by the present data ranges from about $10\,\%$ (when only one isocurvature mode is 
 present on top of the adiabatic mode) up to $40\,\%$ (when two isocurvature modes are allowed 
 together with their cross-correlations). Finally the fraction of isocurvature modes rises 
 to about $60\,\%$ when three isocurvature modes are allowed. Notice, as a remark, that 
 in \cite{moodley2} the authors indeed allowed for general correlations between 
 adiabatic and isocurvature modes while the power spectrum was parametrized 
 by means of a power-law. From the analyses reported so far it seems 
 that the constraints on isocurvature modes are rather dependent 
 upon the possible correlations with the adiabatic modes. 
 In \cite{jussi}, following the analysis of Ref. \cite{hannu2}, the possibility 
 of (correlated) CDM and adiabatic modes has been discussed in light 
 of WMAP and large-scale structure data. Different values of the spectral indices 
 for the power of each mode (and of their correlation) have been scrutinized. 
As a consequence, the bounds reported in \cite{hannu2} became more 
stringent by, roughly, a factor of $1/6$.

 One can also address the question on how some correlated mixture of adiabatic and 
 isocurvature modes may arise in the early Universe. This subject 
 will be also partially addressed in section 6 where 
some general ideas on how isocurvature modes may be 
 excited in the early Universe will be introduced. As an example, 
 consider Ref. \cite{gordonlw} where a model dependent analysis 
 has been performed. In the context of minimal curvaton models (see 
 section 5) the authors found that correlated isocurvature modes seems to 
 be disfavoured with respect to pure adiabatic modes. However, the presence 
 of a (totally correlated) baryon isocurvature mode does not seem to be 
 ruled out (see also \cite{gordonmal,ferrer}).
 
 Various sorts of different initial conditions may also be phenomenologically 
 constrained. In the following we are going to list some possible interesting 
 departure from the generalized (model-independent) analysis of the 
 initial conditions. In the present discussion the dark energy 
 has been always parametrized in terms of a cosmological constant. However, if the 
 quintessence fluid is described in terms of one (or more) scalar degrees of freedom, 
 then the picture may change. Recalling the form 
 of the integrated SW effect, the presence of a quintessence field may 
 induce a time evolution in the longitudinal degrees of freedom of the metric 
 inducing an important integrated contribution (see for instance \cite{luca1}). 
 In \cite{mortak} (see also \cite{gordonhu})  it was argued that 
 if the primordial fluctuation of the quintessence has a correlation with the adiabatic density fluctuations, 
 the large-scale temperature fluctuations may be suppressed. This may have implications 
 for the so-called WMAP anomaly for the low multipoles of the CMB angular power spectrum.
 
 Other possible modifications on the initial conditions include the presence of decaying 
 modes \cite{luca2} as well as the presence of other physical modifications such 
 as a primordial magnetic field \cite{giomagn}. Notice that in the latter 
 case, not only vector and tensor modes are excited \cite{lewisvector1,lewisvector2}
 but also scalar modes. In this case all the solutions described up to now (i.e. 
 the adiabatic mode and the four isocurvature modes) are modified in a 
 computable way \cite{giomagn}. 
 
 \subsection{Simplified views on the problem of initial conditions}

For numerical purposes 
it is safer to set initial conditions in the radiation dominated phase and integrate
across the radiation-matter transition by using, for instance, an interpolating 
solution of the background Einstein equations (\ref{b1})--(\ref{b3}) (like 
the one discussed in Eq. (\ref{interpolation}).   However, since 
 the decoupling time follows the equality time and last scattering 
 occurs when the Universe is dominated by matter, it is also possible to 
 study directly the problem of initial conditions during the matter epoch.
 After introducing some general analytic considerations 
 on the problem of initial conditions during the matter-dominated epoch 
 the adiabatic and isocurvature contributions to the ordinary SW effect 
 will be computed.

The fluctuation in the total pressure density can be connected to the 
total fluctuation of the energy density as 
\begin{equation}
\delta p= c_{\rm s}^2 \delta \rho + \delta p_{\rm nad},
\label{dpnad}
\end{equation}
where 
\begin{equation}
c_{\rm s}^2 = \biggl(\frac{\delta p}{\delta\rho}\biggr)_{\varsigma} = 
\biggl(\frac{p'}{\rho'}\biggr)_{\varsigma},
\label{cs2}
\end{equation}
is the speed of sound computed from the variation 
of the total pressure and energy density at constant specific entropy, i.e. 
$\delta \varsigma =0$. The second term appearing 
in Eq. (\ref{dpnad}) is the pressure density variation 
produced by the fluctuation in the specific entropy 
at  constant energy density, i.e.
\begin{equation}
\delta p_{\rm nad} = 
\biggl( \frac{\delta p}{\delta \varsigma }\biggr)_{\rho} \delta \varsigma,
\label{deltapnaddef1}
\end{equation}
accounting for the 
non-adiabatic contribution to the total pressure perturbation.
 
If only one species is present with equation of state 
$p = w \rho$, then it follows from the 
definition that $ c_{s}^2 =w$ and the 
non-adiabatic contribution vanishes. 
As previously anticipated around Eq. (\ref{SE}), a 
 sufficient condition in order to have $\delta p_{\rm nad} \neq 0$
is that the fluctuation in the specific entropy $\delta \varsigma$ 
is not vanishing. Consider, for simplicity, the case of a plasma 
made of radiation and CDM particles. In this case the speed of 
sound and the non-adiabatic contribution can be easily computed 
and they are:
\begin{eqnarray}
&& c_{\rm s}^2 = \frac{p'}{\rho'} =
 \frac{p_{\gamma}' + p_{\rm c}'}{\rho_{\gamma}' + \rho_{\rm c}'} \equiv 
\frac{4}{3} \biggl( \frac{ \rho_{\gamma}}{3 \rho_{\rm c} + 4 \rho_{\gamma}} \biggr),
\label{cs2ex}\\
&& \varsigma \biggl( \frac{\delta p}{\delta \varsigma}\biggr)_{\rho} = 
\frac{4}{3} \frac{\delta \rho_{\gamma}}{3 \frac{\delta\rho_
{\gamma}}{\rho_{\gamma}} - 4 \frac{\delta\rho_{\rm c}}{\rho_{\rm c}}} \equiv 
\frac{4}{3} \biggl(\frac{ \rho_{\rm c}\rho_{\gamma}}{ 3 
\rho_{\rm c} + 4 \rho_{\gamma}} \biggr) \equiv \rho_{\rm c} c_{\rm s}^2.
\label{entrex} 
\end{eqnarray}
To obtain the final expression appearing at the right-hand-side 
of Eq. (\ref{cs2ex}) the conservation equations for the two species 
(i.e. $\rho_{\gamma}' = - 4 {\cal H} \rho_{\gamma}$ and $\rho_{\rm c}'  = 
- 3 {\cal H} \rho_{\rm c}$) have been 
used.  In Eq. (\ref{entrex}), the first equality follows from the fluctuation 
of the specific entropy computed in Eq. (\ref{SE}); the second 
equality appearing in Eq. (\ref{entrex}) follows from the observation that 
the  increment of the pressure should be computed for constant (total) 
energy density, i.e. $\delta \rho = \delta \rho_{\gamma} + 
\delta \rho_{\rm c} =0$, implying 
$\delta\rho_{\rm c} = - \delta \rho_{\gamma}$.  The third equality appearing
in Eq. (\ref{entrex}) is a mere consequence of the explicit 
expression of $c_{\rm s}^2$ obtained in Eq. (\ref{cs2ex}). 

As in the case of Eq. (\ref{SEgen}), the analysis presented up to now 
can be easily generalized to a mixture of fluids $``{\rm a}"$ and $``{\rm b}"$ with barotropic 
indices $w_{\rm a}$ and $w_{\rm b}$ \cite{KS}.
The generalized speed of sound  is then  given by 
\begin{equation}
c_{\rm s}^2 = \frac{w_{\rm a} (w_{\rm a} + 1) \rho_{\rm a} + w_{\rm b} ( w_{\rm b} + 1) \rho_{\rm b}}{ 
(w_{\rm a } + 1 )\rho_{\rm a} + (w_{\rm b} + 1) \rho_{\rm b}}.
\end{equation}

From Eq. (\ref{cs2ex}) and from the definition of the (total) barotropic index it  
follows that, for a CDM-radiation fluid
\begin{eqnarray}
&& c_{\rm s}^2 = \frac{4}{3} \frac{1}{3 a + 4},
\label{cs2int}\\
&& w = \frac{1}{3} \frac{1}{1 + a}.
\label{wint}
\end{eqnarray}
Using Eqs. (\ref{deltapnaddef1})--(\ref{entrex}) the non-adiabatic contribution to the 
total pressure fluctuation becomes
\begin{equation}
\delta p_{\rm nad} = \frac{4}{3} \rho_{\rm c}  \frac{{\cal S}}{3 a + 4},
\label{nadcdm2}
\end{equation}
where the definition given in Eq. (\ref{SE}), i.e. ${\cal S} = (\delta \varsigma)/\varsigma$, has been used.

Using the splitting of the total pressure density fluctuation into a adiabatic 
and a non-adiabatic parts, Eq. (\ref{p00l}) can be multiplied by a 
factor $c_{\rm s}^2$ and subtracted from Eq. (\ref{B1}). The result 
of this operation leads to a  formally simple expression
for  the evolution of curvature fluctuations in the longitudinal 
gauge, namely:
\begin{equation}
\psi'' + {\cal H}[ \phi' + ( 2 + 3 c_{\rm s}^2 ) \psi'] + 
[ {\cal H}^2 ( 1 + 2 c_{\rm s}^2) + 2 {\cal H}'] \phi 
- c_{\rm s}^2 \nabla^2 \phi + \frac{1}{3}\nabla^2( \phi - \psi) =
4\pi G a^2 \delta p_{\rm nad},
\label{B5}
\end{equation}
which  is indipendent of the specific form of $\delta p_{\rm nad}$. 
The left hand side of Eq. (\ref{B5}) can be written as 
the (conformal) time derivative of a single scalar function whose specific form is,   
\begin{equation}
{\cal R} = - \biggl( \psi+ 
\frac{{\cal H} ( \psi' + {\cal H} \phi)}{{\cal H}^2 
- {\cal H}'} \biggr),
\label{defR2}
\end{equation}
Taking now  the first (conformal) 
time  derivative of ${\cal R}$ as expressed by 
Eq. (\ref{defR2}) and using the definition of $c_{\rm s}^2$ we arrive at the 
following expression 
\begin{equation}
{\cal R}' = -\frac{ {\cal H}}{ 4 \pi G a^2 ( \rho + p)}\{ \psi'' 
+ {\cal H}[ ( 2 + 3 c_{\rm s}^2) \psi' + \phi'] + 
[ 2 {\cal H}' + ( 3 c_{\rm s}^2 + 1 ) {\cal H}^2] \phi \}.
\label{derR}
\end{equation}
Comparing now Eqs. (\ref{derR}) and (\ref{B5}), it is clear that Eq. 
(\ref{derR}) reproduces Eq. (\ref{B5}) but only up to the spatial 
gradients. Hence, using Eq. (\ref{derR}) into Eq. (\ref{B5}) the following 
final expression can be obtained:
\begin{equation}
{\cal R}' = - \frac{{\cal H}}{p + \rho} \delta p_{\rm nad} + 
\frac{k^2 {\cal H}}{ 4 \pi G a^2 ( p + \rho)} \biggl[ \biggl( c_{\rm s}^2 - 
\frac{1}{3} \biggr) \phi+ \frac{1}{3} \psi \biggr].
\label{evolR}
\end{equation}
 There  are slightly different ways to express the terms at the 
right hand side of Eq. (\ref{evolR}).
For instance the term $( \phi - \psi)$ could be replaced 
by the anisotropic stress $\sigma$. However, the physical content of 
the equation remains unchanged.   The specific geometrical meaning 
of ${\cal R}$ will be discussed in detail in section 5 and 
Eq. (\ref{evolR}) should be regarded, for the moment, as a useful mathematical 
simplification allowing, in some cases, a swifter reduction 
to quadratures of the evolution of $\psi$ for $k\tau \ll 1$.

 Disregarding the complication 
 of an anisotropic stress (i.e., from Eq. (\ref{B3}), $\phi =\psi$)
from Eqs. (\ref{delta})--(\ref{theta}) the covariant conservation equations become
 \begin{eqnarray}
 && \delta_{\rm c}' = 3 \psi' - \theta_{\rm c},
 \label{Cdelc}\\
 && \theta_{\rm c}' = - {\cal H} \theta_{\rm c} + k^2 \phi,
 \label{Cthc}\\
 && \delta_{\rm r}' = 4 \psi' - \frac{4}{3} \theta_{\rm r},
 \label{Cdelr}\\
 && \theta_{\rm r}' = \frac{k^2}{4} \delta_{\rm r} + k^2 \phi.
 \label{Cthr}
 \end{eqnarray}
 
 \subsubsection{Adiabatic contribution to the SW effect}
 
Since in the adiabatic case $\delta p_{\rm nad} =0$, Eq. (\ref{B5}) becomes 
 \begin{equation}
  \psi'' + 4 {\cal H} \psi' + \frac{k^2}{3} \psi = 0,
  \label{psirad}
  \end{equation}
  where it has been used that $c_{\rm s}^2 = 1/3$.
  
  Since during radiation $a(\tau)\sim \tau$, Eq. (\ref{psirad}) can be solved as a combination 
  of Bessel functions of order $3/2$ which can be expressed, in turn, as a combination 
  of trigonometric functions weighted by inverse powers of their argument
  \begin{equation}
  \psi(k,\tau) = A_{1}(k) \frac{ y \cos{y} - \sin{y}}{y^3} + B_{1}(k) \frac{y \sin{y}  + \cos{y}}{y^3},
  \label{psiradsol}
  \end{equation}
  where $ y = k\tau/\sqrt{3}$.
 If the solution parametrized by the arbitrary constant $A_{1}(k)$, $\psi\to \psi_{\rm r}$ for $k\tau \ll 1$
 ( where $\psi_{\rm r}$ is a constant). This is the case of purely adiabatic initial conditions. If, on the 
 contrary, $A_{1}(k)$ is set to zero, then $\psi$ will not go to a constant. This second solution is important 
 in the case of the non-adiabatic modes and will be discussed later. 

 In the case of adiabatic fluctuations the constant mode $\psi_{\rm r}$ matches to a constant mode 
 during the subsequent matter dominated epoch. In fact, during the matter dominated epoch 
 and under the same assumptions of absence of anisotropic stresses Eq. (\ref{B5}), in the case $c_{\rm s}^2 =0$,
 leads to the following simple equations 
 \begin{equation}
 \psi'' + 3 {\cal H} \psi' =0,
 \label{psimat}
 \end{equation}
 Since, after equality, $a(\tau)\sim \tau^2$, the solution of Eq. (\ref{psimat}) is then 
 \begin{equation}
 \psi(k,\tau) = \psi_{\rm m} + D_{1}(k) \biggl(\frac{\tau_{\rm eq}}{\tau} \biggr)^{5},
 \label{psimatsol}
 \end{equation}
 where $\psi_{\rm m}$ is a constant. Typilcally the term multiplying $D_{1}(k)$ is 
 of the order of $10^{-4}$ at $\tau = \tau_{\rm dec}$. The values 
 of $\psi_{\rm r}$ and $\psi_{\rm m}$ are different but can be easily connected. In fact 
 we are interested in wave-numbers $k\tau < 1$ after equality. In this limit Eq. 
 (\ref{evolR}) implies that ${\cal R}'=0$. More specifically, if the background is dominated by a 
 perfect fluid with barotropic index $w$, Eq. (\ref{defR2}) leads to 
 \begin{equation}
 {\cal R}= - \frac{ 5 + 3 w}{3 ( 1 + w)} \psi,
 \label{RCONST} 
 \end{equation}
 where Eqs. (\ref{b1})--(\ref{b3}) have been used.
 
 From Eq. (\ref{RCONST}) we obtain that, during radiation (i.e. $ w =1/3$), 
 ${\cal R}_{\rm r} = -(3/2) \psi_{\rm r}$, while during matter (i.e. $w =0$) 
 we have ${\cal R}_{\rm m} = -(5/3) \psi_{\rm m}$. Since, by virtue of Eq. (\ref{evolR}), 
  ${\cal R}$  is conserved in the limit $k\tau \ll 1$, then ${\cal R}_{\rm m} = {\cal R}_{\rm r}$, 
  implying 
 \begin{equation}
 \psi_{\rm m} = \frac{9}{10} \psi_{\rm r}.
 \label{psimpsir}
 \end{equation}
 
 The same result can be also obtained by integrating, from  Eq. (\ref{defR2}), the evolution of 
 $\psi$  directly in terms of the scale factor. From Eq. (\ref{defR2}), changing 
 variable from $\tau$ to $a$  the equation to be solved will be 
 \begin{equation}
 \frac{d\psi}{d a} + \frac{5 a + 6}{2 a ( a + 1)} \psi = \psi_{\rm r} \frac{3 ( 3 a + 4)}{4 a ( a + 1)},
 \label{diffps1}
 \end{equation}
 whose integration, after imposing adiabatic initial conditions 
 for $a\to 0$ ( deep within the radiation-dominated phase),  leads to 
 \begin{equation}
 \psi(a) = \frac{\psi_{\rm r}}{10 a^3 }\{16 ( \sqrt{ a + 1} -1)  + a [ a ( 9 a + 2) -8]\};
 \label{solPSI1}
 \end{equation}
the  limit for $a\to \infty$ (matter-dominated phase)  of the right hand side of Eq. (\ref{solPSI1}) 
leads to $(9/10)\psi_{\rm r}$.

 Combining Eqs. (\ref{Cdelr}) and (\ref{Cthr})  
in the presence of the constant adiabatic mode $\psi_{\rm m}$ and during the 
 matter-dominated phase  
 \begin{equation}
 \delta_{\rm r}'' + k^2 c_{\rm s}^2 \delta_{\rm r} = - 4 c_{\rm s}^2 k^2 \psi_{\rm m}
 \end{equation}
 where  $ c_{\rm s} = 1/\sqrt{3}$. The full solution of Eqs. (\ref{Cthc})--(\ref{Cthr})  will then be 
 \begin{eqnarray}
 && \delta_{\rm c} = - 2 \psi_{\rm m} - \frac{\psi_{\rm m}}{6} k^2 \tau^2 
 \label{deltam}\\
 && \theta_{\rm c} = \frac{k^2 \tau}{3} \psi_{\rm m},
 \label{thetam}\\
 && \delta_{\rm r} = \frac{4}{3} \psi_{\rm m} [ \cos{(k c_{\rm s}\tau)} - 3],
 \label{deltar}\\
 && \theta_{\rm r} = \frac{k \psi_{\rm m}}{\sqrt{3}} \sin{(k c_{\rm s} \tau)}.
 \label{thetar}
 \end{eqnarray}
 As in the case of the adiabatic mode 
 during the radiation-dominated epoch, also in the matter epoch, for 
 $k\tau \ll 1$, $\theta_{\rm c} \simeq \theta_{\rm r}$.
 
 The ordinary SW effect can now be roughly estimated.  Consider Eq. (\ref{SW0}) in the case 
 of the pure adiabatic mode. Since the longitudinal degrees of freedom of the metric are roughly constant
 and equal, inserting the solution of Eqs. (\ref{deltam})--(\ref{thetar}) into  
 Eq. (\ref{SW0}) the following result can be obtained 
  \begin{equation}
 \biggl( \frac{\Delta T}{T}\biggr)^{\rm ad}_{k, {\rm s}} = 
 \biggl(\frac{\delta_{\rm r}}{4} + \psi\biggl)_{\tau\simeq \tau_{\rm dec}}
  \equiv  \frac{\psi_{\rm m}}{3} \cos{(k \,c_{\rm s}\,\tau_{\rm dec})} = \frac{3}{10} \psi_{\rm r}
   \cos{(k\, c_{\rm s} \, \tau_{\rm dec})},
  \label{SWscalad}
 \end{equation}
where the third equality follows from the relation between the constant modes during radiation 
and matter, i.e. Eq. (\ref{psimpsir}).  Concerning Eq. (\ref{SWscalad}) few comments 
are in order:
\begin{itemize}
\item{} for superhorizon modes the baryon peculiar velocity does not contribute to the leading 
result of the SW effect;
\item{} for $k\,c_{\rm s}\,\tau_{\rm dec} \ll 1$ the temperature fluctuations induced by the adiabatic mode are simply $\psi_{\rm m}/3$ ;
\item{} even if more accurate results on the temperature fluctuations on small angular scales 
can be obtained from a systematic expansion in the inverse of the differential optical depth (tight 
coupling expansion, to be discussed in section 8), Eq. (\ref{SWscalad}) suggests that 
the first true peak  in the temperature fluctuations is located at $k c_{\rm s} \eta_{\rm dec} \simeq \pi$.
\end{itemize}
In this 
discussion, the r\^ole of the baryons has been completely neglected. In section 8 a more refined 
picture of the acoustic oscillations will be developed and it will be shown that the inclusion of baryons 
induces a shift of the first Doppler peak.

\subsubsection{Non-adiabatic contribution to the SW effect}
In the simplistic case of CDM-radiation plasma a rather instructive derivation of the 
gross features of the non-adiabatic mode can also be obtained. 
If $\delta p_{\rm nad}\neq 0$, the evolution of ${\cal R}$ is, in the limit $k\tau \ll 1$ becomes:
\begin{equation}
\frac{d{\cal R}}{d a} = -  \frac{4 {\cal S}}{ ( 3 a + 4)^2}.
\label{evolRCDM}
\end{equation}
Eq. (\ref{evolRCDM}) can be easily obtained inserting Eq. (\ref{nadcdm2}) into Eq. (\ref{evolR}) and recalling 
that, in the physical system under consideration, $( p + \rho) = \rho_{\rm c} + (4/3) \rho_{\rm r}$.
In the case of the CDM-radiation isocurvature mode, the non-adiabatic contribution 
is non-vanishing and proportional to ${\cal S}$. Furthermore, it can be easily shown that   
the fluctuations of the entropy density, ${\cal S}$ are roughly constant (up to logarithmic 
corrections) for $k \tau \ll 1$, i.e. for the modes which are relevant for the SW effect after equality.
This conclusion can be easily derived by subtracting Eq. (\ref{Cdelc}) from  $3/4$ of Eq. (\ref{Cdelr}). 
Recalling the definition of ${\cal S}$ the result is 
\begin{equation}
{\cal S}' = - (\theta_{\rm r} -\theta_{\rm c}). 
 \label{entropyevol}
 \end{equation}
Since $\theta_{\rm r}$ and $\theta_{\rm c}$ vanish in the limit $k \tau \ll 1$,  ${\cal S}$ is 
indeed constant. The explicit form of $\theta_{\rm r}$ and $\theta_{\rm c}$ can be obtained (in the limit 
$k \tau \ll 1$) from the asymptotic solution (obtained in the synchronous gauge, 
Eqs. (\ref{isonu})--(\ref{isosigma})) 
applying the gauge transformation (\ref{Tr1}) to re-express the fluctuations in the longitudinal description.  

Eq. (\ref{evolRCDM}) can then be integrated in explicit terms,  across the radiation--matter 
transition
\begin{equation}
{\cal R} = - 4 {\cal S} \int_{0}^{a} \frac{db}{(3 b + 4)^2} \equiv - {\cal S} \frac{a}{3 a + 4},
\label{RCDMsol1}
\end{equation}
implying that ${\cal R} \to 0 $ for $a \to 0$ (radiation-dominated phase)  and that 
${\cal R} \to - {\cal S}/3$ for $a\to \infty$ (matter-dominated phase). 
 
Recalling again the explicit form of ${\cal R}$ in terms of $\psi$, i.e. Eq. (\ref{defR2}),  Eq. (\ref{RCDMsol1}) 
leads to a simple equation giving the evolution of $\psi$ for modes $k \tau \ll 1$, i.e. 
\begin{equation}
\frac{d \psi}{d a } + \frac{ 5 a + 6}{ 2 a ( a + 1)} \psi = \frac{{\cal S}}{2 ( a + 1)}.
\label{PSICDMeq1}
\end{equation}
The solution of Eq. (\ref{PSICDMeq1}) can be simply obtained imposing the isocurvature 
boundary condition, i.e. $\psi(0) \to 0$:
\begin{equation}
\psi(a ) = \frac{{\cal S}}{5 a^3} \{ 16 ( 1 - \sqrt{ a + 1}) + a [ 8 + a ( a - 2)]\}.
\label{PSICDMsol1}
\end{equation}
Eq. (\ref{PSICDMsol1})  is similar to Eq. (\ref{solPSI1}) but with 
few crucial differences. According to Eq. (\ref{PSICDMsol1}) 
(and unlike Eq. (\ref{solPSI1})), $\psi(a)$ vanishes, 
for  $a\to 0$,  as $ {\cal S} a /8$. In the limit $ a \to \infty$ 
$\psi(a) \to {\cal S}/5$. This is the growth of the adiabatic 
mode triggered, during the transition from radiation to matter, by the 
presence of the non-adiabatic pressure density 
fluctuation.  

Having obtained the evolution of $\psi$, the evolution of the total density contrasts
and of the total peculiar velocity field can be immediately obtained by solving 
the Hamiltonian and momentum  constraints of Eqs. (\ref{p00l}) and (\ref{p0il}) 
 with respect to $\delta\rho$ and $\theta$ 
\begin{eqnarray}
&& \delta = \frac{\delta \rho}{\rho} \equiv \frac{\delta_{\rm r} }{ a + 1} + \delta_{\rm c} 
\frac{a}{a + 1} = -2 \biggl(\psi + \frac{d \psi}{ d\ln{a}}\biggr), 
\label{drhonad}\\
&&\theta = \frac{ 2 k^2 ( a + 1)}{( 3 a + 4)} \biggl(\psi +\frac{ d \psi}{d\ln{a}}\biggr),
\label{dtheta}
\end{eqnarray} 
which also implies, for $k \tau \ll 1$, 
\begin{equation}
\theta =- \frac{k^2 (a + 1) }{( 3 a + 4)} \delta.
\label{THsub}
\end{equation}
Equation (\ref{THsub})  is indeed consistent with the result that the total velocity field is negligible 
for modes outside the horizon. Inserting Eq. (\ref{PSICDMsol1}) into Eqs. 
(\ref{drho})--(\ref{theta})
it can be easily argued that the total density contrast goes to zero for $a\to 0$, while, for 
$a \to \infty$ we have  the following relations 
\begin{equation}
\delta_{\rm c} \simeq - \frac{2}{5} {\cal S} \simeq - 2\psi \simeq - \frac{1}{2} \delta_{\rm r}
\label{psitoentr}
\end{equation}
The first two equalities in Eq. (\ref{psitoentr})  follow from the asymptotics of Eq. (\ref{drhonad}), the last equality 
follows from the conservation law (valid for isocurvature modes) which can be 
derived from Eq. (\ref{Cdelr}), i.e. 
\begin{equation}
\delta_{\rm r} \simeq 4 \psi.
\label{drpsi}
\end{equation}

Thanks to the above results, the contribution to the scalar Sachs-Wolfe effect 
can be obtained in the case of the CDM-radiation non-adiabatic mode. From Eq. 
(\ref{SW0}) we have 
 \begin{equation}
 \biggl( \frac{\Delta T}{T}\biggr)^{\rm nad}_{k, {\rm s}} = 
 \biggl(\frac{\delta_{\rm r}}{4} + \psi\biggl)_{\tau\simeq \tau_{\rm dec}} \equiv 2 \psi_{\rm nad}
 \equiv  \frac{2}{5} {\cal S},
 \label{SWscalnad}
 \end{equation}
where the second equality follows from Eq. (\ref{drpsi}) and the third equality 
follows from Eq. (\ref{PSICDMsol1}) in the limit $a\to \infty$ (i.e. $a \gg a_{\rm eq}$).

The following comments are in order:
\begin{itemize}
\item{} as in the case of the adiabatic mode, also in the case of  non-adiabatic mode in the CDM-radiation
system, the peculiar velocity does not contribute to the SW effect; 
\item{} for $k\,c_{\rm s}\,\tau_{\rm dec} \ll 1$ the temperature fluctuations induced by the adiabatic mode are simply $ 2\psi_{\rm nad}$ (unlike the adiabatic case) ;
\item{}  Equation (\ref{SWscalnad}) suggests that 
the first true peak  in the temperature fluctuations is located at $k c_{\rm s} \eta_{\rm dec} \simeq \pi/2$.
\end{itemize}
The last conclusion comes from an analysis similar to the one conducted in the case 
of the adiabatic mode but with the crucial difference that, in the case of the isocurvature 
mode, $\psi$ vanishes as $\tau$ at early times. This occurrence implies the 
presence of sinusoidal (rather than cosinusoidal) oscillations. This point will 
be further discussed in section 8.

\renewcommand{\theequation}{5.\arabic{equation}}
\section{Evolution of metric fluctuations}
\setcounter{equation}{0}

In the following the evolution of the tensor, vector and scalar modes 
of the geometry will be summarized.  The tensor modes 
of the geometry are, as repeatedly stressed, invariant under 
gauge transformation. The vector modes are not gauge-invariant, however, their description 
is rather simple if the Universe expands and in the absence of vector sources. The scalar 
modes are the most difficult ones: they are not gauge-invariant and they 
directly couple to the (scalar) sources of the 
background geometry.  The analysis will be conducted  
in the case of a spatially flat background geometry and in the case 
of different models of early Universe.

\subsection{Evolution of the tensor modes}

For the tensor modes the perturbed 
Einstein equations imply that \footnote{Recall that $\delta_{\rm t}$, as discussed in the appendix, 
denotes the first-order fluctuation with respect to 
the {\em tensor} modes of the geometry. Similarly $\delta_{\rm s}$ and $\delta_{\rm v}$ 
denote, respectively,  the first-order scalar and vector fluctuations.}
$\delta_{\rm t} R_{i}^{j}=0$.  Hence, according to  Eq. (\ref{driccit2}),  the following  
equation should be satisfied, in Fourier space, by the two tensor 
polarizations:
\begin{equation}
{h_{i}^{i}}'' + 2 {\cal H} {h_{i}^{j}}' +k^2 h_{i}^{j} =0. 
\label{T1a}
\end{equation}
Since $\partial_{i} h^{i}_{j} = h_{k}^{k} =0$, 
the direction of propagation can be chosen to lie along the 
third axis and, in this case the two physical polarizations
of the graviton will be
\begin{equation}
h_{1}^{1} = - h_{2}^{2} = h_{\oplus},\,\,\,\,\,\,\,\,\,\,\,\, h_{1}^{2} = 
h_{2}^{1} = h_{\otimes},
\end{equation}
where $h_{\oplus}$ and $h_{\otimes}$ 
obey the same evolution equation (\ref{T1a}) and will be denoted, in the 
remaining part of this section, by $h$.
Equation (\ref{T1a}) can  be written in two slightly different (but mathematically equivalent) forms:  
\begin{eqnarray}
&& ( a^2 h_{k}')' = - k^2 h_{k},
\label{T2a}\\
&& \mu_{k}'' + \biggl[ k^2 - \frac{a''}{a} \biggr] \mu_{k}=0,
\label{T3a}
\end{eqnarray}
where $ \mu_{k} = a h_{k}$.  Equation (\ref{T2a}) can be interpreted 
as a conservation law for the amplitude of the tensor mode.

In the limit $k \tau \ll 1$  Eq. (\ref{T2a}) implies  that the 
amplitude of the gravitational wave evolves as 
\begin{equation}
h_{k}(\tau) \simeq A_{k} + B_{k} \int^{\tau} \frac{ d \tau'}{a^2(\tau')}  = 
A_{k} + B_{k} \int^{t} \frac{ d t'}{a^3(t')},
\label{T4a} 
\end{equation}
where $A_{k}$ and $B_{k}$ are integration constants to be determined 
from the analysis of the (early) initial conditions which will be discussed in section 6.  

By parametrizing the scale factor of a spatially flat Universe as $a(t) \sim t^{\alpha}$, 
the Universe expands and 
accelerates (i.e. $ \dot{a}>0$ and $\ddot{a} >0$) for $ \alpha>1$ and 
decelerates (i.e. $ \dot{a}>0$ and $\ddot{a} <0$) for $ 0<\alpha < 1$. 
Consequently, Eq. (\ref{T4a}) implies
\begin{equation}
h_{k}(t) \simeq A_{k} + B_{k} t^{1 - 3 \alpha}.
\label{T5a}
\end{equation}
According to Eq. (\ref{T5a}), if the Universe inflates 
(i.e. $\alpha >1$), then $h_{k}$ will have a constant mode (proportional 
to $A_{k}$) and a decaying mode (proportional to $B_{k}$). 
If the Universe decelerates the situation can be different since, on top 
of the constant mode, there could be a growing mode (for $\alpha <1/3$). 
The case $\alpha =1/3$ corresponds to the case of a fluid with 
stiff equation of state, namely a fluid where the speed of sound coincides 
with the speed of light \footnote{By solving
 the Friedmann 
equations with barotropic fluid sources ($p = w \rho$), 
$\alpha = 2/(3(w+ 1))$  leading, for $w =1$, to  $\alpha =1/3$.}. In this case 
the solution is not given by Eq. (\ref{T5a}) but rather by 
$h_{k} \simeq A_{k} + B_{k} \ln{t}$ \cite{maxq1}. This case is particularly 
relevant for quintessential inflation.

Eq. (\ref{T3a}) has the same physical content of Eq. (\ref{T2a}).
For $k^2 \gg |a''/a|$ the solutions of Eq. (\ref{T2a}) are oscillatory, and
the amplitude of the tensor modes, $h$, decreases as $1/a$, if the 
background expands. 
For $k^2 \ll |a''/a|$ the solution is non-oscillatory and is given by 
\begin{equation}
\mu_{k}(\tau) \simeq A_{k}a(\tau) + B_{k} a(\tau) \int^{\tau} 
\frac{d\tau'}{a^2(\tau')}.
\end{equation}
For instance, in the case of de Sitter expansion, the previous equation 
implies that $h$ is constant in this regime. If  $k^2 \ll |a''/a|$, the mode is said to
 be super-adiabatically amplified \cite{grgw1,grgw2,grgw3}. The time-dependent 
 function $a''/a$ is customarily named pump field. This terminology 
 is borrowed from quantum optics where a time dependent 
 electric field may produce coherent (or squeezed) states of the 
 electromagnetic field \cite{sq5,sq6,sq7,sq8}.
\begin{figure}[tp]
\centerline{\epsfig{file=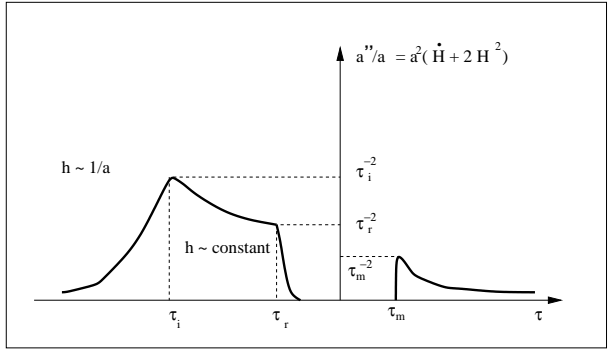, width=8cm}}
\vspace*{9pt}
\caption[a]{The time evolution of the pump field appearing in Eq. (\ref{T3a}) is 
illustrated for a background evolution where the inflationary stage is followed 
by a radiation dominated stage after an intermediate regime. The second barrier 
on the right corresponds to the radiation-matter transition. }
\label{F51}
\end{figure}
The evolution of the tensor modes of the geometry can be deduced from the time 
dependence of the pump field.  An example is illustrated in Fig. \ref{F51} where 
the evolution of $|a''/a|$ is reported for a model where an inflationary 
phase (``i" stage) turns into radiation at a conformal time $\tau_{\rm r}$ 
after an intermediate phase (``s" stage). At $\tau_{\rm m}$ the Universe 
becomes dominated by matter. Both during a de Sitter stage and a matter-dominated 
stage $a''/a = 2 /\tau^2$. During radiation $a''/a=0$ since $a(\tau) \sim \tau$. 
When a given mode with comoving wave-number $k$ gets under the potential barrier 
the corresponding amplitude is amplified during a time inversely proportional to the 
magnitude of $k$. The amplitude  corresponding to $k \sim \tau_{\rm i}^{-1} $ 
will be minimally amplified. The Fourier 
 amplitudes corresponding to $ k < \tau_{\rm m}^{-1}$ will be
maximally amplified. For $ k \gg \tau_{\rm i}^{-1}$ the Fourier amplitudes will not 
be amplified.

When a given mode is in the super-adiabatic regime, the corresponding physical 
frequency is also, grossly speaking, smaller than the Hubble rate: indeed for 
an expanding FRW background $H^2$ is of the order of $\dot{H}$ and $a''/a$ is 
of the order of $a^2 H^2$. 
\begin{figure}[tp]
\centerline{\epsfig{file=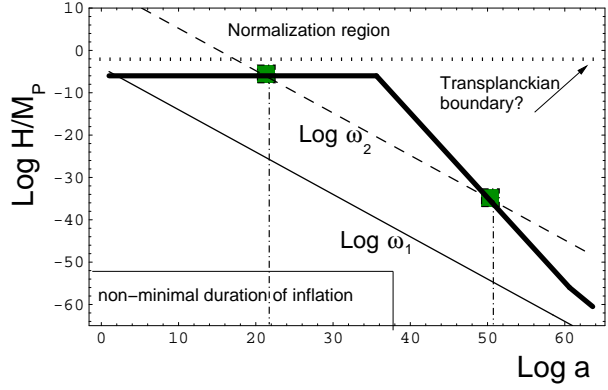, width=8cm}}
\vspace*{9pt}
\caption[a]{The evolution of the Hubble rate in Planck units is illustrated in a 
schematic model of post-inflationary evolution. The 
evolution of two different physical 
frequencies is also illustrated in the case when the duration of inflation is 
non-minimal (about 80 efolds). With the dotted line some fundamental physical scale (smaller than the Planck mass) 
is also illustrated.}
\label{F52}
\end{figure}
In Fig.  \ref{F52} the evolution of the Hubble rate is illustrated as a function of the logarithm (in ten basis)
of the scale factor. The knee corresponds to the inflation-radiation transition. The small ankle 
at the right corresponds to the radiation-matter transition. Superimposed to the plot of the Hubble 
rate the evolution of two physical frequencies is also reported. The first square marks the moment 
when the given mode becomes smaller than the Hubble rate. This moment is called first crossing 
as opposed to the second crossing (second square at the right) when the mode  becomes, again
larger than the Hubble rate. There exist a number of equivalent terminology. For instance, when 
the given mode is smaller than the Hubble rate, its corresponding physical frequency is 
larger than the Hubble radius (i.e. $H^{-1}$) or, in the jargon, superhorizon sized.

\subsubsection{Hamiltonians for the tensor problem}

The variable $\mu_{k} = a h_{k}$ has a special status 
in the theory of initial conditions since it is the canonical variable 
diagonalizing the action perturbed to second-order in the amplitude of 
the tensor fluctuations.  The quadratic action for the tensor modes of the geometry can be
obtained by perturbing the Einstein-Hilbert action to second-order in the 
amplitude of the tensor fluctuations of the metric. To this end, it 
is practical to start  from a form of the gravitational action which 
excludes, from the beginning, the known total derivative. In the case 
of tensor modes this calculation is rather straightforward and the 
result can be written as 
\begin{equation}
S_{\rm gw} = \frac{ 1}{64 \pi G} \int d^4 x a^2 \partial_{\alpha} 
h_{i}^{j} \partial_{\beta} h_{j}^{i} \eta^{\alpha\beta},
\label{Taction1}
\end{equation}
where $\eta_{\alpha\beta}$ is the Minkowski metric.
 
After redefinition of the tensor amplitude through  the Planck mass 
\footnote{Recall that in the present conventions 
 $ \ell_{\rm P} = M_{\rm P}^{-1} = \sqrt{8 \pi G}$.}
\begin{equation}
h = \frac{h_{\oplus}}{\sqrt{2} \ell_{\rm P}} =\frac{h_{\otimes}}{\sqrt{2} \ell_{\rm P}}
\end{equation}
the action (\ref{Taction1}) becomes, for a single tensor polarization,
\begin{equation}
S_{\rm gw}^{(1)} = \frac{1}{2} \int d^{4} x\,\, a^2 \,\,\partial_{\alpha} h \partial_{\beta} h \eta^{\alpha\beta},
\label{TA1}
\end{equation}
whose canonical momentum is simply given by $\Pi= a^2 h'$. The classical
  Hamiltonian associated with Eq. (\ref{TA1}) is then
\begin{equation}
H^{(1)}_{\rm gw}(\tau) =\frac{1}{2} \int d^3 x \biggl[ \frac{ \Pi^2}{ a^2 }
+ a^2 (\partial_{i}h)^2\biggr].
\label{H1t}
\end{equation}
The Hamiltonian (\ref{H1t}) is time-dependent: it is always possible to perform 
time-dependent canonical transformations, leading to a different 
Hamiltonian that will be classically equivalent to (\ref{H1t}). 
Defining the rescaled field, $\mu = a h$, the action (\ref{TA1}) becomes 
\begin{equation}
S^{(2)}_{\rm gw} =\frac{1}{2} \int d^4 x \biggl[ {\mu '}^2 - 2 {\cal H} \mu \mu' 
+ {\cal H}^2 \mu^2 - (\partial_{i}\mu)^2 \biggr],
\end{equation}
while the associated Hamiltonian can be written as 
\begin{equation} 
H^{(2)}_{\rm gw}(\tau) = \frac{1}{2}\int d^{3} x \biggl[  \pi^2 +  2{\cal H} \mu \pi +
 (\partial_{i} \mu)^2 \biggr],
\label{H2t}
\end{equation}
in terms of $\mu$ and of the canonically 
conjugate momentum $\pi = \mu' - {\cal H} \mu$.
A further canonical transformation 
can be performed 
starting from (\ref{H2t}). Defining 
the generating functional of the canonical transformation
 in terms of the 
old fields $\mu$ and of the new momenta $\tilde{\pi}$,  
\begin{equation}
{\cal F}_{2\to 3} ( \mu, \tilde{\pi}, \tau) = \int d^{3} x \biggl( \mu \tilde{\pi}
 - \frac{{\cal H}}{2} \mu^2\biggr),
\label{2to3}
\end{equation} 
the new Hamiltonian can be obtained by taking the partial 
(time) derivative of (\ref{2to3}), with the result
\begin{equation}
H^{(3)}_{\rm gw}(\tau) = \frac{1}{2}\int d^{3} x \biggl[ \tilde{\pi}^2 + 
(\partial_{i} \mu)^2 - ( {\cal H}^2 + {\cal H}')  \mu^2\biggr],
\label{H3t}
\end{equation}
where, recalling the definition of $\pi$, from (\ref{2to3})  
we have $\tilde{\pi} = \mu'$.

The possibility of defining different Hamiltonians 
is connected with the possibility of dropping total (non covariant) derivatives 
from the action of tensor fluctuations.

To illustrate this point consider indeed the action given in Eq. (\ref{TA1}) 
for the single polarization which can be rewritten as:
\begin{equation}
S_{\rm gw} = \frac{1}{2} \int d^{3} x d\tau a^2(\tau) [ {h'}^2 - (\partial_{i} h)^2].
\label{TA2}
\end{equation} 
By looking at Eq. (\ref{TA2}) it is natural to define the canonical field as 
$ \mu = a h$. Recalling that from $\mu = a h$ we also have $ a h' = (\mu'  - {\cal H} \mu)$,
  Eq. (\ref{TA2}) becomes 
\begin{equation}
S_{\rm gw} = \frac{1}{2} \int d^{3} x d\tau [ {\mu'}^2 + {\cal H}^2 \mu^2 - 2 {\cal H} \mu\mu' - 
(\partial_{i} \mu)^2 ].
\label{TA3}
\end{equation}
Recalling now that the action can be always written in terms of an appropriate 
Lagrangian density as 
\begin{equation}
S_{\rm gw} = \int d\tau L(\tau),\,\,\,\,\,\,\,\,\,\,\,\,\,\,\,\,\, L(\tau) = \int d^{3} x {\cal L}(\tau, \vec{x}). 
\label{defL}
\end{equation}
where $L(\tau)$ is the Lagrangian and ${\cal L}(\tau,\vec{x})$ is the Lagrangian density.
In the case of Eq.  (\ref{TA3}), the canonically conjugate momentum is obtained by functional 
derivation of the associated Lagrangian density with respect to $\mu'$ and the 
result will be $\pi = \mu' - {\cal H} \mu$. Hence, the Hamiltonian 
will be 
\begin{equation}
H_{\rm gw}(\tau) = \int d^3 x [ \pi \mu' - {\cal L}_{\rm gw}(\tau,\vec{x})].
\label{defHam}
\end{equation}
Consequently, from Eqs. (\ref{TA3}) and recalling the expression of $\tilde{\pi}$ the Hamiltonian
will exactly be the one defined in Eq. (\ref{H2t}).

Going back to Eq. (\ref{TA3}) and noticing that the term ${\cal H}\mu\mu'$ can be rewritten 
as 
\begin{equation}
{\cal H} \mu\mu'  =-\frac{{\cal H}'}{2} \mu^2 + \frac{d}{d\tau}\biggl[ \frac{{\cal H} \mu^2}{2}\biggr].
\end{equation}
the action
\begin{equation}
\tilde{S}_{\rm gw} = \frac{1}{2}\int d^{4}x [ {\mu'}^2 + ({\cal H}^2 + {\cal H}') \mu^2 - (\partial_{i} \mu)^2 ]
\label{TA4}
\end{equation}
will be classically equivalent to the action of Eq. 
(\ref{TA3}) since the same Euler-Lagrange equations can be obtained from both actions.
Now, however, the conjugate momentum will be $\tilde{\pi} = \mu'$.  Hence, the Hamiltonian 
can be easily obtained, again, from Eq. (\ref{defHam}) but with $\pi$ replaced by $\tilde{\pi}$. After 
performing this simple calculation the result will exactly be the one fiven in Eq. (\ref{H3t}). 

The possibility of defining different Hamiltonians is inherent 
to the time-dependent nature of the problem. The same conclusion also 
applies for the canonical treatment of scalar fluctuations as it will be later illustrated in this 
section. In the Lagrangian approach, this aspect reflects in the possibility 
of defining different classical actions that differ by the addition (or the subtraction) of a total 
time derivative. While at a classical level the equivalence among the different actions 
is complete, at the quantum level this equivalence is somehow broken 
since the minimization of one Hamiltonian or the other may lead to computable differences 
if some recently debated normalization prescriptions are adopted (see section 6).

Up to now the evolution of the tensor modes of the geometry has been discussed 
in the case of expanding FRW backgrounds. Several extensions of this 
analysis have been studied in the last ten years. For instance the present discussion 
can be extended to the context of non-Einsteinian theories of gravity 
like in \cite{BDPBB1,BDPBB2,BDmimoso1,BDhwang}, to the case 
of cosmological models containing compactified 
extra-dimensions \cite{GWint1,GWint2} and higher derivative extension of general 
relativity \cite{GWhigher}. All these techniques are relevant for the 
calculation of the stochastic background of relic gravitons in unconventional 
models of the early Universe such as the pre-big bang \cite{BDPBB1,BDPBB2}
and the ekpyrotic \cite{ekpgw} models.

\subsection{Evolution of the vector modes}

The evolution of the vector modes of the geometry can be obtained 
by perturbing the Einstein equations and the covariant 
conservation equation with respect to the vector modes 
of the geometry, i.e.  
\begin{eqnarray}
&& \delta_{\rm v} {\cal G}_{\mu}^{\nu} = 8\pi G \delta_{\rm v} T_{\mu}^{\nu},
\label{VP1}\\
&& \nabla_{\mu} \delta_{\rm v} T^{\mu\nu} =0.
\label{VP2}
\end{eqnarray}
Recalling  Eqs. (\ref{VCHR})--(\ref{VRICCI}) 
of the appendix,  the explicit form Eqs. (\ref{VP1}) and (\ref{VP2}) 
can be obtained 
\begin{eqnarray}
&& \nabla^2 Q_{i} = - 16\pi G ( p + \rho) a^2 {\cal V}_{i},
\label{VP3}\\
&& Q_{i}' + 2 {\cal H} Q_{i} =0,
\label{VP4}\\
&& {\cal V}_{i}' + (1 - 3 w) {\cal H} {\cal V}_{i}=0,
\label{VP5}
\end{eqnarray}
where ${\cal V}_{i}$ is the divergenceless part of the velocity field, and where
as discussed in section 2, the  gauge has been 
completely fixed by requiring $\tilde{W}_{i}= 0$, implying, according 
to Eqs. (\ref{Q}) and (\ref{W}), that  $\zeta_{i} = -W_{i}$. Notice, furthermore,
that according to the conventions used to represent the velocity field (see 
Eq. (\ref{velfielddef})), $\delta_{\rm v} T_{0}^{i} = ( p + \rho) {\cal V}^{i}$. 

Eqs. (\ref{VP3}), (\ref{VP4}) and (\ref{VP5}) follow, respectively, 
from the $(0i)$--$(ij)$ components of Einstein equations and 
from the spatial component of the covariant conservation equation.
Eqs. (\ref{VP3})--(\ref{VP4}) are not all indipendent. In particular 
it is easy to check that using Eq. (\ref{VP3}), $Q_{i}$ can be 
eliminated from Eq. (\ref{VP4}) and Eq. (\ref{VP5}) is 
easily obtained.

By Fourier transforming Eqs. (\ref{VP3}) and (\ref{VP4}),
the following solution can be found:
\begin{eqnarray}
&& Q_{i}(\tau) = \frac{C_{i}(k)}{a^2 (\tau)},
\label{VP6}\\
&& {\cal V}_{i} (\tau)= \frac{k^2 C_{i}(k)}{16\pi G a^4(p + \rho)},
\label{VP7}
\end{eqnarray} 
where $C_{i}(k) $ is an integration constant.

Two distinct situations are now possible. If the Universe is expanding 
$Q_{i}$ is always decreasing. However, ${\cal V}_{i}$ may also 
increase. Consider, indeed, the case of single barotropic fluid 
$ p = w \rho$. Then, since according to Eq. (\ref{b1})--(\ref{b3}), $\rho \simeq a^{- 3 (w + 1)}$, 
\begin{equation}
{\cal V}_{i} \propto a^{3 w -1}.
\end{equation}
If $w =1$, ${\cal V}_{i}$  
increases as $\tau$ (since $ a(\tau) \sim \sqrt{\tau}$),
while $Q_{i}(\tau)  \simeq \tau^{-1}$, and it is decaying 
for large conformal times. This observation was 
diuscussed, in the past, in connection with the idea that the early 
Universe could be dominated by vorticity \cite{johnvec1,johnvec2}. Recently
a similar observation was also put forward in Ref. \cite{easson1} in the framework 
of the so-called holographic proposal which implies the 
presence of a barotropic fluid with stiff equation of state during 
the post-inflationary evolution \cite{willy1,willy2}.

If the Universe contracts the evolution of the vector modes is  
reversed. In this case $Q_{i}$ may even increase \cite{br1,maxvec1}. 
Finally, the situation may change even 
more radically if the theory is not of Einstein-Hilbert type \cite{maxvec2} or if 
the theory is higher-dimensional \cite{maxvec1}. Notice that if the
energy-momentum tensor of the fluid sources possesses a non-vanishing torque 
force rotational perturbations can be copiously produced \cite{GRV}. 

\subsection{Evolution of the scalar modes}
In analogy with 
the case of the tensor modes of the geometry,  
large-scale scalar fluctuations follow a set of  conservations law 
that are only approximate for $k \tau \ll 1$ and which are  
exact only in the limit $ k \tau \to 0$.  Here, various 
important gauge-invariant quantities will be studied and interpreted 
in light of specific gauge-dependent descriptions.

In section 4  the 
relevant set of equations describing the evolution  
of the metric fluctuations has been derived in the longitudinal gauge.  This 
set of equations is formed by 
Eqs. (\ref{p00l})--(\ref{pijl}) and  Eqs, (\ref{B1})--(\ref{B3}).
Covariant conservation of the evolution of the fluid sources leads to Eqs.
(\ref{delta})--(\ref{theta}). Always in section 4, the CDM-radiation 
isocurvature mode has been treated  by defining the 
non-adiabatic pressure density fluctuation, i.e. $\delta p_{\rm nad}$. 
The introduction of this quantity allowed to derive Eq. (\ref{evolR}), i.e. the evolution 
equation for ${\cal R}$ that can be regarded, in a broad sense,
the scalar analog of Eq. (\ref{T2a}) that holds in the 
case of tensor modes of the geometry. 
Eq. (\ref{T2a}), does not have a source term. On the contrary, Eq. 
(\ref{evolR}) does have a source term, proportional to 
$ \delta p_{\rm nad}$. The following comments are then in order:
\begin{itemize}
\item{} the conservation of ${\cal R}$ is enforced in the $k\to 0$ 
limit provided the non-adiabatic variation of the total 
pressure vanishes;
\item{} the absence of $\delta p_{\rm nad}$ is not guaranteed 
a priori, unless the content of the plasma is only given by a 
single species (i.e. a  single  scalar field, a  single 
fluid); in all other situations $\delta p_{\rm nad}$ will 
not be vanishing and its effects should be taken into account;
\item{} various models share the feature of leading 
to a vanishing $\delta p_{\rm nad}$ like, for instance, the case 
of single field inflationary models.
\end{itemize}

The evolution equations for the fluctuations of the geometry and of the sources 
can be first written without fixing a specific coordinate system. Then, by using the definitions 
of the Bardeen potentials, i.e.  Eqs. (\ref{phigi})--(\ref{vgi}), the wanted gauge-invariant form 
of the equations can be obtained.

Inserting the definitions (\ref{phigi})--(\ref{vgi})
into Eqs. (\ref{00gen}) and (\ref{0igen}), the gauge-invariant 
forms of the Hamiltonian and momentum constraints becomes
\begin{eqnarray}
&& \nabla^2 \Psi - 3 {\cal H} ( \Psi' + {\cal H} \Phi) = 4 \pi a^2 G
\delta \rho_{\rm g},
\label{00gi}\\
&& \nabla^2( \Psi' + {\cal H} \Phi) = - 4 \pi G a^2 ( p + \rho)\Theta,
\label{0igi}
\end{eqnarray}
where, as remarked after Eq. (\ref{vgi}),
 $ \Theta = \partial_{i} V^{i}_{\rm g}$  is the divergence of the 
 gauge-invariant (scalar) peculiar velocity field.
 
Inserting the expressions of the gauge-invariant fluctuations 
given in Eqs. (\ref{phigi})--(\ref{vgi}) into Eqs. (\ref{int1}) 
and (\ref{int2}) the following set of equations
can be also derived:
\begin{eqnarray}
&& \Psi'' + {\cal H} ( \Phi' + 2 \Psi') + ( {\cal H}^2 + 2 {\cal H}') \Phi 
+ \frac{1}{3} \nabla^2 ( \Phi - \Psi) = 4\pi G a^2 \delta p_{\rm g},
\label{ijgi}\\
&& \nabla^2 ( \Phi - \Psi) =  12 \pi G a^2 (p + \rho)\sigma.
\label{difgi} 
\end{eqnarray}
Finally, the gauge-invariant form of the perturbed covariant 
conservation equations can be written as: 
\begin{eqnarray}
&& \delta \rho_{\rm g}' - 3 ( p + \rho) \Psi' + ( p + \rho) \Theta + 
3 {\cal H} ( \delta \rho_{\rm g} + \delta p_{\rm g}) =0,
\label{drgi}\\
&& (p + \rho) \Theta' + [ ( p' + \rho') + 4 {\cal H} ( p + \rho)] \Theta 
+ \nabla^2 \delta p_{\rm g} + ( p + \rho) \nabla^2 \Phi =0.
\label{THgi}
\end{eqnarray}
as it can be checked by using Eqs. (\ref{phi})--(\ref{vgi}) 
into Eqs. (\ref{gencov1})--(\ref{gencov2}) that are written without imposing any gauge condition.

Equations (\ref{00gi})--(\ref{0igi}) and (\ref{ijgi})--(\ref{difgi})
 have the same form of the evolution equations 
in the longitudinal gauge. All the manipulations leading to the 
evolution of ${\cal R}$ (see, in particular, the derivation 
preceding Eq. (\ref{evolR})) can be repeated with the same results.
The explicit expression of ${\cal R}$ in gauge-invariant terms 
can then be written as 
\begin{equation}
{\cal R} = - (\Psi - {\cal H} V_{\rm g}) = - \Psi - \frac{{\cal H} ( \Psi' + 
{\cal H} \Phi)}{ {\cal H}^2 - {\cal H}'}. 
\label{defR4}
\end{equation}

\subsubsection{Physical interpretation of curvature perturbations}
In the comoving  gauge the three velocity 
of the fluid vanishes, i.e. $v_{\rm C} =0$. Since the constant 
(conformal) time hypersurfaces should be orthogonal to the 
four-velocity, we will also have that $B_{\rm C} =0$. In this gauge 
the curvature perturbation can be computed directly from the expressions 
of the perturbed Christoffel connections bearing in mind that, unlike 
the appendix, we want to compute here the fluctuations in the  
spatial curvature, namely
\begin{equation}
\delta R^{(3)}_{\rm C} = \delta_{\rm s} \gamma^{i j} \overline{R}_{ij}^{(3)} 
+ \overline{\gamma}^{i j} 
\delta_{\rm s} R^{(3)}_{ij} \equiv \frac{4}{a^2}\nabla^2 
\psi_{\rm C},
\end{equation}
where the subscript ${\rm C}$ refers to the fact that the calculation has been 
conducted in the on comoving hypersurfaces.  The 
curvature fluctuations of the comoving gauge can be connected to the 
fluctuations in a different gauge characterized 
by a different value of the time coordinate, i.e. 
$ \tau_{\rm C} \to \tau = 
\tau_{\rm C} + \epsilon_0$. Under this shift 
\begin{eqnarray}
&& \psi_{\rm C} \to \psi = \psi_{\rm C} + {\cal H} \epsilon_{0},
\label{CTL1}\\
&& ( v_{\rm C} +B_{C})  \to  B + v  = (v_{\rm C} + B_{\rm C}) + \epsilon_0,
\label{CTL2}
\end{eqnarray}
Since in the comoving orthogonal gauge $ B_{\rm C} + v_{\rm C} =0$,
Eqs. (\ref{CTL1})--(\ref{CTL2}) imply,
in the new coordinate system, 
\begin{equation}
\psi_{\rm C} = \psi - {\cal H} ( v + B) \equiv (\Psi - {\cal H} V_{\rm g}),
\label{calRrat}
\end{equation}
 where the second equality follows from the definitions of gauge-invariant 
fluctuations given in Eqs (\ref{phigi})--(\ref{vgi}). 
From Eq. (\ref{calRrat}) it can be concluded that ${\cal R}$ 
corresponds to the curvature fluctuations of the spatial 
curvature on comoving orthogonal hypersurfaces. 

Other quantities, defined in specific gauges, turn out to have a gauge-invariant
interpretation.
Take, for instance, the curvature fluctuations on constant density 
hypersurfaces, $\psi_{\rm D}$. Under infinitesimal 
gauge transformations 
\begin{eqnarray}
&& \psi_{\rm D}\to \psi= \psi_{\rm D} + {\cal H} \epsilon_{0}, 
\nonumber\\
&& \delta\rho_{\rm D} \to \delta \rho = \delta\rho_{\rm D} - \rho' 
\epsilon_{0}. 
\label{CDH}
\end{eqnarray}
But on constant density hypersurfaces $ \delta\rho_{\rm D} =0$,
by definition of uniform density gauge. Hence, from Eq. (\ref{CDH}),
\begin{equation}
\psi_{\rm D} = \psi + {\cal H} \frac{\delta \rho}{ \rho'} \equiv 
\Psi + {\cal H} \frac{\delta \rho_{\rm g}}{\rho'},
\end{equation}
where, again, the second equality follows from the definitions 
of gauge-invariant fluctuations given in Eqs. (\ref{phigi})--(\ref{vgi}).  
Hence,  the (gauge-invariant) 
 curvature fluctuations on constant density hypersurfaces, can be defined as 
\begin{eqnarray} 
\zeta = - \biggl( \Psi + {\cal H} \frac{\delta\rho_{\rm g}}{\rho'}\biggr). 
\label{ZETAdef}
\end{eqnarray}
 Equation (\ref{ZETAdef}) defines the  
gauge-invariant curvature fluctuations on constant density 
hypersurfaces. The last sentence seems to contain a contradiction, but 
such an  expression simply means that 
$\zeta$ coincides with the curvature fluctuations in the uniform 
density gauge. In a different gauge (for instance the longitudinal 
gauge) $\zeta$ has the same value but does not 
coincide with the curvature fluctuations. 

The values of $\zeta$ and ${\cal R}$ are equal up to terms proportional 
to the laplacian of $\Psi$. This can be shown by using explicitely 
the definitions of ${\cal R}$ and $\zeta$ whose  difference gives 
\begin{equation}
\zeta - {\cal R} = - {\cal H} \biggl( V_{\rm g} + 
\frac{\delta\rho_{\rm g}}{\rho'} \biggr). 
\end{equation} 
Recalling now that from  the Hamiltonian and momentum constraints 
and from the conservation of the energy density of the background 
\begin{eqnarray}
&& V_{\rm g} =- \frac{1}{4\pi Ga^2 ( p + \rho)} ( {\cal H} \Phi + \Psi'),
\nonumber\\
&& \frac{\delta\rho_{\rm g}}{\rho'}=  \frac{1}{4\pi Ga^2 ( p + \rho)} [ 
\nabla^2 \Psi - 3 {\cal H} ( {\cal H} \Phi + \Psi')],
\end{eqnarray}
we obtain 
\begin{equation}
\zeta - {\cal R} = \frac{\nabla^2 \Psi}{12 \pi G a^2 ( p + \rho)}.
\label{Z-R}
\end{equation}

The density contrast on comoving orthogonal 
hypersurfaces  can be also defined as \cite{bardeen,bardeen2}
\begin{equation}
\epsilon_{\rm m} = \frac{\delta \rho_{\rm C}}{\rho} \equiv 
\frac{\delta\rho + \rho' ( v + B)}{\rho} \equiv  \frac{\delta \rho_{\rm g} 
- 3 {\cal H} ( p + \rho) V_{\rm g}}{\rho}, 
\end{equation}
where the second equality follows from the first one by using 
the definitions of gauge-invariant fluctuations.
Again, using the Hamiltonian and momentum constraints, 
\begin{equation}
\epsilon_{\rm m} = \frac{1}{4\pi Ga^2 \rho} \nabla^2 \Psi 
\end{equation}
that means, according to Eq. (\ref{Z-R}), that $(\zeta - {\cal R}) \propto \epsilon_{\rm m}$.
 
\subsubsection{Curvature perturbations induced by scalar fields}
Scalar fields are an almost universal ingredient 
for the discussion of potentially physical effects in  the early Universe since, together 
with relativistic fluids, they have the virtue of leaving unbroken 
the spatial isotropy of the homogeneous 
background geometry.
Consequently scalar fields may appear in diverse (and sometimes unrelated) 
cosmological contexts ranging from conventional inflationary models 
 \footnote{ Various conventional inflationary models have been developed so far. 
A partial list should probably include: old inflation \cite{CONVEN1}, new inflation
\cite{CONVEN2,CONVEN3}, cahotic inflation 
\cite{CONVEN4}, natural inflation \cite{CONVEN5}, eternal inflation \cite{CONVEN6}, 
hybrid inflation\cite{CONVEN7}, quintessential inflation \cite{alex1}. By the word 
conventional we mean that the above models share the feature of being formulated 
in terms of a quasi-de Sitter phase of accelerated expansion driven by a
phenomenological master field generically named inflaton. See also \cite{linde} for a comprehensive 
review of inflationary models.}
to  minimal  \cite{PBB1,PBB2} and non minimal pre-big bang models 
\cite{maxb1,maxb2,maxb3}.  Ekpyrotic models \cite{EKP1,EKP2,SYSK}, together 
with their cyclic extension \cite{EKP3} are formulated in terms of scalar 
degrees of freedom.  Finally, quintessential models of  dark energy \cite{sahni}
are connected with the presence of a scalar degree of freedom 
whose potential is carefully chosen in such a way that the scalar field energy density dominates 
just today, say for $z$ between  $0$ and $3$.

The physical meaning attributed to the scalar degree 
of freedom changes from model to model. For instance, in pre-big 
bang models the scalar fields will be the string theoretic dilaton or the 
Kalb-Ramond axion (which is a pseudo-scalar field in four space-time 
dimensions). In the ekpyrotic model the scalar field will 
parametrize, at least in some versions of the model, the effective  
distance between two colliding domain-walls  defined in five 
dimensions. It is important to remark that the scalar degree of freedom appearing 
in models inspired by string/M theory is the result of the  compactification
(achieved with diverse methods) of a more fundamental degrees of freedom 
which are typically defined either in $10$ or $11$ dimensions. 
It is difficult, sometimes, to attribute  a specific 
meaning to the inflaton field of conventional inflationary models: the inflaton 
is often a  master field (light during the inflationary phase)  that drives 
a phase of quasi-de Sitter expansion.

Some of the considerations developed earlier in this section 
will now be specialized to the case of scalar field  sources characterized, rather 
generically, by a potential $W(\varphi)$. 
In a spatially flat FRW geometry the background Einstein equations read, 
 in the single scalar field case,
\begin{eqnarray}
&&{\cal H}^2 = \frac{8\pi G}{3}\, \biggl( \frac{{\varphi'}^2}{2 } 
+ W a^2 \biggr),
\label{b1s}\\
&&( {\cal H}^2 - {\cal H}')   =  4 \pi G \,\,\varphi'^2 ,
\label{b2s}\\
&& 
\varphi'' + 2 {\cal H} \varphi' + \frac{\partial W}{\partial \varphi} a^2 =0.
\label{b3s}
\end{eqnarray}
The fluctuation of $\varphi$ changes, for infinitesimal coordinate 
transformations, as 
\begin{equation}
\chi \to \tilde{\chi} = \chi - \varphi' \epsilon_{0};
\end{equation}
consequently the associated gauge-invariant scalar field fluctuation is given by 
\begin{equation}
X = \chi + \varphi' ( B - E'). 
\label{chigi}
\end{equation}
The two Bardeen  potentials ($ \Psi$ and $\Phi$) and the 
gauge-invariant scalar field fluctuation $X$ define the coupled system 
of scalar fluctuations of the geometry:
\begin{eqnarray}
&&\nabla^2 \Psi - 3 {\cal H} ( {\cal H} \Phi + \Psi') 
= 4\pi G \biggl[ - \varphi'^2 \Phi + 
\varphi' X' + \frac{\partial W}{\partial \varphi} a^2 X\biggr] ,
\label{00sc}\\
&&  {\cal H} \Phi + \Psi'  = 4\pi G \varphi' X,
\label{0isc}\\
&& \Psi'' + {\cal H} (\Phi' + 2  \Psi') 
 + ({\cal H}^2 + 2 {\cal H}') \Phi  = 4\pi G\biggl[ - \varphi'^2 \Phi + 
\varphi' X'
 - \frac{\partial W}{\partial \varphi} a^2 X \biggr],
\label{ijsc} 
\end{eqnarray}
where Eqs. (\ref{00sc})--(\ref{ijsc}) are, respectively, the perturbed 
$(00)$, $(0i)$ and $(ij)$ components 
of Einstein equations.
Eqs.  (\ref{00sc})--(\ref{ijsc}) are sufficient to determine 
the evolution of the system, however, it is also appropriate to recall
the gauge-invariant form of the perturbed Klein-Gordon equation:
\begin{equation}
X'' + 2 {\cal H} X' - \nabla^2 X + \frac{\partial^2 W}{\partial\varphi^2} a^2 
X + 2 \Phi \frac{\partial W}{\partial\varphi} a^2 - 
\varphi'( \Phi' + 3 \Psi') =0.
\label{KG3}
\end{equation}
Eq. (\ref{KG3}) can be obtained from the perturbed 
Klein-Gordon equation without gauge fixing, reported in 
 Eq. (\ref{KG2}) of the appendix. The gauge-invariant 
fluctuations of the metric of Eqs. (\ref{phigi})--(\ref{psigi}) 
can be inserted into Eq. (\ref{KG2}) together with Eq. (\ref{chigi}): Eq. 
(\ref{KG3}) will then appear as a result of the simplifications.

If the perturbed energy-momentum tensor has vansihing anisotropic stress,
the $(i \neq j)$ component of the perturbed Einstein equations 
leads to $\Phi = \Psi$ \footnote{It would not be correct 
to assume that $\psi=\phi$ already at the leven of Eq. (\ref{KG2}) giving 
the perturbed Klein-Gordon equation without gauge fixing: 
by doing so, the correct form of Eq. (\ref{KG3}) cannot 
be obtained since $ \psi$ and $\phi$ 
change in a different way under infinitesimal coordinate transformations and hence 
lead to different gauge-invariant fluctuations. }.

The gauge-invariant curvature fluctuations on comoving 
orthogonal  hypersurfaces are, for a scalar field source,
\begin{equation}
{\cal R} = - \Psi - {\cal H} \frac{X}{\varphi'}= 
- \Psi - \frac{{\cal H}( {\cal H} \Phi + \Psi')}{{\cal H}^2 - {\cal H}'},
\label{defRchi}
\end{equation}
where the equality follows from the use of Eq. (\ref{0isc})
 and of Eq. (\ref{b2s}).
 
The definition of (\ref{defRchi}) and a linear combination of Eqs. 
(\ref{00sc}) and (\ref{ijsc}) 
leads to the following simple equation
\begin{equation}
{\cal R}' = - \frac{ {\cal H}}{ 4\pi G{\varphi'}^2} \nabla^2 \Psi,
\label{eqR}
\end{equation}
implying the constancy of ${\cal R}$ for modes $k\tau \ll 1$ 
\cite{bardeen,bardeen2,press,lyth}.
The power spectrum 
of the scalar modes amplified during the inflationary phase is customarily 
expressed in terms of ${\cal R}$, which is conserved on super-horizon scales (this 
statement will be made more specific at the end of this section while discussing 
the Weinberg theorem).

Curvature perturbations on comoving spatial hypersurfaces can also 
be simply related to curvature perturbations 
on the constant density hypersurfaces, denoted by $\zeta$
\begin{equation}
\zeta = - \Psi - {\cal H} \frac{\delta\rho_{\varphi}}{\rho_{\varphi}'} \equiv 
- \Psi + \frac{a^2 \delta\rho_{\varphi}}{3 {\varphi'}^2}.
\label{defzeta}
\end{equation}
Taking  the difference in the definitions 
(\ref{defR4}) and (\ref{defzeta}), and using Eq. (\ref{00sc}):
\begin{equation}
\zeta - {\cal R} \equiv {\cal H}\frac{X}{{\varphi}'} + \frac{ a^2 \delta \rho_{\varphi}}{3 {\varphi'}^2} = 
\frac{ 2 M_{\rm P}^2}{3} \frac{\nabla^2 \Psi}{{\varphi '}^2} ,
\end{equation}
${\cal R}$ and $\zeta$ differ by Laplacians of the Bardeen potential.

Taking the time 
derivative of Eq. (\ref{eqR}) and using, repeatedly, Eq. (\ref{defRchi}) and 
Eqs. (\ref{00sc})--(\ref{ijsc}), the following second-order equation can be obtained:
\begin{equation}
{\cal R}'' + 2 \frac{z'}{z} {\cal R}' - \nabla^2 {\cal R} =0,
\label{seceq}
\end{equation}
where the function $z(\tau)$ is given by
\begin{equation}
z(\tau) = \frac{a \varphi'}{\cal H}.
\label{ZDEF}
\end{equation}
Eq. (\ref{seceq}) is {\em formally} similar to Eq. (\ref{T1a}), i.e. the evolution equation 
of a single tensor polarization. This similarity becomes somehow deeper in the 
specific cases when the scale factor has a power-law dependence. In this 
case, let us suppose that the scale factor has a power-law dependence such as 
$a(t) \sim t^{p}$ in cosmic time.

The case of power-law inflation is characterized by a scale factor evolving as 
$a(t) \sim t^{p}$ where $ p >1$. Exact solutions of Eqs. (\ref{b1s})--(\ref{b3s}) 
can be found with exponential potentials  \cite{PL1,PL2,PL3,PL4}. In this case 
the solution can be parametrized, in cosmic time, as
\begin{equation}
a(t) \sim t^{p}, \,\,\,\,\,\,\,\dot{\varphi}= \frac{ \sqrt{2 p} M_{\rm P}}{t}, \,\,\,\,\,\,\,\,\,z(t) = \sqrt{ \frac{2}{p}} M_{\rm P} a(t)
\label{pot}
\end{equation}
where, as usual, the dot denotes a derivation with respect to the cosmic time  coordinate $t$. Clearly, in this 
case $z(t)$ and $a(t)$ are proportional and, therefore, Eqs. (\ref{seceq}) and (\ref{T1a}) are exactly equivalent.
This equivalence surfaces again if we eliminate the first  time derivative appearing in Eq. (\ref{seceq})
by introducing the gauge-invariant variable \cite{luk,muk}
\begin{equation}
q = - z {\cal R} = a X + z \Psi,
\label{qtoR}
\end{equation}
which leads, once inserted into Eq. (\ref{seceq}), to 
\begin{equation}
q'' - \frac{z''}{z} q  - \nabla^2 q=0,
\label{QEQ}
\end{equation}
that is the scalar analog of Eq. (\ref{T3a}).
In the case of power-law inflation given in Eq. (\ref{pot}), Eqs. (\ref{T3a}) and Eq. (\ref{QEQ})
share the same pump field.  This occurrence implies that Fig. \ref{F52} is often 
interpreted, in some vulgarization of the subject, 
as holding both for tensor and scalar modes.
Figure \ref{F52} also describes approximately the evolution of curvature perturbations
in the case of a single scalar field and in the absence of any fluid sources or of any other spectator fields 
which do not drive inflation but whose presence generate a $\delta p_{\rm nad}$, i.e. 
a non-adiabatic fluctuation of the pressure density.
The related caveat is that, unlike the evolution of the tensor modes, the evolution 
of the scalar modes is constrained from both the $(00)$ and $(0i)$ 
components of the perturbed Einstein equations. 

As discussed above, in the case of power-law inflation the simplicity 
of the (exponential) potential leads to a privileged relation 
between the scalar and tensor pump fields that turn out to be 
exactly equal, i.e. $a''/a=z''/z$. For more general inflationary 
potentials (but always in the single field case) this relation is not verified anymore.
The pump fields of the scalar and tensor modes 
for more general potentials leading to a 
quasi-de Sitter expansion of the geometry  can be discussed in terms of the 
slow-roll \footnote{Notice that in \cite{LLKCB} $\eta$ is defined 
with opposite sign with respect to our conventions.  Moreover some authors \cite{SL}
like to 
define $\delta_{\rm t} g_{i j}= - 2  a^2 h_{ij}$ while we define $ \delta_{\rm t}g_{i j}  = - a^2 h_{ij}$.
So the tensor fluctuation defined in \cite{SL} is twice the one defined here. 
This choice is arbitrary provided everything is done consistently 
(for instance, as far as the tensor fluctuation is concerned, the tensor Sachs-Wolfe 
effect will be sensitive to the conventions chosen in order to write the perturbed line 
element). In spite of the arbitrariness of the definition this may lead to confusions
as far as the normalization of the power spectra is concerned (see section 6).
In the following, all the conventions previously established will be carefully followed 
mentioning, when needed, the major differences with the other choices 
made in the literature.}
 parameters $\epsilon$ and 
$\eta$ , i.e. \cite{SL,LLKCB} \begin{equation}
\epsilon= - \frac{\dot{H}}{H^2}, \,\,\,\,\,\,\,\,\,\,\,\,\,\,\,\,\, \eta = \frac{\ddot{\varphi}}{H \dot{\varphi}}.
\label{sl1}
\end{equation}
In the case of power-law potentials the slow-roll parameters have the property of being constants 
and equal in modulus, i.e  $\epsilon =1/p=-\eta$.

To find the expressions of the pump fields in terms of the slow-roll parameters 
it is useful to write the background equations as a first-order set of non-linear 
differential equations \cite{muslim,salopek1,salopek2,lidsey}
\begin{eqnarray}
&& \biggl[ \frac{\partial H}{\partial\varphi}\biggr]^2 - 
\frac{3}{2 M_{\rm P}^2}\,H^2(\varphi) = - 
\frac{W(\varphi)}{2 M_{\rm P}^4},
\label{sl2}\\
&& \dot{\varphi} = - 2 M_{\rm P}^2\frac{\partial H}{\partial\varphi},
\label{sl3}
\end{eqnarray}
where, as usual, the adopted definition for the Planck mass will be the 
one already made explicit in the introduction and after Eq. (\ref{Taction1}).

Bearing in mind Eq. (\ref{sl1}), the explicit relation determining the form of the tensor 
pump field follows from the  chain of equalities 
\begin{equation}
\frac{a''}{a} = {\cal H}^2 + {\cal H}' = 2a^2 H^2 + a^2 \dot{H} \equiv  2 a^2 H^2 
\biggl( 1 - \frac{\epsilon}{2}\biggr),
\label{SLin1}
\end{equation}
where in the last equality, Eq. (\ref{sl1}) has been used. In order to find a suitable 
expression we have to determine $aH$ as a function of $\tau$. To do so we must 
recall that the relation between cosmic and coformal time depends upon $\epsilon$. In fact, if
$\epsilon$ is constant the following equalities hold thanks to integration by parts: 
\begin{equation}
\tau =\int \frac{d t}{a} = \int \frac{da}{ a^{2}H} = - \frac{1}{a H} + \epsilon \int \frac{d a}{a^2 H}.
\label{SLin2}
\end{equation}
The third equality in Eq. (\ref{SLin2}) follows from integration 
by parts and by taking $\epsilon$ to be constant. 
Equation (\ref{SLin2}) allows to determine $a H$ as a function of $\tau$, i.e.  
\begin{equation}
a H= - \frac{1}{\tau ( 1 - \epsilon)}.
\label{aH}
\end{equation}
Equation (\ref{aH}) implies:
\begin{equation}
\frac{a''}{a} = \frac{\mu^2 -1/4}{\tau^2} \equiv \frac{2 -\epsilon}{(1- \epsilon)^2}, \,\,\,\,\,\,\,\,\,\,\mu = 
\frac{3 - \epsilon}{2( 1 - \epsilon)}.
\label{musl}
\end{equation}

Recall now  the definition of $z$ as discussed in Eq. (\ref{ZDEF}). The same calculation 
leading to Eq. (\ref{musl}) implies, for the scalar pump field
\begin{equation}
\frac{z''}{z} = \frac{\nu^2 -1/4}{\tau^2},\,\,\,\,\,\,\,\,\,\,\nu = \frac{3 + \epsilon + 2 \eta}{2( 1 -\epsilon)}.
\label{nuscal}
\end{equation}
The calculation is rather straightforward:  $z''/z$ has to be written in terms of the explicit 
derivatives of ${\cal H}$, $a$ and $\varphi'$ as they appear in Eq. (\ref{ZDEF}); then the conformal 
time expression have to be translated in theur cosmic time analog; then the definitions 
of Eq. (\ref{sl1}) must be used. Ntice that the indices $\mu$ and $\nu$ of Eqs. (\ref{musl}) 
and (\ref{nuscal}) have been introduced for later convenience. Moreover,  
$\mu=\nu$ in the case of power-law inflation when $\epsilon=- \eta$. 
If $\epsilon$ is not constant Eq. (\ref{aH}) is generalized to 
\begin{equation}
\tau= - \frac{1}{aH} \frac{1}{1 - \epsilon} - \frac{2 \epsilon( \epsilon +\eta)}{a H}+....,
\end{equation}
as it can be concluded from integration by parts without assuming the constancy of $\epsilon$; the 
ellipses stand for higher orders in the slow-roll expansion.

To conclude, it is useful to point out that different decoupled equations 
for the gauge-invariant fluctuations can be obtained, in the single scalar field case.
The first equation of this sort is (\ref{QEQ}).
The other equation to be  introduced is a decoupled equation for the Bardeen potential. 
Taking the difference of Eqs. (\ref{ijsc}) and (\ref{00sc}) we obtain rather easily 
\begin{equation}
\Psi'' + 6 {\cal H} \Psi' + ( 2 {\cal H}' + 4 {\cal H}^2) \Psi - \nabla^2 \Psi  = - 
8\pi G \frac{\partial W}{\partial \varphi} a^2  X.
\label{INTPSI}
\end{equation}
To eliminate $X$, the momentum constraint of Eq. (\ref{0isc}) can be used in the form
\begin{equation}
X= \frac{\Psi' + {\cal H} \Psi}{ 4\pi G \varphi'}.
\label{INTPSI2}
\end{equation}
Inserting Eq. (\ref{INTPSI2}) into Eq. (\ref{INTPSI}) and using Eq. (\ref{b3s}) 
to eliminate the derivative of $W$ with respect to $\varphi$ we obtain the 
following decoupled equation:
\begin{equation}
\Psi'' + \biggl( 2 {\cal H} - 2 \frac{\varphi''}{\varphi'}\biggr) \Psi' + 2 \biggl( {\cal H}' - \frac{\varphi''}{\varphi'}\biggr)
\Psi  - \nabla^2 \Psi =0.
\label{INTPSI3}
\end{equation}
The first derivative of a second order equation with time-dependent coefficients can be always 
eliminated by a suitable field re-definition. It is the easy to show that, defining a new variable, be it 
$\overline{\Psi} = (a/\varphi') \Psi$,  Eq. (\ref{INTPSI3}) can be recast in the form
\begin{equation}
\overline{\Psi}'' - \nabla^2 \overline{\Psi} - \biggl(\frac{1}{z}\biggr)'' z \overline{\Psi} =0,
\label{INTPSI4}
\end{equation}
where $z$ has been defined in Eq. (\ref{ZDEF}). Let us remark that, apparently, 
Eqs. (\ref{INTPSI3}) and (\ref{INTPSI4}) are rather simple. However, the price for simplicity 
is, in this case, that the derivation of Eq. (\ref{INTPSI3}) assumes that $\varphi'$ 
never goes to zero, since, by solving the momentum constraint (see Eq. (\ref{INTPSI2}) ) 
we divided by $\varphi'$. The effect of this manipulation is evident from the expressions 
of the pump fields of both Eqs. (\ref{INTPSI3}) and (\ref{INTPSI4}) that contain (explicitly 
or implicitly) inverse powers of $\varphi'$.

Up to now we mainly dealt with the case of a single scalar field. Extensions of some 
of the ideas discussed so far are certainly possible. Consider, for instance, the case 
of two scalar fields $\varphi$ and $\sigma$ characterized by a potential term
$W(\varphi,\sigma)$. In this case the relevant Friedmann equations (\ref{b1s}) and (\ref{b2s}) 
 can be simply obtained by doubling the kinetic terms, i.e. by replacing ${\varphi'}^2$ with 
 $({\varphi'}^2 + {\sigma'}^2)$.  
 The spatial curvature perturbation can be written, in this case (see for instance \cite{mgcurv3,hwa2scal,tar2scal}), as 
\begin{equation}
{\cal R} = - \frac{H}{\dot{\varphi}^2 + \dot{\sigma}^2} 
\bigl[ \dot{\varphi} q_{\varphi}  +\dot{\sigma} q_{\sigma} \bigr],
\label{Rdef1}
\end{equation}
where 
\begin{eqnarray}
&& q_{\varphi} = X_{\varphi} + \frac{\dot{\varphi}}{H} \Psi,
\label{defvph}\\
&& q_{\psi} = X_{\sigma} + \frac{\dot{\sigma}}{H} \Psi,
\label{defvps}
\end{eqnarray}
are, respectively,  the canonically normalized fluctuations of $\varphi$ and $\sigma$, i.e. 
the generalization of the variable $q$ obeying eq. (\ref{QEQ}). 
As discussed in \cite{mgcurv3} the relevant evolution equations 
can be written as 
\begin{eqnarray}
&&\ddot{q}_{\varphi} + 3 H \dot{q}_{\varphi} - \frac{1}{a^2} \nabla^2 q_{\varphi} +
\biggl[ \frac{\partial^2 W}{\partial\varphi^2} - \frac{1}{M_{\rm P}^2 a^3}\frac{\partial}{\partial t} 
\biggl( \frac{a^3}{H} \dot{\varphi}^2\biggr)\biggr] q_{\varphi} 
= \frac{1}{M_{\rm P}^2 a^3}\frac{\partial}{\partial t} 
\biggl( \frac{a^3}{H} \dot{\varphi} \dot{\sigma}\biggr) q_{\sigma},
\nonumber\\
&&\ddot{q}_{\sigma} + 3 H \dot{q}_{\sigma} 
- \frac{1}{a^2} \nabla^2 q_{\sigma} +
\biggl[ \frac{\partial^2 W}{\partial\sigma^2} 
- \frac{1}{M_{\rm P}^2 a^3}\frac{\partial}{\partial t} 
\biggl( \frac{a^3}{H} \dot{\sigma}^2\biggr)\biggr] q_{\sigma} 
=\frac{1}{M_{\rm P}^2 a^3}\frac{\partial}{\partial t} 
\biggl( \frac{a^3}{H} \dot{\varphi} \dot{\sigma}\biggr) q_{\varphi}.
\nonumber\\
\label{qqphisigma}
\end{eqnarray}
Equation (\ref{qqphisigma}), as well as  the whole set of formulae derived in the two-field 
case, can also be generalized to the situation 
of an arbitrary number of scalar fields \cite{anderegg}. This is particularly useful from the point of 
view of the Hamiltonian description of the problem.
The most notable aspect of the previous two equations is that, unlike the single 
field case, the evolution of $q_{\varphi}$ and $q_{\sigma}$ is coupled. This 
already signals a fundamental  physical difference with respect to the single field case.

In fact, in the two-field case the evolution of ${\cal R}$ develops a source term which is 
the analog of the non-adiabatic fluctuation of the pressure density already introduced 
in section 4 when talking about the longitudinal description of isocurvature modes.   
Direct calculation \cite{mgcurv3} shows that the evolution of ${\cal R}$ is, in this case,
\begin{equation}
\dot{\cal R} = - \frac{H}{\dot{\sigma}^2 + \dot{\varphi}^2} 
\delta p_{\rm nad},
\label{zeq}
\end{equation}
where $\delta p_{\rm nad}$ is now
\begin{equation}
\delta p_{\rm nad} = ( c_{\rm s}^2 -1) \Psi ( \dot{\varphi}+ \dot{\sigma}) + 
( 1 - c_{\rm s}^2) ( \dot{\varphi} \dot{X}_{\varphi} + \dot{\sigma} \dot{X}_{\sigma}) 
- ( 1 + c_{\rm s}^2) \biggl( \frac{ \partial W}{\partial\varphi} X_{\varphi} + 
\frac{\partial W}{\partial\sigma} X_{\sigma}\biggr),
\label{dp1}
\end{equation}
where $c_{\rm s}$ is the generalized sound speed, i.e. 
\begin{equation}
c_{\rm s}^2 = \frac{\dot{p}}{\dot{\rho}}=
1 + \frac{2}{3 H( \dot{\varphi}^2 + \dot{\sigma}^2)}\biggl( \frac{\partial W}{\partial\varphi} 
\dot{\varphi} + \frac{\partial W}{\partial \sigma}\dot{\sigma} \biggr).
\label{cs1}
\end{equation}
In the first equality of Eq. (\ref{cs1}) $p$ and $\rho$ are, respectively, the total 
pressure and energy density of the system written in terms of the two background fields $\varphi$ and $\sigma$.

\subsubsection{Hamiltonians for the scalar problem}

In light of possible applications, it is desirable to treat 
the evolution of the scalar fluctuations of the geometry in terms of 
a suitable variational principle. On this basis, Hamiltonians 
for the evolution of the fluctuations can be defined.
By perturbing the action of the scalar fluctuations of the geometry, the 
final form of the action  can be  expressed in terms of the 
curvature fluctuations
\begin{equation}
 S^{(1)}_{\rm scal} = \frac{1}{2} \int d^4 x\,\, z^2 \biggl[{ {\cal R}'}^2 - (\partial_{i} {\cal R})^2\biggr].
\label{Raction}
\end{equation}
Defining now the canonical momentum $\pi_{\cal R} = z^2 {\cal R}'$ the Hamiltonian related 
to the action (\ref{Raction}) becomes
\begin{equation}
H^{(1)}_{\rm scal}(\tau) = \frac{1}{2}\int d^{3}x\,\, 
\biggl[ \frac{\pi_{{\cal R}}^2}{z^2} + z^2  (\partial_{i} {\cal R})^2 \biggr],
\label{hamscal1}
\end{equation}
and the Hamilton equations 
\begin{equation}
 \pi_{\cal R}' =  z^2 \nabla^2 {\cal R},\,\,\,\,\,\,\,\,\,\,\,{\cal R}' = \frac{\pi_{{\cal R}}}{z^2}.
\label{hameq}
\end{equation}
Equations (\ref{hameq}) can be combined in a single second order 
equation so that Eq. (\ref{seceq}) is recovered.

The  canonically conjugate momentum, $\pi_{{\cal R}}$ 
is related to the  
density contrast on comoving hypersurfaces, namely, in the case of a single scalar field 
source\cite{bardeen,press},
\begin{equation}
\epsilon_{\rm m} = \frac{ \delta \rho_{\varphi} + 3 {\cal H} (\rho_{\varphi} + p_{\varphi}) V}{\rho_{\varphi}} = 
\frac{a^2\delta\rho_{\varphi} + 3 {\cal H}\varphi' X}{a^2 \rho_{\varphi}},
\label{epsm}
\end{equation}
where the second equality can be obtained using that $\rho_{\varphi} + p_{\varphi} = {\varphi'}^2/a^2 $ and that the 
effective ``velocity'' field in the case of a scalar field is $ V = X/\varphi'$.
Making now use of Eq. (\ref{0isc}) into Eq. (\ref{00sc}), Eq. 
(\ref{epsm}) can be expressed as  
\begin{equation}
\epsilon_{\rm m} = \frac{ 2 M_{\rm P}^2 \nabla^2 \Psi}{a^2 \rho_{\varphi}} \equiv
 \frac{2}{3}\frac{\nabla^2\Psi}{{\cal H}^2},
\label{epm2}
\end{equation}
where the last equality follows from  Eq. (\ref{b1}).
From Eq. (\ref{epm2}), it also follows that 
\begin{equation}
\pi_{{\cal R}} = z^2 {\cal R}' \equiv - 6 a^2 {\cal H} \epsilon_{\rm m},
\label{piR}
\end{equation}
where Eq. (\ref{epm2}) has been used  together with the expression of ${\cal R}'$ coming from (\ref{eqR}).
Hence, in this description, while the canonical field is the curvature 
fluctuations on comoving spatial hypersurfaces, the canonical momentum is 
the density contrast on the same hypersurfaces.

To bring the second-order action in the simple form of Eq.  (\ref{Raction}),
various (non-covariant) total derivatives have been dropped. Hence, there 
is always the freedom of redeining the canonical fields through time-dependent functions of the background 
geometry. In particular the action (\ref{Raction}) can be rewritten in terms of the variable $q$ defined in 
Eq. (\ref{qtoR}). Then  \cite{muk}
\begin{equation}
S^{(2)}_{\rm scal}
 = \frac{1}{2} \int d^4 x \biggl[ {q'}^2 -2 \frac{z'}{z} q q' - (\partial_{i} q)^2 + \biggl(\frac{z'}{z}\biggr)^2 q^2\biggr],
\label{sca2}
\end{equation}
whose related Hamiltonian and canonical momentum are, respectively
\begin{equation}
H_{\rm scal}^{(2)}(\tau) = \frac{1}{2} \int d^{3} x \biggl[ \pi_{q}^2 + 2 \pi_{q} q + (\partial_{i} q)^2\biggr],~~~~{\rm and}~
\pi = q' - \frac{z'}{z} q.
\label{hamscal2}
\end{equation}
In Eq. (\ref{sca2}) a further total derivative term can be dropped, leading to another action:
\begin{equation}
S^{(3)}_{\rm scal} = \frac{1}{2} \int d^4 x \biggl[ {q'}^2 - (\partial_{i} q)^2 +\frac{z''}{z} q^2\biggr],
\label{sca3}
\end{equation}
and another Hamiltonian
\begin{equation}
H^{(3)}(\tau)_{\rm scal} = \frac{1}{2} \int d^{3} x \biggl[ \tilde{\pi}_{q}^2 +  (\partial_{i} q)^2 - \frac{z''}{z} q^2\biggr].
\label{hamscal3}
\end{equation}
where $\tilde{\pi} = q'$. 
As in the case of the Hamiltonians for the tensor modes,  Eqs. (\ref{hamscal1}), (\ref{hamscal2}) 
and (\ref{hamscal3}) are all related by canonical transformations. Furthermore, notice that, classically, 
the scalar end tensor Hamiltonians have exactly the same form in the case of power-law inflation.

\subsubsection{Weinberg's theorem}

Concerning the  the evolution of scalar fluctuations, 
a plausible question to ask is how many 
adiabatic solutions and how many non-adiabatic solutions 
are compatible, for instance, with the conservation of ${\cal R}$ and 
$\zeta$.  This question, usually approached within the
separate Universe picture \cite{bardeen2} (see also \cite{liddleetal} for a reintroduction 
of some of arguments given in \cite{bardeen2}),
 was recently addressed by Weinberg on a 
more solid ground 
in a series of papers \cite{wein1,wein2,wein3}.  The separate Universe 
picture amounts to stipulate that any portion of the Universe that is larger than 
$H^{-1}$ ($H$ being the Hubble rate) but smaller than the physical wavelength 
on the perturbation $a/k$ will look like a separate unperturbed Universe. 
In the following the separate Universe picture will not be invoked but 
the way of reasoning put forward in refs. \cite{wein1,wein2,wein3}
will be briefly outlined.
The hypotheses of the theorem are the following:
\begin{itemize}
\item{}   suppose that the theory of gravity is, for simplicity, of Einstein-Hilbert type;
\item{} suppose that the background geometry is given by a FRW metric which we can take 
for simplicity to be spatially flat;
\item{} suppose that the Universe is always expanding (possibly with different rates);
\item{} suppose that the sources of the geometry are relativistic fluids.
\end{itemize}
The last assumption is quite general and it allows for the presence of anisotropic stresses. 
From the mentioned assumptions it also follows that Eq. (\ref{evolR}) is fully valid. This 
is true for two separate reasons. The first reason is that Eq. (\ref{evolR}) is deduced 
within an Einsteinian theory of gravity. The second reason is that, since the Universe is 
assumed to be always expanding, the time-dependent coefficients  of the terms at 
the right hand side of Eq. (\ref{evolR}) are all non-singular.
Moreover, since fluids with different barotropic indices  can be simultaneously 
present, the non-adiabatic fluctuation of the pressure density, $\delta p_{\rm nad}$ 
can be present.  

The thesis of the theorem is that the evolution equations, in the longitudinal 
gauge description, have always a pair of physical solutions for which 
$\delta p_{\rm nad} \to 0$ and ${\cal R}$ approaches a constant 
for $k\to 0$. In other words, following the usual terminology, there will be 
always at least a pair of adiabatic solutions with $\delta p_{\rm nad}=0$ 
and ${\cal R}$ constant in the limit $ k \to 0$: one solution with ${\cal R} \neq 0$ and 
the other with ${\cal R}=0$.  In order to understand correctly the thesis 
of the theorem the limit $ k \to 0$ is crucial.  The second point to be borne in mind is 
 that there could also be {\em other} 
solutions with ${\cal R}' \neq 0$. These solutions, if allowed by the 
Hamiltonian constraint,  will be non-adiabatic. 

The theorem can be demonstrated  in general terms, as illustrated in Ref. 
\cite{wein1}.  For purposes 
of presentation it is sufficient to discuss it in the  simple case of a scalar field source. 
In this case, the evolution equations in the longitudinal gauge can be 
simply obtained from Eqs. (\ref{00sc})--(\ref{ijsc}) and (\ref{KG3}) by 
substituting the gauge-invariant fluctuations with  the corresponding longitudinal 
fluctuations, i.e. $\Phi\to \phi$, $\Psi \to \psi$ and $X \to \chi$.

Having done so, consider now the evolution equations in the limit $ k\to 0$ (or,
equivalently, the limit of vanishing spatial gradients). The key observation 
in order to prove the theorem in general terms is that these equations 
have an extra accidental symmetry for a particular class of coordinate 
transformations, i.e. 
\begin{equation}
\tau \to \tilde{\tau} = \tau + \epsilon_{0}(\tau), \,\,\,\,\,\,\,\,\,\,\,\,\, x^{i} \to \tilde{x}^{i} = x^{i} - \lambda x^{i},
\label{TW}
\end{equation}
where the parameter $\lambda$ is a space-time  constant. This coordinate transformation 
clearly induces a transformation in the fluctuations defined in the longitudinal gauge, in particular, 
using Eq. (\ref{T2}) and recalling the definitions of longitudinal fluctuations,
the transformations will be:
\begin{equation}
\phi\to \tilde{\phi} = \phi - {\cal H} \epsilon_{0} - \epsilon_{0}',\,\,\,\,\,\,\,\,\,\,\,\,\,\,\,\,\
\psi\to \tilde{\psi} = \psi + {\cal H} \epsilon_{0} - \lambda,\,\,\,\,\,\,\,\,\,\,\,\,\, \chi \to 
\tilde{\chi} = \chi - \varphi' \epsilon_{0}.
\label{WS0}
\end{equation}
While the gauge transformation for $\phi$ is exactly the one derived in Eq. (\ref{phi}), the 
transformation for $\psi$ derived on the basis of (\ref{TW}) differs from Eq. (\ref{psi}). There is 
no surprise for this occurrence: in fact we are here taking the limit $ k \to 0$. Hence, the part 
of the perturbed metric proportional to $k_{i} k_{j}$ (which would be generically induced by 
a coordinate transformation) is not induced by the specific coordinate transformation 
given in Eq. (\ref{TW}).  Therefore, from Eqs.  (\ref{WS0}) it is clear that a solution 
of the longitudinal gauge equations can be obtained, in the limit $k\to 0$, by setting 
\begin{equation}
\phi = - \epsilon_{0}' - {\cal H} \epsilon_{0}, \,\,\,\,\,\,\,\,\,\,\,\, \psi = {\cal H} \epsilon_{0} - \lambda, 
\,\,\,\,\,\,\,\,\, \chi = - \epsilon_{0} \varphi'.
\label{WS1}
\end{equation}
This solution can be generalized to the case when the sources are more general, namely, for instance, several fluids, several interacting scalar fields and so on.
The solution given in (\ref{WS1}) would simply be a pure gauge. However, it can be 
extended to the case of non-zero $k$ by looking at the conditions implied on $\epsilon_{0}$
by the components of the perturbed Einstein equations which are trivially 
satisfied in the limit $k \to 0$: these are the off-diagonal terms in the $(ij)$ 
equation, implying $\phi = \psi$ (in the absence of an anisotropic stress) 
and the momentum constraint of Eq. (\ref{0isc}) implying $ \psi' + {\cal H} \psi = 4\pi G \varphi' \chi$. 
The condition $\phi = \psi$ implies, according to Eq. (\ref{WS1}) that $\epsilon_{0}$ satisfies 
the following simple equation 
\begin{equation} 
\epsilon_{0}' + 2 {\cal H} \epsilon_{0} = \lambda,
\end{equation}
whose solution can be written as $\epsilon_{0} = \epsilon_{0}^{(1)} + \epsilon_{0}^{(2)}$ where 
\begin{equation}
\epsilon_{0}^{(1)} = \frac{\lambda}{a^2(\tau)} \int^{\tau} a^2(\tau') d\tau', \,\,\,\,\,\,\,\,\,\,\,\,\,\,\,\,\,
\epsilon_{0}^{(2)} = \frac{c_{2}}{a^2(\tau)},
\end{equation}
are, respectively, a particular solution of the inhomoheneous equation and the general solution 
of the homogeneous equation.
The solution expressed by $\epsilon_{0}^{(2)}$ can be also understood from the 
freedom of shifting the lower limit of the integral appearing in $\epsilon_{0}^{(1)}$. 

With this observation, the system of the longitudinal fluctuations can be solved. In order 
to appreciate the significance of the results to be obtained, it is useful  to write down 
the system as a third order differential system in the variables 
$\psi_{k}$, $\chi_{k}$ and $ \chi_{k}' = f_{k}$. 
Using Eqs. (\ref{0isc}) and (\ref{KG3}) the system to be solved is, in Fourier space \footnote{In 
the following the index denoting Fourier transformed quantities will be restored 
since no ambiguity can arise with different subscripts.}, 
\begin{eqnarray}
&& \psi_{k}' = - {\cal H} \psi_{k} + 4 \pi G \varphi' \chi_{k} ,
\label{aex}\\
&& \chi_{k}' = f_{k},
\label{bex}\\
&& f_{k}' = - 2 {\cal H} f_{k} - k^2 \chi_{k} - \frac{\partial^2 W}{\partial\varphi^2} a^2 \chi_{k} - 
2 \psi_{k} \frac{\partial W}{\partial \varphi} a^2 + 4 \psi_{k}' \varphi' .
\label{cex}
\end{eqnarray}
In this case the coefficients of the various terms appearing in the system are 
continuous in a neighborhood of $k=0$. This condition guarantees that the 
would-be-gauge mode can be lifted to the status of physical solution in the limit 
$k \to 0$. In the general case (when the sources are more complicated than a 
single scalar field) this condition may not be satisfied in general but 
it is certainly rather plausible given the typical forms of energy-momentum 
tensors customarily employed in cosmological model building.

After repeated use of the background equations, (and
in particular of Eq. (\ref{b3s})), Eq. (\ref{cex}) implies for $ k =0$ that 
\begin{equation}
\psi_{k} = \frac{1}{a}\biggl( \frac{\chi_{k}}{\varphi'}a\biggr)'.
\label{WS2}
\end{equation}
Now, since 
\begin{equation}
\chi^{(1)}_{k} = - \epsilon^{(1)}_{0} \varphi',\,\,\,\,\,\,\,\,\,\,\, 
\chi^{(2)}_{k} = - \epsilon^{(2)}_{0} \varphi',
\label{chi12}
\end{equation}
 we will also have that
\begin{eqnarray}
&& \psi^{(1)}_{k} = - \lambda  + \lambda {\cal H} \int^{\tau} a^2(\tau') d\tau',
\label{psi1}\\
&& \psi^{(2)}_{k} = c_{2} \frac{\cal H}{a^2}.
\label{psi2}
\end{eqnarray}
Knowing what are $\chi^{(1)}_{k}$, $\chi^{(2)}_{k}$, Eqs. (\ref{psi1}) and (\ref{psi2}) 
allow to determine what is ${\cal R}_{k}$ in the limit $k \to 0$  from the, by now familiar, relation
\begin{equation}
{\cal R}^{(1,2)}_{k} = - \psi_{k}^{(1,2)} - \frac{{\cal H}}{\varphi'}\chi_{k}^{(1,2)}.
\label{familiar}
\end{equation}
In particular, inserting Eqs. (\ref{chi12}) and (\ref{psi1})-(\ref{psi2}) into Eq. (\ref{familiar}) we will have 
\begin{equation}
{\cal R}_{k}^{(1)} = \lambda,\,\,\,\,\,\,\,\,\,\,\,\,\,\,{\cal R}_{k}^{(2)} = 0.
\end{equation}
The result is: for $k \to 0$ there always two adiabatic modes, one with ${\cal R} $ constant and the 
other with ${\cal R}$ vanishing.  Since the system of Eqs. (\ref{aex})--(\ref{cex}) 
is a third-order system ther will be $3$ independent solutions. However, 
the Hamiltonian constraint eliminates one, the non-adiabatic, i.e. the one 
leading to ${\cal R}' \neq 0$ for $ k \to 0$.
If instead of one scalar field there are $N_{\rm s}$ scalar 
degrees of freedom the theorem applies as well. The system of Eqs. (\ref{aex})--(\ref{cex}) 
can be generalized to this case with the result that the number of independent equations will 
be $ (2 N_{\rm s} + 1)$. Of these $(2 N_{\rm s} + 1)$,  $2$ are adiabatic and a third one is eliminated 
by the Hamiltonian constraint.  Thus, the remaining $2 (N_{\rm s} -1)$ are non-adiabatic modes. 

This perspective was further developed in \cite{wein3} (see also \cite{wein2})  with the 
purpose of discussing the fate of non-adiabatic modes in multi-field inflationary models. 
As remarked in \cite{wein4}, in a related context, the conservation laws for ${\cal R}$ 
and its analog for the tensor modes of the geometry bears some similarities with the 
Goldstone theorem \cite{chenglee,gold2} normally employed in relativistic quantum field theories. 
The modes for which ${\cal R}$ or $h_{ij}$ are constant outside the horizon take 
the place here of the Goldstone bosons that become free particles for 
long-wavelength.

As a final side remark, it is interesting to mention that the derivation of the present 
results has been conducted in the longitudinal 
coordinate system. What would happen in a different 
coordinate system, like, for instance, the synchronous? 
The answer to this question is that different gauges may suggest 
that different quantities are constant outside the horizon \cite{wein1}. 
However, this does not imply that different gauges are inequivalent 
as fare as the evolution of physical quantities are concerned. The reason is 
that the limit $k \to 0$ may have different meanings in different gauges \cite{wein1}.

\subsection{Divergences of specific gauge descriptions}

In the study of the evolution of gravitational perturbations a problem often 
encountered is represented by possible ``divergences" in a specific  
set of variables used to describe a given physical system.
This often indicates that the description of the fluctuations 
should be performed in an alternative coordinate system where 
divergences do not necessarily appear.

The first example is the neutrino isocurvature velocity mode. 
As it was analytically shown in section 2, the metric fluctuations 
of the neutrino isocurvature velocity mode are perfectly 
regular in the synchronous gauge description (see Eq. (\ref{metrivm}))
while the longitudinal fluctuations of the metric are singular in the 
limit $k \tau \to 0$ (see Eq. (\ref{divlong1})) for a non-vanishing fractional 
contribution of neutrinos to the radiation background, i.e. $ R_{\nu} \neq 0$.
The neutrino isocurvature velocity mode is perfectly  
physical. However,  the longitudinal coordinate system is not suitable to treat this mode.
As discussed before, a natural set of gauge-invariant variables seems to be, 
the Bardeen potentials $\Phi$ and $\Psi$. In the mentioned example 
they will diverge too. 

Gauge-invariant quantities are useful, however, their 
physical interpretation has always to be discussed within a specific gauge. 
Consider, for instance, the simple case of a  scalar field  (with quadratic potential) 
oscillating  in a FRW geometry with zero spatial curvature. The  background equation for $\varphi$ 
follows from Eq. (\ref{b3s}) and it is 
\begin{equation}
\varphi''  + 2 {\cal H} \varphi' + m^2 a^2  \varphi=0.
\label{oscscal}
\end{equation}
The solution of Einstein equations for an oscillating scalar in a quadratic potential 
are well understood \footnote{Similar problems may arise 
with quartic potentials and with more general polynomial potentials in the 
vicinity of a local minimum. For instance, in the case of a quartic potential 
Eq. (\ref{oscscal}) admits oscillatory solutions which are not sinusoidal 
but rather given by elliptic functions.}
(see, for instance, \cite{turnerc,staro1}). From the solution of the system it  is easy to argue
 that $a^2 (p_{\varphi} + \rho_{\varphi}) = {\varphi'}^2$  goes to zero with  $\varphi'$. 
For conformal times $ {\cal H }< m a $ the solution of Eq. (\ref{oscscal}) 
is simply given by a sinus (or a cosinus) damped by the expansion of the Universe as $a^{3/2}$ 
so that, averaged over many oscillations, $ \langle p_{\varphi} \rangle =0$. This implies that 
the evolution of the scale factor will be $\tau^2$ with oscillating corrections which also 
show up in ${\cal H}$ and ${\cal H}'$. If we would insist to study the fluctuations 
of this system with a decoupled equation for the Bardeen potential (see Eqs. (\ref{INTPSI3})--(\ref{INTPSI4})),
we would soon realize that because of the divergences in the pump field when $ \varphi' \to 0$ 
the solutions will also diverge. Exactly the same occurrence can be verified in the longitudinal gauge where 
the decoupled evolution equation for $\psi$ exactly coincides with Eq. (\ref{INTPSI3}). This problem has been 
discussed in the literature in various frameworks.  In  \cite{kodham1,kodham2} this patology 
of Eq. (\ref{INTPSI3}) has been pointed out and analyzed.  In \cite{taruya1} and \cite{hamaz1,hamaz2} 
the analysis of the fluctuations in the case of multiple oscillating (and interacting) fields has been performed. 
The knowledge of this pathology and of its resolution is now common knowledge in pre-heating studies
\cite{robert1,robert2,staro2,staro3,taruya3,bassett,tsuji1,robert3,tsuji2,gordon6}.   To solve the puzzle, the ``trick"
is to integrate directly the evolution equation for $q$ (see Eq. (\ref{QEQ})) the practical observation is 
then that the pump field of $q$, i.e. $z''/z$, can be re-expressed, by repeated use of the background equations,
in a form which does not contain inverse powers of $\varphi'$ \cite{kodham1,kodham2}. This allows 
easily to compute, for instance, the curvature perturbation, i.e. ${\cal R}$, 
in simple cases like such as the ones  of quadratic or quartic potential \cite{robert1,robert2}.  
Again, the rationale for the resolution of this pathology can be traced back to a gauge choice.
In the longitudinal gauge $\psi$ (and its gauge-invariant generalization, $\Psi$) measures
the curvature fluctuations on zero-shear hypersurfaces. We are then forcing the shear to be zero 
in a case where this choice is questionable.  

Motivated by the divergence of the longitudinal gauge description, let us 
discuss the problem in a different gauge, such as the off-diagonal (or uniform curvature) already 
introduced in section 2.  The evolution equations of the fluctuations in the off-diagonal gauge 
can be simply obtained from the results of the appendix by setting $\psi = E =0$  in Eqs. (\ref{dg00})--(\ref{dg0i}) 
and (\ref{enmomsc1})--(\ref{enmomsc3}). The resulting evolution equations couple then together 
$\phi$, $B$ and $\chi$ which are, respectively, the $(00)$  component of the perturbed metric, the off-diagonal entry 
$(0i)$ of the perturbed metric and the scalar field fluctuation. The relevant set of evolution equations 
can then be written, in the case off a quadratic potential and in Fourier space, as:
\begin{eqnarray}
&& {\cal  H} [ \phi' - \nabla^2 B] = 8\pi G \chi' \varphi',
\label{00+}\\
&& {\cal H}[ \phi' + \nabla^2 B] + 2 ( {\cal H}' + 2  {\cal H}^2) \phi = 8\pi G m^2 a^2 \chi,
\label{00-}\\
&& {\cal H} \phi= 4\pi G \varphi' \chi,
\label{ODC}\\
&& \phi = - B' - 2 {\cal H} B,
\label{ODij}\\
&& \chi'' + 2 {\cal H} \chi' - \nabla^2 \chi + \biggl[ m^2 a^2 + 2 \varphi' \biggl( \frac{{\cal H}'}{{\cal H}^2}
+ 2 \biggr) \biggr] \chi =0.
\label{ODKG}
\end{eqnarray}
Equations (\ref{00+}) and (\ref{00-}) are given, respectively, by the sum and by the 
difference of the $(00)$ and $(ii)$ components of the perturbed Einstein equations. 
Eqs. (\ref{ODC}) and (\ref{ODij}) are the momentum constraint and the $(i\neq j)$ 
component of the perturbed Einstein equations. Finally Eq. (\ref{ODKG}) is 
the perturbed Klein-Gordon equation (see Eq. (\ref{KG2}) where the other equations 
have been used in order to get rid of the dependence upon the metric fluctuations. 

Equation (\ref{ODKG}), being decoupled, 
can be swiftly solved since its pump field does not contain any dependence upon inverse 
powers of $\varphi'$. Then the result can be inserted into Eq. (\ref{ODC}) that 
is, again, free of divergences and this leads to the explicit value of $\phi$. Finally, direct 
integration of Eq. (\ref{ODij}) leads to the evolution of the remaining perturbation 
variable, i.e. $B$. Notice that, in the off-diagonal gauge, $q = a \chi$: this can be 
explicitly verified since by repeated use if the background equations, Eq. (\ref{ODKG}) 
can be brought to the same form of Eq. (\ref{QEQ}).
Thus,  the oscillating scalar field is a perfectly physical problem that requires, however, 
a suitable coordinate system to be treated.

The third example to be discussed is the one of a contracting Universe whose 
dynamics is driven by the kinetic energy of a scalar field. The evolution of the background can be 
easily obtained from Eqs. (\ref{b1s})--(\ref{b3s}) by setting $W(\varphi)=0$. In this case it is
easy to show that ${\cal H}' = - 2 {\cal H}^2$. The scale factor will describe 
then an accelerated contraction for $\tau <0$, i.e. $a(\tau) \simeq (-\tau)^{1/2}$.
Also now, as in the previous two cases, the longitudinal description
breaks down, in particular, the evolution equation of $\psi$ becomes, in this case 
\begin{equation}
\psi'' + 6 {\cal H} \psi' - \nabla^2 \psi=0.
\label{psidiv}
\end{equation}
The solution for $\psi$ diverges as $|\tau|^{-2}$ for $\tau \to 0^{-}$.
This is the typical situation of the case of minimal pre-big bang models in  four 
space-time dimensions.  As in the two previous situations, one can easily argue that 
the description of contracting backgrounds is not unphysical but should be treated 
with some care. A detailed analysis in this direction 
performed in the context of pre-big bang models shows indeed \cite{long2} 
that in the off-diagonal gauge the evolution can be treated perturbatively. Setting
$m^2 =0$ in Eq. (\ref{ODKG}) and recalling that ${\cal H}' = -2 {\cal H}^2$, we have 
that the evolution of the fluctuation of the scalar field 
fluctuation is simply 
\begin{equation}
\chi'' + 2 {\cal H} \chi' - \nabla^2 \chi=0,
\end{equation}
which also implies that cuvature fluctuations are constant (up to logarithmic corrrections).

The example of the accelerated contraction is particularly useful because, from a more accurate 
analysis, it mixes two different problem:
\begin{itemize} 
\item{} the first problem is how (i.e. in which coordinate system) one may describe the 
evolution of the curvature fluctuations in contracting geometries;
\item{} the second problem, more difficult and also more intriguing, cocnerns the singularity 
inherent to contracting backgrounds.
\end{itemize}

To motivate further the second statement we recall that 
in the limit $\tau \to 0^{-}$, the accelerated contraction driven by a minimally coupled scalar field 
leads to a singularity in the curvature.   This aspect can be simply appreciated from Eq. (\ref{b2s}),
implying, in the cosmic time description, that $\dot{H} = - 4\pi G \dot{\varphi}^2$. If the background 
makes a transition from contraction to expansion, $\dot{H}$ has to change sign and this is 
impossible to achieve just with a minimally couples scalar field. 
This kind of problem appeared in the literature thanks to the development 
of pre-big bang models (see \cite{PBB1,PBB2}). 
If the specific evolution of the model is not known in all the details,
it is rather plausible the the perturbations in the different regimes are matched in such a way that 
the super-Hubble conservation of ${\cal R}$ is enforced. Various reprises of these matching conditions
can be found in the literature \cite{DERU1,HV} for the case of conventional inflationary models. 

The same type of conservation laws obeyed in the ordinary inflationary models was also 
suggested for the pre-big bang case \cite{long2}. 
In \cite{maxb1,maxb2}  a specific model for the resolution of the singularity has been studied in detail 
in order to test the arguments put forward in \cite{long2}. 
The model is based on the low-energy  string effective action supplemented by a non-local 
dilaton potential whose effect is to regularize the evolution of the gauge coupling and of the curvature 
at short distances. In spite of the fact that the evolution equations of the background and of the 
linearized fluctuations are different, they reduce to the usual general relativistic equations 
for large times away from the maximal curvature regime. By studying 
both analytically and numerically the evolution of the fluctuations, 
it is found that the evolution of the curvature perturbations 
obeys  an equation that is similar to the one derived in a general relativistic context, i.e. Eq. (\ref{eqR}).
with an important difference:  in ordinary general relativity
 $4\pi G  \varphi'^2  = {\cal H}^2 - {\cal H}'$ while in case of Ref. \cite{maxb1,maxb2} 
 \begin{equation}
 4 ({\cal H}' - {\cal H}^2) + {\varphi'}^2 + \frac{\partial V}{\partial\overline{\varphi}} a^2 e^{\varphi} =0,
 \end{equation}
 in natural string units.
The first and the second term are the analog of the relation stemming from Friedmann 
equations, while the third term arises from the presence of a non local dilaton 
potential $V(\overline{\varphi})$ where $ \overline{\varphi}$ is the shifted dilaton, which 
is invaraint under scale factor duality \cite{PBB2}. Thus, when
$({\cal H}^2 - {\cal H}')=0$, i.e. at the would be singularity, the coefficient at the right-hand side 
of Eq. (\ref{eqR})  does not blow up but it is regular and finite. Thus  curvature perturbations are 
indeed nearly constant as specifically checked in various numerical examples \cite{maxb1,maxb2}.
The evolution of the tensor modes does also obey a generalization 
of the analog conservation law described earlier in this section.
The conservation of super-Hubble curavature perturbations 
has also been checked, recently, in the context of a regularization 
obtained by means of higher derivatives.
Similar results can be obtained 
 \cite{robertpbb1,robertpbb2,robertpbb3} (see also, more recently, \cite{CARTIERB}) 
 if the singularity problem is 
addressed by supplementing the low-energy string effective action with higher 
derivatives terms \cite{high1,high2}.

The ekpyrotic-cyclic models \cite{EKP1,EKP1a,EKP2} 
face situation which is, in some way, similar to the
one of pre-big bang model and some debates have been recorded in the literature
all along the last few years.  The proponents of the ekpyrotic scenario argue 
that curvature perturbations are not continuous across the transition 
from contraction to expansion \cite{EKP1a,gratton1,tolley1}. If curvature 
perturbations are conserved on super -Hubble scales, then, as in the pre-big bang case 
the spectrum of fluctuations of the minimal ekpyrotic model will be violet thought 
with a slightly different spectral index \cite{lythekp1,hwangekp1,lythekp2,hwangekp2} (see section 6). The proponents 
of the ekpyrotic scenario, on the contrary, suggest that the curvature perturbation 
is not the conserved quantity that should be used in order to match across
the transition from contraction to expansion.  More specifically the idea is that 
when branes collide curvature perturbations are not conserved across the bounce 
and they match to the nearly scale-invariant fluctuation of the Bardeen potential.
This type of arguments has been further amplified by some authors \cite{DV1,DV2}. 

As a consequence of the debate on what should be continuous across the transition
various models of general relativistic bounces have been studied.  In general 
relativity there are essentially two ways to get bounces depending on the 
presence (or absence) of spatial curvature. In the presence of (positive) spatial 
curvature Friedmann equations (supplemented by a single scalar field with suitable 
trigonometric potential) may lead to some sort of de Sitter bounce \cite{dsbounce1,dsbounce2} 
where the scale factor, in cosmic time, behaves as $a(t) \sim \cosh{(t/t_0)}$.
In the absence of spatial curvature, the only way to get a bounce is to 
violate the null energy condition. In practical terms $(\rho+ p)$ has to change sign. 
This requirement is impossible to achieve with a single scalar field 
since $(\rho + p) = \dot{\varphi}^2$, which is always positive. 

In the context of the ekpyrotic models some variation on the theme of de Sitter 
bounces has been considered for instance in \cite{gordonturok} and in \cite{peterbounce}.
The path of the violation of the null energy condition has been taken by another set of 
investigations (see, for instance, \cite{wandsallen} and references therein).  Just 
to give an example, in Ref. \cite{wandsallen} the bounce is implemented 
through a system of two minimally coupled scalar fields, say $\varphi_{1}$ and $\varphi_2$. One 
of the two scalars (for instance $\varphi_2$) 
 is a ghost, i.e. it has {\em negative} kinetic energy.  Thus, according to Friedmann equations 
 $\dot{H}$ will now be the difference if two positive definite quantities, i.e. $(\dot{\varphi}_{2}^2 - \dot{\varphi}_1^2)$.
 In this situation $\dot{H}$ changes sign and a bounce is possible provided ghost fields are allowed. 
 Notice that the model studied in \cite{wandsallen} is rather similar to the one discussed in \cite{peterneto}. The conclusions on the evolution of the fluctuations seem, however, to be  different.

\renewcommand{\theequation}{6.\arabic{equation}}
\section{Normalization and amplification of metric fluctuations}
\setcounter{equation}{0}
Cosmological fluctuations 
could be either classical or quantum mechanical.
Classical and quantum cosmological fluctuations  have common 
features hiding physical differences.
Differences and analogies  can be summarized as follows:
\begin{itemize}
\item{} in the linearized approximation, classical and quantum fluctuations obey the  
same evolution equations;
\item{} classical fluctuations are given once forever (on a given space-like hypersurface);
\item{} quantum fluctuations keep on reappearing all the time during the inflationary phase
because of the zero-point fluctuations of the various fields (metric inhomogeneities, perturbations 
of the inflaton, fluctuations of some spectator field);
\item{} the evolution  and normalization of quantum mechanical fluctuations 
can be consistently described both in the Heisenberg and in the 
Schr\"odinger representations.
\end{itemize}
A natural question to ask is wether the fluctuations 
affecting the SW effect are classical or quantum mechanical.
The answer to this question cannot be given in general 
terms but it depends on the model describing the early evolution of the Universe.

Consider, for simplicity, the framework of conventional 
inflationary models driven by a single scalar degree of freedom.
The minimal duration of inflation is   between $60$ and $65$ efolds.
The rationale for such a statement is that, today, the total curvature of the Universe 
receives a leading contribution from the extrinsic curvature and a subleading 
contribution from the intrinsic (spatial) curvature. The ratio between the 
intrinsic and extrinsic curvature (see, for instance, Eq. (\ref{b1ct}))
 goes as $1/\dot{a}^2$. During an epoch of 
decelerated expansion (i.e. $\ddot{a}<0$, $\dot{a} >0$) such as 
the ordinary radiation  and matter-dominated phases, $1/\dot{a}^2$ can become 
very large. The r\^ole of inflation is, in this context, to 
make $1/\dot{a}^2$ very minute at the end of inflation, so that it can easily be 
of order 1 today. The minimal duration of inflation required 
in order to achieve this goal is about $60$-efolds.

If the duration of inflation is minimal (or close to minimal) 
classical fluctuations, which were super-horizon sized at the onset
 of inflation will be affected neither by the inflationary phase nor by the subsequent 
post-inflationary epoch and can have computable large scale effects
like, for instance, the so-called Grishchuk-Zeldovich effect 
\cite{GZ,GZ2}. 
If the fluctuations are   classical,
there are, virtually no ambiguities in  normalizing them: it is sufficient 
to assign the values of the various inhomogeneities over a 
typical scale and at a given time.

\begin{figure}[tp]
\centerline{\epsfig{file=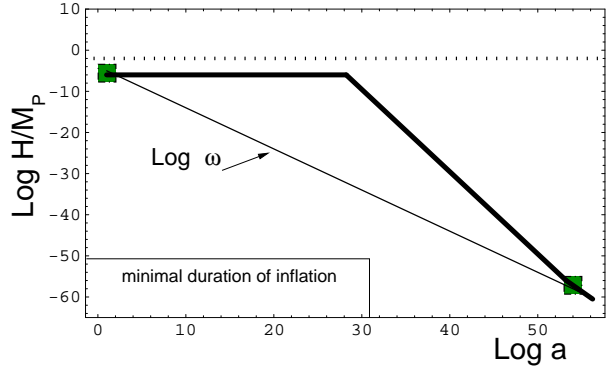, width=8cm}}
\vspace*{8pt}
\caption[a]{The same quantities illustrated in Fig. \ref{F52} are reported in the situation 
when the flat plateau lasts for about $65$ efolds.}
\label{F62}
\end{figure}

When the duration of inflation is 
much longer than $60$-efolds, the large scale 
fluctuations are probably all of quantum-mechanical origin, 
at least in the case of inflationary models driven by a single inflaton field.
Quantum-mechanical fluctuations result 
from the zero-point energy of the metric inhomogeneities (and possibly 
of other quantum fields) 
present during the inflationary epoch. 
The distinction between classical and quantum cosmological fluctuations, 
 can be appreciated by comparing Figs. \ref{F52} and \ref{F62}.

In both figures, a schematic evolution of the logarithm (in ten basis) 
of the Hubble rate (in 
Planck units) is reported both for the case of non-minimal duration of the inflationary 
phase (Fig. \ref{F52}) and in the case of minimal duration (Fig. \ref{F62}).  On the 
horizontal axis the logarithm of the scale factor is reported.
The inflationary region corresponds, in both figures, to the quasi-flat plateau
lasting  $82$ efolds (in Fig. \ref{F52}) and  $65$ efolds (in Fig. \ref{F62}). These 
numerical values are purely illustrative. 

If the duration of inflation is non-minimal
a classical fluctuation present  very close to the beginning of inflation will 
make the second crossing of the Hubble radius in the far future (see Fig. \ref{F52}, thin 
full line). Today, $H/M_{\rm P} \simeq 10^{-61}$ while during inflation $H/M_{\rm P} \simeq 10^{-5}$--$10^{-6}$.
Then it is clear that a hypothetical 
classical fluctuations crossing the horizon close to the beginning of an inflationary phase 
(lasting much more than $60$ efolds) will not cross the horizon the second time 
prior to the present epoch (see Fig. \ref{F52}). Conversely (see Fig. \ref{F62}) 
if the duration of inflation is close to minimal the latter occurrence is not excluded.
In the following, after a detailed discussion of the standard lore, some unconventional 
approaches to the normalization of quantum fluctuation will be reviewed. 

\subsection{The standard lore for the normalization}

The evolution of the metric fluctuations can be described either within
the Schr\"odinger description,  or within the Heisenberg description.
In the Schr\"odinger description,  the evolution can be pictured as the spreading 
of a quantum mechanical wave-functional. The initial wave-functional 
will be constructed as the direct product of states minimizing the 
indetermination relations for the different harmonic oscillators forming the quantum 
field. The initial vacuum state 
will have zero momentum and each Fourier mode of the field will 
evolve into a (two mode) 
squeezed quantum state where the indetermination relations 
are still minimized but in such a way that one of the two canonically 
conjugate operators will have a variance much larger than the quantum limit, 
while the other canonically conjugate operator will have a variance 
much smaller than the quantum uncertainty limit. The description 
of the evolution of cosmological fluctuations by means of the 
squeezed states formalism is rather instructive and has been exploited 
by various authors \cite{sq1,sq3,sq4}. In this formalism, some 
inspiring connections with quantum optics emerge (see for instance \cite{sq7}).  Squeezed 
states are a powerful generalization of the concept of coherent state \cite{sq8}.
As the  pumping action of a background  
electromagnetic field (a laser) is able to produce, under some 
circumstances, squeezed states of photons, the pumping action 
of the classical gravitational field (for instance the curvature) 
is able to produce squeezed states of  gravitons (for the tensor modes) 
or of phonons (in the case of the scalar modes). In the context of the squeezed 
states formalism a natural definition of a coarse grained entropy of the 
quantum fluctuations emerges. This entropy, in the analysis of \cite{sq9,sq10}, is associated 
with the spreading of the wave-functional in an appropriate basis. The coarse 
graining procedure can be performed within different basis with consistent 
results \cite{sq11,sq12,sq13,sq14,sq15,sq16}. 
The concept of entropy of cosmological fluctuations leads naturally 
to the problem of the transition from the quantum to the classical regime which can be 
in turn related to the decoherence of the density matrix of the cosmological 
fluctuations. The Schr\"odinger description proves also very useful for the self-consistent 
evolution of the inflaton and of its fluctuations \cite{sq17}.

In fully equivalent terms, the evolution of the fluctuations can be 
described within the Heisenberg representation. This is 
the description which will be adopted for the purposes of the presentation.
The Hamiltonians 
of the scalar and tensor modes of the geometry will then 
be minimized for  $\tau \to -\infty$, which  is a physical limit, not a mathematical one. 
In fact, inflation cannot last indefinitely in the past.  Since de Sitter space-time 
is not geodesically complete in the past\footnote{The analysis of the past geodesic 
(in)completeness of conventional inflationary models has been discussed, 
in a series of papers, by Borde and Vilenkin \cite{alex2,alex3,alex4} 
and by Borde, Guth and Vilenkin \cite{alex5}. }  , there is no reason to assume 
that $\tau$ should indeed be going to $-\infty$. 

\subsubsection{Large-scale power spectra of the tensor fluctuations}

Consider the Hamiltonian given in Eq. (\ref{H3t}), dropping, for simplicity,
the tildes in the momenta:
\begin{equation}
H(\tau) = \frac{1}{2} \int d^3 x \biggl[ \pi^2 - \frac{a''}{a} \mu^2+ 
(\partial_{i} \mu)^2\biggr],
\label{ham1a}
\end{equation}
where $\mu = a h_{\oplus} M_{\rm P}/\sqrt{2}$ and identically for the other polarization.

After imposing the  commutation relations for the canonically conjugate quantum fields (units $\hbar=1$ are 
adopted)
\begin{equation}
[ \hat{\mu}(\vec{x},\tau), \hat{\pi}(\vec{y},\tau)] = i \delta^{(3)} (\vec{x} - \vec{y}),
\label{cancomma}
\end{equation}
the operator corresponding to the  Hamiltonian (\ref{ham1a}) becomes, then, 
\begin{equation}
\hat{H}(\tau) = \frac{1}{2} \int d^3 x 
\biggl[ \hat{\pi}^2 - \frac{a''}{a} \hat{\mu}^2 
+  (\partial_{i} \hat{\mu})^2\biggr].
\label{ham2a}
\end{equation}
In Fourier space the quantum fields can be written as
\begin{eqnarray}
&&\hat{\mu}(\vec{x},\tau) = \frac{1}{2 (2\pi)^{3/2} } \int d^3 k \biggl[ \hat{\mu}_{\vk} e^{- i \vec{k} \cdot \vec{x} }
+ \hat{\mu}_{\vk}^{\dagger}  e^{ i \vec{k} \cdot \vec{x} }\biggr],
\nonumber\\
&& \hat{\pi}(\vec{x},\tau) = \frac{1}{2 (2\pi)^{3/2} } \int d^3 k \biggl[ \hat{\pi}_{\vk} e^{- i \vec{k} \cdot \vec{x} }
+ \hat{\pi}_{\vk}^{\dagger}  e^{ i \vec{k} \cdot \vec{x} }\biggr],
\label{expansiona}
\end{eqnarray}
Inserting now Eqs. (\ref{expansiona}) into Eq. (\ref{cancomma}) and demanding the validity 
of the latter implies the following canonical commutation relations 
for the Fourier components of the quantum operators:
\begin{eqnarray}
&& [ \hat{\mu}_{\vk}(\tau), \hat{\pi}_{\vp}^{\dagger}(\tau) ] = i \delta^{(3)}(\vec{k} - \vec{p}),
\nonumber\\
&&[ \hat{\mu}_{\vk}^{\dagger}(\tau), \hat{\pi}_{\vp}(\tau) ]= i \delta^{(3)}(\vec{k} - \vec{p}),
\nonumber\\
&& [ \hat{\mu}_{\vk}(\tau), \hat{\pi}_{\vp}(\tau) ]= i \delta^{(3)}(\vec{k} + \vec{p}),
\nonumber\\
&&[ \hat{\mu}_{\vk}^{\dagger}(\tau), \hat{\pi}_{\vp}^{\dagger}(\tau) ]= i \delta^{(3)}(\vec{k} + \vec{p}).
\label{fcomma}
\end{eqnarray}

Inserting now Eq. (\ref{expansiona}) into Eq. (\ref{ham2a}) we get the Fourier space representation
of the quantum Hamiltonian \footnote{ Notice that in order to derive the following equation, 
the relations $\hat{\mu}_{-\vk}^{\dagger} \equiv  \hat{\mu}_{\vk}$ and 
$\hat{\pi}_{-\vk}^{\dagger} \equiv  \hat{\pi}_{\vk}$ should be used .}:
\begin{equation}
\hat{H}(\tau) = \frac{1}{4} \int d^3 k \biggl[(\hat{\pi}_{\vk} \hat{\pi}^{\dagger}_{\vk} + 
\hat{\pi}_{\vk}^{\dagger} \hat{\pi}_{\vk}) + \biggl( k^2 - \frac{a''}{a}\biggr) (\hat{\mu}_{\vk} \hat{\mu}^{\dagger}_{\vk} + 
\hat{\mu}_{\vk}^{\dagger} \hat{\mu}_{\vk}) \biggr],
\label{ham3a}
\end{equation} 
In the Heisenberg representation the field operators obey:
\begin{eqnarray}
i \hat{\mu}' = [\hat{\mu},\hat{H}],
\label{mueq1}\\
i \hat{\pi}' = [\hat{\pi},\hat{H}].
\label{pieq1}
\end{eqnarray}
Using now the mode expansion (\ref{expansiona}) and the Hamiltonian in the form (\ref{ham3a})
the evolution for the Fourier components of the operators is 
\begin{eqnarray}
&&\hat{\mu}_{\vk}'= \hat{\pi}_{\vk} ,
\label{mueq2a}\\
&& \hat{\pi}_{\vk}' = - \biggl( k^2  - \frac{a''}{a}\biggr) \hat{\mu}_{\vk}.
\label{pieq2a}
\end{eqnarray}
The general solution of the system is then 
\begin{eqnarray}
&& \hat{\mu}_{\vk}(\tau) = \hat{a}_{\vk}(\tau_0) f_{k}(\tau) + \hat{a}_{-\vk}^{\dagger}(\tau_0) f^{\ast}_{k}(\tau),
\label{solmu}\\
&& \hat{\pi}_{k}(\tau) = \hat{a}_{\vk}(\tau_0) g_{k}(\tau) + \hat{a}_{-\vk}^{\dagger}(\tau_0)g^{\ast}_{k}(\tau),
\label{solpi}
\end{eqnarray}
where the mode function $f_{k}$ obeys 
\begin{equation}
f_{k}'' + \biggl[ k^2 - \frac{a''}{a}\biggr] f_{k} =0,
\label{fkeq}
\end{equation}
and $g_{k} = f_{k}'$.  In the case when the scale factor has a power dependence, in cosmic 
time, as in Eq. (\ref{pot}) , the scale factor will be, in conformal time 
$a(\tau) = (-\tau/\tau_{1})^{-\beta}$ with $\beta = p/(p -1)$. The solution of Eq. (\ref{fkeq})  is then 
\begin{eqnarray}
f_{k}(\tau) &=& \frac{{\cal N}}{\sqrt{2 k}} \sqrt{-x} H^{(1)}_{\mu}(- x),
\label{fk}\\
g_{k}(\tau) &=& = f_{k}'=  - {\cal N}\sqrt{\frac{k}{2}} \sqrt{-x}\biggl[ H^{(1)}_{\mu -1} (-x) +
\frac{(1 -2 \mu)}{2(-x)} H^{(1)}_{\mu} (-x)\biggr] ,
\label{gk}
\end{eqnarray}
where $ x =  k\tau$ and 
\begin{equation}
{\cal N} = \sqrt{\frac{\pi}{2}} e^{\frac{i}{2}(\mu + 1/2)\pi},~~~~~\mu = \beta+\frac{1}{2}.
\end{equation}
The functions $H^{(1)}_{\mu}(-x) = J_{\mu}(-x)+ iY_{\mu}(-x)$ is the Hankel function \cite{abra1,abra2}
of first kind and the other linearly independent solution will be $H^{(2)}_{\mu} =  {H^{(1)}}^{\ast}_{\mu}$.
Notice that the phases appearing in Eqs. (\ref{fk}) and (\ref{gk}) are carefully selected in such a 
way that for $\tau \to -\infty$, $ f_{k} \to e^{- i\,k\,\tau}/\sqrt{2 k} $ \cite{abra1}.

In the case of more general inflationary potentials the solution of Eq. (\ref{fkeq}) is also 
reasonably simple and it has the same form given in Eq. (\ref{fk}) and (\ref{gk}). However, this 
time, using the slow-roll formalism, the index $\mu$ will be given by Eq. (\ref{musl}).

In Eqs. (\ref{solmu}) and (\ref{solpi}) the operators  $\hat{a}_{\vk}(\tau_{0})$ and $\hat{a}_{-\vk}(\tau_{0}) $
annihilate the vacuum, i.e. the state minimizing the Hamiltonian  (\ref{ham3a}) 
for  $\tau_{0} \to -\infty$ 
which is chosen, as anticipated, close to the onset of inflation. The time $\tau_{0}$ is the same
for different modes $k$. 
The second remark is that, the vacuum, i.e. the state such that $\hat{a}_{-\vk}(\tau_{0})\,\,\hat{a}_{\vk}(\tau_{0})| 0_{\infty} \rangle =0$,  has zero total momentum. This can be understood in terms of a process of pair production
from the vacuum triggered by the pumping action of the gravitational field.  Inserting  Eqs. (\ref{solmu}) and 
(\ref{solpi}) into Eq. (\ref{expansiona}) the following Fourier expansions can be simply obtained 
\begin{eqnarray}
&&\hat{\mu}(\vec{x},\tau) = \frac{1}{ (2\pi)^{3/2} } \int d^3 k \biggl[ \hat{a}_{\vk}(\tau_0) f_{k}(\tau)
 e^{- i \vec{k} \cdot \vec{x} }
+ \hat{a}_{\vk}^{\dagger}(\tau_0) f^{\ast}_{k}(\tau) e^{ i \vec{k} \cdot \vec{x} }\biggr],
\nonumber\\
&& \hat{\pi}(\vec{x},\tau) = \frac{1}{ (2\pi)^{3/2} } \int d^3 k \biggl[ \hat{a}_{\vk}(\tau_0)g_{k}(\tau)
 e^{- i \vec{k} \cdot \vec{x} }
+ \hat{a}_{\vk}^{\dagger}(\tau_0) g^{\ast}_{k}(\tau) e^{ i \vec{k} \cdot \vec{x} }\biggr],
\label{expansion2}
\end{eqnarray}
since, upon integration, 
 $\hat{a}_{-\vk}e^{i \vec{k}\cdot \vec{x} }\equiv \hat{a}_{\vk}e^{- i \vec{k}\cdot \vec{x} }$  and 
 $\hat{a}_{-\vk}^{\dagger}e^{ -i \vec{k}\cdot \vec{x} }\equiv \hat{a}_{\vk}^{\dagger}e^{ i \vec{k}\cdot \vec{x} }$.

The two point function for the tensor fluctuations can then be obtained by computing 
\begin{equation}
\langle 0_{-\infty }| \hat{h}(\vec{x},\eta) \hat{h}(\vec{y},\eta) | 0_{-\infty}\rangle \equiv \frac{2 \ell_{\rm P}^2}{(2 \pi)^3\,\,\,a^2(\tau)} \int d^{3} k \,\,|f_{k}(\tau)|^2 e^{- i \vec{k}\cdot \vec{r}}
\end{equation}
where $ \vec{r} = (\vx - \vy)$. After angular integration, the previous expression 
becomes 
\begin{equation}
\langle 0_{-\infty}| \hat{h}(\vec{x},\eta) \hat{h}(\vec{y},\eta) | 0_{-\infty}\rangle = 
\int d\ln{k} \,\,{\cal P}_{\rm T} \,\,\frac{\sin{k r} }{kr}
\end{equation}
where 
\begin{equation}
{\cal P}_{\rm T} = \frac{ 2 \ell_{\rm P}^2}{a^2(\tau)} \frac{k^3}{2 \pi^2} |f_{k}(\tau)|^2,
\label{PT}
\end{equation}
is the tensor power spectrum.
Taking the small argument limit  \cite{abra2} of the Hankel functions appearing in Eq.  (\ref{PT}),
the square root Eq. (\ref{PT}) becomes   
\begin{equation}
{\cal P}_{\rm T}^{1/2} = \frac{2^{\beta}}{\pi^{3/2}} \Gamma\biggl(\beta + \frac{1}{2}\biggr) |k \tau_{1}|^{(1 - \beta)}
\frac{\ell_{\rm P}^2}{\tau_{1}^2},
\label{PT2}
\end{equation}
where Eq. (\ref{PT}) has been multiplied by $2$ to account for the $2$ poalrizations.

Since, by definition, $k_{1} =| a_1 H_{1}| = \beta/\tau_{1}$,  and bearing in mind 
that $\beta = p/(p-1)$,  Eq. (\ref{PT2}) leads to 
\begin{eqnarray}
&& {\cal P}_{\rm T}^{1/2} =  {\cal C}_{\rm T}(p)  \frac{H_1}{M_{\rm P}} \biggl(\frac{k}{k_{1}}\biggr)^{- 1/(p-1)},
\label{PT3}\\
&& {\cal C}_{\rm T}(p) = \frac{2^{p/(p-1)}}{\pi^{3/2}}
 \Gamma\biggl(\frac{3 p -1}{2(p-1)}\biggr) \biggr[ \frac{p-1}{p}\biggl]^{p/(p-1)}.
\label{CT1}
\end{eqnarray}

In the case of generic slow-roll inflation we will have \footnote{Notice that as a consequence of the 
different (conventional) definitions of the perturbed line element Ref. \cite{SL} arrives at a power spectrum which is 
$1/2$ of our result  (see also previous footnote).}:
\begin{equation}
{\cal P}_{\rm T}^{1/2}  = \frac{2^{\mu -1/2}}{\pi^{3/2}}\Gamma(\mu) ( 1 -\epsilon)^{\mu - 1/2} \biggl(\frac{H}{M_{\rm P}}\biggr)_{aH = k}.
\label{slT}
\end{equation}
where $\mu$ is now determined by Eq. (\ref{musl}).  

In the study of slow-roll dynamics the tensor spectral index is usually defined as 
\begin{equation}
n_{\rm T} = \frac{d \ln{ {\cal P}_{\rm T}}}{d\ln{k}} = -2 \epsilon - ( 3 + c_1) \epsilon^2 - (1 + c_{1}) \epsilon \eta,
\label{NT}
\end{equation}
where the second equality follows from the expansion of the obtained result in terms 
of the slow-roll parameters which are small deep within the inflationary phase; the constant
is biven by $c_{1} = 0.0814$. It is 
an exercise to perform directly the derivative of $\ln{{\cal P}}_{\rm T}$ with respect 
to $\ln{k}$. To obtain the correct result it is useful to employ Eqs. (\ref{sl2}) and (\ref{sl3}).
For instance, recalling that $k= H a$, it can be easily shown that 
\begin{equation}
\frac{d \ln{k}}{d\varphi} = \frac{d \ln{H}}{d\varphi} + \frac{d\ln{a}}{d\varphi} \equiv 
\frac{1}{a} \frac{d a}{d\varphi}- \frac{M_{\rm P}^2}{2} \frac{H}{H_{,\varphi}}
\end{equation}
where the comma followed by $\varphi$ denotes a derivation with respect to $\varphi$ 
and where the second equality follows from Eq. (\ref{sl3}). Introducing the explicit 
definition of slow-roll parameters it can be also shown that 
\begin{equation}
\frac{d \ln{k}}{d\varphi} = \frac{M_{\rm P}}{\sqrt{2 \epsilon}} (\epsilon-1).
\end{equation}

\subsubsection{Large-scale power spectra of the scalar fluctuations}

The calculation of the scalar power spectra follows exactly the same
 algebra discussed in the case of the tensor modes.
In full analogy with the calculation of the tensor power spectrum the Hamiltonian 
given in Eq. (\ref{hamscal3}) can be chosen. Promoting the normal mode 
$q$ and its conjugate momentum to quantum operators, the relevant 
commutation relations are now:
\begin{equation}
[\hat{q}(\tau,\vec{x}), \hat{\pi}_{q}(\tau,\vec{y})]  = i \delta^{(3)}(\vec{x} -\vec{y}).
\end{equation}
Recalling the relation between the normal modes and the gauge-invariant 
curvature fluctuations, obtained in Eq. (\ref{qtoR}), we can then write the scalar 
two-point function as 
\begin{equation}
\langle 0_{-\infty} | \hat{{\cal R}}(\tau,\vec{x}) \hat{{\cal R}}(\tau,\vec{y})| 0_{-\infty}\rangle =
\frac{1}{z^2} \int \frac{d^{3} k}{(2\pi^3)} |f_{{\rm s}\,\,k}(\tau)|^2 e^{-i \vec{k} \cdot \vec{r}},
\label{TWOPSCAL}
\end{equation}
where $f_{k}^{q}(\tau)$ are the mode functions pertaining to the scalar problem and 
obeying the equation
\begin{equation}
f_{{\rm s}\,\,k}'' +\biggl[k^2  - \frac{z''}{z}\biggr] f_{{\rm s}\,\,k}=0.
\label{solfs}
\end{equation}
While in the power-law case the solutions of Eq. (\ref{solfs}) are exactly 
the same as Eq. (\ref{fk}), in the in the generic case $f_{{\rm s}\,\,k}$ will be 
\begin{eqnarray}
f_{{\rm s}\,\,k}(\tau) &=& \frac{{\cal N}_{\rm s}}{\sqrt{2 k}} \sqrt{-x} H^{(1)}_{\nu}(- x),
\label{fksca} \\
{\cal N} &=& \sqrt{\frac{\pi}{2}} e^{\frac{i}{2}(\nu + 1/2)\pi},~~~~~\nu= \frac{3 + \epsilon + 2 \eta}{2( 1 -\epsilon)} .
\end{eqnarray}
where, as usual, $ x =  k\tau$.

After having performed the angular integration in Eq. (\ref{TWOPSCAL}) the scalar power 
spectrum is then 
\begin{eqnarray}
&& {\cal P}_{\cal R}^{1/2} =  {\cal C}_{\cal R}(p)  \frac{H_1}{M_{\rm P}} \biggl(\frac{k}{k_{1}}\biggr)^{- 1/(p-1)}
\label{PR3}\\
&& {\cal C}_{\cal R} (p) = \sqrt{\frac{p}{2}}\frac{2^{1/(p-1)}}{\pi^{3/2}}
 \Gamma\biggl(\frac{3 p -1}{2(p-1)}\biggr) \biggr[ \frac{p-1}{p}\biggl]^{p/(p-1)},
\label{CT2}
\end{eqnarray}
in the case of power-law inflation and 
\begin{equation}
{\cal P}_{{\cal R}}^{1/2}  = \frac{2^{\nu -3/2}}{\pi^{3/2}}\Gamma(\nu) ( 1 -\epsilon)^{\nu - 1/2} \biggl(\frac{H}{\dot{\varphi}M_{\rm P}}\biggr)_{aH = k}.
\label{SLR}
\end{equation}
for the general slow-roll case whose related pump field has been computed in Eq. (\ref{nuscal}).
Notice that, within our notations ${\cal P}^{1/2}_{\rm T}/{\cal P}^{1/2}_{\cal R} = 2\sqrt{2} \epsilon$.

As in the case of the tensor power spectrum, the spectral index is conventionally written 
to lowest order as\footnote{Notice that slightly different definition of the slow-roll parameters 
are possible. Our results follow the conventions previously established in Eq. (\ref{sl1}) 
and are consistent with the ones of \cite{SL}. Some authors like to define, for 
instance, the second slow-roll parameter as $ \overline{\eta} = M_{\rm P}^2 V_{,\varphi\varphi}/V$.  In terms of our definitions we will then have $ \eta = \epsilon - 
\overline{\eta}$. This also implies that $n_{\rm s} = 1 - 6 \epsilon + 2 \overline{\eta}$.}
\begin{equation}
n_{\rm s} -1 = \frac{d \ln{{\cal P}_{\cal R}}}{d\ln{k}} = -4\epsilon - 2\eta .
\end{equation}

By comparing the scalar and tensor power spectra with the tensor spectral index in the 
slow-roll approximation we find that $n_{\rm T}= - (1/4){\cal P}_{\rm T}/{\cal P}_{{\cal R}}$.
These types of relations are often called consistency relations and play a relevant 
r\^ole in the attempts to reconstruct the inflaton pootential in single field 
inflationary models. 
Notice that some authors argue that it is possible to redefine the power spectra \cite{LLKCB} in such a 
way that the ratio of the newly defined spectra indices equals exactly $\epsilon$ (without 
the numerical pre-factor obtained within the present conventions). 

\subsection{Transplanckian problem(s)?}
Consider, for simplicity the case of power inflation parametrized in terms of the conformal
time coordinate $\tau$ as $a(\tau)  \sim (-\tau)^{-\beta}$ with $\beta >0$. If 
the standard normalization prescription is interpreted in strict mathematical 
terms, then it will happen that if the initial normalization time $\tau_{0}$ is sent to $-\infty$, a given physical 
frequency at the time $\tau_{0}$, $\omega = k/a(\tau_0)$ will become much larger than the Planck mass, or 
as often emphasized, transplanckian.  A common theme of various investigations in this direction 
of research is the observation that in the ``transplanckian" regime the precise description 
of the evolution of the metric fluctuations could be foggy.  These statements leave the room to 
various proposals which can be summarized as follows:
\begin{itemize}
\item{} the dispersion relations can be modified 
by transplanckian effect;
 \item{} the indetermination relations are modified;
 \item{} transplanckian physics induces the presence 
 of a new fundamental scale $M$ smaller than the Planck mass;
  observable effects in the scalar and tensor power spectra can be expected.
 \end{itemize}
The modifications 
invoked in the dispersion relations, as often acknowledged by the authors, have always some 
ad-hoc feature.  This possibility was extensively investigated, in a series of papers, 
by Brandenberger and Martin \cite{BM1,BM2,BM3}  as well as by other authors 
\cite{NIEM1,NIEM2,KOW,JAC,GRE1,TMB,BG}. The observation may be summarized 
by saying  that the wave-number 
of the fluctuation is modified in such a way that $k^2 \to k_{\rm eff}^2(k,\tau)$ \cite{BM1,BM2,BM3}.
Such modifications should be derived  from a suitable variational principle \cite{JAC} (but this is 
not completely obvious) and they can well be non-linear  \cite{NIEM2}. 
Some effort has been made in order to justify the existence of modified dispersion relations 
in various frameworks ranging from analog modifications arising in black-hole 
physics \cite{JAC} to quantum-Poincar\'e algebras applied to a cosmological setting \cite{KOW},
to non-commutative geometries \cite{GRE1, TMB}.  

The modifications 
of the indetermination relations are sometimes motivated 
by the scattering of superstrings at Planckian energies \cite{ACV1,ACV2,GRM1,GRM2}. However, 
in that context the modifications in the dispersion relations occurs, in 
critical dimensions ( $26$ in the case of the bosonic string);
the ``position" operator is an impact parameter of two colliding strings. In 
the transplanckian context, on the contrary, the modifications 
of the indetermination relations are studied in a four-dimensional 
context.  Tachyonic instabilities then occur (this is also 
a problem, in some cases, when dispersion relations 
are modified). This approach has been followed by various authors 
\cite{AK1,AK2,HS,Ho}.

One possible approach to the problem would be to assume that 
there is a fundamental scale $M < M_{\rm P}$ and that 
fluctuations should be normalized as soon as they ``exit" from the physical 
regime characterized by the scale $M$. This aspect is partially illustrated in Fig. \ref{F52}
where with the dotted line the new fundamental scale is reported. According to these 
types of proposals \cite{DAN1,DAN2,STA2} the modes of the 
quantum field should be normalized (at a finite conformal time $\tau_{0}$) as soon 
as the physical frequency  (with dashed 
line in Fig. \ref{F52}) crosses the dotted line. This crossing clearly occurs 
at different times for different comoving frequencies. 
In the case of power-law inflation, a given physical frequency $ \omega(\tau) = k/a(\tau)$ 
will ``cross" the scale $M$ at a characteristic time $\tau_{0}(k)$ determined as 
\begin{equation}
\omega(\tau_0) = \frac{k}{a(\tau_0)} \simeq M,\,\, \to\,\,\,\,\,\,\,\,\,\,\tau_{0}(k) = -\tau_{1} \biggl( \frac{M}{k}\biggr)^{\frac{1}{\beta}} \simeq - \tau_{1} \biggl(\frac{M}{k}\biggr)^{1 - \frac{1}{p}},
\label{CROSM}
\end{equation}
where $a(\tau)\sim (-\tau/\tau_{1})^{-\beta}$. The second equality in the second equation 
follows from the relation between the cosmic and conformal times in power-law backgrounds 
(see, for instance, before Eq. (\ref{fk})). In the expanding branch of de Sitter space-time 
$\beta=1$ and $p\to \infty$. Consequently $ k \tau_0$ is constant and roughly given by 
$M/H$ where $H$ is the Hubble rate during the de Sitter phase. Now the claim is that \cite{DAN1,DAN2}
this prescription for setting the initial condition for quantum fluctuations produces observable corrections 
to the tensor power spectrum \footnote{Here we discuss the tensor case since, for {\em exact} 
(expanding) de Sitter space only the tensor modes are excited. However, the same discussion 
hold also for the scalar power spectra in the quasi- de Sitter space-time.}  of the form
\begin{equation}
\overline{{\cal P}}_{{\rm T}} = {\cal P}_{{\rm T}}
\biggl[1 + c_{\gamma}\biggl(\frac{H}{M}\biggr)^{\gamma}+...\biggr],
\label{paramTP}
\end{equation}
where $c_{\gamma}$ is a numerical constant; ${\cal P}_{\rm T} $ is the 
tensor power spectrum obtained through the standard normalization 
prescription discussed earlier in this section. Oscillating terms multiplying 
the ratio $(H/M)^{\gamma}$ have been neglected (see below, however).
The ellipses in Eq. (\ref{paramTP}) stand for terms which are of higher order in $H/M$.
Concerning the parametrization given in Eq. (\ref{paramTP}) two comments are in order:
\begin{itemize}
\item{} it is clear that if the prescription (\ref{CROSM}) is applied the correction 
to the power spectrum has to be in the form (\ref{paramTP}) : the reason is that the power spectrum 
will now depend upon the new scale $x_0= k\tau_0 \sim M/H\gg 1$ that becomes the argument 
of Hankel functions whose limit for $x_0 \gg 1$ must indeed produce a result of the type (\ref{paramTP});
\item{}  the power $\gamma$ is crucial for the possible observational relevance 
of transplanckian effects so it becomes crucial to understand {\em what} controls the power $\gamma$.
\end{itemize}
 Some authors claim that $\gamma$ can be $1$ \cite{DAN1,DAN2}.
In this case the ratio $H/M$ will be larger than $H/M_{\rm P}$ that we know 
can be as large as $10^{-6}$--$10^{-5}$. If $\gamma=2$ or even $\gamma=3$ the correction
is not relevant. Different arguments have been put forward for values $ \gamma >1$
\cite{SHE1,SHE2,GRE2,GRE3,GRE4}.  Notice that often the discussions of these effects 
take place in pure de Sitter space (where, as already pointed out, scalar modes would be absent). In this context 
different vacua can be defined. They are connected by unitary transformations and 
generically indicated as $\alpha$ vacua \cite{alpha1,alpha2,alpha3,alpha4,alpha5}.
Flat space has a global time-like killing vector. This allows to define a Hamiltonian 
whose minimization defines a ``vacuum", i.e. the lowest energy eigenstate.
In de Sitter space there is no global time-like killing vector. So a globally conserved 
energy cannot be defined. It is however still possible to find states which are 
invariant under the connected part of the isometry group of de Sitter space, i.e. 
$SO(4,1)$. In the case of a free scalar field there is an infinite family of invariant 
states called $\alpha$ vacua since they can be distinguished 
by a single complex number $\alpha$.

\subsubsection{Minimization of canonically related Hamiltonians}

In the following the roots of the corrections to the power spectrum 
will be understood in terms of a much more mundane feature 
of the theory of cosmological fluctuations, i.e. the possibility, already 
discussed in section 5, 
of defining different Hamiltonians related by canonical transformations.
More specifically it will be demonstrated that the value of $\gamma$ given 
in Eq. (\ref{paramTP}) depends on  
{\em which Hamiltionian}  one wishes to be minimized at the moment 
when the given physical frequency cosses the new fundamental scale $M$ 
as explained in Eq. (\ref{CROSM}).  
By averaging the energy-momentum pseudo-tensor 
of the fluctuations of the geometry  over the state minimizing 
a given Hamiltonian at  $\tau_{0}(k)$, the energetic  content 
of the fluctuations can be estimated.  The result of such a calculation 
must be {\em smaller} that the energy density of the background geometry.
This chain of calculations has been recently presented in \cite{MGTP2} 
and \cite{MGTP1} and will be briefly reviewed here.

The first step is to show that,   depending on the which Hamiltonian 
is minimized at $\tau_{0}(k)$ (defined as in Eq. (\ref{CROSM}))
a different power of $\gamma$ is obtained in Eq. (\ref{paramTP}).
In order to be specific consider the case of conventional power-law inflationary models. 
 We will first derive the result for the cases of the Hamiltonians (\ref{hamscal1}) (for the scalar 
 fluctuations) and (\ref{H1t}) (for the tensor fluctuations). Then the results 
 of the minimization of the other two Hamiltonians (i.e. Eqs. (\ref{hamscal2})-(\ref{hamscal3}) 
 and Eqs. (\ref{H2t})-(\ref{H3t})) will be given.

Equation (\ref{hamscal1}) implies that the  canonical field is ${\cal R}$, i.e. 
the curvature perturbation. The canonical momentum is the density 
contrast as discussed in Eq. (\ref{piR}).
The Hamiltonian operator will be 
\begin{equation}
\hat{H}(\tau) = \frac{1}{2} \int d^3 x 
\biggl[ \frac{\hat{\pi}_{\cal R}^2}{z^2} + z^2 (\partial_{i} \hat{\cal R})^2\biggr].
\label{ham1ab}
\end{equation}
Repeating the same procedure outlined earlier in this section, 
Eq. (\ref{ham1ab}) can be written as 
\begin{equation}
\hat{H}(\tau) = \frac{1}{4  } \int d^3 k \biggl[\frac{1}{z^2}(\hat{\pi}_{\vk} \hat{\pi}^{\dagger}_{\vk} + 
\hat{\pi}_{\vk}^{\dagger} \hat{\pi}_{\vk}) + k^2 z^2  (\hat{{\cal R}}_{\vk} \hat{{\cal R}}^{\dagger}_{\vk} + 
\hat{{\cal R}}_{\vk}^{\dagger} \hat{{\cal R}}_{\vk}) \biggr].
\label{ham1b}
\end{equation} 
The Hamiltonian (\ref{ham1b}) can be written at $\tau_{0}(k)$ as 
\begin{equation}
\hat{H}(\tau_0) =\frac{1}{4}\int d^{3} k k\biggl[ \hat{Q}^{\dagger}_{\vk} \hat{Q}_{\vk} 
+ \hat{Q}_{\vk} \hat{Q}^{\dagger}_{\vk} +\hat{Q}^{\dagger}_{-\vk} \hat{Q}_{-\vk} 
+ \hat{Q}_{-\vk} \hat{Q}^{\dagger}_{-\vk} \biggr],
\label{min1}
\end{equation}
where
\begin{equation}
\hat{Q}_{\vk}(\tau_0) = \frac{1}{\sqrt{2 k}} \biggl[\frac{\hat{\pi}_{\vk}(\tau_0)}{z(\tau_0)} -
i z(\tau_0) k \hat{{\cal R}}_{\vk}(\tau_0) \biggr],
\label{Q1}
\end{equation}
obeying  $[\hat{Q}_{\vk}, \hat{Q}_{\vp}^{\dagger} ] = \delta^{(3)}(\vec{k} - \vec{p})$.
Consequently, the state minimizing (\ref{ham1b}) at $\tau_0$ is the one annihilated by $\hat{Q}_{\vk}$, i.e. 
\begin{equation}
\hat{Q}_{\vk}(\tau_0) |0^{(1)}\rangle =0,~~~~~~~~~\hat{Q}_{-\vk}(\tau_0) |0^{(1)}\rangle =0.
\label{inst}
\end{equation}
The specific relation between field operators dictated by (\ref{inst}) provides 
initial conditions for the Heisenberg equations
\begin{equation}
i \hat{\cal R}' = [\hat{\cal R},\hat{H}],~~~~~~~~~~~~~
i \hat{\pi}_{\cal R}' = [\hat{\pi}_{\cal R},\hat{H}].
\label{hev}
\end{equation}
The full solution of this equation can be written as
\begin{eqnarray}
&& \hat{\cal R}_{\vk}(\tau) = \hat{a}_{\vk}(\tau_0) f_{k}(\tau) + \hat{a}_{-\vk}^{\dagger}(\tau_0) f^{\ast}_{k}(\tau),
\label{solR}\\
&& \hat{\pi}_{\vk}(\tau) = \hat{a}_{\vk}(\tau_0) g_{k}(\tau) + \hat{a}_{-\vk}^{\dagger}(\tau_0)g^{\ast}_{k}(\tau),
\label{solpiR}
\end{eqnarray}
where,recalling the explicit solution of the equations in the case of the 
exponential potential (\ref{pot}) and defining $ x = k \tau$ 
\begin{eqnarray}
f_{k}(\tau) &=& \frac{\sqrt{\pi}}{4}\frac{e^{\frac{i}{2}(\mu + 1/2)\pi}}{z(\tau)\sqrt{ k}} 
\sqrt{-x} H^{(1)}_{\nu}(- x), ~~~~~~~~~~\nu= \frac{3 p -1}{2 (p -1)} 
\nonumber\\
g_{k}(\tau) &=& -\frac{\sqrt{\pi}}{4}e^{\frac{i}{2}(\mu + 1/2)\pi} z(\tau)\sqrt{k} \sqrt{- x} H^{(1)}_{\nu -1}(-x),
\label{fkgk}
\end{eqnarray}
satisfy the Wronskian normalization condition 
\begin{equation}
f_{k}(\tau) g^{\star}_{k}(\tau) - f^{\star}_{k}(\tau) g_{k}(\tau) = i.
\label{wrcon}
\end{equation}
Notice that the mode functions of Eq. (\ref{fkgk}) are different from the ones previously 
wtitten earlier in this section.
The creation and annihilation operators appearing in (\ref{solpi}) are defined as 
\begin{eqnarray}
\hat{a}_{\vk}(\tau_{0}) &=& \frac{1}{z_0\sqrt{2 k}}  \{ [ g_{k}^{\ast}(\tau_{0}) 
+ i k z_0^2 f_{k}^{\ast}(\tau_0)] \hat{Q}_{\vk}(\tau_0) 
- [ g_{k}^{\ast}(\tau_{0} ) - i k z_0^2 f_{k}^{\ast}(\tau_0)]  \hat{Q}^{\dagger}_{-\vk}(\tau_0)\},
\nonumber\\
\hat{a}_{-\vk}^{\dagger}(\tau_{0}) &=& \frac{1}{z_0\sqrt{2 k}} \{ [ g_{k}(\tau_{0}) 
- i k z_0^2 f_{k}(\tau_0)] \hat{Q}^{\dagger}_{-\vk}(\tau_0)
- [ g_{k}(\tau_{0} ) + i k z_0^2 f_{k}(\tau_0)]  \hat{Q}_{\vk}(\tau_0)\}.  
\label{def}
\end{eqnarray}
So far two sets of creation and annihilation operators have been introduced: 
the operators $\hat{Q}_{\vk}(\tau_0)$ and the 
operators $\hat{a}_{\vk}(\tau_0)$. The state annihilated by $\hat{Q}_{\vk}(\tau_0)$ minimizes 
the Hamiltonian at $\tau_0$ while the state annihilated by $\hat{a}(\tau_0)$ {\em does not} minimize the Hamiltonian at 
$\tau_0$.  It is relevant to introduce these operators 
not so much for the calculation of the two-point function 
but for the subsequent applications to the back-reaction effects. In fact,
in the standard approach to the initial value problem for the 
quantum mechanical fluctuations, the initial state is chosen to be the one annihilated by $\hat{a}_{\vk}(\tau_0)$ for 
$\tau_{0} \to -\infty$. 

The Fourier transform 
of the two-point function, 
\begin{equation}
\langle 0^{(1)},\tau_0| \hat{{\cal R}}(\vx, \tau) \hat{\cal R}(\vy, \tau)| \tau_0, 0^{(1)} \rangle = 
\int \frac{d k}{k} \overline{{\cal P}}_{{\cal R} }  \frac{\sin{ k r}}{k r},~~~~~~~~~~~~~r = |\vx - \vy|,
\label{PSA}
\end{equation}
can now be computed, and the result is   
\begin{eqnarray}
&&\overline{{\cal P}}_{\cal R} = \frac{k^2}{2 \pi^2} 
\Biggl\{ |f_{k}(\tau)|^2 \biggl[ \frac{|g_{k}(\tau_0)|^2}{ z(\tau_0)^2} + k^2 z(\tau_0)^2 | f_{k}(\tau_0)|^2\biggr]
\nonumber\\
&&- \frac{f_{k}(\tau)^2}{2} \biggl[ \frac{{g_{k}^{\ast}(\tau_0)}^2 }{z(\tau_0)^2} + k^2 z(\tau_{0})^2 {f_{k}^{\ast}(\tau_0)}^2\biggr]
\nonumber\\
&& - \frac{{f_{k}^{\ast}(\tau)}^2}{2} \biggl[ \frac{{g_{k}(\tau_0)}^2 }{z(\tau_0)^2} + k^2 z(\tau_{0})^2 {f_{k}(\tau_0)}^2\biggr]\Biggr\}.
\label{ps1}
\end{eqnarray}

The explicit form of $\overline{{\cal P}}_{\cal R}$ and $\overline{{\cal P}}_{\rm T}$  
can be obtained by inserting Eqs. (\ref{fkgk}) into Eq. (\ref{ps1}). 
The results should be expanded for  $|x| = k\tau \ll 1$ 
and  for $ |x_0| = k\tau_0 \gg 1 $.
While $|k\tau|$ measures how much a given mode is outside the horizon,
\begin{equation}
|x_0| = |k\tau_0| \simeq \frac{M}{H(t_{0}(k))} = \frac{M}{H_{\rm ex}}
\label{hex}
\end{equation}
defines the moment at which the given mode crosses the scale $M$.
Notice that Eq. (\ref{hex}) has exactly the same content of Eq. (\ref{CROSM}) 

To have the explicit form of the power spectrum, Eq. (\ref{psi1}) should be expanded for 
$k\tau \ll 1 $ and $|k\tau_0| \gg 1$ and the result is
\begin{equation}
\overline{{\cal P}}_{\cal R}^{1/2} 
= {\cal P}_{{\cal R}}^{1/2}
\biggl[ 1 + \frac{p}{2 (p -1)} \frac{\sin{[ 2 x_0 + p \pi/(p-1)]}}{x_0}\biggr],
\label{scps}
\end{equation}
where ${\cal P}_{{\cal R}}^{1/2}$ is the power spectrum of the scalar fluctutations already 
obtained within the standard lore [i.e. Eqs. (\ref{PR3})--(\ref{CT2})].

The results obtained in the case of Eq. (\ref{hamscal1}) can be also generalized to the case 
of the tensor modes decsribed by the Hamiltonian (\ref{H1t}). In this case the calculation is 
identical with the only difference that $z(\tau)$ is replaced by $a(\tau)$ and ${\cal R}$ 
is replaced by $h$. The result will be :
\begin{equation}
\overline{{\cal P}}_{\rm T}^{1/2} 
= {\cal P}_{\rm T}^{1/2}
\biggl[ 1 + \frac{p}{2 (p -1)} \frac{\sin{[ 2 x_0 + p \pi/(p-1)]}}{x_0}\biggr],
\label{tps}
\end{equation}
where, now, ${\cal P}^{1/2}_{\rm T}$ is given by Eqs. (\ref{PT3})--(\ref{CT1}).

In  Eqs. (\ref{scps}) and (\ref{tps}), on top of the standard (leading) terms there is a correction that goes, roughly, as 
$1/x_0 \sim H_{\rm ex}/M$ where, as discussed in Eq. (\ref{hex}), 
$H_{\rm ex}$ denotes the Hubble parameter evaluated at the 
moment the given scale crosses $M$. If $M\sim M_{\rm P}$, $H_{\rm ex}/M\sim 10^{-6}$. 
This is the correction that would apply 
in the scalar power spectrum if quantum mechanical initial conditions were assigned in such 
a way that the initial state minimizes (\ref{hamscal1}).

Having discussed in detail the results for the case of (\ref{hamscal1}) 
the attention will now be  turned to the case of Eqs. (\ref{hamscal2}) and (\ref{hamscal3}) 
whose  tensor analog are Eqs. (\ref{H2t}) 
and (\ref{H3t}).
In this case the whole procedure of canonical quantization can be repeated 
with the crucial difference that the state minimizing the quantum version of 
(\ref{hamscal2}) will not be the same minimizing (\ref{hamscal1}).
If the Hamiltonian (\ref{hamscal2}) is minimized the result is  
\begin{equation}
\overline{{\cal P}}^{1/2}_{{\cal R}, {\rm T}} ={\cal P}^{1/2}_{{\cal R},{\rm T}} 
\biggl[ 1 -\frac{p}{4 (p -1)} \frac{\cos{[ 2 x_0 + p \pi/(p-1)]}}{x_0^2}\biggr],
\label{tps2}
\end{equation}
where the subscript denotes either the scalar or the tensor power spectrum since 
the correction, in a power-law inflationary background, is the same both for 
scalars and tensors.

 A comparison of Eqs. (\ref{tps}) and (\ref{tps2}) shows two important 
facts. The first is that the leading term of the spectra is  the same. This phenomenon simply reflects the 
occurrence  that different Hamiltonians, connected by canonical transformations, must lead to the same evolution 
and to the same leading term in the power spectra. The second fact to be noticed is that the correction 
to the power spectrum goes as $1/x_0^2$ in the case of (\ref{tps2}). This correction is then much smaller than 
the one appearing in (\ref{tps}). If $M\sim M_{\rm P}$ then the correction will be, in the exact de Sitter case,  
${\cal O}(10^{-12})$, i.e.
six orders of magnitude smaller than the correction appearing in Eq. (\ref{tps}).

Finally the case of  the  Hamiltonians (\ref{hamscal3}) and (\ref{H3t}) will be examined.
Equation (\ref{hamscal3})  can be minimized following the same procedure as
already discussed in the case of Eqs. (\ref{hamscal1}) and (\ref{hamscal2}).  However, again, 
the state minimizing the Hamiltonian of Eq. (\ref{hamscal3}) will be different 
from the states minimizing the Hamiltonians of Eqs. (\ref{hamscal1}) and (\ref{hamscal2}).
The result for the power spectra will then be 
\begin{equation}
\overline{{\cal P}}^{1/2}_{{\cal R},{\rm T}} ={\cal P}^{1/2}_{{\cal R},{\rm T}} 
\biggl[ 1 + \frac{p( 2 p-1)}{(p -1)^2} \frac{\sin{[ 2 x_0 + p \pi/(p-1)]}}{4 x_0^3}\biggr],
\label{tps3}
\end{equation}
In Eq.  (\ref{tps3}) 
 the correction arising from the initial state goes as $1/x_0^3$ and, again, if $M\sim M_{\rm P}$ 
it is ${\cal O}(10^{-18})$, i.e. $12$ orders of magnitude smaller than in the case discussed 
in Eqs. (\ref{scps}) and (\ref{tps}).

\subsubsection{Back-reaction effects} 
In order to select the correct Hamiltonian in a way compatible with the 
idea of assigning initial conditions when a given physical frequency 
crosses the scale $M$ it is desirable  to 
address the issue of back-reaction effects. 
The energetic content of the quantum-mechanical state minimizing 
the given Hamiltonian should be estimated and  compared with the energy density of the background geometry.
The back-reaction 
effects of the different quantum-mechanical states 
minimizing the Hamiltonians will now be computed. 
Without loss of generality, the attention will 
be focused on the tensor modes of the geometry.
The advantage of discussing the gravitons is that they do not couple to 
the sources and, therefore, the form of the energy-momentum pseudo-tensor is simpler 
than in the case of the scalar modes \cite{abramo1,abramo2}.

The appropriate energy-momentum tensor of the fluctuations 
of the geometry will be averaged over the state 
minimizing a given Hamiltonian at $\tau_0(k)$ and the result 
compared with the energy density  of the background geometry,
since he energy density of the perturbations cannot exceed 
that of the background geometry. 

The energy density of the tensor inhomogeneities
 can be computed from 
the energy-momentum pseudo-tensor, written, for simplicity, for one 
of the two polarizations:
\begin{equation}
\langle \hat{{\cal T}}_{0}^{0} \rangle = 
\frac{{\cal H}}{2 a^2} \langle (\hat{h}' \hat{h} + \hat{h} \hat{h}') \rangle  
+ \frac{1}{8 a^2} \langle [ \hat{h}'^2  + (\partial_{i} \hat{h})^2 ] \rangle, 
\label{AVER}
\end{equation}
where $\langle ...\rangle$ denotes the expectation value with respect 
to a quantum mechanical state minimizing a given Hamiltonian and $\hat{h}$ denotes the field 
operator corresponding to a single tensor 
polarization of the geometry. 

If, in Eq. (\ref{AVER}), the average is taken with respect to the state minimizing the Hamiltonian 
(\ref{H1t}), then the energy density is, in the case of exact de Sitter space \cite{MGTP1}:
\begin{equation}
\overline{\rho}_{\rm GW}^{(1)}(\tau,\tau_0) \sim \frac{H^4}{64 \pi^2}\biggl[ x_0^2 + {\cal O} \biggl(\frac{1}{x_0^2}\biggr)\biggr]\simeq 
\frac{H^4}{64 \pi^2}\biggl(\frac{M}{H}\biggr)^2\biggl[ 1 + 
{\cal O}\biggl(\frac{H^2}{M^2}\biggr)\biggr].
\label{en1}
\end{equation}
Since, as already discussed, $ |x_0| = M/H \gg 1$, in the case of de Sitter space, the back-reaction effects 
related to the state minimizing the first Hamiltonian are then large. Recall, in fact, that the energy density of 
the background geometry is ${\cal O} ( H^2 M_{\rm P}^2)$. Hence, if $M\sim M_{\rm P}$ the energy 
density of the fluctuations will be of the same order as that  of the background geometry, which is 
not acceptable since, if this is the case, inflation could not even start. 
Let us now turn our attention to the case of the state minimizing the second Hamiltonian, i.e. Eq. (\ref{H2t}).
The expectation value of the energy-momentum pseudo-tensor will now be taken over the state minimizing 
the second class of Hamiltonians.  Applying the same procedure as before the averaged energy density is 
smaller than the result obtained in Eq. (\ref{en1}) by a factor $(H/M)^2$. 

Finally, in the third and last case,  we have to average the energy-momentum 
pseudo-tensor over the state minimizing the Hamiltonian (\ref{H3t}). In this case the result is \cite{MGTP1}
\begin{equation}
\overline{\rho}_{\rm GW}^{(3)}(\tau,\tau_0) \simeq \frac{27}{256 \pi^2}\,H^4\, 
\biggl(\frac{H}{M}\biggr)^2\biggl[ 1  + {\cal O}
\biggl(\frac{H^4}{M^4}\biggr)\biggr],
\label{en3}
\end{equation}
i.e. even smaller than the result discussed in the case of the second Hamiltonian. In this case, the averaged 
energy density is much smaller 
than that of the background.

The results reported in  \cite{MGTP1} and summarized above suggest the 
following reflections:
\begin{itemize}
\item{} it is physically interesting to set quantum mechanical initial conditions 
as soon as a given physical frequency crosses a fundamental scale $M$;
\item{} to compute ``observables" it is crucial to determine which Hamiltonian 
is minimized as soon as the given physical frequency crosses a fundamental scale $M$;
\item{} large (observable) effects seems to be ruled out if one accepts that the energy density 
of the fluctuations should be smaller than the energy density of the background geometry.
\end{itemize}
Similar considerations, but through different arguments, appeared in \cite{Tanaka,POR1,STA2}. It is here appropriate to mention that the authors 
of Ref. \cite{BaR1,BaR2} claim to 
reach more optimistic conclusions concerning the back-reaction effects associated 
with particular classes of transplanckian effects. In particular, in Ref. \cite{BaR1} 
the analysis has been performed just in terms of the scalar field fluctuation 
(that is not gauge-invariant for infinitesimal diffeomorphisms). It would be interesting 
to repeat the same analysis in terms of the tensor modes of the geometry.
 
\subsection{Spectator fields}

It is rather plausible that the early Universe is not described by a single scalar degree of freedom. 
On the contrary it seem conceivable that various fields are simultaneously 
present.  In the remaining part of the present section the attention will be focused on the possible 
effects of the spectator fields, i.e. fields that are not source of the background geometry but 
whose fluctuations can be amplified with scale-invariant spectra.

Thanks to some of the considerations developed in section 4 and 5, in  the presence of some 
 spectator field, 
a non-adiabatic fluctuation of the pressure density is likely to be generated in the system. 
There are different (but related) concrete examples of such a dynamics. The first 
obvious case is the one where the dynamics of the background is driven by a scalar 
degree of freedom (for instance the dilaton, in pre-big bang models \cite{PBB1,PBB2}). While 
the Universe inflates, some other spectator field (that is not source of the background 
geometry) gets amplified with a quasi-flat spectrum. The r\^ole of spectator field is played, in pre-big 
bang models, by the Kalb-Ramond axion \cite{PBB2}.  When inflation terminates, the large-scale 
fluctuations in the curvature are vanishingly small owing to the steepness of the dilatonic spectrum.
After the dilaton/inflaton decay, the Universe will presumably be dominated by radiation. However, owing 
to the presence of the spectator field,  the dynamics 
of the inhomogeneities will be much richer than the one of a radiation-dominated Universe. Since 
the radiation peculiar velocity and density contrast are present together with the fluctuations 
of the spectator field, a non-adiabatic fluctuation of the pressure density, $\delta p_{\rm nad}$  can 
be expected. Since $\delta p_{\rm nad}$ sources the evolution of the curvature 
perturbations, the quasi-flat spectrum of the spectator/axion
 field may be converted,  under some dynamical requirements,
into curvature fluctuations after axion's decay. 

Another possible example is the one where inflation occurs at relatively low curvature scales, 
i.e. $H \ll 10^{-6}\,\, M_{\rm P}$. In this case one can imagine the situation where the fluctuations 
of a scalar field (that is light during inflation and later on decays), can be efficiently 
converted into adiabatic curvature perturbations.  

In \cite{mollerach1} the possible conversion of isocurvature fluctuations into adiabatic modes was 
investigated in a simple set up where on top of the inflaton field there is only an extra 
spectator field. The chief objective of Ref. \cite{mollerach1} was not the possibility of 
converting isocurvature into adiabatic modes;  on the contrary, there was the hope 
that fluctuations of a pseudo-scalar spectator could indeed give rise to isocurvature 
modes \cite{lindekofman} after the decay (taking place after baryogenesis in the model of \cite{mollerach1}).
The possibility of conversion was also briefly mentioned in Ref. \cite{lindemukhanov}; the main goal 
of \cite{lindemukhanov} was however the analysis of possible non-gaussian effects arising 
in isocurvature models.

In \cite{moroi1} (see also \cite{moroi2}) the r\^ole 
of spectator field was played by some super-string moduli that could be 
amplified during a conventional inflationary phase. The authors correctly 
pointed out the possibility of obtaining, after modulus decay, a correlated mixture 
of adiabatic and isocurvature fluctuations.  In \cite{kari} the conversion of 
isocurvature modes has been analyzed in the context of pre-big bang models.
In fact, while the spectrum the dilaton and of the graviton is rather steep, the 
spectrum of axionic fluctuations may be rather flat or even red. The 
axionic fluctuation amplified during the pre-big bang phase can then be 
converted into adiabatic modes.
In Ref.  \cite{lythwands} the authors called the spectator field curvaton 
and provided further support to this idea.

Consider, for simplicity a scalar field $\sigma$ whose potential is quadratic, i.e. $2 V(\sigma) = m^2 \sigma^2$.
The field $\sigma$, at least initially, is not source of the background geometry.  After inflation its evolution
can be written,  
\begin{eqnarray}
&& M_{\rm P}^2 {\cal H}^2 = \frac{a^2}{3} \biggl[ \rho_{\rm r} + \frac{{\sigma'}^2}{2 a^2} 
+ V\biggr],
\label{ham}\\
&& M_{\rm P} {\cal H}' = - \frac{a^2}{3} \biggl[ \rho_{\rm r} 
+ \frac{ {\sigma'}^2}{a^2} - V\biggr],
\label{dyn}\\
&& \sigma'' + 2 {\cal H} \sigma' + a^2 \frac{\partial V}{\partial\sigma} 
=0, 
\label{kgsigma}\\
&& \rho_{\rm r}' + 4 {\cal H} \rho_{\rm r} =0.
\label{contrad}
\end{eqnarray}
where 
\begin{equation}
\rho_{\sigma} = \frac{{\sigma'}^2}{2 a^2} + V, ~~~~~~~~
p_{\sigma} =    \frac{{\sigma'}^2}{2 a^2} - V.
\label{backdef}
\end{equation}
Since at the onset of the radiation-dominated epoch the 
curvaton is not dominant, its energy density is always 
smaller than the energy density of the background geometry. During 
radiation ${\cal H}a $ is constant; from Eq. (\ref{kgsigma}) it can be easily 
derived that 
\begin{equation}
\sigma' = - \frac{1}{5} \frac{a^2}{\cal H} V_{,\sigma}.
\label{sre}
\end{equation}
The factor of $5$ appearing in Eq. (\ref{sre}) comes from the requirement that 
the potential term dominates in Eq. (\ref{kgsigma}). Thus the approximate evolution 
will follow the equation ${\cal H} \sigma' = -C  a^2 V_{,\sigma}$ where $V_{,\sigma}$
denotes the derivation with respect to $\sigma$ and $C$ is a constant to be determined 
inserting the approximate equation into the full equation and neglecting terms with 
higher derivatives of the potential, i.e. $V_{,\sigma\sigma}$. This requirement 
leads to $ C = 1/5$. Notice that the factor $1/5$ depends on the specific 
evolution of the dominant component of the background geometry.  If the dominant 
component of the background is characterized by a perfect fluid with 
barotropic index $w$, then Eq. (\ref{sre}) becomes 
\begin{equation}
\sigma' = - \frac{2}{{\cal H}[ 6 + 3 ( w + 1)]} V_{,\sigma} a^2
\label{sre2}
\end{equation}
which reproduces Eq. (\ref{sre}) if $w = 1/3$. In the following, on top of the radiation-dominated 
case, the stiff-dominated fluid will be specifically considered. In this case  pre-factor will be $1/6$.

By inspecting the evolution equations of $\sigma$, various relevant scales 
can be derived:
\begin{itemize}
\item{} the moment of oscillations: assuming that sufficiently close 
to the minimum the axion potential is quadratic, coherent 
oscillations of $\sigma$ will start at a typical scale $H_{m} \sim m$;
\item{} the moment of dominance: when the Hubble rate is 
comparable with the axion potential, i.e. $M_{\rm P}
 H_{\sigma}  \sim m \sigma$, the axion starts dominating the 
 background;
 \item{} the moment of decay: being gravitationally coupled, $\sigma$ 
 will decay at a typical scale $H_{\rm d} \sim m^3/M_{\rm P}^2$; this 
 implies that $\sigma$ must decay prior to big-bang nucleosynthesis 
 in order not to jeopardize the production of light elements,
 \end{itemize}
The last requirement implies that $m > 10 \,\,{\rm TeV}$ taking the BBN temperature 
of the order of $0.1$ MeV. This 
bound can be made more restrictive by requiring that the decay takes place prior 
the electroweak epoch (for temperatures of the order of $0.1$ TeV). 

When $H_{\rm m} \sim m$ the oscillations of $\sigma$ commence.  For instance, in the case 
of a quadratic potential the effective equation of state of the oscillating field will 
be the one of dust matter and $ \sigma  $ decreases as $a^{-3/2}$. This 
occurrence, however, depends upon the specific properties of the 
potential. For instance in the case of a quartic potential, $\langle \rho_{\sigma} \rangle 
\simeq 3 \langle  p_{\sigma} \rangle$ \cite{mgcurv1,dim1}.

In the following the main features of the conversion between isocurvature and 
adiabatic modes will be investigated in the two conceptually separate 
but complementary cases of conventional inflationary models and pre-big bang 
models.

\subsubsection{Pre-big bang models} 
During the pre-big bang evolution the Kalb-Ramond axion is amplified with 
quasi-flat spectrum \cite{PBB2}, i.e. 
\begin{equation}
{\cal P}_{\sigma} \simeq \biggl(\frac{H_{1}}{M_{\rm P}}\biggr) \biggl(\frac{k}{k_{1}} 
\biggr)^{n_{\sigma} -1}
\end{equation}
where the spectral index $n_{\sigma}$ depends on the dynamics of the 
internal dimensions \cite{mgcurv1,mgcurv2}. On the contrary, the spectrum of curvature 
fluctuations is characterized by a rather steep spectral index. More precisely, the considerations 
introduced earlier in this section allow to compute the scalar and tensor spectral indices 
\cite{long2}:
\begin{equation}
n_{\rm s} = 4,\,\,\,\,\,\,\,\,\,\,\,n_{\rm T} =3.
\end{equation}
In the context of the ekpyrotic models this estimate leads to 
\begin{equation}
n_{\rm s} = 3,\,\,\,\,\,\,\,\,\,\, n_{\rm T} =2,
\end{equation}
assuming continuity in ${\cal R}$ \cite{lythekp1}.

At the beginning of the post-big bang evolution the background is
characterized by a ``maximal" curvature scale $H_1$, whose finite
value regularizes the big bang singularity of the standard cosmological
scenario, and provides a natural cutoff for the spectrum of
quantum fluctuations amplified by the phase of pre-big bang inflation.  
In string cosmology models such
an initial curvature scale is at most of the order of the string mass
scale, i.e. $H_1 \laq M_{\rm s} \sim 10^{17}$ GeV. 

The evolution of $\sigma$ is different depending upon the 
initial value of $\sigma$ in Planck units, i.e. $\sigma_{\rm i}/M_{\rm P}$. 
If $\sigma_{\rm i} > M_{\rm P}$ the axion dominates earlier since 
the condition of dominance is saturated faster. If $\sigma_{\rm i} < M_{\rm P}$ 
the axion dominates later, i.e. its starts oscillating coherently at 
$H_{m}$ and it will dominate at a redshift $ (a_{m}/a_{\sigma}) \sim (\sigma_{\rm i}/M_{\rm P})^2$. This 
observation also implies that if $\sigma_{\rm i} >1$, the axion dominance 
precedes the oscillation epoch.
For a more specific discussion we refer to \cite{kari,mgcurv1,mgcurv2,martin}. 

Consider then, as an illustration, the case $\sigma_{\rm i} < M_{\rm P}$. Initially
$ 6 H_{\rm inf}^2 M_{\rm P}^2 \gg m^2 \sigma_{\rm i}$ since $\sigma$ is not dominant. Later on 
$\sigma$ oscillates at $ H_{m} \sim m$ and, around this scale 
the approximation of Eq. (\ref{sre2}) breaks down and $\sigma(\tau) \simeq \sigma_{\rm i}
(a_{m}/a)^{3/2}$. At a typical scale $H_{\sigma} \simeq ( \sigma_{\rm i}/M_{\rm P})^4  m/36$,
the field starts dominating the background inducing, possibly, a short period 
of inflationary expansion.  The axion can decay either before being dominant or after. The 
decay of the curvaton has been discussed both analytically and numerically 
by different authors \cite{mgcurv2,curvdec1,curvdec2}. In the following, for sake of simplicity 
the approximation of sudden decay will be adopted. If the decay follows the dominance, i.e. 
$H_{\rm d} < H_{\sigma}$, then 
\begin{equation}
\biggl(\frac{\sigma_{\rm i}}{M_{\rm P}}\biggr) \gg \sqrt{6} \sqrt{\frac{m}{M_{\rm P}}}.
\end{equation}
The opposite inequality holds if $H_{\rm d} > H_{\sigma}$.

The evolution equations for the fluctuations are obtained by considering 
the simultaneous presence of the the fluctuations of $\sigma$ and of 
the fluctuations of the radiation background. The Hamiltonian 
constraint becomes then, in the longitudinal gauge,
\begin{equation}
- k^2 \psi - 3 {\cal H} ( {\cal H} \psi + \psi') = 4\pi G a^2( \rho_{\rm r} \delta_{\rm r} + \delta \rho_{\sigma}),
\label{hamp}
\end{equation}
while the $i=j$ component of the perturbed Einstein equations gives 
\begin{equation}
\psi'' + 3 {\cal H}\psi' + ( {\cal H}^2 + 2 {\cal H}') \psi =  \frac{4 \pi G}{3} a^2  \rho_{\rm r} 
\delta_{\rm r} + 4\pi G a^2  \delta p_{\sigma},
\label{curvij}
\end{equation}
where 
\begin{eqnarray}
&& \delta \rho_{\sigma} = \frac{1}{a^2} \biggl[ - \psi{\sigma'}^2 + \sigma' 
\chi' + \frac{\partial V}{\partial\sigma} a^2\chi \biggr],
\label{deltarhosigma}\\
&& \delta p_{\sigma} = \frac{1}{a^2} \biggl[ - \psi {\sigma'}^2 + 
\sigma'  \chi' - \frac{\partial V}{\partial\sigma} a^2 \chi\biggr].
\label{deltapsigma}
\end{eqnarray}
In all the perturbation equations it has been assumed that anisotropic 
stresses are absent.  Notice that we denoted with $\chi$ the fluctuation of $\sigma$.
Finally the evolution of $\chi$ and of the fluid variables is given by
\begin{eqnarray}
&& \chi'' + 2 {\cal H} \chi' - \nabla^2 \chi + 
\frac{\partial^2 V}{\partial\sigma^2} a^2 \chi - 4 \sigma' \Phi' + 2 
\frac{\partial V}{\partial \sigma }a^2 \Phi =0,
\label{chiredcurv} \\
&& \delta_{\rm r}' = - \frac{4}{3} \theta_{\rm r} + 4 \psi',
\label{deltarcurv}\\
&&\theta_{\rm r}' = \frac{k^2}{4} \delta_{\rm r} + k^2 \psi.
\label{thrcurv}
\end{eqnarray}
We are interested in the solution of the system of Eqs. (\ref{hamp})--(\ref{curvij}) and 
(\ref{chiredcurv})--(\ref{thrcurv}) with non-adiabatic initial conditions, i.e. 
$\psi(k,\tau_{\rm i}) = \delta_{\rm r} (k,\tau_{\rm i}) = 0$ but $\chi(k, \tau_{\rm i}) = 
\chi_{\rm i} (k)$. 

Since for large-scale inhomogeneities $\theta_{\rm r} \ll 1$, Eq. (\ref{deltarcurv}) 
 implies
\begin{equation}
\delta_{\rm r}  \simeq 4 \psi ,
\label{drph}
\end{equation} 
because initially $\psi(\tau_{\rm i}) \to 0$.
Using Eq. (\ref{deltarhosigma}) and (\ref{drph}),   the Hamiltonian
constraint (\ref{hamp}) can be written, for $k \to 0$ as 
\begin{equation}
\frac{ d ( a^3 \psi)}{d\ln{a}} = \frac{ 4 \pi G}{ 3 {\cal H}^2} a^3 [ \sigma' \chi'  + V_{,\sigma} a^2 \chi].
\label{hamint}
\end{equation}
In order to derive Eq. (\ref{hamint}) we used the fact that $(p_{\sigma} +\rho_{\sigma} ) \ll  \rho_{\rm r}$.
Similarly to Eqs. (\ref{sre}) and (\ref{sre2}),  Eq. (\ref{chiredcurv}) implies, with isocurvature initial conditions, 
that 
\begin{equation}
\chi' \simeq - \frac{1}{5 {\cal H}} V_{,\sigma\sigma} a^2 \chi.
\label{chiprcurv}
\end{equation}
Hence, from Eq. (\ref{hamint}), direct integration leads to
\begin{equation}
\psi(\tau) = - \frac{1}{42 M_{\rm P}} \frac{a^2}{{\cal H}^2} 
V_{,\sigma} \chi_{\rm i} + {\cal O}( V_{,\sigma}^2) \simeq 
- \frac{1}{14~\rho_{\rm r}} V_{,\sigma} \chi_{\rm i} 
+ {\cal O}( V_{,\sigma}^2). 
\label{phgen}
\end{equation}  
Inserting Eq. (\ref{phgen}) into Eqs. (\ref{deltarcurv}) and (\ref{thrcurv}) 
\begin{equation}
 \delta_{\rm r}(\tau) = - \frac{2}{21 M_{\rm P}^2} \frac{a^2}{{\cal H}^2} 
V_{,\sigma} \chi_{\rm i} + {\cal O}( V_{,\sigma}^2),
~~~~
\theta_{\rm r}(\tau) = - \frac{1}{105 M_{\rm P}^2} \frac{a^2}{{\cal H}^3} 
V_{,\sigma} \chi_{\rm i} + {\cal O}( V_{,\sigma}^2).
\label{flsol}
\end{equation}

The time evolution
of ${\cal R}$ in the radiation-dominated, slow-roll regime can 
finally be determined from the definition of ${\cal R}$ in terms of $\psi$
\begin{equation}
{\cal R}(\tau) \simeq \frac{1}{4 \rho_{\rm r}} 
\frac{\partial V}{\partial\sigma}
\chi_{\rm i}  + {\cal O}( V_{,\sigma}^2),
\label{calR2}
\end{equation}
so that $\psi$ and ${\cal R}$ obey the approximate relation
$\psi(\tau) \simeq -(2/7) {\cal R}(\tau) + {\cal O}( V_{,\sigma}^2)$.

The same result can be obtained directly by integrating the evolution 
equation for ${\cal R}$ i.e. 
\begin{equation}
\frac{ d {\cal R}}{d \ln a} \simeq - \frac{\delta p_{\rm nad}}{p + \rho},
\label{calR1}
\end{equation}
having approximately determined the form of $\delta p_{\rm nad}$. 
The  exact form of  $\delta p_{\rm nad}$ and of the sound speed 
is, in our system, 
\begin{eqnarray}
&& \delta p_{\rm nad} = \rho_{\rm r} \biggl( \frac{1}{3} - c_{\rm s}^2\biggr) \delta_{\rm r} 
+ \psi ( c_{\rm s}^2 -1) ( p_{\sigma} + \rho_{\sigma}) + \frac{\sigma'\chi'}{a^2} ( 1 - c_{\rm s}^2) 
- V_{,\sigma}\chi ( 1 + c_{\rm s}^2),
\label{dpnadcurv}\\
&& c_{\rm s}^2 = \frac{1}{3} \biggl[ \frac{ 4 \rho_{\rm r} + 9 ( p_{\sigma} + \rho_{\sigma}) +
6 \sigma' /{\cal H} V_{,\sigma}}{ 4 \rho_{\rm r} + 3 ( p_{\sigma} + \rho_{\sigma})}\biggr].
\label{cscurv}
\end{eqnarray}
When the curvaton is subdominant $ c_{\rm s}^2 \simeq 1/3$ and the leading term in 
Eq. (\ref{dpnad}) is the last one. Hence,  to leading order, 
$\delta p_{\rm nad} \simeq - 4 V_{,\sigma} \chi_{\rm i}/3$; inserting the 
approximate form of $\delta p_{\rm nad}$ into Eq. (\ref{calR1}) we obtain again, by 
direct integration Eq. (\ref{calR2}) since 
\begin{equation}
\frac{ d {\cal R}}{d \ln{a}} \simeq \frac{V_{,\sigma}}{\rho_{\rm r}} \chi.
\label{calR3}
\end{equation}

After the initial phase, when the oscillation start, we have to assume a specific form 
of the potential which we take to be quadratic. For $H \leq m$, $ \chi\sim 
\chi_{\rm i} (a_{m}/a)^{3/2}$ and $\sigma \sim \sigma_{\rm i} (a_{m}/a)^{3/2}$ and 
Eq. (\ref{calR3}) then leads to 
\begin{equation}
{\cal R} \simeq \biggl(\frac{\sigma_{\rm i}}{M_{\rm P}}\biggr) \biggl(\frac{\chi_{\rm i}}{M_{\rm P}}\biggr) 
\biggl(\frac{a}{a_{m}}\biggr) = r(a) \frac{\chi_{\rm i}}{\sigma_{\rm i}},,\,\,\,\,\,\,\,\,\,\,\,\,\,\,r(a) =
\biggl(\frac{\sigma_{\rm i}}{M_{\rm P}}\biggr)^2 \biggl(\frac{a}{a_{\rm m}}\biggr)
\label{calR4}
\end{equation}
where we used that, for $ H \leq m$, $ \rho_{\rm r} \sim m^2 M_{\rm P}^2 (a_{m}/a)^4$. The second 
equality follows from the definition of $r(a)$, i.e. the time-dependent ratio of the axion energy density 
over the energy density of the dominant component of the background.
Recalling now that $\chi_{i}(k)$ has a quasi-flat power spectrum, Eq. (\ref{calR4}) 
expresses the conversion of the initial isocurvature mode into the wanted adiabatic mode. The 
expression of ${\cal R}(a)$ will be frozen at the moment of sudden decay taking place at the scale 
$H_{\rm d}$. In the context of pre-big bang models the power spectrum of the axionic 
fluctuations depends on various parameters connected  with the specific model
of pre-big bang evolution. By comparing the generated curvature fluctuations with the 
value experimentally measured various constraints on the parameter of the model  can be obtained \cite{mgcurv1,mgcurv2}. These constraints seem to suggest that the scale $H_{1}$ should be
of the order of $10^{-1} M_{\rm s} $, where  $M_{\rm s} \sim 0.01 M_{\rm P}$ is the string mass scale. 
Notice, furthermore, that we just treated the case where the axion decays before becoming dominant.
However, in the context of pre-big bang models the opposite case is probably the most relevant.
If the axion decays after becoming dominant numerical calculations have to back 
the analytical approach (especially in the case $\sigma_{\rm i} > M_{\rm P}$). 

\subsubsection{Conventional inflationary models}

It is conceivable that the curvaton could also play a r\^ole in the context of 
conventional inflationary models, provided the inflationary scale is sufficiently low 
(i.e. $H_{\rm inf} < 10^{-6} M_{\rm P}$).  In this case, it is argued \cite{liddle}, the 
adiabatic perturbations generated directly by the fluctuations of the inflaton, will be 
too small. In \cite{liddle} a simple model has been discussed where the potentials of the curvaton 
and of the inflaton are both quadratic.  During inflation the curvaton should be effectively massless, 
i.e. $m\ll H_{\rm inf}$.  There are some classes of inflationary models leading naturally 
to a low curvature scale \cite{lyth3} (see also \cite{dim2}). As far as the evolution of the curvaton 
in a radiation dominated epoch is concerned, the results obtained in the case of the pre-big 
bang are also valid provided the scale $H_{1}$ is now interpreted as the curvature scale 
at the end of inflation.

To identify a particle physics candidate acting as a curvaton various proposals have been 
made.  Different authors \cite{kari2,mcdonald,moroi3,postma1} suggest that 
the right-handed sneutrinos (the supersymmetric partners of the righ-handed neutrinos)
 may act as curvatons. Models have been also proposed where the 
curvaton corresponds to a MSSM flat direction \cite{kari3,kari4} and to a MSSM 
Higgs \cite{bastero}. 

In the following we would like to concentrate on one of the besic questions 
arising in low-scale inflationary models: is it possible to lower 
the inflationary scale {\em ad libitum} ? 

In the standard curvaton scenario the energy density of the curvaton
increases with time with respect to the energy density of the 
radiation background. From this aspect of the theoretical construction, a number of constraints 
can be derived; these include an important aspect of the 
inflationary dynamics occurring prior to the curvaton oscillations, namely 
the minimal curvature scale  at the end of inflation compatible with the curvaton 
idea. 
 Suppose, for simplicity, that the curvaton field $\sigma$ 
 has a massive potential 
and that its evolution, after the end of inflation, occurs during a radiation dominated 
stage of expansion. As previously discussed the ratio $r(a)$ increases with time 
during the radiation-dominated oscillations
\begin{equation}
r(a) \simeq \biggl(\frac{\sigma_{\rm i}}{M_{\rm P}}\biggr)^2 \biggl(\frac{a}{a_{\rm m}}\biggr)~~~~~H<H_{\rm m},
\label{r1}
\end{equation}
where $\sigma_{\rm i}$ is the nearly constant value of the curvaton throughout the later stages of inflation.
When $\sigma$ decays the ratio $r$ gets frozen to its value at decay, i.e.  $r(t) \simeq 
r(t_{\rm d}) = r_{{\rm d}}$ for $t > t_{\rm d}$. Equation (\ref{r1}) then implies  
\begin{equation}
m = \frac{\sigma_{\rm i}^2}{ r_{\rm d} M_{\rm P}}. 
\label{mrd}
\end{equation}
The energy density 
of the background fluid just before decay has to be larger
 than the energy density of the decay products, i.e. 
$\rho_{\rm r}(t_{\rm d}) \geq T_{\rm d}^4 $.
Since 
\begin{equation}
\rho_{\rm r}(t_{\rm d}) \simeq m^2 M_{\rm P}^2 \biggl(\frac{a_{\rm m}}{a_{\rm d}}\biggr)^4,
\end{equation}
the mentioned condition implies 
\begin{equation}
\frac{m}{T_{\rm d}} \sqrt{\frac{m}{M_{\rm P}}} > 1,
\end{equation}
which can also be written, using Eq. (\ref{mrd}), as 
\begin{equation}
\biggl( \frac{\sigma_{\rm i}}{M_{\rm P}}\biggr)^3  \geq r_{\rm d}^{3/2} \biggl(\frac{T_{\rm d}}{M_{\rm P}}\biggr).
\label{epscond1}
\end{equation}
Equation (\ref{epscond1}) has to be compared  with the restrictions 
coming from the amplitude of the adiabatic perturbations, which should 
be consistent with observations. If $\sigma$ decays before becoming dominant the curvature 
perturbations at the time of decay  are
\begin{equation}
{\cal R}(t_{\rm d}) \simeq \biggl.\frac{1}{\rho_{\rm r}} \frac{\partial V}{\partial \sigma} 
\chi \biggr|_{t_{\rm d}}\simeq r_{\rm d} \frac{ \chi^{{\rm i}}_{k}}{\sigma_{\rm i}}.
\end{equation}
Recalling that $\chi^{\rm i}_{k} \sim H_{\rm inf}/(2 \pi)$  the power spectrum of curvature perturbations 
\begin{equation}
{\cal P}_{{\cal R}}^{1/2} \simeq \frac{ r_{\rm d} H_{\rm inf}}{ 4 \pi \sigma_{\rm i}} \simeq 5 \times 10^{-5}
\label{pscp}
\end{equation}
implies, using Eq. (\ref{epscond1}) together with Eq. (\ref{mrd}), 
\begin{equation}
\biggl( \frac{H_{\rm inf}}{M_{\rm P}}\biggr) 
\geq 10^{-4} \times r_{\rm d}^{-1/2} \biggl(\frac{T_{\rm d}}{M_{\rm P}}\biggr)^{1/3}.
\label{bound1}
\end{equation}
Recalling now that $ r_{\rm d} \leq 1$ and $m < H_{\rm inf}$, the above inequality 
implies that, at most, $H_{\rm inf} \geq 10^{-12} ~M_{\rm P}$ if $T_{\rm d} \sim 1 ~{\rm MeV}$ is selected.
This estimates is, in a sense, general since the specific relation between $T_{\rm d}$ and $H_{\rm d}$ is not fixed.
The bound (\ref{bound1}) can be even more constraining, for certain 
regions of parameter space, if the condition $ T_{\rm d} \geq \sqrt{H_{\rm d} ~M_{\rm P}}$ is imposed with $H_{\rm d} \simeq m^3/M_{\rm P}^2$.
In this case, Eq. (\ref{bound1}) implies $H_{\rm inf}^2 \geq 10^{-8} m ~M_{\rm P}$,  which is 
more constraining than the previous bound for sufficiently large values of the mass, i.e. $ m \geq 10^{-4} H_{\rm inf}$.
Thus, in the present context, the inflationary curvature scale is bound to be 
in the interval \cite{lyth3,mgcurv4}
\begin{equation}
10^{-12} M_{\rm P} \leq H_{\rm inf} \leq 10^{-6} M_{\rm P} .
\end{equation}

This bound can be relaxed, by a bit, if the evolution of the curvaton takes place, after 
inflation, in a background dominated by the kinetic energy of the 
inflaton field itself.
Consider now the case where the inflationary epoch is not immediately 
followed by radiation. Different models of this kind may be constructed. 
For instance, if the inflaton field is identified with the quintessence field, a long 
kinetic phase occurred prior to the usual radiation-dominated stage of expansion.
The evolution of a massive curvaton field in quintessential inflationary models 
has been recently studied \cite{mgcurv3} in the simplest scenario where the curvaton field
is decoupled from the quintessence field and it is minimally coupled to the metric.
In order to be specific, suppose that,  the inflaton 
potential, $V(\varphi)$ is chosen to be 
a typical power law during inflation and an {\em inverse} power  during 
the quintessential regime:
\begin{eqnarray}
&&V(\varphi) = \lambda ( \varphi^4 + M^4) ,~~~~ \varphi < 0,
\nonumber\\
&& V(\varphi) = \frac{\lambda M^8}{\varphi^4 + M^4}, ~~~~ \varphi \geq 0,
\label{potph}
\end{eqnarray}
where $\lambda$ is the inflaton self-coupling and  $M$ is the typical 
scale of quintessential evolution, i.e. by appropriately fixing $M$, the field $\varphi$ will
be dominant about today. The potential of the curvaton may be 
taken to be, for simplicity, quadratic. In this model the curvaton 
will evolve, right after the end of inflation, in an environnment 
dominated by the kinetic energy of $\varphi$. The curvaton starts oscillating at $H_{\rm m} \sim m$ and 
becomes 
dominant at a typical curvature scale $H_{\sigma} \sim m (\sigma_{\rm i}/M_{\rm P})^2$. 
Due to the different 
evolution of the background geometry, the ratio $r(t)$ will take the form
\begin{equation}
r(t) \simeq m^2 \biggl(\frac{\sigma_{\rm i}}{M_{\rm P}}\biggr)^2 \biggl(\frac{a}{a_{{\rm m}}}\biggr)^3, ~~~~~~H<H_{\rm m}.
\label{kinback}
\end{equation}
to be compared with Eq. (\ref{r1}) valid in the standard case.
For $t>t_{\rm d}$, $r(t)$ gets frozen to the value $r_{\rm d}$ whose relation to 
$\sigma_{\rm i}$ is different from the one obtained previously (see Eq. (\ref{mrd})) and valid 
in the case when $\sigma$ relaxes in a radiation dominated environnment. In fact, from Eq. (\ref{kinback}),
\begin{equation}
m \simeq \frac{\sigma_{\rm i}}{\sqrt{r_{\rm d}}}.
\label{mrdkin}
\end{equation}
From the  requirement
\begin{equation}
\rho_{\rm k}(t_{\rm d}) \simeq m^2 M_{\rm P}^2 \biggl(\frac{a_{\rm m}}{a_{\rm d}}\biggr)^6 \geq T_{\rm d}^4,
\end{equation}
it can be inferred, using (\ref{mrdkin}), that
\begin{equation}
\biggl(\frac{\sigma_{\rm i}}{M_{\rm P}} \biggr)^{3/2} \geq r_{\rm d}^{3/4} \biggl(\frac{T_{\rm d}}{M_{\rm P}}\biggr) 
\end{equation}
Following the analysis reported in \cite{mgcurv3}, the amount of produced fluctuations can be computed.

The calculation exploits Eqs. (\ref{qqphisigma}) which are valid for a two-field model.
Solving Eqs. (\ref{qqphisigma}) with the appropriate initial condition, and using that 
$a^3 \dot{\varphi}^2/H $ is constant during the 
kinetic phase, Eq. (\ref{Rdef1}) can be written as  \cite{mgcurv3}
\begin{equation}
{\cal R}(t) \simeq \frac{H}{\rho_{\rm k}} \dot{\sigma} q_{\sigma} 
\simeq \frac{\chi_{\sigma}}{\rho_{\rm k}} \frac{\partial W}{\partial\sigma}\simeq
 r_{\rm d}\biggl(\frac{\chi^{({\rm i})}_{k}}{\sigma_{\rm i}}\biggr).
\end{equation}
Recalling   that $\chi^{({\rm i})}_{k} \sim H_{\rm inf}/(2 \pi)$, the observed 
value of the power spectrum, i.e.  
$ {\cal P}^{1/2}_{{\cal R}} \sim 5 \times 10^{-5}$, implies 
\begin{equation}
\biggl(\frac{H_{\rm inf}}{M_{\rm P}} \biggr) \geq 10^{-4} r_{\rm d}^{-3/2}
 \biggl(\frac{T_{\rm d}}{M_{\rm P}}\biggr)^{2/3}. 
\label{bound2}
\end{equation}
The same approximations discussed in the standard case can now be applied to the case 
of quintessential inflation. They imply the absolute lower bound of $H_{\rm inf} > 100 {\rm GeV}$
\cite{mgcurv4,postma2}

Up to now we always discussed the case of quadratic potentials. However, the potential 
of $\sigma$ should be allowed to have different profiles. There is only one danger. If the 
potential leads to some attractor solution for the fluctuations, it may happen 
that the fluctuations of the curvaton are erased \cite{mgcurv4,dim1}. If, for some 
reason, the curvaton does not decay completely, it is possible to have a residual 
isocurvature fluctuations which will be completely correlated with the adiabatic component (see, for instance
\cite{wands1}).  The large-scale anisotropy experiments put then interesting 
constraints on this possibility \cite{gordonlw} (see also section 4).

\renewcommand{\theequation}{7.\arabic{equation}}
\section{Spectra of relic gravitons}
\setcounter{equation}{0}

While the present value of the Hubble 
rate, expressed in Hertz, is of the order of $10^{-18}$, relic gravitons of frequencies 
between $100$ Hz and the GHz can be also produced throughout the cosmological evolution. 
In the present section we will compute the stochastic background of relic gravitons in various cosmological models starting with the case of ordinary inflationary models and we will then 
comment on the possible relevance of the obtained signals for experiments 
aimed at the direct detection of backgrounds of relic gravitons. 

\subsection{Relic gravitons from power-law inflation}

In order to estimate the relic graviton background in conventional inflationary 
models in a reasonably generic fashion, we shall be assuming that the evolution of the universe 
consists of three distinct epochs: a power-law inflationary phase, a radiation dominated phase 
and a matter dominated phase. In formulae :
\begin{eqnarray}
&& a_{\rm i} (\tau) = \biggl(-\frac{\tau}{\tau_1}\biggr)^{-\beta},\,\,\,\,\,\,\,\,\,\,\, \tau\leq - \tau_{1}
\label{ai}\\
&& a_{\rm r}(\tau) = \frac{\beta \tau + (\beta + 1) \tau_1}{\tau_{1}}, \,\,\,\,\,\,\,\,\,\, - \tau_{1} \leq \tau \leq \tau_{2}
\label{ar}\\
&& a_{\rm m} (\tau) = 
\frac{[ \beta (\tau + \tau_{2}) + 2 \tau_{1} (\beta + 1)]^2}{4 \tau_1 [ \beta \tau_{2} + (\beta + 1) \tau_1]},\,\,\,\,\,\,\,\,\
\tau > \tau_{2}
\label{am}
\end{eqnarray}
As already discussed, a generic power-law inflationary phase is characterized by a 
power $\beta$. In the case $\beta = 1$ we have the case of the expanding branch of de Sitter
space. During the radiation-dominated epoch the scale factor expands linearly in conformal time 
while during matter it expands quadratically. Notice that the form of the scale factors 
given in Eqs. (\ref{ai})--(\ref{am}) is continuous and differentiable at the transition points, i.e. 
the scale factors and their first derivatives are continuous in $-\tau_1$ and $\tau_{2}$. 
The continuity of the scale factor and its derivative
 prevents the presence of divergences in the pump field, given by $a''/a$. Notice, incidentally
 that Eqs. (\ref{ai})--(\ref{am}) lead to a pump field even simpler than the one 
 reported in Fig. \ref{F51} in the intermediate phase (between inflation and radiation) disappears.  
 Two scales are present in the problem, i.e. 
$\tau_{1}$ and $\tau_{2}$. Hence, the spectrum of relic gravitons will 
be characterized by two branches limited by the comoving wave-numbers $k_{1}\sim 1/\tau_{1}$ and 
$k_{2} \sim 1/\tau_{2}$. For $k \ll k_2$, gravitons will be produced both thanks to the 
transition inflation radiation and to the transition radiation-matter. If, on the contrary, 
$k_2 < k < k_1$, the modes of the field will only feel the pumping action of the inflation-radiation 
transition. Finally, if $k >k_1$, the production of gravitons is absent.

 Consider for instance the case when initial conditions on the tensor modes are set by 
minimizing the Hamiltonian of Eq. (\ref{H3t}). In this case the evolution 
of the field operators in the Heisenberg description can be  written, for $\tau < -\tau_{1}$, as 
\begin{eqnarray}
&& \hat{\mu}_{k}(\tau) = \hat{a}_{\vk}(\tau_0) f_{\rm i}(\tau) + \hat{a}_{-\vk}^{\dagger}(\tau_0) f^{\ast}_{\rm i}(\tau),
\label{solmuIN}\\
&& \hat{\pi}_{k}(\tau) = \hat{a}_{\vk}(\tau_0) g_{\rm i}(\tau) + \hat{a}_{-\vk}^{\dagger}(\tau_0)g^{\ast}_{\rm i}(\tau),
\label{solpiIN}
\end{eqnarray}
where the mode functions during inflation are given by 
\begin{eqnarray}
&& f_{\rm i} (k,\tau)= \frac{{\cal N}}{\sqrt{2 k}} \sqrt{- x} H_{\nu}^{(1)}(- x), \,\,\,\,\,\,\,\,\,\,\,\,\,\,{\cal N} 
=\sqrt{\frac{\pi}{2}} e^{\frac{i}{2}(\nu + 1/2)\pi}.
\nonumber\\
&& g_{\rm r} (k,\tau) =  - {\cal N}\sqrt{\frac{k}{2}} \sqrt{-x}\biggl[ H^{(1)}_{\nu -1} (-x) +
\frac{(1 -2 \nu)}{2(-x)} H^{(1)}_{\nu} (-x)\biggr],
\label{infmode}
\end{eqnarray}
where $\nu = \beta + 1/2$ and, as usual $x = k\tau$. 

For $\tau > - \tau_{1}$ the creation and annihilation operators can be expressed as a linear 
combination of the $\hat{a}_{\vec{k}}$ and $\hat{a}_{\vec{k}}^{\dagger}$ i.e. 
\begin{eqnarray}
&&\hat{b}_{\vk} =  B_{+}(k) \hat{a}_{\vk}(\tau_0) + B_{-}(k)^{\ast}
\hat{a}_{-\vk}^{\dagger}(\tau_0),
\nonumber\\
&& \hat{b}_{\vk}^{\dagger} = B_{+}(k)^{\ast} \hat{a}_{\vk}^{\dagger}(\tau_0) 
+ B_{-}\hat{a}_{-\vk}(\tau_0),
\label{BOG1}
\end{eqnarray}
and analogously for $\vk\to -\vk$. 
During radiation the pump field changes since, according to Eq. (\ref{ar}),
$a_{\rm r}''/a_{\rm r}=0$; the mode functions are then plane waves, i.e.  
\begin{eqnarray}
&& \tilde{f}_{\rm r}(k,\tau) = \frac{1}{\sqrt{2 k} } e^{ - i y},
\nonumber\\
&&  \tilde{g}_{\rm i}(k,\tau) = -i  \sqrt{\frac{k}{2}}  e^{- i y},
\label{radmode}
\end{eqnarray}
where $ y = k [\beta \tau + (\beta + 1) \tau_{1}]$. The field operators can then be written 
in terms of the new vacuum annihilated by $\hat{b}_{\vk}$ and $\hat{b}_{-\vk}$, i.e. 
\begin{eqnarray}
&& \hat{\mu}_{\vk} = \hat{b}_{\vk} \tilde{f}_{\rm r} + \hat{b}_{-\vk}^{\dagger}
\tilde{f}_{\rm r}^{\ast},
\nonumber\\
&&\hat{\pi}_{\vk} = \hat{b}_{\vk} \tilde{g}_{\rm r} + \hat{b}_{-\vk}^{\dagger}
\tilde{g}_{\rm r}^{\ast}.
\label{secsol}
\end{eqnarray}
Inserting Eq. (\ref{BOG1}) into Eq. (\ref{secsol}) we can get the expression 
of the field operators in terms of the ``new" mode functions and of the 
``old" creation and annihilation operators. The result is 
simply 
\begin{eqnarray}
&& \hat{\mu}_{\vk}(\tau) = \hat{a}_{\vk} [   B_{+}(k) f_{\rm r} + B_{-}(k) 
f_{\rm r}^{\ast}]+ \hat{a}_{-\vk}^{\dagger}[   B_{+}(k)^{\ast} 
f_{\rm r}^{\ast} + B_{-}(k) ^{\ast}f_{\rm r}],
\label{solmurad}\\
&& \hat{\pi}_{k}(\tau) = \hat{a}_{\vk} [  B_{+}(k) g_{\rm r} + B_{-} g_{\rm r}^{\ast}  ]  
+ \hat{a}_{-\vk}^{\dagger}  [  B_{+}(k)^{\ast} g_{\rm r}^{\ast} + B_{-}^{\ast} g_{\rm r}  ] .
\label{solpirad}
\end{eqnarray}
Since the evolution of the canonical fields must be continuous, Eqs. (\ref{solmuIN})--(\ref{solpiIN}) 
and Eqs. (\ref{solmurad})--(\ref{solpirad}) imply 
\begin{eqnarray}
&& f_{\rm i}(-\tau_1) = B_{+}(k) f_{\rm r}(-\tau_1) + B_{-}(k)f_{\rm r}^{\ast}(-\tau_1),
\nonumber\\
&& g_{\rm i}(- \tau_1)  =  B_{+}(k) g_{\rm r}(-\tau_1) + B_{-}(k) g_{\rm r}^{\ast}(-\tau_1),
\label{match1}
\end{eqnarray}
which allows to determine the coefficients of the Bogoliubov transformation $B_{\pm}(k)$.
Inserting Eqs. (\ref{infmode}) and (\ref{radmode}) into Eq. (\ref{match1}) we get 
\begin{equation}
B_{\pm}(k) = \frac{\cal N}{2}\,\, e^{\pm i x_1}\,\,\sqrt{x_1} \biggl\{ H_{\nu}^{(1)}(x_1) 
\biggl[ 1 \pm \frac{2 i}{x_1}\biggl(\nu - \frac{1}{2}\biggr)\biggr] \mp i H_{\nu -1}^{(1)}(x_1)\biggr\}.
\end{equation}
Notice that because of the unitarity of the evolution $|B_{+}(k)|^2 - |B_{-}(k)|^2 =1$. Since 
$B_{+}(k)$ and $B_{-}(k)$ are complex, the unitary evolution implies that the 
Bogoliubov transformation is completely determined by three real numbers.

The mean number of produced pairs of gravitons
 with momentum $k_{2} < k <k_{1}$
 will then be given by 
\begin{equation}
\overline{n}(k)  = \langle 0 | \hat{N} | 0 \rangle = |B_{-}|^2,
\end{equation}
where $ \hat{N} = [\hat{b}_{\vk}^{\dagger} \hat{b}_{\vk} +  
 \hat{b}_{-\vk}^{\dagger} \hat{b}_{-\vk}]$. Since $k <k_1 \sim\tau_{1}^{-1}$, we also 
 have that $x_{1}<1$. Hence we have that the expression of the mean 
 number of produced particles can be written for $|x_{1}|\ll1$, with 
 the result that 
 \begin{equation}
\overline{n}(k) = \frac{2^{2\nu -5}}{\pi} \Gamma(\nu)^2
(1 - 2\nu)^2 |k\tau_{1}|^{- 2\nu -1}, \,\,\,\,\,\,\,\,\,\,\,\,\,\, k_{2} < k < k_{1}.
\end{equation}
 In the case of pure de Sitter space, $\beta =1$, $\nu = 3/2$, 
 $\overline{n}(k) = 1/(4 x_1^4)$. 
 
 The energy density of the relic gravitons 
 can be written as 
 \begin{equation}
 d \rho_{\rm GW} = 2 \overline{n}(k) \frac{d^{3}k}{(2\pi)^3},
 \end{equation}
 where the factor $2$ accounts for the two polarizations.
 Hence, the logarithmic energy spectrum can be written as 
 \begin{equation}
 \frac{d \rho_{\rm GW}}{d \ln{k}} = \frac{k^4}{\pi^2} \overline{n}(k).
\label{logspec}
 \end{equation}
The logarithmic energy spectrum can be usefully presented as a fraction of the critical 
energy. Dividing Eq. (\ref{logspec}) by the energy density of the radiation background 
at $\tau_1$, i.e. $\rho_{\gamma} \simeq 3  H_{1}^2 M_{\rm P}^2 (a_1/a)^4$ 
we obtain 
\begin{equation}
\Omega_{\rm GW}(\omega) \simeq \biggl(\frac{H_{1}}{M_{\rm P}}\biggr)^2
 \biggl(\frac{\omega}{\omega_1}\biggr)^4 \overline{n}(\omega) 
\sim \biggl(\frac{H_{1}}{M_{\rm P}}\biggr)^2   \biggl(\frac{\omega}{\omega_1}\biggr)^{3 - 2\nu}, 
\,\,\,\,\,\,\,\,\,\,\,\,\,\,\,\omega_{2} <\omega < \omega_{1},
\label{OM1}
\end{equation}
where we switched from comoving to physical frequencies.

Equation (\ref{OM1}) shows that $\Omega_{\rm GW}$ produced in power-law inflation 
(and in all conventional inflationary models) can be (at most) flat. The completely flat spectrum 
arises in the pure de Sitter case (i.e. $\beta =1$). In the case of standard power law and slow-roll
inflation $\beta >1$ and $\Omega_{\rm GW}$ decreases.
Discounting for a possible variation of the relativistic degrees of freedom, 
the  maximal frequency of the spectrum, 
$\omega_{1} \sim 100 \sqrt{H_{1}/M_{\rm P}}\,\,{\rm GHz}$., while $\omega_{2} \sim 10^{-16} \,\,\, {\rm Hz}$.

Let us finally compute the spectrum in the infrared branch, i.e. for comoving frequencies $k < k_{2}$.
During the matter-dominated stage, 
the evolution of the mode functions is given by 
\begin{eqnarray}
&& f_{\rm m}(k,\tau) = \frac{1}{\sqrt{2 k}} \biggl( 1 - \frac{i}{z}\biggr) e^{- i z}
\nonumber\\
&& g_{\rm m}(k,\tau) =  - i \sqrt{\frac{k}{2}} \biggl[ 1 - \frac{i}{z} - \frac{1}{z^2}\biggr] e^{- i z}
\label{matmode}
\end{eqnarray}
where $ z = [\beta (\tau + \tau_{2}) + 2 (\beta + 1) \tau_1]/2$. Hence, repeating the same steps 
outlined above, the mode functions during radiation can be matched, in $\tau_2$ to the mode 
functions during matter by means of two (complex) mixing coefficients, i.e. 
\begin{eqnarray}
&& f_{\rm r}(\tau_2) = C_{+} f_{\rm m}(\tau_2) + C_{-} f^{\ast}_{\rm m}(\tau_2),
\nonumber\\
&& g_{\rm r}(\tau_2) = C_{+} g_{\rm m}(\tau_2) + C_{-} g^{\ast}_{\rm m}(\tau_2).
\end{eqnarray}
The mixing coefficients are then determined to be 
\begin{eqnarray}
&&C_{\pm} = \frac{- B_{\mp} e^{\pm 2 i [x_1 + (x_1 +x_2)\beta]} }{2 [ x_1+ (x_1 + x_2)\beta]^2}
\nonumber\\
&& +\frac{ B_{\pm}\{-1 + 2 [x_1 + (x_1 + x_2)\beta][ x_1 \pm i +  
(x_1 + x_2)\beta]\}}{2 [ x_1+ (x_1 + x_2)\beta]^2}
\end{eqnarray}
The mean number of gravitons will then be, for $x_{2} \ll 1$
\begin{equation}
\overline{n}(k) \simeq \frac{(2\,i \,\beta -1 )\left(  B_{+}+ B_{-} \right)   - 4\,i \,B_{-}\,\beta }{2\,x_2\,{\beta }^2}.
\end{equation}
This result implies that for $k <k_{2}$ the logarithmic energy spectrum decreases faster than 
in the high-frequency branch. 
\begin{equation}
\Omega_{\rm GW} \propto \biggl( \frac{\omega}{\omega_1}\biggr)^{3 - 2\nu} \biggl(\frac{\omega_{2}}{\omega}\biggr)^2.
\end{equation}
\begin{figure}[tp]
\centerline{\epsfig{file=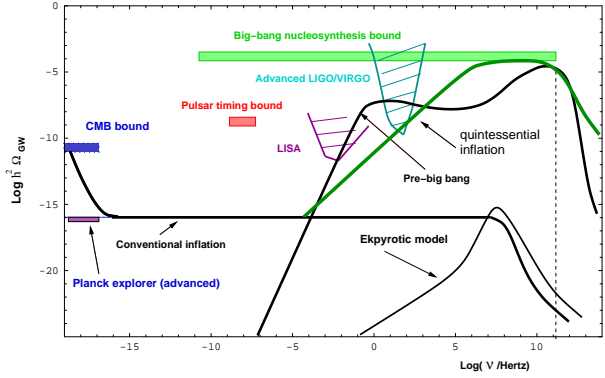, width=8cm}}
\vspace*{8pt}
\caption[a]{The logarithmic energy spectrum of relic gravitons is illustrated in different models 
of the early Universe as a function of the present frequency, $\nu$.}
\label{F71}
\end{figure}
The spectra of relic gravitons arising in conventional inflationary models, as well as in other models
are reported in Fig. \ref{F71} as a function of the present frequency. Notice that in the study 
of stochastic GW backgrounds the logarithmic energy spectrum is given typically as a function of the 
present physical frequency $\nu$ or as a function of the present physical wave-number $\omega$. 
The relation between the two quantities is $\omega = 2 \pi \nu$. Notice, furthermore, that 
$ \Omega_{\rm GW}$  enters the expression of the signal-to-noise 
ratio for the correlation between two detectors (say two  wide-band interferometers). 
This formula (see below) depends on the value of $h$, i.e. the present indetermination 
on the Huble rate.  So sometimes it is useful to discuss directly $ h^2 \Omega_{\rm GW}$.
In terms of the physical frequency we have that the logarithmic energy spectrum 
of relic gravitons in the pure de Sitter case can be  written as 
\begin{eqnarray}
&&\Omega_{GW}(\nu,\tau_0) \simeq \Omega_{\gamma}(\tau_0) 
\biggl(\frac{H_1}{M_{P}}\biggr)^2,~~~~~~~~~~~\nu_{{\rm dec}} < \nu < \nu_{1} 
\nonumber\\
&& \Omega_{GW}(\nu,\tau_0) \simeq \Omega_{\gamma}(\tau_0) 
\biggl(\frac{H_1}{M_{P}}\biggr)^2 
\biggl(\frac{\nu_{{\rm dec}}}{\nu}\biggr)^2,
~~~\nu_{0}< \nu < \nu_{{\rm dec}} 
\end{eqnarray}
where $ \Omega_{\gamma} \simeq 7.95 \times 10^{-5}$ is the 
present fraction of critical energy density attributed to radiation and $H_{1}$ is the 
curvature scale of the inflationary phase (strictly constant for this estimate).

\subsection{Constraints on the spectra of relic gravitons}
If an inflationary phase 
is followed by a radiation dominated phase 
preceding the  matter dominated epoch, 
the amplitude of the produced gravitons background can be computed  
and the result is illustrated  in Fig. \ref{F71} with the flat plateau.
The idea that stochastic backgrounds of relic gravitons can be produced in the early Universe 
goes back to the works of Grishchuk \cite{gr1,gr2} (see also \cite{thorne1} for a classical review 
and \cite{grrev} for a more recent one).

The spectrum consists of two branches 
a soft branch ranging  between 
$\nu_0 \simeq 10^{-18}~h$ Hz and 
$\nu_{{\rm dec}}\simeq  10^{-16}$ Hz. 
For 
$\nu >\nu_{{\rm dec}}$ we have instead the  hard
 branch consisting of high frequency gravitons mainly produced 
thanks to the transition from the inflationary regime to radiation. In the 
soft branch $\Omega_{{\rm GW}}(\nu,\tau_0) \sim \nu^{-2}$. 
In the hard branch $\Omega_{{\rm GW}}(\nu,\tau_0) $
 is  constant in frequency (or slightly decreasing in the quasi-de Sitter case). 
The soft branch was computed for the first 
time in \cite{RUB,FP,AB}. The hard branch has been 
computed originally in \cite{STAR} (see also \cite{ALLEN,SAHNI,SOL,GGGW}).  

The large-scale observation of the first multipole moments 
of the temperature anisotropy imply a bound for the relic graviton background. 
The rationale for this statement is very simple since relic gravitons contribute 
to the integrated Sachs-Wolfe effect as discussed in section 2.  
The gravitational wave contribution to the Sachs-Wolfe 
integral 
cannot be larger than the (measured) amount of anisotropy directly
detected. The soft branch 
of the  spectrum is then constrained and  the bound reads \cite{ALLEN2}
\begin{equation}
\Omega_{{\rm GW}}(\nu,\tau_0)h^2 \laq~ 6.9~ \times 10^{-11},
\label{cobe}
\end{equation}
for  $\nu\sim \nu_{0} \sim 10^{-18} {\rm Hz}$.
Moreover, the very small size of the fractional timing error in the 
arrivals of the millisecond plusar's pulses imply that also the hard
branch is bounded according to 
\begin{equation}
\Omega_{{\rm GW}}(\nu, \tau_0)~ \laq~ 10^{-8},
\label{puls}
\end{equation}
for $\nu\sim 10^{-8}$ Hz corresponding, roughly, to the inverse of the 
observation time during which the various millisecond pulsars have been 
monitored \cite{taylor} (see also \cite{SCHUTZ}).

The two constraints of Eqs. (\ref{cobe}) and (\ref{puls}) are reported 
in Fig. \ref{F71}, at the two relevant frequencies.
The Sachs-Wolfe and millisecond pulsar constraints are  differential
since they limit, locally, the logarithmic derivative of the 
gravitons energy density. There exists also an  integral 
bound coming from standard BBN analysis \cite{schw} 
and constraining the integrated graviton energy spectrum: 
\begin{equation}
h^2\int_{\nu_{{\rm n}}}^{\nu_{{\rm max}}} \Omega_{{\rm GW}}(\nu,\tau_0) d\ln{\nu}
\laq ~ 0.2\,\,\,h^2 \Omega_{\gamma} \simeq 8\times 10^{-6}
\label{NS}
\end{equation}
for $h \simeq 0.72$, $\Omega_{\gamma} \sim 7.95 \times 10^{-5}$.
In Eq. (\ref{NS})  $\nu_{{\rm max}}$ 
corresponds to the (model dependent) ultra-violet cut-off 
of the spectrum and $\nu_{{\rm n}}$ is the frequency corresponding 
to the horizon scale at nucleosynthesis.
 Notice 
that the BBN 
constraint of Eq. (\ref{NS}) has been derived in the 
context of the simplest BBN model, namely, 
assuming that no inhomogeneities and/or matter anti--matter domains 
are present at the onset of nucleosynthesis. In the presence of  
matter--antimatter domains for scales comparable with the 
neutron diffusion scale \cite{mam2,mam3,HKS} this bound is relaxed. 

From Fig. \ref{F71} we see that also  the global bound of 
 Eq. (\ref{NS})  is satisfied and the typical amplitude of the 
logarithmic energy spectrum in critical units 
for frequencies $\nu_{I} \sim 100 $ Hz (and larger)
cannot exceed $10^{-14}$. This amplitude 
has to be compared with the LIGO sensitivity to a flat
$\Omega_{{\rm GW}}(\nu_{I}, \tau_0)$ 
which could be {\em at most} of the order of 
$h^2 \Omega_{{\rm GW}}(\nu_{I},\tau_0)
= 5\times 10^{-11}$ after four months of
 observation with $90\% $ confidence \cite{ALLEN3}.

\subsection{Relic gravitons from a stiff phase}

Suppose now, as a toy example, that the ordinary inflationary phase is not immediately 
followed by a radiation dominated phase but by a quite long phase 
expanding slower than radiation \cite{maxq1}. This speculation  is theoretically 
plausible since we ignore what was the thermodynamical history of the Universe 
prior to BBN. If the Universe expanded slower than radiation the equation of state 
of the effective sources driving the geometry had to be, for some time, 
stiffer than radiation. This means that the effective speed of sound $c_s$ 
had to lie in the range $1/\sqrt{3} < c_{s} \leq 1$.
Then the resulting logarithmic energy spectrum,  for the 
modes leaving the horizon during the inflationary phase and 
re-entering during the stiff phase, is tilted towards large 
frequencies with typical (blue) slope given by \cite{maxq1}
\begin{equation}
 \frac{d \ln{\Omega_{\rm GW}}}{d\ln{\nu}} = \frac{ 6 c_s^2 - 2 }{3 c_s^2 +1},
\end{equation}
which is always limited between $0$ (matter) and $1$ stiff fluid.
A situation very similar to the one we just described occurs in 
quintessential inflationary models \cite{alex1}. In this case the 
tilt is maximal  and a more precise calculation shows 
the appearance of logarithmic corrections in the logarithmic 
energy spectrum which becomes \cite{maxq2,maxq3} , 
$\Omega_{\rm GW} \propto \nu \ln^2{\nu}$ (see Fig. \ref{F71}).

The maximal frequency $\nu_{\rm max}(\tau_0)$ is of the order of $100$ GHz
(to be compared with the $100$ MHz of ordinary inflationary models)
and it corresponds to 
the typical frequency of a  spike in the GW background. In quintessential 
inflationary models the relic graviton background will then have 
the usual infra-red and flat branches supplemented, at high 
frequencies (larger than the mHz and smaller than the GHz) by a true 
spike \cite{maxq3} whose peak can be, in terms of $h^2 \,\,\, 
\Omega_{\rm GW}$, 
 of the order of $10^{-6}$, compatible 
with the BBN bound and  roughly eight orders of magnitude larger than the 
signal provided by ordinary inflationary models. 

An interesting aspect of this class of models is that the maximal signal 
occurs in a frequency region between the MHz and the GHz. 
Microwave cavities can be used as GW detectors precisely in the 
mentioned frequency range (see below).  This signal is certainly a candidate 
for this type of devices. The sensitivity of wide-band interferometers 
to quintessential gravitons has been computed in \cite{gioquint}.

\subsection{Relic gravitons from pre-big bang}

In  pre-big bang models 
$h^2 \,\,\,\Omega_{\rm GW}$ can 
be as large as $10^{-7}$--$10^{-6}$ for frequencies ranging between $1$ Hz and 
$100$ GHz \cite{BDPBB1,BDPBB2,GGGW}. 
The logarithmic energy spectrum can be 
either blue or violet depending upon the given mode of the spectrum. If the mode
under consideration 
left the horizon during the dilaton-dominated epoch the typical 
slope will be violet (i.e. $ \Omega_{\rm GW} \sim \nu^3 $ up to logarithmic corrections).
If the given mode left the horizon during the stringy phase the slope can 
be also blue (i.e. less steep). An example is reported in Fig. \ref{F71}. This 
behaviour is representative of the minimal string cosmological 
scenarios. However, in the non-minimal case the spectra can also be non 
monotonic. Recently the sensitivity of a pair of VIRGO detectors
to string cosmological gravitons was specifically analyzed \cite{gio10} with the 
conclusion that a VIRGO pair, in its upgraded stage, will certainly be able to probe 
wide regions of the parameter space of these models. If we  maximize the 
overlap between the two detectors \cite{gio10} or 
if we would  reduce (selectively) the pendulum and pendulum's internal modes
contribution to the thermal noise of the instruments \cite{gio11}, the 
visible region (after one year of observation and with SNR equal to one)
of the parameter space will get even larger. Unfortunately, as in the 
case of the advanced LIGO detectors, also in the case of the advanced VIRGO 
detector the sensitivity to a flat spectrum will be irrelevant for 
ordinary inflationary models.

\subsection{Detectors of relic gravitons}

 GW detectors can be divided in three broad classes: resonant 
mass detectors, interferometers and microwave cavities.
There are five (cryogenic) resonant mass detectors which are 
now operating: NIOBE \cite{niobe} (Perth, Australia), 
ALLEGRO \cite{allegro} (Baton Rouge, Lousiana, USA), AURIGA 
\cite{auriga} (Legnaro, Italy), EXPLORER \cite{explorer}
(Geneva, Switzerland) and NAUTILUS \cite{nautilus} 
(Frascati, Italy). They all have 
cylindrical shape (the are ``bars''). They are all made in Aluminium 
(except NIOBE which is made of Niobium). Their approximate mass is of the 
order of $2200$ kg (except NIOBE whose mass is of the order of $1500$ kg).
Their mode frequencies range from $694$ Hz (in the case of NIOBE) to 
the $912$ Hz of AURIGA.

There are, at the moment, four Michelson-Morley interferometers 
being built. They are 
GEO \cite{geo} (Hannover, Germany), TAMA \cite{tama} (Tokyo, Japan), 
VIRGO \cite{virgo} (Cascina, Italy),
and the two LIGO \cite{ligo} 
(in Hanford [Washington], and Livingston [Lousiana], USA).
The arms of the instruments range from the 400 m of TAMA up to the 
three km of VIRGO and to the 4 km of LIGO. The effective optical path of 
the photons in the interferometers is greatly enhanced by the use of 
Fabry-P\'erot cavities.

Microwave cavities have been originally proposed as GW detectors in the 
GHz--MHz region of the spectrum \cite{mw1}. A first prototype has been 
built in MIT in 1978 showing that this idea could be 
actually implemented in order to detct small harmonic displacements \cite{mw2}. 
It is not unreasonable to think that sensitive measurements could 
be performed in the near future. In particular improvements in the quality 
factors of the cavities (if compared 
with the prototypes of \cite{mw1}) could be foreseen.
Two experiments (in Italy \cite{paco} and in England \cite{bir})
 are now trying to achieve this goal with slightly different technologies.

\renewcommand{\theequation}{8.\arabic{equation}}
\section{Radiative transfer equations}
\setcounter{equation}{0}

The properties of the 
radiation field at decoupling are affected by the presence of metric fluctuations and 
a preliminary analysis of  this problem has been presented in section 3. 
The effect of metric inhomogeneities on the properties  
of the radiation field will now be analyzed using the radiative transfer (or 
radiative transport) equations that can be integrated 
numerically even if useful analytical 
results can also be obtained. 
A classical preliminary reference  
 is the textbook of Chandrasekar
\cite{CHAN1} (see in particular chapter 1 in light of the calculation 
of the collision term of  Thompson scattering that is quite relevant for the present ends).
Another recent reference is \cite{PER1}. In broad terms the radiative transfer 
equations describe the evolution of the Stokes parameters of the radiation field through 
some layer of matter which could be, for instance, the stellar atmosphere or, in the present case, 
the primeval plasma prior to the recombination of free electrons and protons.

In the present framework the radiative transfer equations have a further complication with 
respect to the flat space case:  the 
collisionless part of the Boltzmann equation  is modified by the inhomogeneities 
of the geometry. These 
inhomogeneities induce a direct coupling of the Boltzmann equation 
to the perturbed Einstein equations.  An interesting system 
of equations naturally emerges: the Einstein-Boltzmann
system of equations which is, in some approximation,  exactly 
what has been described in section 4 for the discussion of the initial 
conditions of CMB anisotropies. In that case the perturbed Einstein equations 
were coupled to a set of fluid equations for the density contrasts 
and for the peculiar velocities. These are, indeed, the first two terms (i.e. the 
monopole and the dipole)
in the Boltzmann hierarchy.  Truncated 
Boltzmann hierarchies are a useful tool for the analysis of
initial conditions, but  their limitations have been already 
emphasized in connection  with the description of collisionless 
particles.

While the general  conventions established  in the previous sections 
will be consistently enforced, further conventions related to the specific way  the brightness 
perturbations  are defined \footnote{See Eqs. (\ref{deBRf1})--(\ref{deBRf4}) below for the 
definition of brightness perturbation.}.   Denoting by $\Delta$ a  brightness 
perturbation (related generically to one of the four Stokes parameters 
of the radiation field), the expansion of $\Delta$ in terms of Legendre 
polynomials will be written, in this paper, as 
\begin{equation}
\Delta(\vec{k}, \hat{n}, \tau) = \sum_{\ell} (-i)^{\ell}\,\, (2 \ell + 1)\,\, \Delta_{\ell} (\vec{k},\tau)\,\,P_{\ell}(\hat{k}\cdot\hat{n}).
\label{CONV}
\end{equation}
where $\vec{k}$  is the momentum of the Fourier expansion, $\hat{k}$ its direction; $ \hat{n}$ 
is the direction of the photon momentum; $P_{\ell}(\hat{k}\cdot\hat{n})$ are the Legendre polynomials. The same 
expansion will be consistently employed for the momentum averaged phase-space 
density perturbation (see below Eq. (\ref{reddist})). This quantity will be also 
called, for short, reduced phase-space density and it is related to the brightness 
perturbation by a numerical factor. The conventions 
of \cite{HS1,HS2,HS3} (see also \cite{HS4})  are such that the factor $(2\ell +1)$  {\em  is not} 
included in the expansion.  Furthermore, in \cite{HS1,HS2,HS2a} the metric fluctuations 
are parametrized in terms of the Bardeen potential while in \cite{HS3} 
the treatment follows the conformally Newtonian gauge. 
Finally, in \cite{MB}  the conventions are the same as the ones of Eq. (\ref{CONV})
but  the metric convention is mostly plus (i.e. 
$-,+, +, +$) and the definition of the longitudinal 
degrees of freedom is inverted (i. e.  Ref.  \cite{MB} calls $\psi$ what we call $\phi$ and viceversa).
In \cite{HZ1,HZ2,HZ3} (see also \cite{kos2,KL})  the expansion of the brightness perturbation is different with respect 
to Eq. (\ref{CONV}) since the authors {\em do not} include the factor $(-i)^{\ell}$ in the expansion. In the latter case
the collision terms are modified by a sign difference in the 
dipole terms (involving a mismatch of $(-i)^2$ with respect to the conventions fixed by Eq. (\ref{CONV})). 

\subsection{Collisionless Boltzmann equation}
If the space-time would be 
homogeneous the position variables $x^{i}$ and  the conjugate momenta $P_{j}$ could 
constitute a practical set of pivot variables for the analysis of Boltzmann equation in curved 
backgrounds. However, since, in the 
present case, the space-time 
is not fully homogeneous, metric perturbations 
do affect the definition of conjugate momenta.
Hence,  for practical reasons, the approach usually followed is to write
the Boltzmann equations in terms of the proper moementa, i.e. 
the momentum measured by an observer at a fixed value of the spatial coordinate. 

Consider, for simplicity, the case of massless particles (like photons or massless neutrinos).
Their mass-shell condition can be written, in a curved background, as 
\begin{equation}
g_{\alpha\beta}P^{\alpha} P^{\beta}=0,
\label{MS}
\end{equation}
where  $g_{\alpha\beta}$  is now  the full
metric tensor (i.e. background plus inhomogeneities).
Equation (\ref{MS}) implies, quite trivially, 
\begin{equation}
g_{00} P^{0} P^{0} = - g_{ij} P^{i} P^{j} \equiv \delta_{ij} p^{i} p^{j},
\end{equation}
where the second equality is the definition of the physical
three momentum $p_{i}$.  Recalling that, to first order and in the 
longitudinal gauge, 
$g_{00}= a^2( 1 + 2\phi)$ and   $g_{i j} = - a^2 ( 1 - 2 \psi) \delta_{ij}$, then 
the relation between the conjugate momenta and the physical three-momenta 
can be easily obtained by expanding the obtained expressions 
for small $\phi$ and $\psi$. The result is simply 
\begin{eqnarray}
&& P^{0} = \frac{p}{a} ( 1 - \phi) = \frac{q}{a^2} ( 1 - \phi),\,\,\,\,\,\,\,\,\,\,\, P_{0} = a p ( 1 + \phi) = q ( 1 + \phi),
\nonumber\\
&& P^{i} =\frac{p^{i}}{a} ( 1 + \psi) = \frac{q^i}{a^2} ( 1 + \psi),\,\,\,\,\,\,\,\,\,\,\, P_{i} = - a p_{i} ( 1 - \psi) =- q_{i} ( 1 - \psi).
\label{CtoP}
\end{eqnarray}
The quantity $q_{i}$ defined in Eq. (\ref{CtoP}) are nothing but the 
comoving three-momenta, i.e.  $ p_{i} a = q_i$, while 
 $q =p a $ is the modulus of the comoving three-momentum. 
 Generalization of Eq. (\ref{CtoP}) is trivial since, in the 
 massive case, the mass-shell condition implies that 
 $g_{\alpha\beta}P^{\alpha} P^{\beta} = m^2$ and, for instance 
 $P^{0} = \sqrt{q^2 + m^2 a^2} ( 1 - \phi)$.
 
In terms of the  modulus and direction of the comoving three-momentum \cite{bond1}, i.e. 
\begin{equation}
q_{i} = q n_{i}, \,\,\,\,\,\, n_{i}n^{i} = n_{i} n_{j} \delta^{ij} =1,
\label{C3m}
\end{equation}
the Boltzmann equation can be written as  
\begin{equation}
\frac{D f}{D \tau} = \frac{\partial f}{\partial \tau} + \frac{\partial x^{i}}{\partial\tau} \frac{\partial f}{\partial x^{i}} + 
\frac{\partial f}{\partial q} \frac{\partial q}{\partial \tau} + \frac{\partial f}{\partial n_{i}} \frac{\partial n^{i}}{\partial \tau} = 
{\cal C}_{\rm coll},
\label{BZ1}
\end{equation}
where a generic collision term, ${\cal C}_{\rm coll}$ has  been included for future convenience. 

Eq. (\ref{BZ1}) can now be perturbed around a configuration of local 
thermodynamical equilibrium by writing
\begin{equation}
f( x^i, q, n_{j}, \tau)= f_{0}(q) [ 1 + f^{(1)}( x^i, q, n_{j}, \tau)],
\label{expBZ}
\end{equation}
where $f_0(q)$ is the Bose-Einstein (or Fermi-Dirac in the case of fermionic 
degrees of freedom) distribution. Notice that $f_{0}(q)$  does not depend on
$n^{i}$ but only on $q$.

Inserting Eq. (\ref{expBZ}) into Eq. (\ref{BZ1}) the first-order form of the perturbed 
Boltzmann equation  can be readily obtained 
\begin{equation}
f_{0}(q)\,\, \frac{\partial f^{(1)}}{\partial \tau} + f_0(q)\,\,\frac{\partial f^{(1)}}{\partial x^{i}} \,\,n^{i}+ 
\frac{\partial f_0}{\partial q} \frac{\partial q}{\partial \tau}= 
{\cal C}_{\rm coll},
\label{BZ2}
\end{equation}
by appreciating  that a pair of terms 
\begin{equation}
\frac{\partial f^{(1)}}{\partial q} \frac{\partial q}{\partial \tau}, \,\,\,\,\,\,\,\,\,\,\,\,\,\,\,\,\,
 \frac{\partial f^{(1)}}{\partial n_{i}} \frac{\partial n^{i}}{\partial \tau},
\end{equation}
are of higher order (i.e. ${\cal O}(\psi^2)$) and have been neglected to first-order.

Dividing by $f_{0}$ Eq. (\ref{BZ2}) can also be written as 
\begin{equation}
\frac{\partial f^{(1)}}{\partial \tau} +   \frac{\partial f^{(1)}}{\partial x^{i}}  \,n^{i}+ 
\frac{\partial \ln{f_0}}{\partial q} \frac{\partial q}{\partial \tau}= 
\frac{1}{f_{0}}{\cal C}_{\rm coll}.
\label{BZ3}
\end{equation}
Notice that in Eq. (\ref{BZ2})--(\ref{BZ3})  the generalization  of known 
special relativistic expressions 
\begin{equation}
\frac{ d x^{i}}{d\tau} = \frac{P^{i}}{P^{0}} = \frac{q^{i}}{q} = n^{i},
\end{equation}
has been used.

To complete the derivation, $dq/d\tau$ must be written 
in explicit terms.  The geodesic 
equation gives essentially the first time derivative of the 
conjugate momentum, i.e. 
\begin{equation}
\frac{d P^{\mu}}{d s } = P^{0} \frac{d P^{\mu}}{d \tau} = - \Gamma^{\mu}_{\alpha\beta} P^{\alpha} P^{\beta},
\label{geod1a}
\end{equation}
where $s$ is the affine parameter.
As before in this section, $\Gamma^{\mu}_{\alpha\beta}$ denotes the full Christoffel connection (background 
plus fulctuations).  Using the values of the perturbed connections in the longitudinal gauge 
(take Eq. (\ref{SCHR}) of the appendix and set $B=E=0$),  Eq. (\ref{geod1a}) becomes  
\begin{equation}
\frac{d P^{i}}{d\tau} = - \partial^{i} \phi P^{0} + 2 \psi' P^{i} - 2 {\cal H} P^{i} - 
\frac{P^{j} P^{k}}{P^{0}} [ \partial^{i} \psi \delta_{jk} - \partial_{k} \psi \delta_{j}^{i} - \partial_{j} \psi \delta^{i}_{k}].
\label{geod2a}
\end{equation}
Recalling now that $q = q_{i} n^{i}$, the explicit form of $d q/d\tau$ will be 
\begin{equation}
\frac{d q}{d\tau} = \biggl[ \frac{\partial P^{i}}{\partial \tau} a^2 ( 1 - \psi) + 2 {\cal H} a^2 (1 - \psi ) P^{i} - a^2 \psi' P^{i} \biggr] n_{i}  - P^{i} a^2 \partial_{j}  \psi n^{j} n_{i}. 
\label{qtau}
\end{equation}
Inserting now Eq. (\ref{geod2a}) into Eq. (\ref{qtau}) 
the explicit form of $d q/d\tau$ becomes\footnote{To derive Eq. (\ref{qtau2}) from Eq. (\ref{qtau}),
the factors $P^{i}$ and $P^{0}$ appearing at the right hand side 
 Eq. (\ref{geod2a}) have to be replaced with their first-order 
expression in terms of the comoving-three momentum $q^{i}$ (and $q$) 
 as previously discussed in Eqs. (\ref{CtoP}).}
\begin{equation}
\frac{d q }{ d\tau} = q \psi' - q n_{i} \partial^{i} \phi.
\label{qtau2}
\end{equation}
Finally, using Eq. (\ref{qtau2}) into Eq.  (\ref{BZ3}) to eliminate $dq/d\tau$ the final form of the 
Boltzmann equation for massless particles becomes:
\begin{equation}
\frac{\partial f^{(1)}}{\partial \tau} +  n^{i} \frac{\partial f^{(1)}}{\partial x^{i}} + 
\frac{\partial \ln{f_0}}{\partial\ln q} [\psi' -  n_{i} \partial^{i} \phi] = 
\frac{1}{f_{0}}{\cal C}_{\rm coll},
\end{equation}
which can be also written, going to Fourier space, as 
\begin{equation}
\frac{\partial f^{(1)}}{\partial \tau} +  i k\mu f^{(1)} + 
\frac{\partial \ln{f_0}}{\partial\ln q} [\psi' -  i k\mu \phi] = 
\frac{1}{f_{0}}{\cal C}_{\rm coll},
\label{BZ4}
\end{equation}
where we have denoted, according to the standard notation, $k$ as the Fourier mode and 
$\mu = \hat{k}\cdot \hat{n} $ as the projection of the Fourier mode along 
the direction of the photon momentum \footnote{Notice that here there may be, in principle, a 
clash of notations since, in section 5 (see for instance Eq. (\ref{QEQ})) 
we denoted with $q$ and $\mu$ the 
normal modes for the scalar and tensor actions; in the present section $q$ and $\mu$ 
denote, on the contrary the comoving three-momentum and the cosinus between the Fourier 
mode and the photon direction. The two sets of variables never appear together and there should not
be obvious confusion.}. Clearly,  given the axial symmetry of the problem 
it will be natural to identify  the direction of $\vec{k}$ with the $\hat{z}$ direction in which 
case $\mu =\cos{\theta}$.

The result obtained so far can be easily generalized to the case of massive particles 
\begin{equation}
\frac{\partial f^{(1)}}{\partial \tau} +  i \alpha(q,m) k\mu f^{(1)} + 
\frac{\partial \ln{f_0}}{\partial\ln q} [\psi' -  i  \alpha(q,m) k\mu \phi] = 
\frac{1}{f_{0}}{\cal C}_{\rm coll},
\label{BZ5}
\end{equation}
where $\alpha(q,m) = q/\sqrt{q^2 + m^2 a^2}$ and where, now, the appropriate mass dependence 
has to appear in the equilibrium distribution $f_{0}$. 

\subsection{Boltzmann hierarchy for massless neutrinos}
The Boltzmann equations derived in Eqs. (\ref{BZ4}) and (\ref{BZ5}) are general.  
In the following, two relevant cases will be discussed, namely the case of massless 
neutrinos and the case of photons. 
In order to proceed further with the case of massless neutrinos 
let us define the reduced phase-space distribution as 
\begin{equation}
{\cal F}_{\nu}( \vec{k}, \hat{n}, \tau) = \frac{\int q^{3} d q f_{0} f^{(1)}}{\int q^{3} d q f_{0}}.
\label{reddist}
\end{equation}
Eq. (\ref{BZ4}) becomes, in the absence of collision term,
\begin{equation}
\frac{\partial {\cal F}_{\nu}}{\partial \tau} + i k\mu {\cal F}_{\nu} = 4 (\psi' - i k \mu \phi).
\label{bz3} 
\end{equation}
 The factor $4$ appearing in Eq. (\ref{bz3}) follows 
from the explicit expression of the equilibrium Fermi-Dirac distribution and observing that integration by parts 
implies 
\begin{equation}
\int_{0}^{\infty} q^{3} d q \frac{\partial f_{0}}{ \partial \ln{q}} = - 4  \int_{0}^{\infty} q^{3} d q f_{0}.
\label{4fact}
\end{equation}
The reduced phase-space distribution of Eq. (\ref{reddist})  can be expanded in series of Legendre 
polynomials as defined in Eq. (\ref{CONV})
\begin{equation}
{\cal F}_{\nu}( \vec{k}, \hat{n}, \tau) = \sum_{\ell} (-i)^{\ell} ( 2 \ell + 1) {\cal  F}_{\nu\ell}(\vec{k},\tau) P_{\ell}(\mu).
\label{expF}
\end{equation}
Equation (\ref{expF}) will now be inserted  into Eq. (\ref{bz3}). 
The orthonormality relation for Legendre polynomials,
\begin{equation}
\int_{-1} ^{1}  P_{\ell}(\mu) P_{\ell'}( \mu) d\mu  = \frac{2}{2 \ell + 1} \delta_{\ell\ell'},
\label{norm}
\end{equation}
together with the well-known recurrence relation 
\begin{equation}
(\ell + 1) P_{\ell + 1}(\mu) = (2 \ell +1) \mu P_{\ell}(\mu) - \ell P_{\ell -1}(\mu),
\label{rec1}
\end{equation}
allows to get a hierarchy of differential equations coupling together 
the various multipoles.  After having multiplied
each of the terms of Eq. (\ref{bz3}) by $\mu$, integration of the obtained 
quantity will be performed over $\mu$ (varying between $-1$ and $1$); in formulae:
\begin{eqnarray}
&& \int_{-1}^{1} P_{\ell'}(\mu) {\cal F}_{\nu} d\mu = 2 (-i)^{\ell'} {\cal F}_{\nu\ell'},
\label{r1a}\\
&& ik \int_{-1}^{1} \mu P_{\ell'}(\mu) {\cal F}_{\nu} 
d\mu = 2 i k \biggl[ (-i)^{\ell' + 1} \frac{\ell' +1}{ 2 \ell' + 1}  {\cal F}_{\nu(\ell' +1)}  
+ (-i)^{\ell' -1}\frac{\ell'}{2\ell' + 1} {\cal F}_{\nu (\ell' -1)}\biggr],  
\label{r2}\\
&& 4\int_{-1}^{1}  \psi' P_{\ell'}(\mu) d\mu = 8\psi' \delta_{\ell'0} ,\,\,\,\,\,\,\,\,\,\,\,,\,- 4 i \phi \int_{-1}^{1} \mu P_{\ell'}(\mu) d\mu  = - \frac{8}{3} i k \phi \delta_{\ell' 1}.
\label{r4}
\end{eqnarray}
Equation (\ref{r2}) follows from the relation
\begin{equation}
\int_{-1}^{1} \mu P_{\ell}(\mu) P_{\ell'}(\mu) d\mu = \frac{2}{2\ell + 1} \biggl[ \frac{\ell' +1 }{2\ell' + 1} \delta_{\ell,\ell'+1} 
+ \frac{\ell'}{2\ell' + 1} \delta_{\ell,\ell' -1} \biggr]
\end{equation}
that can be easily derived using Eqs. (\ref{rec1}) and (\ref{norm}).
Inserting Eqs. (\ref{r1a})--(\ref{r4}) into Eq. (\ref{bz3}) the first example of Boltzmann hierarchy can be derived:
\begin{eqnarray}
&& {\cal F}_{\nu 0}' = - k {\cal F}_{\nu 1} + 4 \psi',
\label{mom0}\\
&& {\cal F}_{\nu 1}' = \frac{k}{3} [ {\cal F}_{\nu 0} - 2 {\cal F}_{\nu 2}] + \frac{4}{3} k \phi,
\label{mom1}\\
&& {\cal F}_{\nu\ell}' = \frac{k}{2\ell +1} [ \ell {\cal F}_{\nu,(\ell-1)}  - (\ell+1) {\cal F}_{\nu (\ell+1)}].
\label{mom2}
\end{eqnarray}
Equation (\ref{mom2}) holds for $\ell \geq 2$. Eqs. (\ref{mom0}) and (\ref{mom1}) 
are nothing but the evolution equations for the density contrast and for the neutrino 
velocity field. This aspect can be easily appreciated by computing, in explicit terms, 
the components of the energy-momentum tensor as a function 
of the reduced neutrino phase-space density. In general terms, the energy-momentum 
tensor can be written, in the kinetic approach, as 
\begin{equation}
T_{\mu}^{\nu} = - \int \frac{d^{3} P}{\sqrt{-g}} \frac{P_{\mu}P^{\nu}}{P^{0}} f(x^{i}, P_{j}, \tau).
\label{defFTmunu}
\end{equation}
According to Eq. (\ref{CtoP}), establishing the connection between conjugate momenta 
and comoving three-momenta, the $(00)$ component of Eq. (\ref{defFTmunu}) becomes, for a 
completely homogeneous distribution,
\begin{equation}
\rho_{\nu} = \frac{1}{a^4} \int d^{3} q q  f_{0}(q),
\end{equation}
i.e. the homogeneous energy density.  Using instead the first-order phase space density, 
the density contrast, the peculiar velocity field and the neutrino anisotropic stress 
are connected, respectively, to the monopole, dipole and quadrupole moments of the 
reduced phase-space distribution:
\begin{eqnarray}
&& \delta_{\nu} = \frac{1}{4\pi} \int d \Omega {\cal F}_{\nu}(\vec{k},\hat{n},\tau) = {\cal F}_{\nu 0},
\label{def1a}\\
&& \theta_{\nu} = \frac{3i}{16\pi} \int d\Omega (\vec{k}\cdot \hat{n}) {\cal F}_{\nu}(\vec{k}, \hat{n},\tau) 
= \frac{3}{4} k {\cal F}_{\nu 1},
\label{def0a}\\
&& \sigma_{\nu} = -\frac{3}{16\pi }\int d\Omega \biggl[ (\vec{k}\cdot \hat{n})^2 - \frac{1}{3}\biggr] {\cal F}_{\nu}(\vec{k},\hat{n},\tau) =  \frac{{\cal F}_{\nu 2}}{2}.
\label{def2a}
\end{eqnarray}
Inserting Eqs. (\ref{def1a}) and (\ref{def2a}) into Eqs. (\ref{mom0})--(\ref{mom2}), the system following from 
the perturbation of the covariant conservation equations can be partially recovered 
\begin{eqnarray}
&& \delta_{\nu}' = - \frac{4}{3} \theta_{\nu} + 4\psi',
\label{mom4a}\\
&& \theta_{\nu}' = \frac{k^2}{4} \delta_{\nu} - k^2 \sigma_{\nu} + k^2 \phi,
\label{mom4b}\\
&& \sigma_{\nu}' = \frac{4}{15} \theta_{\nu} - \frac{3}{10} k {\cal F}_{\nu 3},
\label{mom4}
\end{eqnarray}
with the important addition of the quadrupole (appearing in Eq. (\ref{mom4a})) and of the whole 
Eq. (\ref{mom4}), which couples the quadrupole, the peculiar velocity field, and the octupole ${\cal F}_{\nu3}$.
For the adiabatic mode, after neutrino decoupling, ${\cal F}_{\nu 3}=0$. 
The problem of dealing with neutrinos 
while setting initial conditions for the evolution of the CMB anisotropies can be now fully
 understood. The fluid approximation 
implies that the dynamics of neutrinos can be initially described, after neutrino decoupling, by 
the evolution of the monopole and dipole of the neutrino phase space distribution. However, in order 
to have an accurate description of the initial conditions one should solve an infinite 
hierarchy of equations for the time derivatives of higher order moments 
of the photon distribution function. 

Eqs. (\ref{mom0})--(\ref{mom2}) hold for massless neutrinos but a similar hierarchy 
can be derived also in the case of the photons or, more classically, in the case of the 
brightness perturbations of the radiation field to be discussed below. The spatial 
gradients of the longitudinal fluctuations of the metric are sources of the equations 
for the lowest multipoles, i.e. Eqs. (\ref{mom0}) and (\ref{mom1}). For $\ell >2$, each multipole 
is coupled to the preceding (i.e. $(\ell -1)$) and to the following (i.e. $(\ell +1)$) multipoles. To 
solve numerically the hierarchy one could truncate the system at a certain $\ell_{\rm max}$.
This is, however, not the best way of dealing with the problem since \cite{MB} the effect 
of the truncation could be an unphysical reflection of power down through the lower 
(i.e. $\ell <\ell_{\rm max}$) multipole moments. This problem can be efficiently 
addressed with the method of line-of-sight integration (to be discussed later in this section) that 
is also a rather effective in the derivation of approximate expressions, for instance, of the polarization 
power spectrum. The method of line-of-sight integration is the one used, for instance, in CMBFAST 
\cite{CMBF,zaldsel}.

\subsection{Brightness perturbations of the radiation field}
Unlike  neutrinos, photons are a collisional species, so the generic collision term appearing in 
Eq. (\ref{BZ5}) has to be  introduced.  With this warning in mind, all the results derived so
far can be simply translated to the case of photons (collisionless part of Boltzmann equation, 
relations between the moments of the reduced phase-space and the components of the energy-momentum tensor...)
provided the Fermi-Dirac equilibrium distribution is replaced by the Bose-Einstein distribution.
 
Thompson scattering leads to a collision term  that depends both on the baryon velocity 
field \footnote{Since the electron-ion collisions are sufficiently rapid, it is 
normally assumed, in analytical estimates of CMB effects, that electrons 
and ions are in kinetic equilibrium at a common temperature $T_{\rm eb}$}
and on the direction cosinus $\mu$ \cite{CHAN1}.  
The collision term is different for the brightness function describing 
the fluctuations of the total intensity of the radiation field (related to the 
Stokes parameter $I$) and for the brightness functions describing the degree 
of polarization of the scattered radiation (related to the Stokes parameters 
$U$ and $V$).

The conventions 
for the Stokes parameters and their well known properties will now be summarized: they 
can be found in standard electrodynamics textbooks
\cite{jackson} ( see also \cite{kos1,zalexp,huwhpol}  for phenomenological 
introduction to the problem of CMB polarization and \cite{kos2} for a more theoretical
perspective). Consider, for simplicity, 
a monochromatic  radiation field decomposed according to its linear polarizations and 
travelling along the $z$ axis:
\begin{equation}
\vec{E} = [ E_{1} \hat{e}_{x} + E_{2} \hat{e}_{y}] e^{i ( k z - \omega t)}.
\end{equation}
The decomposition according to circular polarizations can be written as 
\begin{equation}
\vec{E}= [ \hat{\epsilon}_{+} E_{+} + \hat{\epsilon}_{-} E_{-}] e^{i ( k z - \omega t)},
\end{equation}
where 
\begin{eqnarray}
 &&\hat{\epsilon}_{+} = \frac{1}{\sqrt{2}}( \hat{e}_{x} + i \hat{e}_{y}),\,\,\,\,\,\,\,\,\,
 \label{posh}\\
 && \hat{\epsilon}_{-} = \frac{1}{\sqrt{2}}( \hat{e}_{x} - i \hat{e}_{y}).
\label{neghel}
\end{eqnarray}
Eq. (\ref{posh}) is defined to be, conventionally, a {\em positive} helicity, while 
Eq. (\ref{neghel}) is the {\em negative} helicity. Recalling that  $E_{1}$ and $E_{2}$ can be written as 
\begin{equation}
E_{1} = E_{x} e^{i \delta_{x}},\,\,\,\,\,\,\,\,\,\,\,\,\,\,E_{2} = E_{y} e^{i \delta_{y}},
\end{equation}
the polarization properties of the radiation field can be described  in terms of $4$ real numbers
given by the projections of the radiation field over the linear and circular polarization unit vectors, i.e.
\begin{equation}
(\hat{e}_{x} \cdot \vec{E}),\,\,\,\,\,\,(\hat{e}_{y} \cdot \vec{E}),\,\,\,\,\,\,
(\hat{\epsilon}_{+} \cdot \vec{E}),\,\,\,\,\,\,(\hat{\epsilon}_{-} \cdot \vec{E}).
\end{equation}
The four  Stokes parameters are then, in the linear polarization basis
\begin{eqnarray}
&& I = |\hat{e}_{x} \cdot \vec{E}|^2 +  |\hat{e}_{y} \cdot \vec{E}|^2= E_{x}^2 + E_{y}^2,
\label{Idef}\\
&& Q = |\hat{e}_{x} \cdot \vec{E}|^2  - |\hat{e}_{y} \cdot \vec{E}|^2= E_{x}^2 - E_{y}^2 ,
\label{Qdef}\\
&& U = 2 {\rm Re}[(\hat{e}_{x} \cdot \vec{E})^{\ast} (\hat{e}_{y} \cdot \vec{E})]= 2 E_{x} E_{y} \cos{(\delta_y - \delta_{x})},
\label{Udef}\\
&& V= 2 {\rm Im}[(\hat{e}_{x} \cdot \vec{E})^{\ast} (\hat{e}_{y} \cdot \vec{E})]= 2 E_{x} E_{y} \sin{(\delta_y - \delta_{x})}.
\label{Vdef}
\end{eqnarray}

Stokes parameters are not all invariant under rotations. Consider a two-dimensional (clock-wise)
rotation of the coordinate system, namely
\begin{eqnarray}
&& \hat{e}_{x}' = \cos{\varphi} \hat{e}_{x}  + \sin{\varphi} \hat{e}_{y},
\nonumber\\
&&  \hat{e}_{y}' = -\sin{\varphi} \hat{e}_{x}  + \cos{\varphi} \hat{e}_{y}.
\label{rotation}
\end{eqnarray}
Inserting Eq. (\ref{rotation}) into Eqs. (\ref{Idef})--(\ref{Vdef}) it can be easily shown that 
$I'= I$ and $V'= V$ where the prime denotes the expression of the Stokes parameter 
in the rotated coordinate system. However, the remaining two parameters 
mix, i.e.
\begin{eqnarray}
&& Q' = \cos{2\varphi} Q + \sin{2\varphi} U,
\nonumber\\
&& U' = -\sin{2 \varphi} Q + \cos{2\varphi} U.
\label{clockwise}
\end{eqnarray}
From the last expression it can be easily shown that the polarization degree $P$ is invariant
\begin{equation}
P= \sqrt{Q^2 + U^2}= \sqrt{{Q'}^2 + {U'}^2}, 
\label{DP}
\end{equation}
while $U/Q= \tan{2\alpha}$ transform as $U'/Q' = \tan{2(\alpha -  \varphi)}$.

Stokes parameters are not independent (i.e. it holds that 
$I^2 = Q^2 + U^2 + V^2$ ), they only depend on the difference of the phases (i.e. $(\delta_{x} - \delta_{y})$) 
but not on their sum (see Eqs. (\ref{Idef})--(\ref{Vdef}) ).
 Hence the polarization tensor of the electromagnetic field can be written in matrix notation as 
\begin{equation}
\rho = \left(\matrix{I + Q 
& U - i V &\cr
U + i V  & I - Q &\cr}\right) \equiv \left(\matrix{E_{x}^2 
& E_{x} E_{y} e^{ - i \Delta} &\cr
 E_{x} E_{y} e^{ i \Delta} & E_{y}^2 &\cr}\right),
\label{matrixpol}
\end{equation}
where $\Delta = (\delta_{y} - \delta_{x})$. If the radiation field would be treated in a second 
quantization approach, Eq. (\ref{matrixpol})  can be promoted to the status of density matrix 
of the radiation field \cite{kos2}.

The evolution equations for the brightness functions will now be derived. 
Consider, again Eq. (\ref{BZ4}) written, this time, in the case of photons. As in the case 
of neutrinos we can define a reduced phase space distribution ${\cal F}_{\gamma}$, just changing $\nu$ 
with $\gamma$ in Eq. (\ref{reddist}) and using the Bose-Einstein 
instead of the Fermi-Dirac equlibrium distribution. 
The reduced photon phase-space density describes 
the fluctuations of the intensity of the radiation field (related to the Stokes parameter 
$I$); a second reduced  phase-space distribution, be it  ${\cal G}_{\gamma}$, can be 
defined  for the difference of the two intensities
(related to the stokes parameter Q). The equations for ${\cal F}_{\gamma}$ and ${\cal G}_{\gamma}$ 
can be written as 
\begin{eqnarray}
&&\frac{\partial F_{\gamma}}{\partial \tau} + i k\mu F_{\gamma} - 4 (\psi' - i k \mu \phi) = {\cal C}_{I},
\nonumber\\
&&\frac{\partial G_{\gamma}}{\partial \tau} + i k\mu G_{\gamma}  = {\cal C}_{Q},
\label{FGC}
\end{eqnarray}
The collision terms for these two equations are different \cite{bes1,bes2} and can be 
obtained following the derivation reported in the chapter 1 of Ref. \cite{CHAN1} or by 
following the derivation of Bond (with different notations) in the appendix C of Ref. \cite{bond} (see from p. 638).
Another way of deriving the collision 
terms for the evolution equations of the brightness perturbations is by employing 
the total angular momentum method \cite{tot1} that will be swiftly discussed in connection 
with CMB polarization.

Before writing the explicit form of the equations, including the collision terms,  it is useful to pass 
directly to the brightness perturbations. 
For the fluctuations of the total intensity  of the radiation field
the brightness perturbations is simply given by 
\begin{equation}
f( x^{i}, q, n_{j}, \tau) = f_{0}\biggl( \frac{q}{1 + \Delta_{\rm I}}\biggr).
\label{deBRf1}
\end{equation}
Recalling  now that, by definition,
\begin{equation}
f_{0}\biggl( \frac{q}{1 + \Delta_{\rm I}}\biggr) = f_{0}(q) + \frac{\partial f_{0}}{\partial q} [ q( 1 - \Delta_{\rm I}) - q],
\label{deBRf2}
\end{equation}
the perturbed phase-space distribution and the brightness perturbation must satisfy:
\begin{equation}
f_0(q) [ 1 + f^{(1)}(x^{i}, q,n_{j}, \tau) ] = f_{0}(q) \biggl[1 - \Delta_{{\rm I}}(x^{i}, q,n_{j}, \tau)\frac{\partial \ln{f_{0}}}{\partial\ln{ q}}\biggr],
\label{deBRf3}
\end{equation}
that also implies 
\begin{equation}
\Delta_{\rm I} = - f^{(1)} \biggl(\frac{\partial \ln{f_0}}{\partial \ln{q}} \biggr)^{-1},\,\,\,\,\,\,\,\,\,\,\,\,
F_{\gamma} = - \Delta_{\rm I} \frac{\int q^{3} d q  
f_{0}\frac{\partial f_{0}}{\partial \ln{q}}}{\int q^{3} d q f_{0}} = 
4 \Delta_{\rm I},
\label{deBRf4}
\end{equation}
where the second equality follows from integration by parts as in Eq. (\ref{4fact}).

The Boltzmann equations for the perturbation of the brightness are then
\begin{eqnarray}
&& \Delta_{\rm I}' + i k\mu ( \Delta_{\rm I} + \phi) = \psi' + 
\epsilon' \biggl[ - \Delta_{\rm I} + \Delta_{{\rm I}0 } +  \mu v_{b} - 
\frac{1}{2} P_{2}(\mu) S_{\rm Q}\biggr],
\label{BRI}\\
&& \Delta_{\rm Q}' + i k\mu  \Delta_{\rm Q}  =  \epsilon' \biggl\{ - \Delta_{\rm Q}  + 
\frac{1}{2} [1- P_{2}(\mu)] S_{\rm Q}\biggr\},
\label{BRQ}\\
&& \Delta_{\rm U}' + i k\mu  \Delta_{\rm U}  = - \epsilon'  \Delta_{\rm U} ,
\label{BRU}\\
&& \Delta_{\rm V}' + i k \mu \Delta_{\rm V} = - \epsilon' \biggl[ \Delta_{\rm V} +
\frac{3}{2} \,\,i\mu\,\, \Delta_{{\rm V}1}\biggr],
\label{BRV}
\end{eqnarray}
where we defined, for notational convenience and for homogeneity with the notations of other authors \cite{HZ1}
\begin{equation}
v_{\rm b} = \frac{\theta_{\rm b}}{i k}
\label{defvb}
\end{equation}
and 
\begin{equation}
S_{\rm Q}=  \Delta_{{\rm I}2} + \Delta_{{\rm Q}0} + \Delta_{{\rm Q}2}.
\end{equation}
In Eqs. (\ref{BRQ})--(\ref{BRU}), $P_{2}(\mu) = (3 \mu^2 -1)/2$ is the Legendre 
polynomial of second order, which appears in the collision operator of the Boltzmann 
equation for the photons 
due to the directional nature of Thompson scattering. Eq. (\ref{BRV}) is somehow decoupled from the system. So 
if, initially, $\Delta_{\rm V}=0$ it will also vanish at later times.

In Eqs. (\ref{BRI})--(\ref{BRU}) the function $\epsilon'$ denotes the differential 
optical depth for Thompson scattering 
\begin{equation}
\epsilon' = x_{\rm e} n_{\rm e} \sigma_{\rm T} \frac{a}{a_0} =\frac{x_{\rm e} n_{\rm e} \sigma_{\rm T} }{z + 1},
\label{OPD1}
\end{equation}
having denoted with $x_{\rm e}$ the ionization fraction and $z = a_{0}/a -1$ the redshift. 
Defining with $\tau_{0}$ the time at which the 
signal is received, the optical depth will then be 
\begin{equation}
\epsilon(\tau,\tau_0) = \int_{\tau}^{\tau_0} x_{\rm e} n_{\rm e} \sigma_{\rm T} \frac{a(\tau)}{a_0} d\tau.
\label{OPD2}
\end{equation}
There are two important limiting cases. In the optically thin limit $\epsilon \ll 1$ absorption along the 
ray path is negligible so that the emergent radiation is simply the sum of the contributions 
along the ray path.  In the opposite case $\epsilon \gg 1$ the plasma is said to be 
optically thick.  Notice that the mean free path of CMB photons discussed in Eq. (\ref{MFPphot}) 
is nothing but the inverse of the differential optical depth, i.e. $\ell_{\gamma} \sim 1/\epsilon'$.

To close the system the evolution of the baryon velocity field can be 
rewritten as 
\begin{equation}
 v_{\rm b}' + {\cal H} v_{\rm b} + i k\phi  + \frac{\epsilon'}{\beta}
 \biggl( 3 i \Delta_{{\rm I}1} + v_{b} \biggr) =  0,
\label{vb}
\end{equation}
having defined with $\beta$ the ratio between the baryon and photon energy densities, i.e.
\begin{equation}
\beta = \frac{3}{4} \frac{\rho_{\rm b}}{\rho_{\rm r}} \simeq \frac{700}{1 + z} \biggl(\frac{h^2 \Omega_{\rm b}}{0.023}\biggr);
\label{DEFBETA}
\end{equation}
at the decoupling epoch occurring for $z \simeq 1100$, $\beta \sim 7/11$ for a typical
baryonic content of $h^2 \Omega_{\rm b} \sim 0.023$.
Notice that the photon velocity field has been eliminated, in Eq. (\ref{vb}) with the 
corresponding expression involving the monopole of the brightness function.

As pointed out  in Eq. (\ref{DP}), while $Q$ and $U$ change under rotations, the 
degree of linear polarization is invariant. Thus, it is sometimes useful to combine Eqs. (\ref{BRI}) 
and (\ref{BRQ}). The result of this combination is 
\begin{eqnarray}
&& \Delta_{\rm P}' + ( i k\mu + \epsilon')  \Delta_{\rm P}  =  \frac{3}{4} \epsilon' ( 1 - \mu^2) S_{\rm P},
\nonumber\\
&& S_{\rm P}=   \Delta_{{\rm I}2} + \Delta_{{\rm P}0} + \Delta_{{\rm P}2}.
\label{BRP}
\end{eqnarray}
With the same notations Eq. (\ref{BRI}) can be written as 
\begin{equation}
\Delta_{\rm I}' + (i k \mu + \epsilon')\Delta_{\rm I} = 
\psi' - i k\mu \phi  + \epsilon'[ \Delta_{{\rm I}\,0}   + \mu v_{\rm b}- \frac{1}{2} P_{2}( \mu) S_{\rm P}].
\label{BRIPA}
\end{equation}
By adding a $\phi'$ and $\epsilon'\phi$ both at the left and right hand sides of Eq. (\ref{BRIPA}), 
the equation for the temperature fluctuations can also be written as:
\begin{equation}
(\Delta_{\rm I} + \phi)' + (i k \mu + \epsilon') ( \Delta_{\rm I} + \phi) = 
(\psi' + \phi') + \epsilon'[ (\Delta_{{\rm I}\,0} + \phi)  + \mu v_{\rm b}- \frac{1}{2} P_{2}( \mu) S_{\rm P}].
\label{BRIP}
\end{equation}
 This form of the equation is relevant in order to 
find formal solutions of the evolution of the brightness equation (see below the discussion 
of the line of sight integrals). 

\subsubsection{Visibility function}

An important function appearing naturally in various subsequent expressions is the 
so-called {\em visibility function}, ${\cal K}(\tau)$, giving the probability that a CMB photon was last 
scattered between $\tau$ and $\tau + d\tau$; the definition of ${\cal K}(\tau)$ is 
\begin{equation}
{\cal K}(\tau) = \epsilon' e^{- \epsilon(\tau-\tau_{0})},
\label{visibilityf}
\end{equation}
usually denoted by $ g(\tau)$ in the literature. The function ${\cal K}(\tau) $ is a rather important 
quantity since it is sensitive to the whole ionization history of the Universe.
The visibility function is strongly peaked around the decoupling time $\tau_{\rm dec}$ and 
can be approximated, for analytical purposes, by a Gaussian with variance of the order of few
$\tau_{\rm dec}$ \cite{wyse}.  In ${\rm Mpc}$  the width of the visibility function is about $70$.
A relevant limit is the so-called sudden decoupling 
limit  where the visibility function 
can be approximated by a Dirac delta function and its integral, i.e. the optical depth, can be approximated 
by a step function; in formulae:
\begin{equation}
{\cal K}(\tau) \simeq \delta(\tau -\tau_{\rm dec}),\,\,\,\,\,\,\,\,\,\,\,\, e^{- \epsilon(\tau,\tau_{0})} \simeq 
\theta( \tau - \tau_{\rm dec}).
\label{suddendec}
\end{equation}
This approximation will be used, below, for different applications and it is justified since the free electron density 
diminishes suddenly at decoupling.  In spite of this occurrence  there are 
 convincing indications that, at some epoch after decoupling, the 
Universe was reionized. The ionization fraction $x_{\rm e}$ becomes again of order $1$ and the CMB 
photons can be rescattered.
The WMAP results seem to imply that the reionization 
occurred already between redshift $z\simeq 15$ and $z\simeq 20$ \cite{spergeletal}.
Temperature anisotropies on scales smaller than  the 
angle subtended by the horizon at reionization can be affected \cite{tegsilk} and 
polarized fluctuations can be generated. The angular scale corresponding 
to the horizon at reionization can be estimated as 
$\vartheta_{\rm reio} \sim \sqrt{\Omega_{\rm m}/z_{\rm reio}}$ in the case of a spatially flat Universe
(see \cite{tegsilk} and references therein). The apparent detection of early reionization 
detected by WMAP (for $15 \leq z \leq 20$) seems to be compatible with a prolonged 
reionization process. 
This occurrence would  be supported by the detection of traces of smoothly 
distributed neutral hydrogen around $z\sim 6$ via Gunn-Peterson throughs 
in the spectra of high-redshift quasars \cite{GP1,GP2}.

\subsubsection{Line of sight integrals}

 Equations  (\ref{BRP}) and (\ref{BRIPA})--(\ref{BRIP})  can be formally written as 
 \begin{equation}
 {\cal M}(\vec{k}, \tau)' + ( i k \mu + \epsilon') {\cal M}(\vec{k},\tau) = {\cal N}(\vec{k},\tau),
\label{FORM1}
 \end{equation}
 where ${\cal M}(\vec{k},\tau)$ are appropriate functions changing from case to case and 
 ${\cal N}(\vec{k},\tau)$ is a source term which also depends on the specific equation 
 to be integrated.

The formal solution of the class of equations parametrized in the form (\ref{FORM1}) can be written as  
\begin{equation}
{\cal M}(\vec{k},\tau_0) = e^{- A(\vec{k},\tau_0)} \int_{0}^{\tau_{0}} e^{A(\vec{k},\tau)} {\cal N}(\vec{k},\tau) d \tau,
\label{FORM2}
\end{equation}
where the boundary term for $\tau\to 0$ can be dropped since it is unobservable \cite{HS1,HZ2}. The 
function $A(\vec{k},\tau)$ determines the solution of the homogeneous equations and it is:
\begin{equation}
A(\vec{k},\tau) 
= \int_{0}^{\tau} (i k \mu + \epsilon') d\tau  = i k\mu \tau + \int_{0}^{\tau} x_{\rm e} n_{\rm e} \sigma_{\rm T} 
\frac{a}{a_0} d\tau.
\label{FORM3}
\end{equation}
Using the results of  Eqs. (\ref{FORM1})--(\ref{FORM3}),  the solution of Eqs. (\ref{BRP}) and (\ref{BRIP})
 can be formally  written  as 
\begin{eqnarray}
(\Delta_{\rm I} + \phi)(\vec{k},\tau_0) &=& \int_{0}^{\tau_{0}} \,\,d\tau \,\,e^{- i k\mu \Delta \tau - \epsilon(\tau,\tau_{0})} 
(\phi' + \psi') 
\nonumber\\
&+& \int_{0}^{\tau_{0}} \,\,d\tau\,\,\,
 {\cal K}(\tau)\biggl[ ( \Delta_{{\rm I}\,0} + \phi + \mu v_{\rm b}) - \frac{1}{2} P_{2 }(\mu)
S_{\rm P}(k,\tau) \biggr],
\label{LSI}
\end{eqnarray}
and as
\begin{equation}
\Delta_{\rm P}( \vec{k}, \tau_{0}) =\frac{3}{4} \int_{0}^{\tau_{0}} {\cal K}(\tau)  e^{- i k \mu \Delta\tau} (1-\mu^2)
S_{\rm P}(k, \tau) d\tau,
\label{LSP}
\end{equation}
where $\epsilon(\tau,\tau_{0})$ is the optical depth already introduced in Eq. (\ref{OPD2})  and $\Delta \tau = (\tau_{0} 
-\tau)$ is the (conformal time) increment between the reception of the signal (at $\tau_0$)
and the emission (taking place for $\tau\simeq \tau_{\rm dec}$). In Eqs.
(\ref{LSI}) and (\ref{LSP}) the visibility function ${\cal K}(\tau)$, already defined in Eq. (\ref{visibilityf}), 
has been explicitly introduced.

Equations (\ref{LSI}) and (\ref{LSP}) are called for short line of sight integral solutions. There are at least two important applications of Eqs. (\ref{LSI}) 
and (\ref{LSP}). The first one is numerical and will be only swiftly described. The second one is 
analytical and will be exploited both in the present section and in the following. 

Equation (\ref{LSI}) and (\ref{LSP}) can be used to get an integral solution of  the Boltzmann hierarchy.
Let us start from Eq.  (\ref{LSP}) 
and let us multiply both sides of the equation by $P_{\ell}(\mu)$. Then the right and left hand sides 
can be integrated over $\mu$ between $-1$ and $1$. Recalling the conventions for the 
expansion of the brightness perturbations, i.e.  Eq. (\ref{CONV}), the left hand side will simply give, up to a 
numerical factor, the $\ell$-th multipole of $\Delta_{\rm P}$, i.e. $\Delta_{{\rm P}\,\ell}$. The right
hand side will instead give an explicit expression.  Using repeatedly the orthonormality and 
the recurrence relations, i.e. Eqs. (\ref{norm}) and (\ref{rec1}),  the result of the (tedious) calculation can 
be expressed as
\begin{equation}
\Delta_{{\rm P}\,\ell}(k,\tau_0) = \frac{3}{4} \int_{0}^{\tau_{0}} {\cal K}(\tau) S_{\rm P}(k,\tau)
[{\cal U}_{\ell} j_{\ell}( k \Delta\tau)
+ {\cal U}_{\ell -2} j_{\ell -2}( k \Delta\tau) + {\cal U}_{\ell + 2} j_{\ell + 2}( k \Delta\tau) ] d\tau,
\label{LSPell}
\end{equation}
where 
\begin{equation}
{\cal U}_{\ell} = 2 \frac{\ell^2 + \ell -1}{(2\ell -1)(2 \ell + 3)},\,\,\,\,\,\,\,\,\,\,\, {\cal U}_{\ell -2} = \frac{\ell(\ell -1)}{(2 \ell + 1)
(2\ell -1)}, \,\,\,\,\,\,\,\,\,\,\,\, {\cal U}_{\ell + 2} = \frac{(\ell + 2) (\ell + 3)}{(2 \ell + 1) ( 2 \ell + 3)}.
\label{UL}
\end{equation}
In Eq. (\ref{LSPell}), $j_{\ell}( k \Delta \tau)$ 
 are the spherical Bessel functions \cite{abra1,abra2} which can be related to the ordinary 
 Bessel functions as:
 \begin{equation}
 j_{\ell}( k \Delta\tau) = \sqrt{ \frac{\pi}{2 k \Delta\tau}} J_{\ell + 1/2}( k\Delta\tau).
 \label{BESSEL}
 \end{equation}
The $j_{\ell}( k\Delta\tau)$ emerge 
 as a consequence of the expansion in a series of Legendre polynomials  of the plane wave 
\begin{equation}
e^{ - i \mu\,k\,\Delta\tau} = \sum_{\ell} (-i)^{\ell}(2 \ell + 1) j_{\ell}(k \Delta\tau) P_{\ell}(\mu)
\label{PLW}
\end{equation}
appearing in Eq. (\ref{LSP}).  

The same calculation performed in the case of Eq. (\ref{LSP}) can now be performed for 
Eq. (\ref{LSI}) with the result 
\begin{eqnarray}
&&\Delta_{\rm I}(\vec{k},\tau_0) = \int_{0}^{\tau_{0}} e^{- \epsilon(\tau,\tau_0)}\,\,d\tau \,\,\biggl\{ \biggr[ (\psi' + \phi') + 
\epsilon'(\Delta_{{\rm I}\,0} + \phi) 
\nonumber\\
&-& \frac{\epsilon'}{4}\frac{\ell(\ell + 1)}{(2 \ell + 3) (2\ell -1)}
S_{\rm P}(k,\tau)\biggr] 
j_{\ell}(k \Delta \tau)+
 i v_{\rm b}\epsilon' \biggl[ \frac{\ell}{2\ell + 1} j_{\ell -1}( k \Delta\tau) - \frac{\ell + 1}{2\ell +1}
 j_{\ell + 1}( k \Delta\tau)\biggr]
 \nonumber\\
 &+& \frac{3}{4}\epsilon' \biggl[ \frac{\ell (\ell + 1)}{(2 \ell + 1)(2\ell -1)} j_{\ell -2}( k\Delta\tau) + 
 \frac{(\ell + 1)(\ell + 2)}{(2 \ell + 1) (  2 \ell + 3)} j_{\ell + 2}( k\Delta\tau)\biggr]S_{\rm P}(k,\tau) \biggr\}.
 \label{LSIell}
 \end{eqnarray}
 
 In the tight coupling approximation (which will be one of the subjects of the 
 forthcoming analysis) it is possible to determine with reasonable accuracy the 
 evolution of the monopole and of the dipole. Thanks to the recurrence relations of the Boltzmann 
 hierarchy, higher multipoles can be also estimated. Then, the integral 
 solutions derived above can be used in order to estimate the full angular power spectrum.
 
 Equation (\ref{LSIell}) can be used in order to compute (numerically)  the $\Delta_{{\rm I}\ell}$ without 
 solving the Boltzmann hierarchy for too high $\ell$. In this 
 approach it is enough to compute accurately 
 the source terms appearing in the integrand at a moderately high value value of $\ell$.
  
Bessel functions appearing in Eq. (\ref{LSIell}) 
have a maximum for $ k \Delta \tau \sim \ell$. This occurrence provides 
a relation between the wave-number and the multipole so that  a multipole 
$\ell$ is sensitive to a typical inhomogeneity $ k \sim \ell/\Delta\tau$. More precisely, the relation 
between a given multipole and the comoving wave-number, can be expressed, in the 
spatially flat case, as 
\begin{equation}
k \simeq \frac{h}{6000\,\,{\rm Mpc}} \,\,\,\ell
\end{equation}
where $h$ is the indetermination in the Hubble parameter.

There is another (related) way of obtaining simplified expressions for the integral solutions 
of the Boltzmann equations in the form (\ref{BRP})-- (\ref{BRIPA}) and (\ref{BRIP}). 
The formal solution  of Eq. (\ref{BRP}) can be written in a different form 
if the term $\mu^2$ is integrated by parts (notice, in fact, that the $\mu$ enters also the exponential). 
The boundary terms arising as a result of the integration by parts can be dropped because they are 
vanishing in the limit $\tau\to 0$ and are irrelevant for $\tau=\tau_{0}$ (since only an unobservable monopole 
is induced). The result the integration by parts of the $\mu^2$ term in Eq. (\ref{LSP}) can be expressed as 
\begin{eqnarray}
&& \Delta_{\rm P}(\vec{k},\tau_0) = \int_{0}^{\tau_{0}} e^{- i k\mu \Delta\tau} {\cal N}_{\rm P}(k,\tau)\,\, d\tau,
\label{LSPb}\\
&& {\cal N}_{\rm P}(\vec{k},\tau) = \frac{3}{4 k^2} [ {\cal K}( S_{\rm P}'' + k^2 S_{\rm P}) + 2 
{\cal K}' S_{\rm P}' + S_{\rm P} {\cal K}''],
\label{sourcePb}
\end{eqnarray}
where, as usual $\Delta\tau = (\tau_{0} -\tau)$.
The same exercise can be performed in the case of  Eq. (\ref{BRIPA}).
Before giving the general result, let us just integrate by parts the term $-ik\mu \phi$ 
appearing at the right hand side of Eq. (\ref{BRIPA}). The result of this manipulation
is 
\begin{eqnarray}
&&\Delta_{\rm I}(\vec{k},\tau_{0}) = \int_{0}^{\tau_{0}}e^{i k \mu (\tau - \tau_0) - \epsilon(\tau,\tau_{0})} (\psi' + \phi') \,\,d\tau
\nonumber\\
&& + \int_{0}^{\tau_{0}} {\cal K}(\tau) 
e^{i k \mu (\tau- \tau_{0})}\,\,d\tau\,\biggl[ \Delta_{{\rm I}\,0} + \phi + \mu v_{\rm b} - 
\frac{1}{2} P_{2}(\mu) S_{\rm P} \biggr].
\label{SWBZ1}
\end{eqnarray}
Let us now exploit the sudden decay approximation illustrated around Eq. (\ref{suddendec}) 
and assume that the (Gaussian) visibility function ${\cal K}(\tau)$ is indeed a Dirac delta function 
centered around $\tau_{\rm dec}$ (consequently the optical depth $\epsilon(\tau,\tau_{0})$ will be 
a step function). Then Eq. (\ref{SWBZ1}) becomes 
\begin{equation}
\Delta_{{\rm I}}(\vec{k}, \tau_{0} ) = \int_{\tau_{\rm dec}}^{\tau_{0}} e^{i k \mu (\tau - \tau_0)}[ \psi' + \phi'] d\tau + 
e^{i k \mu(\tau_{\rm dec} - \tau_{0})} [ \Delta_{{\rm I}\,0} + \phi + \mu v_{\rm b}]_{\tau_{\rm dec}},
 \label{SWBZ2}
 \end{equation}
 where the term $S_{\rm P}$ has been neglected since it is subleading at large scales. Equation (\ref{SWBZ2}) 
 is exactly (the Fourier space version of) Eq. (\ref{scalSW}) already derived with a different chain of arguments and we can directly recognize the integrated SW term (first term at the right hand side), the ordinary SW effect (proportional 
 to \footnote{Recall, in fact, that 
because of the relation between brightness and perturbed energy-momentum tensor, i.e. Eqs. (\ref{def1a}) 
and (\ref{deBRf4}), $4 \Delta_{{\rm I}\,0} = \delta_{\gamma}$.} 
 $ (\Delta_{{\rm I}0} + \phi)$) and the Doppler term receiving contribution from the 
 peculiar velocity of the observer and of the emitter. 
 
 If all the $\mu$ dependent terms appearing in Eq. (\ref{BRIPA}) are integrated by parts the result will be 
\begin{eqnarray}
&& \Delta_{\rm I}(\vec{k},
\tau_0) = \int_{0}^{\tau_{0}} e^{- i k\mu \Delta\tau - \epsilon(\tau,\tau_0)} ( \psi' + \phi')\,\, d\tau + 
 \int_{0}^{\tau_{0}} e^{- i k\mu \Delta\tau} {\cal N}_{\rm I}(k,\tau) d \tau,
 \label{LSIb}\\
 && {\cal N}_{\rm I}(k,\tau) = \biggl\{{\cal K}(\tau)\biggl[ \Delta_{{\rm I}, 0} + \frac{S_{\rm P}}{4} + \phi + 
 \frac{i}{k} v_{\rm b}' + \frac{3}{4 k^2} S_{\rm P}''\biggr] 
 \nonumber\\
 &&+ {\cal K}' \biggl[ \frac{i}{k} v_{\rm b} + \frac{3}{2 k^2} S_{\rm P}' \biggr] + \frac{3}{4 k^2} {\cal K}'' S_{\rm P}\biggr\}.
 \label{sourceIb}
 \end{eqnarray}

\subsubsection{Angular power spectrum and observables}

Equations (\ref{SWBZ2}) together with the results summarized in section 4 for the initial conditions 
of the metric fluctuations after equality allow the estimate of the angular power spectra in the case 
of adiabatic and isocurvature initial conditions.
The brightness perturbation can be Fourier transformed \footnote{Notice that as 
in the definition of the Fourier transforms of all the quantities introduced so far the factor 
$(2 \pi)^{3/2}$ has been always incuded. Some authors, however, choose, consistently, 
not to include it.}
\begin{equation}
 \Delta_{\rm I}(\vec{x}, \hat{n}, \tau) = \frac{1}{(2\pi)^{3/2}} 
\int d^{3} k e^{i \vec{k}\cdot\vec{x}} \Delta_{\rm I}(\vec{k}, \hat{n}, \tau). 
\end{equation}
Assuming that the receiver is located at some conformal time $\tau$ (eventually 
coinciding with $\tau_0$) and at  $\vec{x}=0$ 
the previous formula can be also expanded in spherical harmonics, i.e. 
the following chain of equalities holds 
\begin{equation}
\Delta_{\rm I}(\hat{n},\tau ) = \sum_{\ell=0}^{\infty}
 \sum_{m=-\ell}^{\ell} a_{\ell m} Y_{\ell m}(\hat{n}) = \frac{1}{(2\pi)^{3/2}} 
\int d^{3} k \Delta_{\rm I}(\vec{k}, \hat{n}, \tau),
\label{ANIS}
\end{equation}
where $a_{\ell m}$ are the coefficients to be determined. Now, as previously done in the present section,  
$ \Delta_{\rm I}(\vec{k}, \hat{n}, \eta) $  can be represented as  series of Legendre polynomials 
\begin{equation}
 \Delta(\vec{k}, \hat{n}, \tau) = \sum_{\ell=0}^{\infty} (- i)^{\ell} ( 2 \ell + 1) \Delta_{{\rm I}\ell}( \vec{k}, \tau) P_{\ell}(\hat{k}\cdot \hat{n}).
\label{LEGser}
\end{equation}
An important property of the $Y_{\ell m}$  is expressed by the theorem of addition 
of spherical harmonics which stipulates that 
\begin{equation}
P_{\ell }(\hat{k}\cdot \hat{n})  = \frac{4\pi}{2\ell + 1}\sum_{m=-\ell}^{\ell} Y_{\ell m}^{\ast}(\hat{k}) Y_{\ell m} (\hat{n}).
\label{ADD}
\end{equation}
Inserting now Eq. (\ref{ADD}) into Eq. (\ref{LEGser}) and plugging the obtained result
into the second equality of  Eq. (\ref{ANIS}), the coefficients $a_{\ell m}$ are determined as
 \begin{equation}
a_{\ell m} = \frac{(4\pi)}{(2\pi)^{3/2}} 
(-i)^{\ell} \int d^{3} k Y_{\ell m }^{\ast} (\hat{k}) 
\Delta_{{\rm I} \ell}(\vec{k}, \tau).
\label{AELLM}
\end{equation}

The two-point temperature correlation function on the sky 
between two directions conventionally denoted by $\hat{n}_{1}$ and $\hat{n}_{2}$, can be written as 
$C(\vartheta) = \langle \Delta_{\rm I}(\hat{n}_{1},\tau_{0})  \Delta_{\rm I}(\hat{n}_{2},\tau_{0}) \rangle$
where $C(\vartheta)$ does not depend on the azimuthal angle because of isotropy of the background 
space-time and where the angle brackets denote a theoretical ensamble average. 
Since the background space-time is isotropic, the ensamble average of the $a_{\ell m}$ will only depend 
upon $\ell$, not upon $m$, i.e. 
\begin{equation}
\langle a_{\ell m}a^{\ast}_{\ell' m'} \rangle = C_{\ell} \delta_{\ell\ell'} \delta_{m m'},
\label{AVALM}
\end{equation}
where $C_{\ell}$ is the angular power spectrum.
Thus, using the first equality of Eq. (\ref{ANIS}) and exploiting Eq. (\ref{ADD}), the relation 
(\ref{AVALM}) implies 
\begin{equation}
C(\vartheta) = \langle \Delta_{\rm I}(\hat{n}_{2},\tau_{0})  \Delta_{\rm I}(\hat{n}_{1},\tau_{0}) \rangle \equiv 
\frac{1}{4\pi} \sum_{\ell} (2 \ell + 1) C_{\ell } P_{\ell}(\hat{n}_{1} \cdot \hat{n}_{2}).
\end{equation}

The $C_{\ell}$ spectrum will now be derived for few interesting examples. 
Consider, for instance, the adiabatic mode.
In this case Eq. (\ref{SWBZ2}) (or, Eq. (\ref{scalSW})) has vanishing integrated contribution and vanishing
Doppler contribution at large scales (as discussed in section 4). Using then the result of Eq. (\ref{SWscalad}), 
the adiabatic contribution to the temperature fluctuations can be written as \footnote{Recall that there are models 
where the Universe gets dominated by the cosmological constant when, approximately, $a/a_{0} > (\Omega_{\rm m}/\Omega_{\Lambda})^{1/3}$. This generates a time-evolution in the longitudinal fluctuations 
of the geometry: it is the integrated SW effect which will not be specifically addressed here but which has been 
studied in connection with various problems \cite{luca1,mortak,gordonhu} previously addressed in this paper.} 
\begin{equation}
\Delta^{\rm ad}_{{\rm I}}(\vec{k}, \tau_{0} ) = 
e^{- i k \mu \tau_{0}} [ \Delta_{{\rm I}\,0} + \phi ]_{\tau_{\rm dec}}
 \simeq e^{- i k \mu \tau_{0}} \frac{1}{3} \psi_{\rm m}^{\rm ad}(\vec{k}),
 \label{ADSW3}
\end{equation}
noticing that, in the argument of the plane wave $\tau_{\rm dec}$ can be dropped since 
$\tau_{\rm dec} \ll \tau_{0}$.
The plane wave appearing in Eq. (\ref{ADSW3}) can now be expanded in series of Legendre polynomials 
and, as a result,
\begin{equation}
\Delta^{\rm ad}_{{\rm I},\ell}(\vec{k},\tau_{0}) = \frac{j_{\ell}(k\tau_{0})}{3} \psi_{\rm m}^{\rm ad}(\vec{k}),
\label{deltaAD}
\end{equation}
where $j_{\ell}(k\tau_{0})$ are defined as in Eq. (\ref{BESSEL}) with their appropriate argument.

Assuming now that $\psi_{\rm m}^{\rm ad}(\vec{k})$ are the Fourier components of a Gaussian and isotropic 
random field (as, for instance, implied by some classes of inflationary models) then 
\begin{equation}
\langle \psi_{\rm m}^{\rm ad}(\vec{k})  \psi_{\rm m}^{\rm ad}(\vec{k}') \rangle = 
\frac{2\pi^2}{k^3} {\cal P}^{\rm ad}_{\psi}(k) \delta^{(3)}(\vec{k} -\vec{k}'),\,\,\,\,\,\,\, {\cal P}^{\rm ad}_{\psi}(k) = \frac{k^3}{2\pi^2} |\psi^{\rm ad}_{\rm m}(k)|^2 ,
\label{CORRAN}
\end{equation}
where ${\cal P}^{{\rm ad}}_{\psi}(k)$ is the power spectrum of the longitudinal fluctuations of the metric after equality.
Then, Eq. (\ref{deltaAD}) can be inserted into Eq. (\ref{AELLM}): the obtained result has to 
be  plugged into Eq. (\ref{AVALM}) and from Eq. (\ref{CORRAN}) (together with the orthogonality 
of spherical harmonics)
In this case 
\begin{equation}
C_{\ell}^{({\rm ad})} = \frac{4\pi}{9}\int_{0}^{\infty} \frac{d k}{k} {\cal P}^{\rm ad}_{\psi}(k) j_{\ell}(k\tau_{0})^2.
\label{CL1}
\end{equation}
To perform the integral it is customarily assumed that the power spectrum of adiabatic fluctuations 
has a power-law dependence characterized by a single spectral index $n$ 
\begin{equation}
{\cal P}^{\rm ad}_{\psi}(k) = \frac{k^3}{2\pi^2} |\psi_{k}|^2 = A_{\rm ad} \biggl( \frac{k}{k_{\rm p}}\biggr)^{n -1}.
\label{ANSPPSI}
\end{equation}
Notice that $k_{\rm p}$ is a typical pivot scale which is conventional since the whole dependence 
on the parameters of the model is encoded in $A_{\rm ad}$ and $n$. For instance, the WMAP 
collaboration \cite{spergeletal,VERDE}, chooses to normalize $A$ at 
\begin{equation}
k_{\rm p} = k_1 = 0.05 \,\,{\rm Mpc}^{-1},
\end{equation}
while the scalar-tensor ratio (defined in section 6) is evaluated at a scale 
\begin{equation}
k_{0} = 0.002\,\,\, {\rm Mpc}^{-1} \equiv 6.481 \times 10^{-28}\,\,{\rm cm}^{-1} = 1.943\times 
10^{-17}\,\,{\rm Hz},
\end{equation}
recalling that $1\,\,{\rm Mpc} = 3.085\times 10^{24} \,\,{\rm cm}$.

Inserting Eq. (\ref{ANSPPSI}) into Eq. (\ref{CL1}) and recalling the explicit form of the spherical Bessel 
functions in terms of ordinary Bessel functions 
\begin{equation}
C^{({\rm ad})}_{\ell} = \frac{2\pi^2}{9} (\tau_{0} \,\,k_{\rm p})^{1 - n}\,\, A_{\rm ad}\,\,\int_{0}^{\infty} dy  y^{n -3}
 J^2_{\ell + 1/2}(y),
\label{CLMAP}
\end{equation}
where $ y = k \tau_{0}$.
The integral appearing in Eq. (\ref{CLMAP}) can be performed for 
$-3 < n< 3$ with the result
\begin{equation}
\int_{0}^{\infty} dy  y^{n -3}
 J^2_{\ell + 1/2}(y)  = \frac{1}{ 2 \sqrt{\pi}} \frac{ 
\Gamma\biggl( \frac{3 -n}{2}\biggr)
 \Gamma\biggl(\ell + \frac{n}{2} - \frac{1}{2} \biggr)}{\Gamma\biggl(\frac{4 - n}{2}\biggr) \Gamma\biggl( \frac{5}{2} + \ell  - \frac{n}{2} \biggr)}.
\label{CL3}
\end{equation}
To get the standard form of the $C_{\ell}$ use now the 
duplication formula for the $\Gamma$ function, namely in our case 
\begin{equation}
\Gamma\biggl( \frac{3 - n}{2}\biggr) = \frac{\sqrt{2\pi} \Gamma( 3 - n) }{
2^{5/2 -n} \Gamma\biggl( \frac{4 - n}{2}\biggr) }.
\label{rel2a}
\end{equation}
Insert now Eq. (\ref{rel2a}) into Eq. (\ref{CL3}); inserting then 
Eq. (\ref{CL3}) into Eq. (\ref{CLMAP}) we do get 
\begin{eqnarray}
&& C^{({\rm ad})}_{\ell} = \frac{\pi^2 }{36}  A_{\rm ad} {\cal Z}(n, \ell)
\nonumber\\
&&{\cal  Z}(n,\ell) = (\tau_{0} \,\,k_{\rm p})^{1 - n}\,\,2^{n}\frac{\Gamma( 3 - n) 
\Gamma\biggl( \ell + \frac{n}{2} - 
\frac{1}{2}\biggr)}{\Gamma^2\biggl(\frac{4 - n}{2}\biggr)
 \Gamma\biggl( \frac{5}{2} + \ell - \frac{n}{2} \biggr) },
\label{CLUS}
\end{eqnarray}
where the function ${\cal Z}(n,\ell)$ has been introduced for future convenience. Notice, as 
a remark, that for the approximations made in the evaluation of the SW effects, Eq. (\ref{CLUS}) 
holds at large angular scales, i.e. $\ell < 30$.

The $C^{({\rm ad})}_{\ell}$ have been given in the case 
of the spectrum of $\psi$.  There is a specific relation between the spectrum 
of $\psi$ and the spectrum of curvature perturbations which has been derived in section 
4 (see discussion after  Eq. (\ref{RCONST}))  which implies, quite trivially, 
${\cal P}^{\rm ad}_{\cal R} = (25/9) {\cal P}^{\rm ad}_{\psi}$. Finally, the spectrum of the longitudinal 
fluctuations of the geometry may also be related to the spectrum of the same quantity but computed 
before equality: this entails the $(9/10)$ factor discussed in Eq. (\ref{psimpsir}). 

The same calculation performed in the case of the adiabatic mode can be repeated, with minor 
(but relevant) modifications for the CDM-radiation non-adiabatic mode. In the specific case 
of this non-adiabatic mode, Eq. (\ref{deltaNAD}) is modified as 
\begin{equation}
\Delta^{({\rm nad})}_{{\rm I},\ell}(\vec{k},\tau_{0}) = 2 j_{\ell}(k\tau_{0})\psi_{\rm m}^{\rm nad}(\vec{k}),
\label{deltaNAD}
\end{equation}
as it follows directly from Eq. (\ref{SWBZ2}) in the case of non-adiabatic initial conditions after equality 
(see also Eq. (\ref{SWscalnad})).  Performing the same computation 
Eq. (\ref{CLUS}) becomes 
\begin{equation}
C^{({\rm nad})}_{\ell} =\pi^2   A_{\rm nad} {\cal Z}(n_{\rm nad}, \ell),
\end{equation}
where the power spectrum of non-adiabatic fluctuations has been defined as 
\begin{equation}
{\cal P}_{\psi}^{\rm nad}= A_{\rm nad} \biggl(\frac{k}{k}_{\rm p}\biggr)^{n_{\rm nad} -1}. 
\end{equation}
 Again, 
following the considerations reported in Eqs. (\ref{psitoentr}) and (\ref{SWscalnad}), the 
spectrum of non-adiabatic fluctuations can be directly expressed in terms of the fluctuations 
of ${\cal S}$, i.e. the fluctuations of the specific entropy 
 (see Eqs. (\ref{SE}) and (\ref{SWscalnad})), with 
the result that ${\cal P}_{\psi} = (1/25){\cal P}_{{\cal S}}^{\rm nad}$. Of course 
the major difference between adiabatic and non-adiabatic fluctuations will be much more 
dramatic at smaller angular scales (i.e. say between $\ell \sim 200$ and $\ell \sim 350$) 
where the patterns of acoustic oscillations have a crucial phase difference (this aspect will be 
discussed in the context of the tight coupling expansion).

There could be physical situations where adiabatic and non-adiabatic modes are simultaneously present with some 
degree of correlation. In this case the derivations given above change qualitatively, but not crucially.
The contribution to the SW effect will then be the sum of the adiabatic and non adiabatic contributions 
(weighted by the appropriate coefficients) i.e. 
\begin{equation}
\Delta^{\rm tot}_{{\rm I}}(\vec{k}, \tau_{0} ) 
 \simeq e^{- i k \mu \tau_{0}} \biggl[ \frac{1}{3} \psi_{\rm m}^{\rm ad}(\vec{k}) + 2 \psi_{\rm m}^{\rm nad}(\vec{k})\biggr].
 \label{ADSW4}
\end{equation}
While taking expectation values, there will not only be the adiabatic and non-adiabatic 
power spectra, i.e. ${\cal P}_{\psi}^{\rm ad}(k)$ and ${\cal P}^{\rm nad}_{\psi}(k)$, but also 
the power spectrum of the correlation between the two modes arising from 
\begin{equation}
\langle \psi_{\rm ad}(\vec{k}) \psi_{\rm nad}(\vec{k}') \rangle = \frac{2\pi^2}{k^3} {\cal P}_{\psi}^{\rm cor}(k) \delta^{(3)}(\vec{k} -\vec{k}'), \,\,\,\,\,\,\,\,\,\, {\cal P}_{\psi}^{\rm cor}(k)= \sqrt{A_{\rm ad} A_{\rm nad}} \biggl(\frac{k}{k_{\rm p}}\biggr)^{n_{\rm c} -1}
\cos{\alpha_{\rm c}},
\label{CORRPWS}
\end{equation}
where the angle $\alpha_{\rm c}$ parametrizes the degree of correlation between the adiabatic and non-adiabatic mode. The total angular power spectrum will then be given not only 
by the adiabatic and non-adiabatic contributions, but also by their correlation, i.e.  
\begin{equation}
C_{\ell}^{({\rm cor})} = \frac{\pi^2}{3} \sqrt{ A_{\rm ad} A_{\rm nad} } \cos{\alpha_{\rm c} } {\cal Z}(n_{\rm c},\ell).
\end{equation}

Consider, finally, the specific case of adiabatic fluctuations with Harrison-Zeldovich, i.e. the case $n=1$ 
in eq. (\ref{CLUS}). In this case
\begin{equation}
\frac{\ell (\ell +1)}{2\pi} C^{({\rm ad})}_{\ell} = \frac{A_{\rm ad}}{9}.
\label{CLAD}
\end{equation}
If the fluctuations were of purely adiabatic nature, then large-scale anisotropy experiments 
(see Fig. \ref{F13}) imply\footnote{To understand fully the quantitative features 
of Fig. \ref{F13} it should be borne in mind that sometimes the $C_{\ell}$ are given not in absolute units 
(as implied in Eq. (\ref{CLAD}) but they are  multiplied by the CMB temperature. To facilitate 
the conversion recall that the CMB temperature is $T_{0} = 2.725 \times 10^{6}
\,\,\mu {\rm K}$. For instance the WMAP collaboration normalizes the power 
spectrum of the curvature fluctuations at the pivot scale $k_{\rm p}$ 
as ${\cal P}_{{\cal R}} =(25/9)\times (800\pi^2/T_{0}^2) \times \tilde{A}$
where $\tilde{A}$ is not the $A$ defined here but it can be easily related to it.}
$A\sim 9 \times 10^{-10}$.
Notice that in Fig. \ref{F13} the quantity $C_{\ell}\,\ell(\ell+1)/(2\pi)$ is directly plotted: as it follows 
from the approximate equality
\begin{equation}
\sum_{\ell} \frac{2\ell + 1}{4\pi} C_{\ell} \simeq \int \frac{\ell(\ell +1)}{2\pi} C_{\ell} d \ln{\ell},
\end{equation}
$C_{\ell}\,\ell(\ell+1)/(2\pi)$ is roughly the power per logarithmic interval of $\ell$.

Up to now the large angular scale anisotropies have been treated. In the following 
the analysis of the smaller angular scales will be introduced in the 
framework of the tight coupling approximation.

\subsection{Tight coupling expansion}

The tight coupling approximation has been already implicitly used in section 4 where 
it has been noticed that, prior to recombination, 
 for comoving scales shorter than the mean free path of CMB photons, the baryons 
 and the photons evolve as a single fluid (see Eq. (\ref{MFPphot})). 

If tight coupling is  exact,  photons and baryons 
are synchronized so well that the photon phase-space distribution 
is isotropic in the baryon rest frame. In other words since the typical time-scale 
between two collisions is set by $\tau_{\rm c} \sim 1/\epsilon'$,  the scattering 
rate is rapid enough to equilibrate the photon-baryon fluid.  Since the photon distribution is 
isotropic, the resulting radiation is not polarized. The idea is then to tailor a systematic expansion
in $\tau_{\rm c} \sim 1/\epsilon'$ or, more precisely, in 
$k \tau_{\rm c} \ll 1$ and $\tau_{\rm c}{\cal H} \ll 1$.  
 
Recall the expansion of the brightness perturbations:
\begin{eqnarray}
&& \Delta_{\rm I}(\vec{k}, \hat{n}, \tau) = \sum_{\ell} (- i)^{\ell}( 2 \ell + 1) \Delta_{{\rm I}\ell}(\vec{k},\tau) P_{\ell}(\mu), 
\nonumber\\
&& \Delta_{\rm Q}(\vec{k}, \hat{n}, \tau) = \sum_{\ell} (- i)^{\ell}( 2 \ell + 1)  \Delta_{{\rm Q}\ell}(\vec{k},\tau) P_{\ell}(\mu), 
\label{expD}
\end{eqnarray}
$\Delta_{{\rm I}\ell}$ and $  \Delta_{{\rm Q}\ell}$ being the  $\ell$-th multipole of the brightness 
function $\Delta_{{\rm I}}$ and $\Delta_{{\rm Q}}$.

The idea is now to expand Eqs. (\ref{BRI}) and (\ref{BRQ}) 
in powers of the small parameter $\tau_{\rm c}$.  Before doing the expansion, it is useful to derive the hierarchy for the brightness 
functions in full analogy with what is  discussed in the appendix for the case of the neutrino phase-space 
distribution. 
To this aim, each side of Eqs. (\ref{BRI})--(\ref{BRQ}) and (\ref{vb}) will be multiplied 
by the various Legendre polynomials  and the  integration  over $\mu$ will be performed.
Noticing that, from the orthonormality relation for Legendre polynomials (i. e. Eq. (\ref{norm})),
\begin{equation}
\int_{-1}^{1} P_{\ell}(\mu) \Delta_{\rm I} d\mu = 2 (-i)^{\ell} \Delta_{{\rm I} \ell},\,\,\,\,\,\,\,\,\,\,\,
\int_{-1}^{1} P_{\ell}(\mu) \Delta_{\rm Q} d\mu = 2 (-i)^{\ell} \Delta_{{\rm Q} \ell},
\label{INTPL}
\end{equation}
and recalling that
\begin{equation}
P_{0}(\mu) =1,\,\,\,\,\,P_{1}(\mu) = \mu,\,\,\,\,\,\,\,\,P_{2}(\mu)= \frac{1}{2}(3 \mu^2 -1),\,\,\,\,\,\, P_{3}(\mu) = \frac{1}{2}( 5\mu^3 -3 \mu),
\end{equation}
Eqs. (\ref{BRI})--(\ref{BRQ}) and (\ref{vb}) 
allow the determination of the first three sets of equations for  the hierarchy of the brightness.
More specifically, multiplying Eqs. (\ref{BRI})--(\ref{BRQ}) and (\ref{vb}) by $P_{0}(\mu)$ and integrating over $\mu$, the following relations can be obtained
\begin{eqnarray}
&& \Delta_{{\rm I}0}' + k \Delta_{{\rm I}1} =  \psi',
\label{L01}\\
&& \Delta_{{\rm Q}0} '+k \Delta_{{\rm Q}1} = \frac{\epsilon'}{2} [ \Delta_{{\rm Q}2} + \Delta_{{\rm I}2} - \Delta_{{\rm Q}0} ],
\label{L02}\\
&& v_{b}' + {\cal H} v_{\rm b} = - i k \phi - \frac{\epsilon'}{\beta} ( 3 i \Delta_{{\rm I}1} + v_{b} ).
\label{L03}
\end{eqnarray}
If Eqs. (\ref{BRI})--(\ref{BRQ}) and (\ref{vb}) are multiplied by  $P_{1}(\mu)$, both at right and left-hand sides, 
the integration  over $\mu$ of the various terms implies, using Eq.  (\ref{INTPL}):
\begin{eqnarray}
&& - \Delta_{{\rm I} 1}' - \frac{2}{3}k \Delta_{{\rm I}2} + \frac{k}{3} \Delta_{{\rm I}0} = - \frac{k}{3}  \phi + \epsilon' \biggl[ \Delta_{{\rm I} 1} + 
\frac{1}{3 i} v_{\rm b}\biggr],
\label{L11}\\
&& - \Delta_{{\rm Q}1}' - \frac{2}{3} k \Delta_{{\rm Q}2} + \frac{k}{3} \Delta_{{\rm Q}0} = \epsilon' \Delta_{{\rm Q} 1},
\label{L12}\\
&& v_{b}' + {\cal H} v_{b} = - i k \phi - \frac{\epsilon'}{\beta} ( 3 i \Delta_{{\rm I}1} + v_{b} ).
\label{L13}
\end{eqnarray}
The same  procedure, using $P_{2}(\mu)$, leads to
\begin{eqnarray}
&& - \Delta_{{\rm I} 2}' - \frac{3}{5} k \Delta_{{\rm I}3} + \frac{2}{5} k \Delta_{{\rm I} 1} = \epsilon'\biggl[ \frac{9}{10} \Delta_{{\rm I}2} - \frac{1}{10} (\Delta_{{\rm Q}0} + 
\Delta_{{\rm Q} 2} )\biggr],
\label{L21}\\
&&  - \Delta_{{\rm Q} 2}' - \frac{3}{5} k \Delta_{{\rm Q}3} + \frac{2}{5} k \Delta_{{\rm Q} 1} = \epsilon'\biggl[ \frac{9}{10} \Delta_{{\rm Q}2} - \frac{1}{10} (\Delta_{{\rm Q}0} + 
\Delta_{{\rm I} 2} )\biggr],
\label{L22}\\
&& v_{b}' + {\cal H} v_{b} = - i k \phi - \frac{\epsilon'}{\beta} \biggl( 3 i \Delta_{{\rm I}1} + v_{b} \biggr).
\end{eqnarray}
For $\ell\geq 3$ the hierarchy of the brightness can be determined in general terms by using 
the recurrence relation for the 
Legendre polynomials reported in Eq. (\ref{rec1}):
\begin{eqnarray}
&&\Delta_{{\rm I}\ell}' + \epsilon' \Delta_{{\rm I}\ell} 
= \frac{k}{2 \ell + 1} [ \ell \Delta_{{\rm I}(\ell-1)} - (\ell + 1) \Delta_{{\rm I}(\ell + 1)}],
\nonumber\\
&& \Delta_{{\rm Q}\ell}' + \epsilon' \Delta_{{\rm Q}\ell} 
= \frac{k}{2 \ell + 1} [ \ell \Delta_{{\rm Q}(\ell-1)} - (\ell + 1) \Delta_{{\rm Q}(\ell + 1)}].
\end{eqnarray}

\subsubsection{Zeroth order in the tight coupling expansion: acoustic oscillations}
We are now ready to compute the evolution of the various terms to a given order in the tight-coupling expansion parameter $\tau_{\rm c} = |1/\epsilon'|$. 
After expanding the various moments of the brightness function and the velocity field in $\tau_{\rm c}$ 
\begin{eqnarray}
&&\Delta_{{\rm I}\ell} = \overline{\Delta}_{{\rm I}\ell} + \tau_{\rm c} \delta_{{\rm I}\ell},
\nonumber\\
&& \Delta_{{\rm Q}\ell} = \overline{\Delta}_{{\rm Q}\ell} + \tau_{\rm c} \delta_{{\rm Q}\ell},
\nonumber\\
&&v_{\rm b} = \overline{v}_{\rm b} + \tau_{\rm c} \delta_{v_{\rm b}},
\label{DEFEXP}
\end{eqnarray}
the obtained expressions can be inserted  into Eqs. (\ref{L01})--(\ref{L13}) and the evolution of the various moments of the brightness 
function can be found order by order.

To zeroth order in the tight-coupling approximation, the evolution equation for the baryon velocity field, i.e. Eq. (\ref{L03}), leads to:
\begin{equation}
\overline{v}_{b} = - 3 i  \overline{\Delta}_{{\rm I}1},
\label{vb1}
\end{equation} 
while Eqs. (\ref{L02}) and (\ref{L12}) lead, respectively, to
\begin{equation}
\overline{\Delta}_{{\rm Q}0} = \overline{\Delta}_{{\rm I}2} + \overline{\Delta}_{{\rm Q}2},\,\,\,\,\,\,\,\,\,\,\, \overline{\Delta}_{{\rm Q}1} =0.
\label{int1Q}
\end{equation}
Finally Eqs. (\ref{L21}) and (\ref{L22}) imply
\begin{equation}
9\overline{\Delta}_{{\rm I}2}  = \overline{\Delta}_{{\rm Q}0} + \overline{\Delta}_{{\rm Q}2},
\,\,\,\,\,\,\,\,\,\,\,\,\, 9\overline{\Delta}_{{\rm Q}2}  = \overline{\Delta}_{{\rm Q}0} + \overline{\Delta}_{{\rm I}2}.
\label{int2Q}
\end{equation}
Taking together the four conditions expressed by Eqs. (\ref{int1Q}) and (\ref{int2Q}) we have, to zeroth order in the 
tight-coupling approximation:
\begin{equation}
\overline{\Delta}_{{\rm Q}\ell} =0,\,\,\,\,\,\,\,\,\,\,\ell\geq 0,\,\,\,\,\,\,\,\,\overline{\Delta}_{{\rm I}\ell} =0,\,\,\,\,\,\, \ell \geq 2.
\label{int3}
\end{equation}
Hence, to zeroth order in the tight coupling, the relevant equations are 
\begin{eqnarray}
&& \overline{v}_{b} = - 3i \overline{\Delta}_{{\rm I}1},
\label{zerothorder1}\\
&& \overline{\Delta}_{{\rm I}0}' + k \overline{\Delta}_{{\rm I}1} =  \psi'.
\label{zerothorder2}
\end{eqnarray}
This means, as anticipated, that to zeroth order in the tight-coupling expansion the CMB is not polarized 
since $\Delta_{\rm Q}$ is vanishing.

A decoupled evolution equation for the monopole can be derived. Summing up
Eq. (\ref{L11}) (multiplied by $ 3 i$) and  Eq. (\ref{L13})  (multiplied by $\beta$) we get, to zeroth 
order in the tight coupling expansion:
\begin{equation}
\beta \overline{v}_{\rm b}' - 3 i \overline{\Delta}_{{\rm I}1}' + i k \phi (\beta + 1) -
2i k \overline{\Delta}_{{\rm I}2} + i k \overline{\Delta}_{{\rm I}0} + \beta {\cal H} \overline{v}_{\rm b} =0.
\label{MON1}
\end{equation}
Recalling now
 Eq. (\ref{zerothorder1})  to eliminate $\overline{v}_{\rm b}$  from Eq. (\ref{MON1}),
 the following equation can be obtained
\begin{equation}
(\beta+ 1) \overline{\Delta}_{{\rm I}1}' + {\cal H} \beta 
\overline{\Delta}_{{\rm I} 1} - \frac{k}{3} \overline{\Delta}_{{\rm I}0} 
= 0.
\label{int4}
\end{equation}
Finally, the dipole term can be eliminated from Eq. (\ref{int4}) using Eq. (\ref{zerothorder2}). By doing so, Eq. (\ref{int4}) leads to  the wanted 
decoupled equation for the monopole:
\begin{equation}
\overline{\Delta}_{{\rm I}0}''  + \frac{\beta'}{\beta + 1} \overline{\Delta}_{{\rm I}0}' + 
k^2 c_{{\rm s,b}}^2 \overline{\Delta}_{{\rm I}0} = 
 \biggl[ \psi'' + \frac{\beta'}{\beta +1} \psi' - \frac{k^2}{3} \phi \biggr],
 \label{monopole}
\end{equation}
where
\begin{equation}
c_{{\rm s,b}}= \frac{1}{\sqrt{3 (\beta+1)}}, 
\label{CSB}
\end{equation}
is the sound speed which includes the effect of baryons.
The term  $k^2 c_{\rm s}^2 \overline{\Delta}_{{\rm I}0}$ is the photon pressure.
Defining, from Eq. (\ref{CSB}), the sound horizon as 
\begin{equation}
r_{\rm s}(\tau) = \int_{0}^{\tau} c_{\rm s}(\tau') d \tau',
\label{Shor}
\end{equation}
the photon pressure cannot be neglected for modes $k r_{\rm s}(\tau) \geq 1$.

At the right hand side of Eq. (\ref{monopole}) several forcing terms appear.  The 
term $\psi''$ dominates, if present, on super-horizon scales and causes 
a dilation effect on $\overline{\Delta}_{{\rm I}0}$. The term containing $k^2 \phi$ leads 
to the adiabatic growth of the photon-baryon fluctuations and becomes 
important for $k \tau \simeq 1$. 

In Eq. (\ref{monopole}) the damping term arises from the redshifting of the baryon momentum 
in an expanding Universe, while photon pressure provides the restoring force which is weakly suppressed by  the additional inertia of the baryons.

\subsubsection{Solutions of the evolution of monopole and dipole}
Equation (\ref{monopole}) can be solved under different approximations (or even exactly \cite{HS1}).
The first brutal approximation would be to set $\beta'= \beta =0$, implying the the r\^ole of the baryons 
in the acoustic oscillations is totally neglected. As a consequence, in this case $c_{{\rm s,b}}\equiv 1/\sqrt{3}$
which is nothing but the sound speed discussed in Eqs. (\ref{deltam})--(\ref{thetar}) for the fluid
analysis of the adiabatic mode. In the case of the adiabatic mode, neglecting 
neutrino anisotropic stress, $\psi = \phi = \psi_{\rm m}$ and $\psi' =0$. Hence, the solution 
for the monopole and the dipole to zeroth order in the tight coupling expansion 
follows by solving Eq. (\ref{monopole}) and by inserting the obtained result 
into Eqs. (\ref{zerothorder1}) and (\ref{zerothorder2}), i.e. 
\begin{eqnarray}
&& \overline{\Delta}_{{\rm I}0}(k,\tau) = \frac{\psi_{\rm }}{3} [ \cos{(k c_{{\rm s, b}}\tau)} -3],
\nonumber\\
&& \overline{\Delta}_{{\rm I}1}(k,\tau) = - \frac{\psi_{\rm m}}{3} k c_{{\rm s, b}} \sin{(k c_{\rm s, b}\tau)},
\label{beta=0}
\end{eqnarray}
which is exactly the solution discussed in section 4 if we recall Eq. (\ref{zerothorder1}) and 
the definition (\ref{defvb}).

If $\beta' =0$ but $\beta\neq 0$, then the solution of Eqs. (\ref{zerothorder1})--
(\ref{zerothorder2}) and (\ref{monopole})  becomes, in the case of the adiabatic mode, 
\begin{eqnarray}
&& \overline{\Delta}_{{\rm I}0}(k,\tau) = \frac{\psi_{\rm m}}{3} (\beta + 1) [ \cos{(k c_{\rm s, b} \tau)} -3],
\nonumber\\
&& \overline{\Delta}_{{\rm I}1}(k,\tau) = \frac{\psi_{\rm m}}{3} \sqrt{\frac{\beta + 1}{3}} \sin{(k c_{\rm s, b} \tau)}.
\label{betaneq0}
\end{eqnarray}
Equation (\ref{betaneq0}) shows that the presence of the baryons increases the amplitude of the monopole 
by a factor $\beta$. This phenomenon can be verified also in the case of generic time-dependent $\beta$.
In the case of $\beta\neq 0$ the shift in the monopole term is $(\beta +1)$ with respect to the case $\beta =0$.
This phenomenon produces a modulation of the height of the acoustic peak that depends on the baryon content 
of the model.

Consider now the possibility of setting directly initial conditions for the Boltzmann hierarchy during 
the radiation dominated epoch. During the radiation dominated epoch
and for modes which are outside the horizon, the initial conditions for the monopole 
and the dipole are fixed as 
\begin{eqnarray}
&&\Delta_{{\rm I}0}(k,\tau) =  - \frac{\phi_{0}}{2} - \frac{525 + 188 R_{\nu} + 16 R_{\nu}^2}{180( 25 + 2 R_{\nu}) } \phi_{0}
k^2 \tau^2 ,
\nonumber\\
&& \Delta_{{\rm I}1}(k,\tau)= \frac{\phi_{0}}{6} k\tau - \frac{65 + 16 R_{\nu}}{108( 25  + 2 R_{\nu})} \phi_{0} k^3 \tau^3
\label{RADBZ}
\end{eqnarray}
where $\phi_{0}$ is the constant value of $\phi$ during radiation. This result is fully 
compatible with the results obtained in Eqs. (\ref{ad1})--(\ref{solsig1}). The constant value 
of $\psi$, i.e. $\psi_{0}$ will be related to $\phi_{0}$ through $R_{\nu}$, i.e. the fractional contribution
of the neutrinos to the total density (see Eqs. (\ref{DEFRNU}) and (\ref{sigma01})).

It is useful to observe that in terms of the quantity $\Delta_{0} = (\overline{\Delta}_{{\rm I}0} -\psi)$,
Eq. (\ref{monopole})  becomes
\begin{equation}
\Delta_{0}'' + k^2 c_{{\rm s,b}}^2 \Delta_{0} = - k^2 \biggl[ \frac{\phi}{3} + c_{{\rm s,b}}^2 \psi\biggr].
\label{DEL0}
\end{equation}
The initial conditions for $\Delta_{0}$ are easily obtained from its definition in terms 
of $\Delta_{{\rm I}0}$ and $\psi$.
 
The same strategy can be applied to more realistic
 cases, such as the one where the scale factor interpolates between 
a radiation-dominated phase and a matter-dominated phase, as discussed in Eq. (\ref{interpolation}). In this case 
the solution of Eq. (\ref{monopole}) will be more complicated but always analytically tractable.
Equation (\ref{monopole}) can indeed be solved  in general terms. The general solution 
of the homogeneous equation is simply given, in the WKB 
approximation, as 
\begin{equation}
\overline{\Delta}_{{\rm I}0}  = \frac{1}{(\beta + 1)^{1/4}} [ A \cos{k r_{\rm s}} + B \sin{k r_{\rm s}}].
\label{GEN}
\end{equation}
 For adiabatic fluctuations, $k^2 \phi$ contributes primarily to the cosinus. The reason is that , in this 
 case, $\psi$ is constant until the moment of Jeans scale crossing at which moment it begins to decay.
 Non-adiabatic fluctuations, on the contrary, have vanishing gravitational potential at early 
 times and their monopole is dominated by sinusoidal harmonics.  Consequently, the peaks 
 in the temperature power spectrum will be located, for adiabatic fluctuations, 
 at a scale $k_{n}$ such that 
 $ k_{n} r_{\rm s} (\tau_{\ast}) = n \pi$.  Notice that, according  to Eq. (\ref{zerothorder2}) 
 the dipole, will be anticorrelated with the monopole. So if the monopole is 
 cosinusoidal , the dipole will be instead sinusoidal. Hence the ``zeros" of the cosinus
 (as opposed to the maxima) will be filled by the monopole. 
  The solution of Eq. (\ref{monopole})  can then be obtained by supplementing the 
 general solution of the homogeneous equation (\ref{GEN}) with a particular 
 solution of the inhomogeneous equations that can be found easily with the 
 usual Green's function methods \cite{HS1}. 
 
 The amplitude of the monopole term shifts as $(1 + \beta)^{-1/4}$. Recalling 
 Eq. (\ref{DEFBETA}), it can be argued that the height of the Doppler peak 
 is weakly sensitive to $h^2 \Omega_{\rm b}$ in the $\Lambda$CDM model 
 where $\Omega_{\rm b} \ll \Omega_{\rm m}$ and $\beta(\tau_{\rm dec}) < 1$.

In $\ell$ space the position of the first peak for adiabatic and isocurvature modes is 
 given, respectively, by 
 \begin{eqnarray}
 && \ell^{(n)} = n\,\pi \frac{(\tau_{\rm dec} - \tau_{0})}{ \,\,r_{\rm s}(\tau_{\rm dec})},
 \label{ADdop}\\
 && \ell^{(n)} = \biggl(n + \frac{1}{2}\biggr)
 \,\pi \frac{(\tau_{\rm dec} - \tau_{0})}{r_{\rm s}(\tau_{\rm dec})},
 \label{ISOdop}
 \end{eqnarray}
 where, in the spatially flat case 
 \begin{equation}
 \tau_{\rm dec} -\tau_{0} = \frac{1}{H_{0} a_0} \int_{0}^{z_{\rm dec}} \frac{d z }{\sqrt{\Omega_{\rm r} ( 1 + z)^4 + \Omega_{\rm m}( 1 + z)^3 
 + \Omega_{\Lambda}}},
 \label{INT1}
 \end{equation}
 and $a_{\rm dec}(\tau_{\rm dec} - \tau_{0})$ is the angular diameter distance.
In the case $\Omega_{\Lambda} =0 $ and for $h^2 \Omega_{\rm b} =0.023$ (whose value  affects the sound horizon)
the position of the first peak will be about $220$ for adiabatic modes and about $330$ isocurvature modes.  The position of the first peak detected in large scale CMB experiments 
is in good agrrement with Eq. (\ref{ADdop}) (see the introduction).
The predicted position of the first adiabatic peak is in good agreement with the value inferred from the 
data reported in Fig. \ref{F13} .
Notice that when integrating up to $z_{\rm dec} \sim 1100$ the radiation density (i.e. $\Omega_{\rm r} \sim 8 \times 
10^{-5}$) does affect the result since it is multiplied by a factor $(1+z)^4$.  Notice, conversely, that the 
region of small redshifts (i.e. $z\leq 10$) is sensitive to the presence of a cosmological constant.

It is difficult to obtain general analytic formulas for the position of the peaks. Degeneracies among the 
parameters may appear \cite{BE}. In \cite{W1s} a semi-analytical expression for the integral giving 
the angular diameter distance has been derived for various cases of practical interest.

Once the evolution of the lowest multipoles is known, the obtained expressions can be used in the 
integral solutions of the Boltzmann equation (like Eq. (\ref{LSIell})) and the angular power spectrum can be 
computed analytically. Recently Weinberg in a series of papers \cite{Wbz1,Wbz2,Wbz3} 
computed the temperature fluctuations in terms of a pair of generalized form factors 
related, respectively, to the monopole and the dipole. This set of calculations were conducted in the synchronous 
gauge (see also \cite{old1,old2,old3} for earlier work on this subject). Reference \cite{MUKa} also presents 
analytical estimates for the angular power spectrum exhibiting explicit dependence 
on the cosmological parameters in the case of the concordance model.

The results of the tight coupling expansion hold for $k \tau_{\rm c} \ll 1$. Thus 
the present approximation scheme breaks down, strictly speaking, for 
wave-numbers  $k > \tau_{\rm c}^{-1}$.
 Equation (\ref{monopole})  holds to zeroth-order in the tight coupling expansion, i.e. 
 it can be only applied on scales much larger than the photon mean free path. By comparing 
 the rate of the Universe expansion with the rate of dissipation we can estimate that 
 $\tau_{\rm c} k^2 \sim  \tau^{-1}$ defines approximately the scale above which 
 the wave-numbers will experience damping.  From these considerations the typical 
 damping scale can be approximated by 
 \begin{equation}
 k_{\rm d}^{-2} \simeq 0.3 ( \Omega_{\rm m} h^2)^{-1/2} 
 (\Omega_{\rm b} h^2)^{-1} ( a/a_{\rm dec})^{5/2}\,\,\,{\rm Mpc}^2.
 \end{equation}
 The effect of diffusion is to damp the photon and baryon oscillations exponentially by the time
  of last scattering on comoving scales smaller than $3$ Mpc. For an experimental 
  evidence of this effect see \cite{DICK} and references therein.
 
 In order to have some qualitative estimate for the damping scale in the framework 
 of the tight coupling approximation, it is necessary to expand the temperature, polarization and 
 velocity fluctuations to second order in $\tau_{\rm c}$. Since for very small scales the r\^ole 
 of gravity is not important the longitudinal fluctuations 
 of the metric can be neglected. The result of this analysis \cite{HZ2} shows that 
 the monopole behaves approximately as 
 \begin{equation}
 \Delta_{{\rm I}0} \simeq e^{\pm i k r_{\rm s}} e^{ -(k/k_{\rm d})^2},
 \end{equation}
 where 
 \begin{equation}
 \frac{1}{k_{\rm d}^2} = \int_{0}^{\tau} \frac{\tau_{\rm c}}{6 (\beta + 1)^2} \biggl[ \beta^2 
+ \frac{16}{15} ( 1 + \beta)\biggr].
\label{KD}
\end{equation}
The  factor $16/15$ arises when the polarization fluctuations are taken consistently 
into account in the derivation \cite{HZ2}.

\subsubsection{First order in tight coupling expansion: polarization}

To first order in the tight-coupling limit, the relevant equations can be obtained by keeping all terms of order 
$\tau_{\rm c} $ and by using the 
first-order relations to simplify the expressions. From Eq. (\ref{L12}) the condition $\delta_{{\rm Q}1} =0$ can be derived. From Eqs. (\ref{L02}) and  (\ref{L21})--(\ref{L22}), the following remaining conditions are obtained respectively:
\begin{eqnarray}
&& - \delta_{{\rm Q}0} + \delta_{{\rm I} 2} + \delta_{{\rm Q} 2} =0,
\label{firstord1}\\
&& \frac{9}{10} \delta_{{\rm I}2} - \frac{1}{10} [ \delta_{{\rm Q}0} + \delta_{{\rm Q}2} ]= \frac{2}{5} k \overline{\Delta}_{{\rm I}1},
\label{firstord2}\\
&&  \frac{9}{10} \delta_{{\rm Q}2} - \frac{1}{10} [ \delta_{{\rm Q}0} + \delta_{{\rm I}2} ]=0.
\label{firstord3}
\end{eqnarray}
Equations  (\ref{firstord1})--(\ref{firstord3}) are a set of algebraic conditions 
implying that the  relations to be satisfied are:
\begin{eqnarray}
&& \delta_{{\rm Q}0} = \frac{5}{4} \delta_{{\rm I}2},
\label{Cond1}\\
&& \delta_{{\rm Q}2} = \frac{1}{4} \delta_{{\rm I}2},
\label{Cond2}\\
&& \delta_{{\rm I}2} = \frac{8}{15} k \overline{\Delta}_{{\rm I}1}.
\label{Cond3}
\end{eqnarray}
Recalling the original form of the expansion of the quadrupole as defined in Eq. (\ref{DEFEXP}),
 Eq. (\ref{Cond3}) can be also written 
\begin{equation}
\Delta_{{\rm I} 2} = \tau_{\rm c} \delta_{{\rm I}2}= \frac{8}{15} k\tau_{\rm c} \overline{\Delta}_{{\rm I}1},
\label{Cond4}
\end{equation}
since to zeroth order the quadrupole vanishes and the first non-vanishing effect comes 
from the first-order quadrupole whose value is determined from the zeroth-order monopole.

Now, from Eqs. (\ref{Cond1}) and (\ref{Cond2}), the quadrupole moment of $\Delta_{\rm Q}$ is proportional to the quadrupole of $\Delta_{\rm I}$, which is, in turn, 
proportional to the dipole evaluated to first order in $\tau_{\rm c}$. But $\Delta_{\rm Q}$ measures exactly 
the degree of  linear polarization of the radiation field. So, to first order in the tight-coupling expansion, the CMB  is linearly  polarized. Notice that the same derivation performed in the case of the equation for $\Delta_{\rm Q}$ 
can be more correctly performed in the case of the evolution equation of $\Delta_{\rm P}$ with the same 
result \cite{HZ2}.   Using the definition of $S_{\rm P}$ (i.e. Eq. (\ref{BRP})),
 and recalling Eqs. (\ref{Cond1})--(\ref{Cond3}),  we have that the source term 
 of Eq. (\ref{LSP}) can be approximated as
 \begin{equation}
 S_{\rm P} \simeq  \frac{4}{3} k \tau_{\rm c} \overline{\Delta}_{{\rm I}1}.
 \end{equation}
 Since $\tau_{\rm c}$ grows very rapidly during recombination, in order to have 
 quantitative estimates of the effect we have to know the evolution of $S_{\rm P}$ with 
 better accuracy. In order to achieve this goal, let us go back to the (exact) system 
 describing the coupled evolution of the various multipoles and, in particular, to Eqs. 
 (\ref{L02}) and (\ref{L21})--(\ref{L22}).  Taking the definition 
 of $S_{\rm P}$ (or $S_{\rm Q}$) and performing a first 
 time derivative we have 
 \begin{equation}
 S_{\rm P}' = \Delta_{{\rm I}2}' + \Delta_{{\rm P}2}' + \Delta_{{\rm P}0}'.
 \end{equation}
 Then, from Eqs. (\ref{L02}) and (\ref{L21})--(\ref{L22}), the time
 derivatives of the two quadrupoles and of the monopole can be expressed 
 in terms of the monopoles, quadrupoles and octupoles.  Simplifying 
 the obtained expression we get the following 
 evolution equation for $S_{\rm P}$, i.e.  \cite{HZ2}
 \begin{equation}
 S_{\rm P}' + \frac{3}{10} \epsilon' S_{\rm P} = k \biggl[ \frac{2}{5} \Delta_{{\rm I}1} - 
 \frac{3}{5}\biggl( \Delta_{{\rm P}1}+ \Delta_{{\rm P}3} + \Delta_{{\rm I}3}\biggr)\biggr].
 \end{equation}
This equation can be solved by evaluating the right hand side to zeroth-order 
in the tight coupling expansion, i.e. 
\begin{equation}
S_{\rm P}(\tau) =\frac{2}{5} k \int_{0}^{\tau}  d x \overline{\Delta}_{{\rm I}1} e^{-\frac{3}{10} \epsilon(\tau,x)}.
\end{equation}
This equation, giving the evolution of $S_{\rm P}$, can be inserted back into Eq. (\ref{LSP}) in order 
to obtain $\Delta_{\rm P}$. The result of this procedure is \cite{HZ2}
\begin{eqnarray}
&&\Delta_{\rm P} \simeq ( 1 - \mu^2) e^{i k\mu(\tau_{\rm dec} -\tau_{0})} {\cal D}(k),
\nonumber\\
&& {\cal D}(k)\simeq (0.51) \,\,k \,\,\sigma_{\rm dec}
 \overline{\Delta}_{{\rm I}1}(\tau_{\rm dec}),
 \label{IMPROVED}
\end{eqnarray}
where $\sigma_{\rm dec}$ is the width of the visibility function. This result allows to estimate 
with reasonable accuracy the angular power spectrum of the cross-correlation between 
temperature and polarization (see below), for instance, in the case of the adiabatic mode.

While the derivation of the polarization dependence of Thompson scattering 
has been conducted within the framework of the tight coupling approximation, it is useful to recall 
here that these properties follow dirrectly from the polarization dependence of Thompson 
scattering whose differential cross-section can be written as 
\begin{equation}
\frac{d\sigma}{d\Omega} = r_{0}^2 | \epsilon^{(\alpha)}\cdot \epsilon^{(\alpha')}|^2 \equiv 
 \frac{3 \sigma_{\rm T}}{8\pi} | \epsilon^{(\alpha)}\cdot \epsilon^{(\alpha')}|^2.
\end{equation}
where $\epsilon^{(\alpha)}$ is the incident polarization and $\epsilon^{(\alpha')}$ is the scattered polarization;
$r_{0}$ is the classical radius of the electron and $\sigma_{\rm T}$ is the 
total cross-section both defined in Eq. (\ref{thompson}).

Suppose that the incident radiation is  not polarized, i.e. $U = V = Q=0$; then we can write 
\begin{equation}
Q = {\cal I}_{x} - {\cal I}_{y}=0,\,\,\,\,\,\,\, {\cal I}_{x} = {\cal I}_{y} = \frac{{\cal I}}{2}.
\label{nonpol}
\end{equation}
Defining the incoming and outgoing polarization vectors as 
\begin{eqnarray}
&&\epsilon_{x} = (1,\,\,\,\,0,\,\,\,0),\,\,\,\,\,\,\,\,\,\,\epsilon_{y} = (0,\,\,\,\,1,\,\,\,0),\,\,\,\,\,\,\,\,\,\, \hat{k}=(0,\,\,\,\,0,\,\,\,1).
\nonumber\\
&& \epsilon_{x}' = (-\sin{\varphi},\,\,\,\,-\cos{\varphi},\,\,\,0),\,\,\,\,\,\,\,\,\,\,
\epsilon_{y}' = (\cos{\vartheta}\cos{\varphi},\,\,\,\,-\cos{\vartheta}\sin{\varphi},\,\,\,-\sin{\vartheta}),
\label{geom1}
\end{eqnarray}
the explicit form of the scattered amplitudes will be 
\begin{eqnarray}
&& {\cal I}_{x}'= \frac{3 \sigma_{T}}{8\pi} \biggl[ |\epsilon_{x}\cdot\epsilon_{x}'|^2 {\cal I}_{x} + |\epsilon_{y}\cdot\epsilon_{x}'|^2 {\cal I}_{y} \biggr] =\frac{3 \sigma_{T}}{16\pi} {\cal I},
\nonumber\\
&& {\cal I}_{y}'= \frac{3 \sigma_{T}}{8\pi} \biggl[ |\epsilon_{x}\cdot\epsilon_{y}'|^2 {\cal I}_{x} + |\epsilon_{y}\cdot\epsilon_{y}'|^2 {\cal I}_{y} \biggr]=\frac{3 \sigma_{T}}{16\pi} {\cal I} \cos^2{\vartheta}.
\label{geom2}
\end{eqnarray}

Recalling the definition of Stokes parameters:
\begin{eqnarray}
&& I'= {\cal I}_{x}' + {\cal I}_{y}' = \frac{3}{16 \pi } \sigma_{T}  {\cal I} ( 1 + \cos^2{\vartheta}),
\nonumber\\
&& Q' =  {\cal I}_{x}' - {\cal I}_{y}' = \frac{3}{16 \pi } \sigma_{T}  {\cal I}\sin^2{\vartheta}.
\end{eqnarray}
Even if $U'=0$ the obtained $Q$ and $U$ must be rotated 
to a common coordinate system:
\begin{equation}
Q'= \cos{2\varphi} Q,\,\,\,\,\,\,\,\,\,\,\,\,\,\,\,U'= -\sin{2\varphi} Q.
\end{equation}
So the final expressions  for the Stokes parameters of the scattered radiation are:
\begin{eqnarray}
&& I' =\frac{3}{16 \pi } \sigma_{\rm T}  {\cal I} ( 1 + \cos^2{\vartheta}),
\nonumber\\
&& Q'= \frac{3}{16 \pi } \sigma_{\rm T}  {\cal I} \sin^2{\vartheta} \cos{2\varphi},
\nonumber\\
&& U' = - \frac{3}{16 \pi } \sigma_{\rm T}  {\cal I} \sin^2{\vartheta} \sin{2\varphi}.
\end{eqnarray}
We can now expand the incident intensity in spherical harmonics:
\begin{equation}
{\cal I}(\theta,\varphi) = \sum_{\ell m} a_{\ell m} Y_{\ell m }(\vartheta,\varphi).
\label{CALI}
\end{equation}
So, for instance, $Q'$ will be
\begin{equation}
Q' = \frac{3}{16 \pi }\sigma_{\rm T} \int \sum_{\ell m} Y_{\ell m}(\vartheta,\varphi) a_{\ell m} \sin^2{\vartheta} \cos{2\varphi} d\Omega.
\label{HEU}
\end{equation}
By inserting the explicit form of the spherical harmonics into Eq. (\ref{HEU}) it can be easily shown that 
$Q'\neq0$ provided the term $a_{22}\neq 0$ in the expansion of Eq. (\ref{CALI}).  

\begin{figure}[tp]
\centerline{\epsfig{file=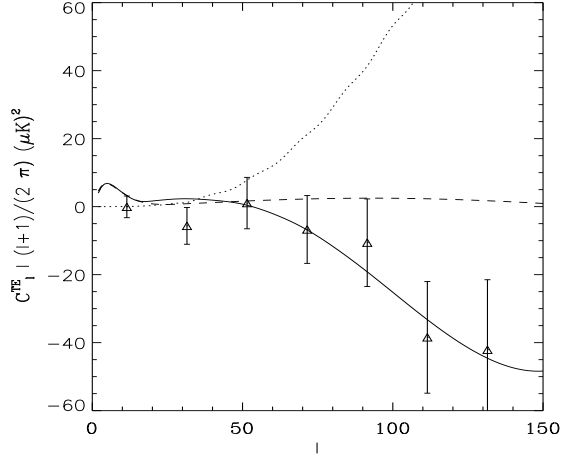, width=8cm}}
\vspace*{8pt}
\caption[a]{Temperature-polarization correlation measured by the WMAP collaboration 
(adapted from Ref. \cite{spergeletal}.}
\label{F81}
\end{figure}

Recall now that under clockwise rotations the Stokes parameters $Q$ and $U$ 
transform as in Eq. (\ref{clockwise}). As a consequence 
\begin{equation}
(Q \pm i U)^{\prime} = e^{\mp 2i \varphi}(Q \pm iU)
\end{equation} 
where $\varphi$ is the rotation angle.
From this observation it follows that the combinations 
\begin{equation}
(\Delta_{\rm Q} \pm i \Delta_{\rm U})(\hat{ n}) = \sum_{\ell m} a_{\pm 2,\ell m}
\,\,\,\,{_{\pm2}}{\cal Y}_{\ell m}(\hat{n})
\end{equation}
can be expanded in terms of the  spin-2 spherical
harmonics, i.e.   $_{\pm 2}{\cal Y}_{\ell}^{m}(\hat{n})$ \cite{sphharm1,sphharm2}.

The expansion coefficients are 
\begin{equation}
a_{\pm 2,\ell m}=\int d\Omega \,\,\,\,\,\,{_{\pm2}} {\cal Y}_{\ell m}^{*}(\Delta_{\rm Q} \pm i \Delta_{\rm U})(\hat{n}).
\end{equation}
In Ref. \cite{sphharm1,sphharm2}, the authors introduce the following 
linear combinations of 
$a_{\pm 2,  \ell m}$ to circumvent the impasse that the Stokes 
parameter are not invariant under rotations;
\begin{equation}
a_{E,\ell m}=-\frac{1}{2}(a_{2, \ell m}+a_{-2,\ell m})  \nonumber \\
a_{B, \ell m}= \frac{i}{2}(a_{2,\ell m}-a_{-2,\ell m}).
\end{equation}
These newly defined variables are expanded in terms of ordinary spherical
harmonics, $Y_{\ell m}(\hat{n})$,
\begin{equation}
E(\hat{n})= \sum_{lm} a_{E,\ell m} Y_{lm}(\hat{n}), \,\,\,\,\,\,\,\,\,\,\,\,\,
B(\hat{n})= \sum_{lm} a_{B,\ell m} Y_{lm}(\hat{ n}).
\end{equation}
The spin-zero spherical harmonics, $Y_{lm}(\hat{n})$, is free from
 the ambiguity with the rotation of the coordinate system,
and therefore $E$ and $B$ are rotationally invariant quantities.
The  $E$-mode  has $(-1)^{\ell}$ parity and the  $B$
mode $(-1)^{(\ell+1)}$ parity in analogy with electric and magnetic
fields.
Scalar perturbations generate only the $E$ mode \cite{kamionkowski}. 
While the scalar fluctuations only generate an $E$ mode, tensor fluctuations also 
generate a B mode. The Boltzmann equation for the tensor modes can be 
easily derived  following Refs. \cite{polnarev,crit} (see 
also \cite{kos2,knox}). 

Consider now, specifically, the adiabatic mode.
While the temperature fluctuation $\Delta_{\rm I}$ oscillates like $\cos{(kc_{\rm s,b}
\tau_{\rm dec})}$, the polarization is proportional to the dipole and oscillates like the sinus 
of the same argument. The correlation function of the temperature and polarization, i.e. 
$\langle \Delta_{\rm I} \Delta_{\rm P}\rangle$ will then be 
proportional to $\sin{(k c_{\rm s,b} \tau_{\rm dec})} \cos{(kc_{\rm s,b} \tau_{\rm dec})}$.  
An analytical prediction for this quantity can be inferred from Eq. (\ref{IMPROVED})  (see \cite{HZ2}). 
The spectrum of the cross correlation must then have a peak for 
$k  c_{\rm s,b} \tau_{\rm dec} \sim \,\,3\pi/4$, corresponding to $\ell \sim 150$. This is the 
result suggested by Fig. \ref{F81} and taken from Ref. \cite{spergeletal} reporting the 
measurement of the WMAP collaboration. In  Fig. \ref{F81} the temperature-polarization angular
power spectrum is reported for adiabatic models (solid line) and for isocurvature models 
(dashed line).  

We should mention here that a  rather effective method in order to treat on equal footing the scalar, vector 
and tensor radiative transfer equations is the total angular momentum method 
\cite{tot1,tot2,tot3}. Within this approach, the collision terms couple only the quadrupole moments 
of the distributions and each moment corresponds directly to observable patterns in the microwave sky. 
In this language the analysis of the polarization of the radiation field becomes somehow more transparent.

\section*{Acknowledgments}
The author wishes to express his gratitude to D. Babusci,  M. Gasperini,
 H. Kurki-Suonio, E. Keih\"anen,  G. Veneziano, A. Vilenkin  for discussions and collaborations leading, directly or indirectly, 
 to some of the results summarized in the present article. Various discussions presented here 
 grew through the years as a consequence of lectures and seminars presented in different contexts. The gratitude 
 of the author is also for those stimulating questions and comments of students and colleagues.

\newpage 

\begin{appendix}
\renewcommand{\theequation}{A.\arabic{equation}}
\setcounter{equation}{0}
\section{Metric fluctuations}
In this appendix, some technical aspects concerning the scalar, vector and 
tensor fluctuations of the geometry will be summarized with the purpose of 
making the present review self-contained. The fluctuations of the geometry and 
of the sources are parametrized as in section 2.  The background geometry will be 
assumed to be spatially flat. 
To first order in the amplitude of the metric fluctuations, the 
fluctuations of the Christoffel connections can be computed
\begin{equation}
\delta \Gamma_{\alpha\beta}^{\mu} = \frac{1}{2}  \overline{g}^{\mu\nu}( 
-\partial_{\nu} \delta g_{\alpha\beta} + \partial_{\beta} \delta g_{\nu\alpha} 
+ \partial_{\alpha} \delta g_{\beta\nu} ) + 
\frac{1}{2} \delta g^{\mu\nu} ( -\partial_{\nu}  
\overline{g}_{\alpha\beta} + \partial_{\beta}  \overline{g}_{\nu\alpha} 
+ \partial_{\alpha}  \overline{g}_{\beta\nu} ),
\label{CHRF}
\end{equation}
where the overline denotes that the corresponding quantity is to 
be evaluated on the background geometry and the usual convention 
of summation over repeated indices is adopted. The first-order fluctuation of the Ricci scalar can also be computed as 
\begin{equation}
\delta R_{\mu\nu} = \partial_{\alpha} \delta \Gamma_{\mu\nu}^{\alpha}
- \partial_{\nu} \delta \Gamma_{\mu \beta}^{\beta} + 
\delta \Gamma_{\mu\nu}^{\alpha}\, \overline{\Gamma}_{\alpha\beta}^{\beta} 
+ \overline{\Gamma}_{\mu\nu}^{\alpha}\,\delta \Gamma_{\alpha\beta}^{\beta}
- \delta \Gamma_{\alpha\mu}^{\beta} \overline{\Gamma}_{\beta\nu}^{\alpha} 
- \overline{\Gamma}_{\alpha\mu}^{\beta} \delta \Gamma_{\beta\nu}^{\alpha}.
\label{RICF}
\end{equation}
The fluctuations of the Ricci tensors with one controvariant 
index and one covariant index, as well as the fluctuations 
of the Ricci scalar,  are then
\begin{eqnarray}
&& \delta R_{\mu}^{\nu} = \delta g^{\nu\alpha} \overline{R}_{\alpha\mu}  + 
\overline{g}^{\nu\alpha} \delta R_{\alpha\mu},
\label{MIXIN}\\
&& \delta R = \delta g^{\alpha\beta} \overline{R}_{\alpha\beta} + 
\overline{g}^{\alpha\beta} \delta R_{\alpha\beta}.
\label{RS}
\end{eqnarray}
Finally, the fluctuations of the components 
of the Einstein tensor can be easily deduced once Eqs. (\ref{MIXIN})--(\ref{RS}) are known 
in explicit terms: 
\begin{equation}
\delta {\cal G}_{\mu}^{\nu} = \delta R_{\mu}^{\nu} - \frac{1}{2} 
\delta_{\mu}^{\nu} \delta R.
\end{equation}
Formally, the 
perturbation of the covariant conservation equation is:
\begin{equation}
\partial_{\mu} \delta T^{\mu\nu} + 
\overline{\Gamma}_{\mu\alpha}^{\mu} \delta T^{\alpha\nu} 
+ \delta \Gamma_{\mu\alpha}^{\mu}\overline{T}^{\alpha\nu} 
+ \overline{\Gamma}_{\alpha\beta}^{\nu} \delta T^{\alpha\beta} 
+ \delta \Gamma_{\alpha\beta}^{\nu} \overline{T}^{\alpha\beta} =0.
\label{GENCOVCONS1}
\end{equation}
To obtain explicit formulae for scalar, vector and tensor modes it 
is necessary to have the values of the Christoffel connections of the 
background:
\begin{equation}
\overline{\Gamma}_{00}^{0} = {\cal H},\,\,\,\,\,\, \overline{\Gamma}_{i j}^{0} 
={\cal H} \delta_{i j}, \,\,\,\,\,\,\, \overline{\Gamma}_{0i}^{j} = 
{\cal H} \delta_{i}^{j}.
\label{Chback}
\end{equation}
The components of the Ricci tensor and the Ricci scalar 
of the background are then:
\begin{eqnarray}
&& \overline{R}_{00} = - 3 {\cal H}', \,\,\,\,\,\,\, \overline{R}_{0}^{0} = - 
\frac{3}{a^2} {\cal H}',
\nonumber\\
&& \overline{R}_{i j} = ( {\cal H}' + 2 {\cal H}^2) \delta_{i j},
\,\,\,\,\,\,\, \overline{R}_{i}^{j} = - \frac{1}{a^2} ( {\cal H}' + 2 {\cal H}^2) \delta_{i}^{j},
\nonumber\\
&& \overline{R} = - \frac{6}{a^2}( {\cal H}^2 + {\cal H}').
\end{eqnarray}

\subsection{The scalar modes}
The scalar fluctuations of the geometry with covariant
and controvariant indices can be written to first-order , as
\begin{eqnarray}
&& \delta_{\rm s} g_{00} = 2 a^2 \phi,\,\,\,\,\,\,\,\,\,\,\,\,\,\,\,\,\,\,\,\,\,\,\,\, 
\delta_{\rm s} g^{00} = - \frac{2}{a^2} \phi
\nonumber\\
&&  \delta_{\rm s} g_{ij} = 
2 a^2 (\psi \delta_{i j} - \partial_{i} \partial_{j} E),\,\,\,\,\,
\delta_{\rm s} g^{ij} = 
-\frac{2}{ a^2} (\psi \delta^{i j} - \partial^{i} \partial^{j} E),
\nonumber\\
&&  \delta_{\rm s} g_{0i} = - a^2  \partial_{i} B,\,\,\,\,\,\,\,\,\,\,\,\,\,\,\,\,\,\,\,\,\,\,\,\,
\delta_{\rm s} g^{0i} = - \frac{1}{a^2} \partial^{i} B.
\label{SF}
\end{eqnarray}
Inserting Eqs. (\ref{SF}) into Eq. (\ref{CHRF}) 
the various first-order components of the (scalar) Christoffel 
fluctuations can be easily computed:
\begin{eqnarray}
&& \delta_{\rm s} \Gamma^{0}_{00} = \phi' ,
\nonumber\\
&& \delta_{\rm s} \Gamma_{i j}^{0} = - 
[ \psi' + 2 {\cal H}( \phi + \psi)]\delta_{ij} + \partial_{i } 
\partial_{j} [ (E' + 
2 {\cal H} E) - B],
\nonumber\\
&& \delta_{\rm s} \Gamma_{i0}^{0} = \delta_{\rm s} \Gamma_{0i}^{0}= 
\partial_{i}( \phi + {\cal H} B),
\nonumber\\
&&\delta_{\rm s} \Gamma^{i}_{00} = \partial^{i}( \phi + B' + {\cal H} B),
\nonumber\\
&& \delta_{\rm s} \Gamma_{ij}^{k} = ( \partial^{k} \psi \delta_{ij} - 
\partial_{i} \psi \delta^{k}_{j} - \partial_{j}\psi \delta^{k}_{i})
+ \partial_{i}\partial_{j}\partial^{k} E - {\cal H}\partial^{k} B \delta_{i j},
\nonumber\\
&& \delta_{\rm s} \Gamma_{0 i}^{j} = - \psi' \delta_{i}^{j} + \partial_{i} 
\partial^{j}E'.
\label{SCHR}
\end{eqnarray}
The first-order fluctuations of the Ricci tensors are 
\begin{eqnarray}
 \delta_{\rm s} R_{00} &=& \nabla^2[ \phi + (B- E')' + {\cal H}( B - E')]
+ 3 [ \psi'' + {\cal H}( \phi' + \psi') ],
\nonumber\\
 \delta_{\rm s} R_{0 i} &=& \partial_{i} [ ( {\cal H}' + 2 {\cal H}^2) B + 
2 ( \psi' + {\cal H} \phi)],
\nonumber\\
 \delta_{\rm s} R_{i j} &=& - \delta_{i j}\{ [\psi'' + 2 ( {\cal H}' + 2 {\cal H}^2) 
(\psi + \phi)  + {\cal H} ( \phi' + 5 \psi') - \nabla^2 \psi ]+ {\cal H} 
\nabla^2 ( B - E') \},
\nonumber\\
&+&\partial_{i} \partial_{j}[ (E' - B)' + 2 ( {\cal H}' + 2 {\cal H}^2 )E + 
2 {\cal H} (E' -B) + ( \psi - \phi) ]
\label{SRICCI}
\end{eqnarray}
while  the first-order fluctuation of the Ricci scalar are determined to be, 
\begin{equation}
\delta_{\rm s}
 R = \frac{2}{a^2}\{ 3 \psi'' + 6 ( {\cal H}' + {\cal H}^2) \phi 
+ 3 {\cal H} ( \phi' + 3 \psi') +
\nabla^2[ ( \phi - 2 \psi) + ( B - E')' + 3 {\cal H} ( B - E') ]\}.
\end{equation}
Using Eqs. (\ref{SRICCI}) into Eqs. (\ref{MIXIN}) and (\ref{RS}), 
the fluctuations of the components of the Ricci tensor with 
mixed indices will then be:
\begin{eqnarray}
\delta_{\rm s} R_{0}^{0} &=& \frac{1}{a^2} \{ \nabla^2 [ \phi + ( B - E')' + 
{\cal H}( B - E') + 
3 [\psi'' + {\cal H} ( \phi' + \psi') + 2 {\cal H}' \phi]\},
\nonumber\\
\delta_{\rm s} 
 R_{i}^{j} &=& \frac{1}{a^2} [ \psi'' + 2 ( {\cal H}' + 2 {\cal H}^2)
 \phi + {\cal H} ( \phi' + 5 \psi') - \nabla^2 \psi + 
{\cal H} \nabla^2 ( B - E') ] \delta_{i}^{j} 
\nonumber\\
&-& \frac{1}{a^2}\partial_{i}\partial^{j} [ ( E' - B)' + 2 {\cal H} ( E' - B) 
+ ( \psi -\phi)],
\nonumber\\
\delta_{\rm s} R_{i}^{0} &=& \frac{2}{a^2} \partial_{i}[ \psi' + {\cal H} \phi],
\nonumber\\
\delta_{\rm s} R_{0}^{i} &=& \frac{2}{a^2} \partial^{i}
[ - ( \psi' + {\cal H} \phi) + ({\cal H}' - {\cal H}^2)B].
\label{mixedR}
\end{eqnarray}

Finally the fluctuations of the components of the Einstein tensor
with mixed indices are computed to be 
\begin{eqnarray}
\delta_{\rm s} {\cal G}_{0}^{0} &=& \frac{2}{a^2} \{ \nabla^2 \psi - 
{\cal H} \nabla^2( B - E') - 3 {\cal H} ( \psi' + {\cal H} \phi)\},
\label{dg00}\\
\delta_{\rm s} {\cal G}_{i}^{j} &=& \frac{1}{a^2} \{ [- 2 \psi'' - 
2 ({\cal H}^2 + 2 {\cal H}') \phi - 2 {\cal H} \phi' - 4 {\cal H}\psi']
\nonumber\\
&-& 
\nabla^2 [ ( \phi - \psi) + ( B - E')' + 2 {\cal H} ( B - E')]\} \delta_{i}^{j}
\nonumber\\
&-& \frac{1}{a^2}
\partial_{i}\partial^{j} [ ( E' - B)' + 2 {\cal H} ( E' - B) + 
(\psi - \phi) ],
\label{dgij}\\
\delta_{\rm s} {\cal G}_{i}^{0} &=& \delta R_{i}^{0}.
\label{dg0i}
\end{eqnarray} 

\subsection{The vector modes}
The vector fluctuations of the geometry can be written as 
\begin{eqnarray}
&& \delta_{\rm v} g_{0 i} = - a^2 Q_{i},\,\,\,\,\,\,\,
\delta_{\rm v} g^{0 i} = - \frac{ Q^{i}}{a^2},
\nonumber\\
&& \delta_{\rm v} g_{i j} = 
a^2 ( \partial_{i} W_{j} + \partial_{j} W_{i}),\,\,\,\,\,
 \delta_{\rm v} g^{i j} = -
\frac{1}{a^2}( \partial^{i} W^{j} + \partial^{j} W^{i}),
\label{VF}
\end{eqnarray}
and the related fluctuations of the Chrristoffel connections 
can be deduced, after some algebra, by inserting Eqs. (\ref{VF}) 
into Eq. (\ref{CHRF}). The result is:
\begin{eqnarray}
&&\delta_{\rm v} \Gamma_{i0}^{0} = {\cal H} Q_{i},
\nonumber\\
&& \delta_{\rm v} \Gamma_{ij}^{0} = -\frac{1}{2} ( \partial_{i} Q_{j} + 
\partial_{j} Q_{i}) - {\cal H} (\partial_{i} W_{j} + \partial_{j} W_{i}) 
- \frac{1}{2}(\partial_{i} W_{j}' + \partial_{j} W_{i}') ,
\nonumber\\
&& \delta_{\rm v} \Gamma_{00}^{i} = { Q^{i}}' + {\cal H} Q^{i},
\nonumber\\
&& \delta_{\rm v} \Gamma_{i j}^{k} = - {\cal H} Q^{k} \delta_{i j} 
+ \frac{1}{2} \partial^{k}(\partial_{i} W_{j} + \partial_{j} W_{i}) - 
\frac{1}{2}\partial_{j}( \partial^{k} W_{i} + \partial_{i} W^{k}) 
- \frac{1}{2} \partial_{i}(\partial_{j}W^{k} + \partial^{k}W_{j}),
\nonumber\\
&& \delta_{\rm v} \Gamma_{i0}^{j} = \frac{1}{2}( \partial_{i} Q^{j} 
- \partial^{j} Q_{i}) - \frac{1}{2}(\partial^{j} W_{i}' 
+ \partial_{i}{W^{j}}'),
\label{VCHR}
\end{eqnarray}
where, the conditions of Eq. (\ref{div1}) on $Q_{i}$ and $W_{i}$
 have been extensively used.

The fluctuations of the Ricci tensors are then 
\begin{eqnarray}
 \delta_{\rm v} R_{0 i} &=& [ {\cal H}' + 2 {\cal H}^2 ] Q_{i} 
 - \frac{1}{2} \nabla^2 Q_{i}  - \frac{1}{2} \nabla^2 W_{i}'
\nonumber\\
\delta_{\rm v} R_{i j} &=& 
- \frac{1}{2} \biggl\{ ( \partial_{i} Q_{j} + \partial_{j} Q_{i})' + 2 {\cal H}   ( \partial_{i} Q_{j} + \partial_{j} Q_{i})\biggr\}
- \frac{1}{2}( \partial_{i}W_{j}'' + \partial_{j}W_{i}'') 
\nonumber\\
&-& {\cal H}  (\partial_{i}W_{j}' + \partial_{j}W_{i}')
- \frac{1}{2} (\partial_{i}W_{j} + \partial_{j}W_{i}) 
( 2 {\cal H}' + 4 {\cal H}^2 ) ,
\label{VRICCI}
\end{eqnarray}

\subsection{The Tensor modes}

Consider now the case of the tensor modes of the geometry, i.e., according to 
Eq. (\ref{div1}), the two polarization of the graviton:
\begin{equation}
\delta_{\rm t} g_{i j} = - a^2 h_{ij},\,\,\,\,\,\,\,\,\,\,\,
\delta_{\rm t}  g^{i j} = \frac{h^{ij}}{a^2}.
\end{equation}
From Eq. (\ref{CHRF}) the tensor contribution to the 
fluctuation of the connections can be expressed as 
\begin{eqnarray}
&& \delta_{\rm t} \Gamma_{ij}^{0} =  \frac{1}{2} ( h_{ij}' + 2 {\cal H}
h_{i j}),
\nonumber\\
&&  \delta_{\rm t} \Gamma_{i0}^{j} = \frac{1}{2} {h_{i}^{j}}',
\nonumber\\
&& \delta_{\rm t} \Gamma_{ij}^{k} = \frac{1}{2} [ \partial_{j} h_{i}^{k} + 
\partial_{i} h_{j}^{k} - \partial^{k} h_{ij}].
\label{TCHR}
\end{eqnarray}
Inserting these results into the perturbed expressions 
of the Ricci tensors it is easy to obtain:
\begin{eqnarray}
&& \delta_{\rm t} R_{i j} = \frac{1}{2}[ h_{i j}'' + 2 {\cal H} 
h_{ij} +  2 ( {\cal H}' + 2 {\cal H}^2) h_{i j} - \nabla^2 h_{ij} ],
\label{driccit1}\\
&& \delta_{\rm t} R_{i}^{j} = - \frac{1}{2 a^2}[ {h_{i}^{i}}'' + 2 {\cal H} 
{h_{i}^{j}}' - \nabla^2 h_{i}^{j}].
\label{driccit2}
\end{eqnarray}

\subsection{Fluctuations of the energy momentum tensor}

The energy-momentum tensors usually discussed in the 
theory of cosmological perturbations range from 
the fluid case to the case of scalar fields and 
possibly vector fields (i.e. electromagnetic energy-momentum
tensor). The background energy-momentum tensor of a perfect 
fluid is 
\begin{equation}
T_{\mu \nu} = ( p + \rho) u_{\mu} u_{\nu} - p g_{\mu \nu}.
\end{equation}
From the normalization condition $g_{\mu\nu} u^{\mu} u^{\nu} = 1$ 
it can be concluded $ u_0 = a$ and $ \delta u^{0} = - \phi/a$.
Hence,
\begin{equation}
\delta_{\rm s} T_{0}^{0} = \delta\rho, \,\,\,\,\,\,\,\,\,\,\,\,\,
\delta_{\rm s} T^{00} = \frac{1}{a^2} ( \delta \rho - 2 \rho \phi),
\label{dT00}
\end{equation}
and 
\begin{eqnarray}
&&\delta_{\rm s} T_{i}^{j} = - \delta p \delta_{i}^{j},
\label{dTij1}\\
&&\delta_{\rm s} T^{i j} = \frac{1}{a^2} [ \delta p \delta^{i j} + 2 p ( \psi \delta^{i j} - \partial^{i} \partial^{j} E)],
\label{dTij2}\\
&& \delta_{\rm s} T_{0}^{i} = ( p + \rho) v^{i},
\label{dT0i}\\
&&  \delta_{\rm s} T^{0i} 
= \frac{1}{a^2} [ ( p + \rho) v^{i} + \partial^{i} B].
\end{eqnarray}
where we have chosen to define $ \delta u^{i} = v^{i}/a$. Notice that the velocity 
field can be split into  divergenceless and  divergencefull parts, i.e. 
\begin{equation}
v^{i} = \partial^{i} v + {\cal V}^{i},\,\,\,\,\,\,\,\,\partial_{i} {\cal V}^{i} =0
\label{velfielddef}
\end{equation}
Notice that in the case of neutrinos (and possibly also in the 
case of other collisionless species) the anisotroopic 
stress has to be considered both in the Einstein 
equations and in the covariant conservation equations. 
The anisotropic stress is the introduced as 
\begin{equation}
\delta_{s} T_{i}^{j} = - \delta p \delta_{i}^{j} + \Pi_{i}^{j},
\end{equation}
where $\Pi_{i}^{j} = T_{i}^{j} - \delta_{i}^{j} T_{k}^{k}/3$. 
In the case of scalar fluctuations it is 
practical to adopt the following notation
\begin{equation}
\partial_{j} \partial^{i} \Pi_{i}^{j} =  -( p + \rho)\nabla^2 \sigma,
\end{equation}
which is equaivalent, in Fourier space, to the following 
identity 
\begin{equation}
( p + \rho) \sigma = \biggl( \hat{k}_{j} \hat{k}^{i} - 
\frac{1}{3} \delta_{j}^{i}\biggr) \Pi_{i}^{j}.
\end{equation}
There could be a  potential confusion in the definition of the 
perturbed velocity field. It is plausible to define the peculiar velocity 
in two slightly different ways, namely
\begin{eqnarray}
&& \delta_{\rm s} T_{0}^{i} = ( p + \rho) u_{0} \delta u^{i} \equiv 
( p + \rho) v^{i} \equiv ( p + \rho) \partial^{i} v
\label{def1}\\
&& \delta_{\rm s} T_{i}^{0} = ( p + \rho) u^{0} \delta u_{i} = 
(p + \rho)  \overline{v}_{i} 
\equiv (p + \rho) \partial_{i} \overline{v},
\label{def2}
\end{eqnarray}
where, from the normalization condition $g_{\mu\nu} u^{\mu} u^{\nu} =1$, 
$u_{0}= a$ and $u^{0} =1/a$.
The velocity fields $\overline{v}$ and $v$ defined in Eqs. (\ref{def1}) 
and (\ref{def2}) are not equivalent. In fact $\delta u_{i} 
= a \overline{v}_{i}$ and $\delta u^{i} = v^{i}/a$. Recall, now, that 
\begin{equation}
\delta u^{i} = \delta( g^{i\alpha} u_{\alpha}) \equiv 
\delta g^{i 0} u_{0} + g^{i k} \delta u_{k}.
\end{equation}
Inserting now the explicit definitions of $ \delta u^{i}$ and $\delta u_{k}$ 
in terms of $v^{i}$ and $\overline{v}_{k}$ and recalling that 
$ \delta g^{i0} = - \partial^{i} B/a^2$, we have 
that 
\begin{equation}
v_{i} = - \overline{v}_{i} - \partial_{i} B.
\end{equation}
This difference in the definition of the velocity field reflects 
in the gauge transformations which are different for $v_{i}$ and 
$ \overline{v}_{i}$, namely
\begin{eqnarray}
&& v_{i} \to \tilde{v}_{i} = v_{i} + \partial_{i} \epsilon',
\nonumber\\
&& \overline{v}_{i} \to \tilde{\overline{v}}_{i} = \overline{v}_{i} 
- \partial_{i}\epsilon_{0}.
\end{eqnarray}
In this paper we always define the velocity field as in Eq. (\ref{def1}) 
however, this remark should be borne in mind since 
different authors use different definitions which may only be 
equivalent, up to a sign, in specific gauges (like the class of gauges where $B=0$).

Let us now consider the fluctuations of the energy-momentum tensor of 
a scalar field $ \varphi$ characterized by a potential $W(\varphi)$:
\begin{equation}
T_{\mu\nu} = \partial_{\mu} \varphi \partial_{\nu} \varphi - g_{\mu\nu} 
\biggl[\frac{1}{2} g^{\alpha\beta} \partial_{\alpha} \varphi \partial_{\beta} 
\varphi 
- W(\varphi)\biggr].
\end{equation}
Denoting with $ \chi$ the first-order fluctuation of the scalar field 
$\varphi$ we will have
\begin{eqnarray}
&& \delta_{\rm s} T_{\mu\nu} = 
\partial_{\mu}\chi \partial_{\nu} \varphi + 
\partial_{\mu} \varphi \partial_{\nu} \chi
\nonumber\\
&& - 
\delta_{\rm s} 
g_{\mu\nu} \biggl[ \frac{1}{2} g^{\alpha\beta} \partial_{\alpha} 
\varphi \partial_{\beta} \varphi 
- W\biggr] - g_{\mu\nu} \biggl[ \frac{1}{2} \delta_{\rm s} g^{\alpha\beta} 
\partial_{\alpha} \varphi \partial_{\beta} \varphi
+ g^{\alpha\beta} \partial_{\alpha} \chi \partial_{\beta}
 \varphi - \frac{\partial W}{\partial\varphi} \chi\biggr], 
\end{eqnarray}
and , in explicit terms, 
\begin{eqnarray}
&& \delta_{\rm s} T_{00} = \chi' \varphi' + 2 a^2 \phi W + a^2 
\frac{\partial W}{\partial \varphi} \chi,
\nonumber\\
&& \delta_{\rm s} T_{0i} = \varphi' \partial_{i} \chi + 
a^2 \partial_{i}B \biggl[ \frac{{\varphi'}^2}{2 a^2 } - W \biggr],
\nonumber\\
&& \delta_{\rm s} T_{i j} = 
\delta_{ij} \biggl[ \varphi' \chi' - 
\frac{\partial W}{\partial\varphi} \chi a^2 - ( \phi + \psi) {\varphi'}^2 + 
2 a^2 W \psi\biggr] 
\nonumber\\
&& + 
2 a^2  \biggl[ \frac{{\varphi'}^2}{2 a^2} - W\biggr]
\partial_{i} \partial_{j} E.
\end{eqnarray}
Recalling that 
\begin{equation}
\overline{T}_{00} = \frac{{\varphi'}^2}{2 } +a^2 W, \,\,\,\,\,\,\,\,\,\,\,\,
\overline{T}_{ij} = \biggl[\frac{{\varphi'}^2}{2 } -a^2 W \biggr] 
\delta_{ij},
\end{equation}
the perturbed components of the scalar field energy-momentum tensor 
with mixed (one covariant the other controvariant) indices 
can be written as 
\begin{equation}
\delta_{\rm s} T_{\mu}^{\nu} = \delta_{\rm s} T_{\alpha\mu}
 \overline{g}^{\alpha\nu} + \overline{T}_{\alpha\mu} \delta_{\rm s} 
g^{\alpha\nu}, 
\end{equation}
i.e., in explicit terms, 
\begin{eqnarray}
&& \delta_{\rm s} T_{0}^{0}= \frac{1}{a^2} \biggl( - \phi {\varphi'}^2 
+ \frac{\partial W}{\partial \varphi} a^2 \chi+ \chi' \varphi'\biggr),
\label{enmomsc1}\\
&& \delta_{\rm s} T_{i}^{j}= \frac{1}{a^2}\biggl( \phi {\varphi'}^2 
+ \frac{\partial W}{\partial \varphi} a^2\chi - \chi' \varphi'\biggr) 
\delta_{i}^{j}, 
\label{enmomsc2}\\
&& \delta_{\rm s} T_{0}^{i} = - \frac{1}{a^2} \varphi' \partial^{i} \chi 
- \frac{{\varphi'}^2}{a^2} \partial^{i} B . 
\label{enmomsc3}
\end{eqnarray}

The covariant conservation of the energy-momentum tensor implies, in the 
scalar field case, the validity of the Klein-Gordon equation
which can be written as 
\begin{equation}
g^{\alpha\beta} \nabla_{\alpha} \nabla_{\beta} \varphi + 
\frac{\partial W}{\partial \varphi} a^2 =0. 
\label{KG1}
\end{equation}
From Eq. (\ref{KG1}) the perturbed Klein-Gordon equation can be 
written as 
\begin{equation}
\delta_{\rm s} g^{\alpha\beta} [ \partial_{\alpha} \partial_{\beta} \varphi 
- \overline{\Gamma}_{\alpha\beta}^{\sigma} \partial_{\sigma} \varphi] 
+ \overline{g}^{\alpha\beta}[ \partial_{\alpha}\partial_{\beta} \chi - 
\delta_{\rm s} \Gamma^{\sigma}_{\alpha\beta} \partial_{\sigma} \varphi - 
\overline{\Gamma}_{\alpha\beta}^{\sigma} \partial_{\sigma} \chi] 
+ \frac{\partial^2 W}{\partial\varphi^2} =0,
\end{equation}
which becomes, in explicit form, 
\begin{equation}
\chi'' + 2 {\cal H} \chi' - \nabla^2 \chi +
\frac{\partial^2 W}{\partial\varphi^2}a^2 \chi + 
2 \phi \frac{\partial W}{\partial\varphi}a^2 - 
\varphi'( \phi' + 3 \psi') + \varphi' \nabla^2( E' - B) =0.
\label{KG2}
\end{equation}
 
 \subsection{Generalized scalar perturbation equations without gauge fixing}
 
 Different gauges are suitable for different problems. Hence, it is practical 
 to collect here the generalized evolution equations of the scalar fluctuations obtained without 
 fixing a particular coordinate system. The sources will be assumed to be 
 barotropic fluids but this is not a severe limitation since, with the appropriate identifications, this 
 set of equations can even describe more general situations.
 
 From Eqs. (\ref{dg00}),(\ref{dT00}) and from Eqs. (\ref{dg0i}), (\ref{dT0i})
the general expressions for the Hamiltonian and momentum constraints reads
\begin{eqnarray}
&& \nabla^2 \psi - {\cal H} \nabla^2 ( B- E') - 3 {\cal H} ( \psi' + {\cal H}
\phi) = 4 \pi G a^2 \delta\rho,
\label{00gen}\\
&& \partial^{i}[ ( \psi' + {\cal H} \phi ) + ( {\cal H}^2 - {\cal H}')B ]
= - 4\pi G a^2 ( p + \rho) v^{i},
\label{0igen}
\end{eqnarray}

From Eqs. (\ref{dgij}) and (\ref{dTij2}) the generalized 
expression of the $(ij)$ component of the perturbed 
equations is:
\begin{eqnarray}
&& [ \psi'' + ( {\cal H}^2 + 2  {\cal H}') \phi + {\cal H}
(\phi' + 2 \psi') ] \delta_{i}^{j} + 
\frac{1}{2} \delta_{i}^{j} \nabla^2[ (\phi - \psi) + 
( B - E' )' + 2 {\cal H} ( B- E') ]
\nonumber\\
&& - \frac{1}{2} \partial_{i}\partial^{j} [ (\phi - \psi) + 
( B - E' )' + 2 {\cal H} ( B- E') ] = 4 \pi G a^2 [ \delta p \delta_{i}^{j} 
- \Pi_{i}^{j}].
\label{ijgen}
\end{eqnarray}
Separating, in Eq. (\ref{ijgen}) the tracefull from the traceless part,
we obtain, respectively, 
\begin{equation}
\psi'' + {\cal H} ( \phi' + 2 \psi') + 
({\cal H}^2 + 2 {\cal H}') \phi + 
\frac{1}{3} \nabla^2[ ( \phi - \psi) + ( B- E')' + 2 {\cal H}( B - E')] 
= 4 \pi G a^2 \delta p.
\label{int1}
\end{equation}
and 
\begin{equation}
\nabla^2 [ ( \phi - \psi) + ( B- E')' + 2 {\cal H} ( B - E') ] = 12 
\pi G a^2 ( p + \rho) \sigma.
\label{int2}
\end{equation}
In general  terms the  the $(0)$ and $(i)$ components of  
 Eq. (\ref{GENCOVCONS1}) can be written, for scalar 
 fluctuations as 
 \begin{eqnarray}
&&\partial_{0} \delta_{\rm s} T^{00} + \partial_{j} \delta_{\rm s} T^{j0} + 
( 2 \delta_{\rm s} \Gamma_{00}^{0} + \delta_{\rm s} \Gamma_{k_0}^{k} ) 
\overline{T}^{00} 
\nonumber\\
&& + 
(2 \overline{\Gamma}_{00}^{0}+ \overline{\Gamma}_{k 0}^{k}) \delta_{\rm s} 
 T^{00}
+ \overline{\Gamma}_{ij}^{0} \delta_{\rm s} T^{ij} + 
\delta_{\rm s} \Gamma_{ij}^{0} \overline{T}^{ij} =0,
\label{T0a}\\
&& \partial_{0} \delta_{\rm s} T^{0 j} + \partial_{k} \delta_{\rm s} T^{ k j}
+ ( \delta_{\rm s} \Gamma_{0 k}^{0} + \delta_{\rm s} \Gamma_{m k}^{m}) 
\overline{T}^{ k j} 
\nonumber\\
&& + ( \overline{\Gamma}_{00}^{0} + \overline{\Gamma}_{k 0}^{k}) 
\delta_{\rm s} T^{0 j} 
+ \delta_{\rm s} \Gamma_{00}^{j} \overline{T}^{00}  + \delta_{\rm s} 
\Gamma_{k m}^{j} \overline{T}^{k m} + 
2 \overline{\Gamma}_{0 k}^{j} \delta_{\rm s}  T^{0 k} =0.
\label{T0b}
\end{eqnarray}
Inserting now the specific form of the perturbed connections of Eqs. (\ref{SCHR}) 
into Eqs. (\ref{T0a}) and (\ref{T0b}) the following 
result can be, respectively, obtained:
\begin{equation}
 \delta \rho' - 3 \psi' ( p + \rho) + ( p + \rho) \theta + 
3 {\cal H} ( \delta \rho + \delta p) + 
( p + \rho) \nabla^2 E' =0,
\label{gencov1}
\end{equation}
for the  $(0)$ component, and  
\begin{eqnarray}
&& ( p + \rho ) \theta' + \theta [ ( p' + \rho') + 4 {\cal H}( p + \rho)]
+ (p + \rho) \nabla^2 B' ,
\nonumber\\
&&
+ [ p' + {\cal H} ( p + \rho) ] \nabla^2 B + 
\nabla^2 \delta p + ( p + \rho) \nabla^2 \phi =0,
\label{gencov2}
\end{eqnarray}
for the $(i)$ component.
In the above equations, as explained in the text, the 
divergence of the velocity field, i.e. $ \theta = \partial_{i}v^{i} = \partial_{i}\partial^{i} v$, has been 
directly introduced. In the text, several gauge-dependent discussions are 
present.  The longitudinal gauge equations are obtained, for instance, by setting everywhere 
in the above equations $E=B=0$. The off-diagonal gauge equations are obtained 
from the above equations by setting everywhere $E =\psi =0$. The sychronous gauge equations 
can be derived by setting everywhere in the above equations $ \phi= B=0$, and so on 
for the gauge that is mostly suitable in a given calculation.

\end{appendix}


\newpage

\begin{thebibliography}{999}

\bibitem{feynman} R. Feynman, 
{\it Feynman Lectures on Gravitation},  (Addison-Wesley, US, 1995),  p. 186.

\bibitem{weinberg}  S. Weinberg,  {\it Gravitation and Cosmology},  (John Wiley \& Sons, New York, 
US, 1972).

\bibitem{KT} E. W. Kolb and M. S. Turner,  {\it The Early Universe}, ( Addison-Wesley, US, 1990).

\bibitem{patridge} R. B. Patridge, {\it 3 K: the Cosmic Microwave 
Background Radiation} (Cambridge University Press, Cambridge, UK, 1995).

\bibitem{zel1} Ya. B. Zeldovich, Sov. Phys. Usp. {\bf 6}, 475 (1964) 
[ Usp. Fiz. Nauk. {\bf 80}, 357 (1963)].

\bibitem{LK} E. M. Lifshitz and I. M. Khalatnikov, Sov. Phys. Usp. {\bf 6}, 
495 (1964) [Usp. Fiz. Nauk. {\bf 80}, 391 (1964)].

\bibitem{friedmann} A. Friedmann, Z. Phys. {\bf 10}, 377 (1922); {\bf 
21}, 326 (1924).

\bibitem{lemaitre1} G. Lema\^itre,  Ann.\ Soc.\  Sci.\  Bruxelles {\bf 47A}, 49 
(1927); Rev. Questions Sci. {\bf 129}, 129 (1958).

\bibitem{lemaitre2} G. Lema\^itre, Nature {\bf 128}, 700 (1931).

\bibitem{eddington} A. S. Eddington, Month. Not. R. Astron. Soc. {\bf 90}, 672 (1930).

\bibitem{KS}H.~Kodama and M.~Sasaki, Prog.\ Theor.\ Phys.\ Suppl.\  {\bf 78}, 1 (1984).

\bibitem{MFB} V. F. Mukhanov, H. A. Feldman, and R. H.  Brandenberger, Phys. Rept.{\bf 215}, 203 (1992).

\bibitem{sachs} R. K. Sachs and A. M. Wolfe, Astrophys.\ J.\  {\bf 147}, 73  (1967).

\bibitem{sakharov} A. D. Sakharov, Sov. Phys. JETP {\bf 22}, 241 (1965)  [Zh. Eksp. Teor. 
Fiz.{\bf 49}, 345 (1965)].

\bibitem{silk1} J. Silk, Nature {\bf 215}, 1155 (1967); Astrophys. J. {\bf 151} 459 (1968).

\bibitem{PeeblesYu} P. J. E. Peebles and J. T. Yu, Astrophys. J.  {\bf 162}, 815 (1970).

\bibitem{doro} A. G. Doroshkevich, Ya. B. Zeldovich, and R. A. Sunyaev, Sov. Astron. {\bf 22}, 523 (1978).

\bibitem{wilson} M. L. Wilson and J. Silk, Astrophys. J. {\bf 243},  14 (1981).

\bibitem{cobe} C. L. Bennet et al., Astrophys. J. Lett. {\bf 464}, L1 (1996).

\bibitem{cob2} M. Tegmark, Astrophys. J. Lett. {\bf 464}, L35 (1996).

\bibitem{boom1}  de Bernardis {\it et al}. Astrophys. J. {\bf 564}, 559 (2002).

\bibitem{boom2} C. B. Netterfield et al., Astrophys. J. {\bf 571}, 604 (2002).

\bibitem{dasi} N. W. Halverson et al., Astrophys. J. {\bf 568}, 38 (2002).

\bibitem{maxima} A. T. Lee et al., Astrophys. J. {\bf 561}, L1 (2001).

\bibitem{archeops}  A. Beno\^it {\it et al.},  Astron. Astrophys. {\bf 399}, L19 (2003).

\bibitem{silk} E. Gawiser and  J. Silk, Phys. Rept. {\bf 333}, 245 (2000).

\bibitem{WMAP02}  L. Page {\it et al}, Astrophys. J. Suppl.{\bf 148}, 223 (2003).

\bibitem{DICK} C. Dickinson {\it et al}  arXiv:astro-ph/0402498, (2004).

\bibitem{mason} B.S. Mason {\it et al} Astrophys.~J. {\bf 591}, 540 (2003).

\bibitem{pearson} T.J. Pearson {\it et al} Astrophys.~J. {\bf 591}, 556 (2003).

\bibitem{kuo} C.L. Kuo {\it et al}  Astrophys.~J.  {\bf 600}, 32 (2004).

\bibitem{WMAP01} C. L. Bennett {\it et al.}, Astrophys. J. Suppl., {\bf 148}, 1.

\bibitem{LSS01}  M. Tegmark {\it et al.}, Astrophys. J. {\bf 606}, 702  (2004).

\bibitem{CL01} N. Bachall {\it et al.}, Astrophys. J. {\bf 585}, 182 (2003).

\bibitem{BBN01} B. Fields and S. Sarkar, Phys. Lett. B {\bf 592}, 202 (2004).

\bibitem{SN01} J. Tonry {\it et al.}, Astrophys. J. {\bf 594}, 1 (2003). 

\bibitem{white} M.~J.~White, D.~Scott and J.~Silk, Ann.\ Rev.\ Astron.\ Astrophys.\  {\bf 32}, 319 (1994).

\bibitem{dodelson}  W.~Hu and S.~Dodelson, Ann.\ Rev.\ Astron.\ Astrophys.\  {\bf 40}, 171 (2002).

\bibitem{bond} J. R. Bond, {\it Cosmology and Large Scale Structure}  (Les Houches, Session LX, 1993 ), eds. R. Schaeffer, J. Silk, M. Spiro and J. Zinn-Justin, p. 469.

\bibitem{carroll} S.~M.~Carroll, W.~H.~Press and E.~L.~Turner,
Ann.\ Rev.\ Astron.\ Astrophys.\  {\bf 30}, 499 (1992).

\bibitem{pad} T. Pdmnabhan, Phys. Rep. {\bf 380},  235 (2003).

\bibitem{sahni} V.~Sahni, {\it Dark matter and dark energy},  arXiv:astro-ph/0403324.

\bibitem{bardeen}  J. M. Bardeen, Phys. Rev. D {\bf 22}, 1882 (1980).

\bibitem{PV1} W. Press and E. Vishniac, Astrophys. J. {\bf 239}, 1 (1980).

\bibitem{PV2} W. Press and E. Vishniac, Astrophys. J. {\bf 236}, 323 (1980).

\bibitem{MB} C.-P. Ma and E. Bertschinger, Astrophys. J. {\bf 455}, 7 (1995).

\bibitem{bardeen2} J. Bardeen, P. Steinhardt, M. Turner, Phys. Rev. D {\bf 28}, 679 (1983).

\bibitem{GRV}  L. P. Grishchuk, Phys. Rev. D {\bf 48}, 5581 (1993).

\bibitem{br1}  T.~J.~Battefeld and R.~Brandenberger, arXiv:hep-th/0406180.

\bibitem{maxvec1}M.~Giovannini, Phys.\ Rev.\ D {\bf 70}, 103509 (2004).

\bibitem{maxvec2} M.~Giovannini, arXiv:hep-th/0410094.

\bibitem{harr1} E. Harrison, Rev. Mod. Phys. {\bf 39}, 862 (1967).

\bibitem{long2} R.~Brustein, M.~Gasperini, M.~Giovannini, V.~F.~Mukhanov and G.~Veneziano,
Phys.\ Rev.\ D {\bf 51}, 6744 (1995).

\bibitem{hwang1} J. Hwang, Phys. Rev. D {\bf 53}, 762 (1995).

\bibitem{hwang2} J. Hwang, Class. Quantum Grav. {\bf 11},  2505  (1994).
 
\bibitem{hwang3} J. Hwang, Phys. Rev. D  {\bf 48},  3544  (1993) .

\bibitem{hwang4} J. Hwang, Astrophys. J.  {\bf 375}, 443 (1990).

\bibitem{whitehu1}  M. White and W. Hu, Astron. Astrophys. {\bf 321}, 8 (1997).

\bibitem{jackson}  J. D.  Jackson, {\it Classical Electrodynamics}, (Wiley, New York, US, 1975).

\bibitem{Peebles1} P. J.  E.  Peebles,  Nature {\bf 327}, 210 (1987).

\bibitem{Peebles2}  P. J.  E.  Peebles,  Astrophys. J. {\bf 315}, L73 (1987).

\bibitem{Peebles3} R.~y.~Cen, J.~P.~Ostriker and P.~J.~E.~Peebles, Astrophys.\ J.\  {\bf 415}, 423 (1993).

\bibitem{Peebles4} W.~Hu and P.~J.~E.~Peebles, Astrophys.\ J.\  {\bf 528}, L61 (2000).

\bibitem{turokiso}  M.~Bucher, K.~Moodley and N.~Turok, Phys.\ Rev.\ D {\bf 62}, 083508 (2000).

\bibitem{challinoriso} A. Challinor and A. Lasenby, Astrophys. J. {\bf 513}, 1 (1999).

\bibitem{hannu2}  K.~Enqvist, H.~Kurki-Suonio and J.~Valiviita, Phys.\ Rev.\ D {\bf 62}, 103003 (2000).

\bibitem{hannu1} K.~Enqvist and H.~Kurki-Suonio, Phys.\ Rev.\ D {\bf 61}, 043002 (2000).

\bibitem{hannu3} K.~Enqvist, H.~Kurki-Suonio and J.~Valiviita, Phys.\ Rev.\ D {\bf 65}, 043002 (2002).

\bibitem{turok1} M.~Bucher, K.~Moodley and N.~Turok, Phys.\ Rev.\ Lett.\  {\bf 87}, 191301 (2001).

\bibitem{turok2}  M.~Bucher, K.~Moodley and N.~Turok, Phys.\ Rev.\ D {\bf 66} ,  023528 (2002).

\bibitem{moodley1}  K.~Moodley, M.~Bucher, J.~Dunkley, P.~G.~Ferreira and C.~Skordis, Phys.\ Rev.\ D {\bf 70}, 103520 (2004).

\bibitem{moodley2} M.~Bucher, J.~Dunkley, P.~G.~Ferreira, K.~Moodley and C.~Skordis, Phys.\ Rev.\ Lett.\  {\bf 93}, 081301 (2004).

\bibitem{jussi}  H.~Kurki-Suonio, V.~Muhonen and J.~Valiviita, astro-ph/0412439.

\bibitem{gordonlw} C.~Gordon and A.~Lewis, Phys.\ Rev.\ D {\bf 67}, 123513 (2003).

\bibitem{gordonmal} C.~Gordon and K.~A.~Malik, Phys.\ Rev.\ D {\bf 69}, 063508 (2004).

\bibitem{ferrer} F.~Ferrer, S.~Rasanen and J.~Valiviita, JCAP {\bf 0410}, 010 (2004).

\bibitem{luca1}L.~Amendola, Phys.\ Rev.\ D {\bf 69}, 103524 (2004).

\bibitem{mortak} T.~Moroi and T.~Takahashi, Phys.\ Rev.\ Lett.\  {\bf 92}, 091301 (2004).

\bibitem{gordonhu}  C.~Gordon and W.~Hu, Phys.\ Rev.\ D {\bf 70}, 083003 (2004).

\bibitem{luca2} L.~Amendola and F.~Finelli, arXiv:astro-ph/0411273.

\bibitem{lewisvector1} A.~Lewis, Phys.\ Rev.\ D {\bf 70}, 043518 (2004).

\bibitem{lewisvector2} A.~Lewis, Phys.\ Rev.\ D {\bf 70}, 043011 (2004).

\bibitem{giomagn}  M.~Giovannini,   Phys.\ Rev.\ D {\bf 70}, 123507 (2004).

\bibitem{maxq1}  M.~Giovannini, Phys.\ Rev.\ D {\bf 58}, 083504 (1998).

\bibitem{grgw1}  L.~P.~Grishchuk, Phys.\ Rev.\ D {\bf 48}, 3513 (1993).

\bibitem{grgw2} L.~P.~Grishchuk, Class.\ Quant.\ Grav.\  {\bf 10}, 2449 (1993).

\bibitem{grgw3} L.~P.~Grishchuk and M.~Solokhin, Phys.\ Rev.\ D {\bf 43} 2566 (1991).

\bibitem{sq5} B. L. Shumaker, Phys. Rep. {\bf 135}, 317 (1986).

\bibitem{sq6} J. Grochmalicki and M. Lewenstein, Phys. Rep. {\bf 208}, 189 (1991).

\bibitem{sq7} R. Loudon, {\it The Quantum Theory of Light}, (Oxford University 
Press, 1991).

\bibitem{sq8}  L. Mandel and E. Wolf, {\it Optical Coherence and Quantum 
optics}, (Cambridge University Press, Cambridge, England, 1995).

\bibitem{BDPBB1} M.~Gasperini and M.~Giovannini, Phys.\ Rev.\ D {\bf 47}, 1519 (1993).

\bibitem{BDPBB2}   R.~Brustein, M.~Gasperini, M.~Giovannini and G.~Veneziano,  
Phys.\ Lett.\ B {\bf 361}, 45 (1995).

\bibitem{BDmimoso1} J.~D.~Barrow, J.~P.~Mimoso and M.~R.~de Garcia Maia, Phys.\ Rev.\ D {\bf 48}, 3630 (1993)
[Erratum-ibid.\ D {\bf 51}, 5967 (1995)].

\bibitem{BDhwang} J.~c.~Hwang, Class.\ Quant.\ Grav.\  {\bf 15}, 1401 (1998).

\bibitem{GWint1} M.~Gasperini and M.~Giovannini, Class.\ Quant.\ Grav.\  {\bf 9}, L137 (1992).

\bibitem{GWint2}  M.~Giovannini, Phys.\ Rev.\ D {\bf 55}, 595 (1997).

\bibitem{GWhigher}  M.~Gasperini, Phys.\ Rev.\ D {\bf 56}, 4815 (1997).

\bibitem{ekpgw} L.~A.~Boyle, P.~J.~Steinhardt and N.~Turok, Phys.\ Rev.\ D {\bf 69}, 127302 (2004).

\bibitem{johnvec1} J. Barrow, Mon. Not. R. Astron. Soc. {\bf 178}, 625 (1977).

\bibitem{johnvec2} J. Barrow, Mon. Not. R. Astron. Soc. {\bf  179}, 47 (1977).

\bibitem{easson1} T.~J.~Battefeld and D.~A.~Easson, Phys.\ Rev.\ D {\bf 70}, 103516 (2004).

\bibitem{willy1} T.~Banks and W.~Fischler, arXiv:hep-th/0111142.

\bibitem{willy2} T.~Banks and W.~Fischler, arXiv:hep-th/0310288.

\bibitem{CONVEN1}  A. Guth, Phys. Rev. D {\bf 23}, 347 (1981).

\bibitem{CONVEN2}  A. Linde,  Phys. Lett. {\bf 108B} (1982).

\bibitem{CONVEN3} A. Albrecht and P. J. Steinhardt, Phys.  Rev. Lett. {\bf 48}, 1220 (1982).

\bibitem{CONVEN4} A. Linde,  Phys. Lett. {\bf 129B}, 177 (1983).

\bibitem{CONVEN5} K. Freese, J.  Frieman, and A. Olinto, Phys. Rev. Lett. {\bf 65},  3233 (1990).

\bibitem{CONVEN6}  A.~Vilenkin, Phys.\ Rev.\ D {\bf 27}, 2848 (1983).

\bibitem{CONVEN7} A. Linde,  Phys. Rev. D {\bf 49}, 748 (1994).

\bibitem{alex1}  P.~J.~E.~Peebles and A.~Vilenkin, Phys.\ Rev.\ D {\bf 59}, 063505 (1999).

\bibitem{linde}  A. Linde, {\it Particle Physics and Inflationary Cosmology},  (Hardwood, Chur, Switzerland, 1990).

\bibitem{PBB1} G. Veneziano, Phys.\ Lett.\ B {\bf 265}, 287 (1991).

\bibitem{PBB2} M. Gasperini and G. Veneziano,  Phys.\ Rept.\  {\bf 373}, 1 (2003).

\bibitem{maxb1} M.~Gasperini, M.~Giovannini and G.~Veneziano, Phys.\ Lett.\ B {\bf 569}, 113 (2003).

\bibitem{maxb2} M.~Gasperini, M.~Giovannini and G.~Veneziano, Nucl.\ Phys.\ B {\bf 694}, 206 (2004).

\bibitem{maxb3}  M.~Giovannini, Class.\ Quant.\ Grav.\  {\bf 21}, 4209 (2004).

 \bibitem{EKP1} J.~Khoury, B.~A.~Ovrut, P.~J.~Steinhardt and N.~Turok,
Phys.\ Rev.\ D {\bf 64}, 123522 (2001).

\bibitem{EKP1a}  J.~Khoury, B.~A.~Ovrut, P.~J.~Steinhardt and N.~Turok,
Phys.\ Rev.\ D {\bf 66}, 046005 (2002).

\bibitem{EKP2} J.~Khoury, B.~A.~Ovrut, N.~Seiberg, P.~J.~Steinhardt and N.~Turok,
Phys.\ Rev.\ D {\bf 65}, 086007 (2002).

\bibitem{SYSK} S.~Rasanen, arXiv:astro-ph/0208282.

\bibitem{EKP3} P.~J.~Steinhardt and N.~Turok, Phys.\ Rev.\ D {\bf 65}, 126003 (2002).

\bibitem{press} R. Brandenberger, R. Kahn, and W. Press, Phys. Rev. D {\bf 28}, 1809 (1983).

\bibitem{lyth} D. H. Lyth, Phys. Rev. D {\bf 31}, 1792 (1985). 

\bibitem{PL1} L. F. Abbott and M. B. Wise, Nucl. Phys. B {\bf 244} , 541 (1984).

\bibitem{PL2} F. Lucchin and S. Matarrese, Phys. Rev. D {\bf 32}, 1316 (1985).

\bibitem{PL3} F. Lucchin and S. Matarrese,  Phys. Lett. B {\bf 164}, 282 (1985).

\bibitem{PL4} D. Lyth and E. Stewart, Phys. Lett. B {\bf 274}, 168 (1992).

\bibitem{luk} V.~N.~Lukash, 
Sov.\ Phys.\ JETP {\bf 52}, 807 (1980) [Zh.\ Eksp.\ Teor.\ Fiz.\  {\bf 79},  (1980)]

\bibitem{muk} V.~F.~Mukhanov,
Sov.\ Phys.\ JETP {\bf 67}, 1297 (1988) [Zh.\ Eksp.\ Teor.\ Fiz.\  {\bf 94N7}, 1 (1988)].

\bibitem{SL} E. Stewart and D. Lyth, Phys. Lett. B {\bf 302}, 171 (1992).

\bibitem{LLKCB} J. Lidesy et al., Rev.  Mod.  Phys.  {\bf 69}, 373 (1997).

\bibitem{muslim} A. G. Muslimov, Class. Quant. Grav., {\bf 7}, 231 (1990).

\bibitem{salopek1} D. Salopek and J. R. Bond, Phys. Rev. D {\bf 42}, 3936 (1990).

\bibitem{salopek2} D. Salopek and J. R. Bond, Phys. Rev. D {\bf 43}, 1005 (1991).

\bibitem{lidsey} J. E. Lidsey, Phys. Lett. B {\bf 273} , 42 (1991).

\bibitem{mgcurv3}M.~Giovannini, Phys.\ Rev.\ D {\bf 67}, 123512 (2003).

\bibitem{hwa2scal} J.~c.~Hwang, Astrophys.\ J.\  {\bf 375}, 443 (1991).

\bibitem{tar2scal}  A.~Taruya and Y.~Nambu, Phys.\ Lett.\ B {\bf 428}, 37 (1998).

\bibitem{anderegg}
S.~Anderegg and V.~Mukhanov, Phys.\ Lett.\ B {\bf 331}, 30 (1994).

\bibitem{liddleetal} D. Wands {\it et al.}, Phys. Rev. D {\bf 62}, 043527 (2000).

\bibitem{wein1} S.~Weinberg, Phys.\ Rev.\ D {\bf 67}, 123504 (2003).

\bibitem{wein2}  S.~Weinberg, Phys.\ Rev.\ D {\bf 70}, 043541 (2004).

\bibitem{wein3} S.~Weinberg, arXiv:astro-ph/0405397.

\bibitem{wein4} S.~Weinberg, Phys.\ Rev.\ D {\bf 69}, 023503 (2004).

\bibitem{chenglee}  T.-P. Cheng and L.-F. Lee, {\it Gauge theory of elementary particle physics}, 
(Clarendon Press, Oxford, UK, 1984).

\bibitem{gold2} J. Goldstone, A. Salam and S. Weinberg, Phys. Rev. {\bf 127} (1962).

\bibitem{turnerc}  M. S. Turner, Phys. Rev. D {\bf 28}, 1243 (1983).

\bibitem{staro1} L.~Kofman, A.~D.~Linde and A.~A.~Starobinsky, Phys.\ Rev.\ Lett.\  {\bf 73}, 3195 (1994).

\bibitem{kodham1} H.~Kodama and T.~Hamazaki, Prog.\ Theor.\ Phys.\  {\bf 96}, 949 (1996).

\bibitem{kodham2} T.~Hamazaki and H.~Kodama, Prog.\ Theor.\ Phys.\  {\bf 96}, 1123 (1996).

\bibitem{taruya1} Y.~Nambu and A.~Taruya, Prog.\ Theor.\ Phys.\  {\bf 97}, 83 (1997)

\bibitem{hamaz1}  T.~Hamazaki, Nucl.\ Phys.\ B {\bf 698}, 335 (2004).

\bibitem{hamaz2}  T.~Hamazaki, Phys.\ Rev.\ D {\bf 66}, 023529 (2002). 

\bibitem{robert1} F.~Finelli and R.~H.~Brandenberger, Phys.\ Rev.\ D {\bf 62}, 083502 (2000).

\bibitem{robert2}  F.~Finelli and R.~H.~Brandenberger, Phys.\ Rev.\ Lett.\  {\bf 82}, 1362 (1999).

\bibitem{staro2} L.~Kofman, A.~D.~Linde and A.~A.~Starobinsky, Phys.\ Rev.\ D {\bf 56}, 3258 (1997).

\bibitem{staro3} P.~B.~Greene, L.~Kofman, A.~D.~Linde and A.~A.~Starobinsky, Phys.\ Rev.\ D {\bf 56}, 6175 (1997)

\bibitem{taruya3}  A.~Taruya, Phys.\ Rev.\ D {\bf 59}, 103505 (1999)

\bibitem{bassett} B.~A.~Bassett, D.~I.~Kaiser and R.~Maartens, Phys.\ Lett.\ B {\bf 455}, 84 (1999).

\bibitem{tsuji1}  S.~Tsujikawa, Phys.\ Rev.\ D {\bf 61}, 083516 (2000).

\bibitem{robert3} J.~P.~Zibin, R.~H.~Brandenberger and D.~Scott, Phys.\ Rev.\ D {\bf 63}, 043511 (2001).

\bibitem{tsuji2} S.~Tsujikawa and B.~A.~Bassett, Phys.\ Rev.\ D {\bf 62}, 043510 (2000).

\bibitem{gordon6}  C.~Gordon, arXiv:astro-ph/0112523.

\bibitem{DERU1} N.~Deruelle and V.~F.~Mukhanov, Phys.\ Rev.\ D {\bf 52}, 5549 (1995).

\bibitem{HV} J.-C. Hwang and E. Vishniac,  Astrophys. J. {\bf 382}, 363 (1991).

\bibitem{high1} I. Antoniadis, J. Rizos, and K. Tamwakis, Nucl. Phys. B {\bf 415}, 497 (1994).

\bibitem{high2} C. Cartier, E. Copeland and R. Madden, JHEP {\bf 0001}, 035 (2000).

\bibitem{robertpbb1} F.~Finelli and R.~Brandenberger, Phys.\ Rev.\ D {\bf 65}, 103522 (2002).

\bibitem{robertpbb2} S.~Tsujikawa, R.~Brandenberger and F.~Finelli, Phys.\ Rev.\ D {\bf 66}, 083513 (2002).

\bibitem{robertpbb3} F.~Di Marco, F.~Finelli and R.~Brandenberger, Phys.\ Rev.\ D {\bf 67}, 063512 (2003).

\bibitem{CARTIERB}  C.~Cartier, arXiv:hep-th/0401036.

\bibitem{gratton1}
S.~Gratton, J.~Khoury, P.~J.~Steinhardt and N.~Turok,
Phys.\ Rev.\ D {\bf 69}, 103505 (2004).

\bibitem{tolley1} A.~J.~Tolley, N.~Turok and P.~J.~Steinhardt,
Phys.\ Rev.\ D {\bf 69}, 106005 (2004).

\bibitem{lythekp1} D.~H.~Lyth, Phys.\ Lett.\ B {\bf 524}, 1 (2002).

\bibitem{hwangekp1} J.~c.~Hwang, Phys.\ Rev.\ D {\bf 65}, 063514 (2002).

\bibitem{lythekp2} D.~H.~Lyth, Phys.\ Lett.\ B {\bf 526}, 173 (2002).

\bibitem{hwangekp2} J.~Hwang and H.~Noh, Phys.\ Lett.\ B {\bf 545}, 207 (2002).

\bibitem{DV1}
R.~Durrer and F.~Vernizzi, Phys.\ Rev.\ D {\bf 66}, 083503 (2002).

\bibitem{DV2}  C.~Cartier, R.~Durrer and E.~J.~Copeland,
Phys.\ Rev.\ D {\bf 67}, 103517 (2003).

\bibitem{dsbounce1}
G.~F.~R.~Ellis and M.~S.~Madsen,
Class.\ Quant.\ Grav.\  {\bf 8}, 667 (1991).

\bibitem{dsbounce2} G.~F.~R.~Ellis and R.~Maartens,
Class.\ Quant.\ Grav.\  {\bf 21}, 223 (2004).

\bibitem{gordonturok} C.~Gordon and N.~Turok, Phys.\ Rev.\ D {\bf 67}, 123508 (2003).

\bibitem{peterbounce}  J.~Martin and P.~Peter, Phys.\ Rev.\ Lett.\  {\bf 92}, 061301 (2004).

\bibitem{wandsallen} L.~E.~Allen and D.~Wands, Phys.\ Rev.\ D {\bf 70}, 063515 (2004).

\bibitem{peterneto} P.~Peter and N.~Pinto-Neto, Phys.\ Rev.\ D {\bf 66}, 063509 (2002).

\bibitem{GZ} L. P. Grishchuk and Ya. B. Zeldovich,
Astron. Zh. {\bf 55}, 209 (1978)[ Sov. Astron. {\bf 22}, 125 (1978)].

\bibitem{GZ2} L. P. Grishchuk, Phys. Rev. D {\bf 45}, 4717 ( 1992).

\bibitem{sq1} L.~P.~Grishchuk and Y.~V.~Sidorov, Phys.\ Rev.\ D {\bf 42}, 3413 (1990).

\bibitem{sq3} M.~Gasperini and M.~Giovannini, Class.\ Quant.\ Grav.\  {\bf 10}, L133 (1993).

\bibitem{sq4} M.~Gasperini, M.~Giovannini and G.~Veneziano, Phys.\ Rev.\ D {\bf 48}, 439 (1993).

\bibitem{sq9} M.~Gasperini and M.~Giovannini, Phys.\ Lett.\ B {\bf 301}, 334 (1993).

\bibitem{sq10} R. Brandenberger, V. Mukhanov and T. Prokopec, Phys. Rev. Lett. {\bf 69}, 3606 (1992).

\bibitem{sq11} M.~Kruczenski, L.~E.~Oxman and M.~Zaldarriaga, Class.\ Quant.\ Grav.\  {\bf 11}, 2317 (1994).

\bibitem{sq12}
D.~Koks, A.~Matacz and B.~L.~Hu, Phys.\ Rev.\ D {\bf 55}, 5917 (1997)
[Erratum-ibid.\ D {\bf 56}, 5281 (1997)].

\bibitem{sq13}
D.~Polarski and A.~A.~Starobinsky, Class.\ Quant.\ Grav.\  {\bf 13}, 377 (1996).

\bibitem{sq14}
I.~L.~Egusquiza {\it et al.}, Class.\ Quant.\ Grav.\  {\bf 15}, 1927 (1998).

\bibitem{sq15} S.~Hirai, Class.\ Quant.\ Grav.\  {\bf 20}, 1673 (2003).

\bibitem{sq16} E.~Keski-Vakkuri and M.~S.~Sloth,
JCAP {\bf 0308}, 001 (2003).

\bibitem{sq17} D.~Boyanovsky, H.~J.~de Vega and R.~Holman, Phys.\ Rev.\ D {\bf 49}, 2769 (1994).

\bibitem{sq18} D.~Boyanovsky {\it et al.},  Phys.\ Rev.\ D {\bf 56}, 1939 (1997).

\bibitem{abra1}  M. Abramowitz and I. A. Stegun, {\it Handbook of 
Mathematical Functions} (Dover, New York, 1972).

\bibitem{abra2}  A. Erdelyi, W. Magnus, F. Obehettinger, and F. Tricomi,  {\it Higher Trascendental Functions} (Mc Graw-Hill, New York, 1953).

\bibitem{alex2}   A.~Borde and A.~Vilenkin, Phys.\ Rev.\ Lett.\  {\bf 72}, 3305 (1994).

\bibitem{alex3}  A.~Borde and A.~Vilenkin, Int.\ J.\ Mod.\ Phys.\ D {\bf 5}, 813 (1996).

\bibitem{alex4}  A.~Borde and A.~Vilenkin, Phys.\ Rev.\ D {\bf 56}, 717 (1997).

\bibitem{alex5}   A.~Borde, A.~H.~Guth and A.~Vilenkin, Phys.\ Rev.\ Lett.\  {\bf 90}, 151301 (2003).

\bibitem{BM1} J.~Martin and R.~H.~Brandenberger,   Phys.\ Rev.  {\bf D63}, 123501 (2001).

\bibitem{BM2}  R.~H.~Brandenberger and J.~Martin,  Mod.\ Phys.\ Lett. {\bf A16}, 999 (2001).

\bibitem{BM3} J.~Martin and R.~H.~Brandenberger, Phys.\ Rev.\ D {\bf 65}, 103514 (2002).

\bibitem{NIEM1} J.~C.~Niemeyer, Phys.\ Rev.\ D {\bf 63}, 123502 (2001).

\bibitem{NIEM2} J.~C.~Niemeyer and R.~Parentani, Phys.\ Rev.\ D {\bf 64}, 101301 (2001).

\bibitem{KOW} J.~Kowalski-Glikman, Phys.\ Lett.\ B {\bf 499}, 1 (2001).

\bibitem{JAC}  T.~Jacobson and D.~Mattingly, Phys.\ Rev.\ D {\bf 63}, 041502 (2001).

\bibitem{GRE1} C.~S.~Chu, B.~R.~Greene and G.~Shiu, Mod.\ Phys.\ Lett.\ A {\bf 16}, 2231 (2001).

\bibitem{TMB} S.~Tsujikawa, R.~Maartens and R.~Brandenberger, Phys.\ Lett.\ B {\bf 574}, 141 (2003).

\bibitem{BG} M.~Bastero-Gil, P.~H.~Frampton and L.~Mersini, Phys.\ Rev.\ D {\bf 65}, 106002 (2002).

\bibitem{ACV1} D.~Amati, M.~Ciafaloni and G.~Veneziano, Phys.\ Lett. {\bf B197}, 81 (1987).

\bibitem{ACV2} D.~Amati, M.~Ciafaloni and G.~Veneziano, Phys.\ Lett.\ B {\bf 216}, 41 (1989).

\bibitem{GRM1}  D.~J.~Gross and P.~F.~Mende,  Nucl.\ Phys. {\bf B303}, 407 (1988).

\bibitem{GRM2}D.~J.~Gross and P.~F.~Mende, Phys.\ Lett.\ B {\bf 197}, 129 (1987).

\bibitem{AK1} A.~Kempf, Phys.\ Rev.\ D {\bf 63}, 083514 (2001).

\bibitem{AK2} A.~Kempf and J.~C.~Niemeyer, Phys.\ Rev. {\bf D64}, 103501 (2001).

\bibitem{HS}S.~F.~Hassan and M.~S.~Sloth, Nucl.\ Phys.\ B {\bf 674}, 434 (2003).

\bibitem{Ho}  R.~Brandenberger and P.~M.~Ho, Phys.\ Rev. {\bf D66}, 023517 (2002).

\bibitem{DAN1} U.~H.~Danielsson, Phys.\ Rev.\ D {\bf 66}, 023511 (2002).

\bibitem{DAN2} U.~H.~Danielsson, JHEP {\bf 0207}, 040 (2002).

\bibitem{STA1} A.~A.~Starobinsky,
{\it Pisma Zh.\ Eksp.\ Teor.\ Fiz.}  {\bf 73}, 415 (2001) [{\it JETP Lett.}  {\bf 73}, 371 (2001)].

\bibitem{SHE1}
N.~Kaloper, M.~Kleban, A.~E.~Lawrence and S.~Shenker,
Phys.\ Rev.\ D {\bf 66}, 123510 (2002).

\bibitem{SHE2}
N.~Kaloper, M.~Kleban, A.~E.~Lawrence and S.~Shenker,
Phys.\ Rev.\ D {\bf 66}, 123510 (2002).

\bibitem{GRE2}
R.~Easther, B.~R.~Greene, W.~H.~Kinney and G.~Shiu,
Phys.\ Rev. {\bf D64}, 103502 (2001).

\bibitem{GRE3}
R.~Easther, B.~R.~Greene, W.~H.~Kinney and G.~Shiu,
Phys.\ Rev.\ D {\bf 67}, 063508 (2003).

\bibitem{GRE4}
R.~Easther, B.~R.~Greene, W.~H.~Kinney and G.~Shiu,
Phys.\ Rev.\ D {\bf 66}, 023518 (2002).

\bibitem{alpha1}
N.~A.~Chernikov and E.~A.~Tagirov,   Annales Poincare Phys.\ Theor.\ A {\bf 9}, 109 (1968).

\bibitem{alpha2} E.~A.~Tagirov,
Annals Phys.\  {\bf 76}, 561 (1973).

\bibitem{alpha3}
K.~Goldstein and D.~A.~Lowe, Nucl.\ Phys.\ B {\bf 669}, 325 (2003).

\bibitem{alpha4}
H.~Collins and R.~Holman, Phys.\ Rev.\ D {\bf 70}, 084019 (2004).

\bibitem{alpha5}H.~Collins, R.~Holman and M.~R.~Martin, Phys.\ Rev.\ D {\bf 68}, 124012 (2003).

\bibitem{MGTP2} V.~Bozza, M.~Giovannini and G.~Veneziano, JCAP {\bf 0305}, 001 (2003).

\bibitem{MGTP1} M.~Giovannini, Class.\ Quant.\ Grav.\  {\bf 20}, 5455 (2003).

\bibitem{abramo1} V.~F.~Mukhanov, L.~R.~W.~Abramo and R.~H.~Brandenberger,
Phys.\ Rev.\ Lett.\  {\bf 78}, 1624 (1997).

\bibitem{abramo2} L.~R.~W.~Abramo, R.~H.~Brandenberger and V.~F.~Mukhanov,
Phys.\ Rev.\ D {\bf 56}, 3248 (1997).

\bibitem{Tanaka} T.~Tanaka, arXiv:astro-ph/0012431.

\bibitem{POR1} M.~Porrati, Phys.\ Lett.\ B {\bf 596}, 306 (2004).

\bibitem{STA2} A.~A.~Starobinsky and I.~I.~Tkachev,
JETP Lett.\  {\bf 76}, 235 (2002) [Pisma Zh.\ Eksp.\ Teor.\ Fiz.\  {\bf 76}, 291 (2002)].

\bibitem{BaR1}
B.~R.~Greene, K.~Schalm, G.~Shiu and J.~P.~van der Schaar, hep-th/0411217.

\bibitem{BaR2} R. Brandenberger and J. Martin, hep-th/0410223.

\bibitem{mollerach1} S. Mollerach, Phys. Rev. D {\bf 42}, 313 (1990).

\bibitem{lindekofman} L.  Kofman and A. Linde, Nucl. Phys. B {\bf 282}, 555 (1987).

\bibitem{lindemukhanov} A. Linde and V. Mukhanov, Phys. Rev. D  {\bf 56}, 535 (1997).

\bibitem{moroi1}  T. Moroi and T. Takahashi, Phys. Lett. B {\bf 522}, 215 (2001)
 [Erratum, {\it ibid} {\bf 539}, 303 (2002)].
 
\bibitem{moroi2} T. Moroi and T. Takahashi, Phys. Rev. D {\bf 66}, 063501 (2002).

\bibitem{kari}  K.~Enqvist and M.~S.~Sloth, Nucl.\ Phys.\ B {\bf 626}, 395 (2002).

\bibitem{lythwands} D.~H.~Lyth and D.~Wands, Phys.\ Lett.\ B {\bf 524}, 5 (2002).

\bibitem{mgcurv1} V.~Bozza, M.~Gasperini, M.~Giovannini and G.~Veneziano,  Phys.\ Lett.\ B {\bf 543}, 14 (2002).

\bibitem{mgcurv2} V.~Bozza, M.~Gasperini, M.~Giovannini and G.~Veneziano, Phys.\ Rev.\ D {\bf 67}, 063514 (2003).

\bibitem{martin} M. Sloth, Nucl. Phys. B {\bf 656}, 239 (2003).

\bibitem{dim1} K. Dimopoulos {\it et al.}, J. High Energy Phys. {\bf 05}, 057 (2003).

\bibitem{curvdec1} K.~A.~Malik, D.~Wands and C.~Ungarelli, Phys.\ Rev.\ D {\bf 67}, 063516 (2003).

\bibitem{curvdec2} S.~Gupta, K.~A.~Malik and D.~Wands, Phys.\ Rev.\ D {\bf 69}, 063513 (2004).

\bibitem{liddle} N. Bartolo and A. Liddle, Phys. Rev. D {\bf 65},  121301 (2002).

\bibitem{lyth3}D.~H.~Lyth, Phys.\ Lett.\ B {\bf 579}, 239 (2004).

\bibitem{dim2} K.~Dimopoulos and D.~H.~Lyth, Phys.\ Rev.\ D {\bf 69}, 123509 (2004).

\bibitem{kari2} K.~Enqvist and A.~Mazumdar, Phys.\ Rept.\  {\bf 380}, 99 (2003).

\bibitem{mcdonald} J.  McDonald, Phys. Rev. D {\bf 68}, 043505 (2003).

\bibitem{moroi3} T. Moroi and H. Murayama, Phys. Lett. B {\bf 553}, 126 (2003).

\bibitem{postma1} M. Postma, Phys. Rev. D {\bf 67}, 063518 (2003).

\bibitem{kari3}  K. Enqvist, A. Jokinen, S. Kasuya, A. Mazumdar, Phys. Rev. D {\bf 68}, 103507 (2003).

\bibitem{kari4} K.~Enqvist, Mod.\ Phys.\ Lett.\ A {\bf 19}, 1421 (2004)

\bibitem{bastero} M. Bastero-Gil, V. Di Clemente, and S. F. King, Phys. Rev. D {\bf 67},
103516 (2003).

\bibitem{mgcurv4}  M.~Giovannini, Phys.\ Rev.\ D {\bf 69}, 083509 (2004).

\bibitem{postma2} M.~Postma, JCAP {\bf 0405}, 002 (2004).

\bibitem{wands1} D. Wands, New Astron.Rev. {\bf 47}, 781 (2003). 

\bibitem{gr1}  L. P. Grishchuk,  Sov. Phys. JETP {\bf 40}, 409 (1975) [ Zh. \'Eksp. Teor. Fiz. {\bf 67},  825 (1974)].

\bibitem{gr2} L. P. Grishchuk,  JETP Lett. {\bf 23}, 293 
(1976) [Pis'ma Zh. Eksp. Teor. Fiz. {\bf 23}, 326 (1976)].

\bibitem{thorne1}  K. S. Thorne, in {\it 300 years of gravitation}, edited by S. Hwaking and W. Israel, (Cambridge University 
press, Cambridge, UK, 1987) p 330.

\bibitem{grrev} L.~Grishchuk, {\it et al.}  Phys.\ Usp.\  {\bf 44}, 1 (2001)
[Usp.\ Fiz.\ Nauk {\bf 171}, 3 (2001)].

\bibitem{RUB} V. A. Rubakov, M. V. Sazhin and A. V. Veryaskin, Phys. Lett.
{\bf 115B}, 189 (1982).

\bibitem{FP} R. Fabbri and M. D. Pollock, Phys. Lett. {\bf 125B}, 445 (1983).

\bibitem{AB} L. F. Abbott and M. B. Wise, Nucl. Phys. {\bf 224}, 541 (1984).

\bibitem{STAR} A. A. Starobinsky, JETP Lett. {\bf 30}, 682 (1979) [Pis'ma Zh. \'Eksp. Teor. Fiz. {\bf 30}, 719 (1979)]

\bibitem{ALLEN} B. Allen, Phys. rev. D {\bf 37}, 2078 (1988).

\bibitem{SAHNI} V. Sahni, Phys. Rev. D {\bf 42}, 453 (1990).

\bibitem{SOL} L. P. Grishchuk and M. Solokhin, Phys.Rev. D {\bf 43}, 2566 (1991).

\bibitem{GGGW} M. Gasperini and M. Giovannini, Phys.Lett.B {\bf 282},  36 (1992).

\bibitem{ALLEN2} B. Allen, in {\ Proceedings of the Les Houches School on
Astrophysical Sources
of Gravitational Waves}, edited by J. Marck and J.P. Lasota
(Cambridge University Press, Cambridge England, 1996).

\bibitem{SCHUTZ} B. Schutz, Class. Quantum Grav. {\bf 16}, A131 (1999).

\bibitem{taylor} V. Kaspi, J. Taylor, and M. Ryba, Astrophy. J.  {\bf 428}, 713 (1994).

\bibitem{schw} V. F. Schwartzman,  JETP Lett {\bf 9}, 184 (1969) [Pis'ma Zh. \'Eksp. Teor. Fiz, 
{\bf 9}, 315 (1969)].

\bibitem{mam2} J. Rehm and K.  Jedamzik  Phys. Rev. Lett. {\bf 81},  3307 (1998).

\bibitem{mam3} H.  Kurki-Suonio and E. Sihvola,  Phys. Rev. Lett.  {\bf 84}, 3756  (2000).

\bibitem{HKS} M. Giovannini, H. Kurki-Suonio,  E. Sihvola   Phys.Rev.D {\bf 66},  043504
(2003).

\bibitem{ALLEN3} B. Allen and J. D. Romano, Phys. Rev. D {\bf 59}, 102001 (1999).

\bibitem{maxq2}  M.~Giovannini, Phys.\ Rev.\ D {\bf 60}, 123511 (1999).

\bibitem{maxq3}  M.~Giovannini, Class.\ Quant.\ Grav.\  {\bf 16}, 2905 (1999).

\bibitem{gioquint} D.~Babusci and M.~Giovannini,
Phys.\ Rev.\ D {\bf 60}, 083511 (1999).
 
\bibitem{gio10} D.~Babusci and M.~Giovannini, Int.\ J.\ Mod.\ Phys.\ D {\bf 10}, 477 (2001).

\bibitem{gio11} D.~Babusci and M.~Giovannini, Class.\ Quant.\ Grav.\  {\bf 17}, 2621 (2000).
 
\bibitem{niobe} D. Blair {\it et al.},  Phys. Rev. Lett. {\bf 74}, 1908 
(1995).

\bibitem{allegro} E. Manuceli {\it et al.},  Phys. Rev. D {\bf 54}, 1264 (1996).

\bibitem{auriga}  M. Cerdonio, {\it  et al.},   Class. Quantum Grav.  {\bf 14}, 
1491 (1997).

\bibitem{explorer} P. Astone, {\it et al.}, Phys. Rev. D {\bf 47}, 362 (1993).

\bibitem{nautilus} P. Astone,  {\it et al.}, Astroparticle Physics, {\bf 7}, 231 (1997).

\bibitem{tama} K. Tsubono,  {\it Gravitational Wave Experiments}, Proceedings of the E. 
Amaldi Conference, edited by Coccia E., Pizzella G. , and Ronga F., ( World Scientific, 
Singapore, 1995), p. 112.

\bibitem{geo} K. Danzmann, {\it et al.}, Class. Quantum Grav. {\bf 14}, 1471 (1997).

\bibitem{virgo}B.  Caron,  {\it et al.},  Class. Quantum Grav. {\bf 14}, 1461 (1997).

\bibitem{ligo} Abramovici, A. {\it et al.}, Science {\bf 256}, 325 (1992).

\bibitem{mw1} F. Pegoraro, L. A. Radicati,  
Ph. Bernard ,  and E. Picasso, Phys. Lett. A {\bf 68}, 165 (1978).


\bibitem{mw2} C.  Caves, Phys. Lett. B {\bf 80}, 323 (1979).

\bibitem{paco} Ph.~Bernard, G.~Gemme, R.~Parodi and E.~Picasso,
Rev.\ Sci.\ Instrum.\  {\bf 72}, 2428 (2001).

\bibitem{bir}  A. Cruise, Class. Quantum Grav. {\bf 17}, 2525 (2000); 
 Mon. Not. R. Astron. Soc {\bf 204}, 485 (1983).

\bibitem{CHAN1} S.  Chandrasekar, {\it Radiative Transfer}, (Dover, New York, US, 1966).

\bibitem{PER1} A. Peraiah, {\it An Introduction to Radiative Transfer}, (Cambridge University Press, 
Cambridge, UK, 2001).

\bibitem{HS1} W.~Hu and N.~Sugiyama, Astrophys.\ J.\  {\bf 444}, 489 (1995).

\bibitem{HS2} W.~Hu and N.~Sugiyama,  Phys.\ Rev.\ D {\bf 51}, 2599 (1995).

\bibitem{HS2a} W.~Hu and N.~Sugiyama, Phys.\ Rev.\ D {\bf 50}, 627  (1994).

\bibitem{HS3}  W.~Hu and N.~Sugiyama,  Astrophys.\ J.\  {\bf 471}, 542 (1996).

\bibitem{HS4} W.~Hu, D.~Scott, N.~Sugiyama and M.~J.~White, Phys.\ Rev.\ D {\bf 52}, 5498 (1995).

\bibitem{HZ1} D.~D.~Harari and M.~Zaldarriaga, Phys.\ Lett.\ B {\bf 319}, 96 (1993).

\bibitem{HZ2}  M.~Zaldarriaga and D.~D.~Harari, Phys.\ Rev.\ D {\bf 52}, 3276 (1995).

\bibitem{HZ3}D.~D.~Harari, J.~D.~Hayward and M.~Zaldarriaga, Phys.\ Rev.\ D {\bf 55}, 1841 (1997).

\bibitem{kos2} A.~Kosowsky, Annals Phys.\  {\bf 246}, 49 (1996).

\bibitem{KL} A.~Kosowsky and A.~Loeb, Astrophys.\ J.\  {\bf 469}, 1 (1996).

\bibitem{bond1} J.~R.~Bond and A.~S.~Szalay,  Astrophys.\ J.\  {\bf 274}, 443 (1983).

\bibitem{kos1}  A.~Kosowsky, New Astron.\ Rev.\  {\bf 43}, 157 (1999).

\bibitem{CMBF} See, for instance, http://www.cmbfast.org.

\bibitem{zaldsel} U. Seljak and M. Zaldarriaga, Astrophys. J. {\bf 469}, 437 (1996).

\bibitem{Huint} W.~Hu, Annals Phys.\  {\bf 303}, 203 (2003).

\bibitem{zalexp} M.~Zaldarriaga, Astrophys.\ J.\  {\bf 503}, 1 (1998)

\bibitem{huwhpol}  W.~Hu and M.~J.~White, New Astron.\  {\bf 2}, 323 (1997).

\bibitem{tot1} W. Hu and M. White, Phys. Rev. D {\bf 56}, 596 (1997).

\bibitem{bes1}  J.~R.~Bond and G.~Efstathiou,
Mon.\ Not.\ Roy.\ Astron.\ Soc.\  {\bf 226}, 655 (1987).

\bibitem{bes2} J.~R.~Bond and G.~Efstathiou, Astrophys.\ J.\  {\bf 285}, L45 (1984).

\bibitem{wyse}  B. Jones and R. Wyse, Astron. Astrophys. {\bf 149}, 144 (1985).

\bibitem{spergeletal} D. Spergel {\it et al.},  Astrophys. J. Suppl. {\bf 148}, 175 (2003).

\bibitem{tegsilk}  M. Tegmark and J. Silk, Astrophys. J. {\bf 441}, 458 ( 1995).

\bibitem{GP1} S. Djorgovski {\it et al.}, Astrophys. J. Lett. {\bf 560}, 5 (2001).

\bibitem{GP2} R. Becker {\it et al.}, Astronon. J. {\bf 122}, 2850 (2001).

\bibitem{VERDE} L. Verde {\it et al.}, Astrophy. J. Suppl. {\bf 148}, 195 ( 2003).

\bibitem{BE} G. Esftathiou and J. Bond, Mon. Not. Roy. Astron. Soc. {\bf 304}, 75 (1999).

\bibitem{W1s} S.~Weinberg,  Phys.\ Rev.\ D {\bf 62}, 127302 (2000).

\bibitem{Wbz1} S.~Weinberg, Astrophys.\ J.\  {\bf 581}, 810 (2002).

\bibitem{Wbz2} S.~Weinberg, Phys.\ Rev.\ D {\bf 64}, 123512 (2001).

\bibitem{Wbz3} S.~Weinberg, Phys.\ Rev.\ D {\bf 64}, 123511 (2001).

\bibitem{old1} F. Atrio-Barandela, A. Doroshkevich, and A. Klypin, Astrophys. J. {\bf 378}, 1 (1991).

\bibitem{old2} P. Naselsky and I. Novikov, Astrophys. J. {\bf 413}, 14 (1993).

\bibitem{old3} H. Jorgensen, E. Kotok, P. Naselsky, and I Novikov, Astron. Astrophys. {\bf 294}, 639 (1995).

\bibitem{MUKa} V. Mukhanov, e-print arXiv: astro-ph/0303072.

\bibitem{sphharm1} U. Seljak and M. Zaldarriaga, Phys. Rev. Lett
  {\bf 78}, 2054 (1997).
  
 \bibitem{sphharm2} M. Zaldarriaga and U. Seljak, Phys. Rev.
  {\bf D 55}, 1830 (1997).
  
\bibitem{hu} W. Hu and M. White, Phys. Rev. {\bf D 56} (1997) 596

\bibitem{kamionkowski} M. Kamionkowski, A. Kosowsky and A. Stebbins,
   Phys. Rev. {\bf D 55} , 7368 (1997).

\bibitem{polnarev} A. Polnarev, Sov. Astron. {\bf 29}, 607 (1985).

\bibitem{crit} R. Crittenden, J. Bond, R. Davis, G. Efsthathiou,  and P. Steinhardt, Phys. Rev. Lett. {\bf 71}, 324 (1993).

\bibitem{knox} L. Knox, Y.-S. Song, Phys. Rev. Lett. {\bf 89}, 011303 (2002).

\bibitem{tot2} W. Hu, U. Seljak, M. White, and M. Zaldarriaga, Phys. Rev. D {\bf 57}, 3290 (1998). 

\bibitem{tot3} U. Seljak and M. Zaldarriaga, Astrophys. J. {\bf 469}, 437 (1996).


\end{thebibliography}
\end{document}